\documentclass[12pt,a4paper]{article} 
\pdfoutput=1

\usepackage{jheppub}
\usepackage{epstopdf}
\usepackage{graphicx}
\usepackage{epsfig}
\usepackage{dcolumn}  
\usepackage{amssymb} 
\usepackage{amsmath}
\usepackage{amsfonts}
\usepackage{etex}
\usepackage{slashed}  
\usepackage{youngtab}
\usepackage[mathscr]{euscript}
\usepackage{epsfig}
\usepackage[utf8]{inputenc}
\usepackage{xcolor}
\usepackage{mathrsfs}
\usepackage{cases}
\usepackage{bbding}
\usepackage{pifont}
\usepackage{dcolumn}
\usepackage{graphicx}
\usepackage[all]{xy}
\usepackage{titlesec}
\usepackage{ytableau}
\usepackage{collectbox}
\usepackage{centernot}
\usepackage[utf8]{inputenc}
\usepackage{color}
\usepackage{young}
\usepackage[most]{tcolorbox}

\newcounter{savefootnote}

\usepackage{tikz}
\usepackage{tikz-cd}
\usetikzlibrary{tikzmark,calc,arrows,shapes,decorations.pathmorphing,decorations.pathreplacing}
\usetikzlibrary{arrows}
\usetikzlibrary{automata}

\makeatletter
\newcommand{\sqbox}{%
    \collectbox{%
        \@tempdima=\dimexpr\width-\totalheight\relax
        \ifdim\@tempdima<\z@
            \fbox{\hbox{\hspace{-.5\@tempdima}\BOXCONTENT\hspace{-.5\@tempdima}}}%
        \else
            \ht\collectedbox=\dimexpr\ht\collectedbox+.5\@tempdima\relax
            \dp\collectedbox=\dimexpr\dp\collectedbox+.5\@tempdima\relax
            \fbox{\BOXCONTENT}%
        \fi
    }%
}
\makeatother

\titleclass{\subsubsubsection}{straight}[\subsection]

\newcounter{subsubsubsection}[subsubsection]
\renewcommand\thesubsubsubsection{\thesubsubsection.\arabic{subsubsubsection}}

\titleformat{\subsubsubsection}
  {\normalfont\normalsize\bfseries}{\thesubsubsubsection}{1em}{}
\titlespacing*{\subsubsubsection}
{0pt}{3.25ex plus 1ex minus .2ex}{1.5ex plus .2ex}

\makeatletter
\renewcommand\paragraph{\@startsection{paragraph}{5}{\z@}%
  {3.25ex \@plus1ex \@minus.2ex}%
  {-1em}%
  {\normalfont\normalsize\bfseries}}
\renewcommand\subparagraph{\@startsection{subparagraph}{6}{\parindent}%
  {3.25ex \@plus1ex \@minus .2ex}%
  {-1em}%
  {\normalfont\normalsize\bfseries}}
\def\toclevel@subsubsubsection{4}
\def\toclevel@paragraph{5}
\def\toclevel@paragraph{6}
\def\l@subsubsubsection{\@dottedtocline{4}{7em}{4em}}
\def\l@paragraph{\@dottedtocline{5}{10em}{5em}}
\def\l@subparagraph{\@dottedtocline{6}{14em}{6em}}
\makeatother

\newdimen\tableauside\tableauside=1.0ex
\newdimen\tableaurule\tableaurule=0.4pt
\newdimen\tableaustep
\def\phantomhrule#1{\hbox{\vbox to0pt{\hrule height\tableaurule width#1\vss}}}
\def\phantomvrule#1{\vbox{\hbox to0pt{\vrule width\tableaurule height#1\hss}}}
\def\sqr{\vbox{%
		\phantomhrule\tableaustep
		\hbox{\phantomvrule\tableaustep\kern\tableaustep\phantomvrule\tableaustep}%
		\hbox{\vbox{\phantomhrule\tableauside}\kern-\tableaurule}}}
\def\squares#1{\hbox{\count0=#1\noindent\loop\sqr
		\advance\count0 by-1 \ifnum\count0>0\repeat}}
\def\tableau#1{\vcenter{\offinterlineskip
		\tableaustep=\tableauside\advance\tableaustep by-\tableaurule
		\kern\normallineskip\hbox
		{\kern\normallineskip\vbox
			{\gettableau#1 0 }%
			\kern\normallineskip\kern\tableaurule}%
		\kern\normallineskip\kern\tableaurule}}
\def\gettableau#1 {\ifnum#1=0\let\next=\null\else
	{{\tiny\yng(1)}}s{#1}\let\next=\gettableau\fi\next}

\newcommand{\includeCroppedPdf}[2][]{%
    \IfFileExists{./#2-crop.pdf}{}{%
        \immediate\write18{pdfcrop #2 #2-crop.pdf}}%
    \includegraphics[#1]{#2-crop.pdf}}
        
\newcommand\Kappa{\mathrm{K}}

\colorlet{mygreen}{green!10}

\graphicspath{{./figures/}}

\title{Quiver Yangian from Crystal Melting}

\author{Wei Li$^a$ and Masahito Yamazaki$^b$} 
\affiliation{$^a$Institute of Theoretical Physics, Chinese Academy of Sciences,\\
\hspace*{0.3cm}100190 Beijing, P.R.\ China} 
\affiliation{$^b$Kavli Institute for the Physics and Mathematics of the Universe (WPI),\\
\hspace*{0.3cm}University of Tokyo, Kashiwa, Chiba 277-8583, Japan}

\emailAdd{weili@mail.itp.ac.cn, masahito.yamazaki@ipmu.jp}

\abstract{
We find a new infinite class of infinite-dimensional algebras acting on BPS states for non-compact toric Calabi-Yau threefolds. 
In Type IIA superstring compactification on a toric Calabi-Yau threefold, the D-branes wrapping holomorphic cycles represent the BPS states, and the fixed points of the moduli spaces of BPS states are described by statistical configurations of crystal melting. 
Our algebras are ``bootstrapped" from the molten crystal configurations, hence they act on the BPS states. 
We discuss the truncation of the algebra and its relation with D4-branes. We illustrate our results in many examples, with and without compact $4$-cycles.}

\begin{document}


\maketitle

\makeatletter
\g@addto@macro\bfseries{\boldmath}
\makeatother

\section{Introduction}

The counting of Bogomol'nyi-Prasad-Sommerfield (BPS) states \cite{Bogomolny:1975de, Prasad:1975kr} has been one of the most central questions in quantum field theories, black holes, and string theory. 
Toric Calabi-Yau manifolds provide an ideal setup for addressing this problem --- the geometry of a toric Calabi-Yau manifold in itself is described by the combinatorial data of the toric diagram, and the BPS state counting problem can be recast as the statistical counting problem of crystal melting \cite{Okounkov:2003sp, Iqbal:2003ds}. 

The original crystal melting model of \cite{Okounkov:2003sp} counts three-dimensional plane partitions, and counts BPS states on the simplest Calabi-Yau geometry, $\mathbb{C}^3$.
The crystal melting configuration was subsequently generalized to arbitrary toric Calabi-Yau threefold \cite{Ooguri:2008yb},\footnote{
This crystal is different from that obtained from the topological vertex formalism \cite{Aganagic:2003db}. The two crystal descriptions are related by BPS wall crossing.}
based on earlier works \cite{Szendroi, MR2592501}.
This description also accommodates BPS wall crossing, where the wall crossing is described by a change of the crystal configuration \cite{Young:2008hn, MR2836398, MR2999994, Jafferis:2008uf, Chuang:2008aw, Nagao:2009ky, Nagao:2009rq, Sulkowski:2009rw, Aganagic:2010qr} (see e.g.\ \cite{Yamazaki:2010fz} for a summary).

Despite the success of the BPS state counting program for toric Calabi-Yau manifolds, there remained one unsatisfactory aspect of the program.
While the BPS counting problem generates an infinite set of numbers (BPS degeneracies), there are clearly some structures in them,
and it has long been expected that there is an underlying algebra, the algebra of BPS states acting on BPS states \cite{Harvey:1996gc}.
One hopes such an algebra will provide a better organizing principle for the BPS state counting problem.
There was, however, little discussion of this algebra, in particular not for general toric Calabi-Yau threefolds.

Recently there have been impressive developments in this direction.
For the case of the $\mathbb{C}^3$-geometry, it was found that we can define an action of $Y(\widehat{\mathfrak{gl}_1})$ (the affine Yangian of  $\mathfrak{gl}_1$, which is equivalent with the universal enveloping algebra of $\mathcal{W}_{1+\infty}$-algebra \cite{SV,Maulik:2012wi,Tsymbaliuk,Tsymbaliuk:2014fvq,Prochazka:2015deb,Gaberdiel:2017dbk}) on the set of plane partitions \cite{feigin2012, Tsymbaliuk:2014fvq, Prochazka:2015deb} (see also \cite{MR2793271}), and hence on the BPS states contributing to the BPS degeneracy.
The BPS partition function (which is the MacMahon function) is
identified with the character of the affine Yangian of $\mathfrak{gl}_1$.
In other words, the problem of explicitly constructing the algebra for the $\mathbb{C}^3$ has now been solved.

The natural question is then whether similar algebras exist for other toric Calabi-Yau geometries.
Namely, can we explicitly construct an infinite-dimensional algebra such that it acts on the BPS crystal configurations of \cite{Ooguri:2008yb}?\footnote{Related problems were posed recently in \cite{Rapcak:2018nsl}.}

The goal of this paper is to provide an answer to this question.
We explicitly define an infinite-dimensional algebra $\mathsf{Y}$ for an {\it arbitrary} toric Calabi-Yau threefold, and show that we can define a representation of the algebra in terms of the statistical model of BPS crystal melting. 
Indeed, for each toric Calabi-Yau threefold $M$, its associated BPS algebra $\mathsf{Y}$ is ``bootstrapped" from the set of the colored crystals that describes the BPS states of the IIA string in $M$, by demanding that the algebra acts on the corresponding set of crystals appropriately. 
Our algebra and representation reduce to $Y(\widehat{\mathfrak{gl}_1})$ and its plane-partition representation for the special case of $\mathbb{C}^3$.

The colored crystals furnish the representations that we use to ``bootstrap" the algebras. 
However for the purpose of defining the algebra, once the algebras are obtained, one can forget about the details of the crystal configurations. 
Namely, what enters the definition of the BPS algebra are only the basic ingredients that defines the crystal, which is a pair $(Q,W)$ of a quiver diagram $Q$ and a superpotential $W$, which are determined by the toric Calabi-Yau geometry. 
For this reason one can denote our algebra as $\mathsf{Y}_{(Q,W)}$, and call it the {\it BPS quiver Yangian}, or simply the quiver Yangian.

In our discussion it is crucial to keep track of the orientations of the quiver, and also to have closed loops in the quiver diagram $Q$; the quiver is in general chiral.
The existence of loops in the quiver is the necessary ingredient for the existence of a non-zero superpotential $W$, which in itself is an independent data.
In this respect our discussion seems to be more general than similar discussions of infinite-dimensional algebra in the literature, e.g.\ the work of \cite{Maulik:2012wi} where the Yangian associated with the quiver acts on the cohomologies of quiver varieties. 
It would be interesting to fully understand the relation with \cite{Maulik:2012wi} and other works, e.g.\  \cite{Kimura:2015rgi}, as we will discuss further in section \ref{sec:summary}. 
Let us also mention that during the preparation of this manuscript we have been notified of the ongoing work \cite{RSYZ},
who studies cohomological Hall algebras \cite{MR2851153} for some toric Calabi-Yau manifolds.\footnote{
The quiver Yangian is conjectured to be the Drinfeld double of CoHA; or inversely, the CoHA captures the positive part of the quiver Yangian, i.e.\ the creation part instead of the annihilation part.}

The rest of this paper is organized as follows. 
We begin with a review of the BPS crystal melting (section \ref{sec:review_crystal}) and affine Yangian of $\mathfrak{gl}_1$ (section \ref{sec-Yangian-W-PP}). We introduce the BPS quiver Yangian in section \ref{sec-quiver-Yangian}.
In order to motivate this definition, in section \ref{sec:bootstrap_gl1} we first go back to the plane partitions discussed in section \ref{sec-Yangian-W-PP} and bootstrap the affine Yangian of $\mathfrak{gl}_1$. 
Then in section \ref{sec:bootstrap_general} we repeat a similar analysis for a general quiver corresponding to a toric Calabi-Yau threefold, to obtain our BPS quiver Yangian. 
We discuss the truncation of the algebra and the relation with D4-branes in section \ref{sec:truncation}.
We present many examples both for toric Calabi-Yau threefolds without compact $4$-cycles (section \ref{sec:generalno4cycle}) and with compact $4$-cycles (section \ref{sec:general4cycle}). These examples will provide useful illustrations of many of the general results of the previous sections. 
The final section \ref{sec:summary} is devoted to a summary and discussions.

\bigskip

\section{Review: BPS Crystal Melting} \label{sec:review_crystal}

\subsection{Quiver Diagram and Superpotential}

Let us first briefly summarize the BPS crystal melting for general toric Calabi-Yau threefolds.
For a more complete discussion, see \cite{Ooguri:2008yb, Yamazaki:2010fz}.

\begin{figure}[htbp]
\centering\includegraphics[scale=0.5]{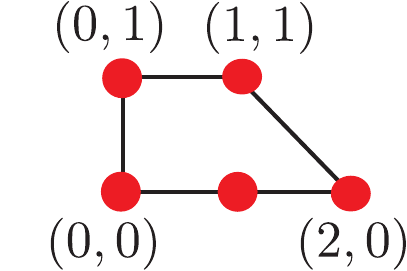}
\caption{The toric diagram for a toric Calabi-Yau threefold, the so-called Suspended Pinched Point geometry $x y=z^2 w$.}
\label{fig.SPP_quiver}
\end{figure}

Let us consider type IIA string theory compactified on a non-compact toric Calabi-Yau threefold $X$.
Combinatorially, the choice of $X$ is encoded in the so-called toric diagram $\Delta$, a lattice convex polytope in $\mathbb{Z}^2$, see Figure \ref{fig.SPP_quiver} for an example.

The BPS states of the theory are described by D-branes (D0/D2/D4-branes) wrapping holomorphic cycles (0/2/4-cycles) inside the Calabi-Yau threefold $X$.
The effective theory on the D-branes is a supersymmetric quiver quantum mechanics, and the moduli space of BPS states can be identified with the vacuum moduli space of the quiver quantum mechanics.

The quiver quantum mechanics generically has four supercharges, and can be thought of as the dimensional reduction of a four-dimensional $\mathcal{N}=1$ supersymmetric quiver gauge theory.
The theory is specified by a pair $(Q,W)$, where $Q$ is a quiver diagram and $W$ is a holomorphic superpotential. 

A quiver diagram $Q=\{Q_0, Q_1\}$ is given by a set of vertices $Q_0$ and a set of arrows $Q_1$ between vertices.
In the following we use the notation
\begin{equation}
Q_0=\{a\}_{a\in Q_0} \qquad \textrm{and} \qquad Q_1=\{I\}_{I\in Q_1} \;,
\end{equation}
namely we use $a,b,\dots$ to label the vertices, and $I, J,\dots$ to label the edges. 
We denote the number of vertices and arrows by $|Q_0|$ and $|Q_1|$, respectively.
The source and the target of an arrow $I \in Q_1$ will be denoted by $s(I)\in Q_0$ and $t(I)\in Q_0$, respectively.

In quiver quantum mechanics $Q_0$ and $Q_1$ specify the gauge groups and the bifundamental matter fields: we have a vector multiplet $V_a$ for each vertex $a \in Q_0$ and a bifundamental matter chiral multiplet $\Phi_I$ for each arrow $I \in Q_1$.

The superpotential $W$ specifies the interactions between the matter.
Given $W$, we can write down the so-called F-term relations $\partial W/\partial \Phi_I=0$ for each bifundamental matter $\Phi_I$ corresponding to the arrow $I$.
In the mathematics literature $W$ is known as a potential \cite{Ginzburg:2006fu} and defines the so-called ``path algebra for quiver with relations" $A_{(Q,W)}$. 
This is defined to be the Jacobian algebra $\mathbb{C}Q/(\partial W)$, where $\mathbb{C}Q$ is the path algebra generated by the set of (in general open) paths on the quiver diagram, with multiplication defined by concatenations of the paths.
The infinite-dimensional algebra $A_{(Q,W)}$ underlies the definition of the crystal melting model, and is closely related to our infinite-dimensional algebra $\mathsf{Y}_{(Q,W)}$.

\begin{figure}[htbp]
\centering\includegraphics[scale=0.4]{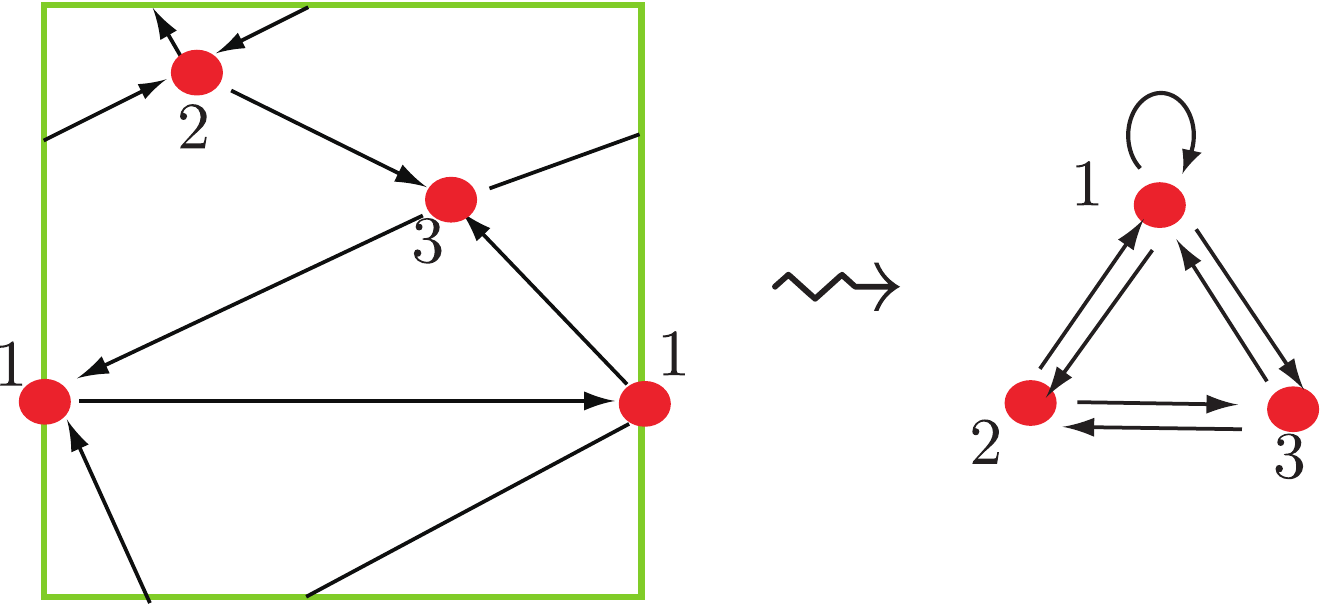}
\caption{The periodic quiver for the Suspended Pinched Point geometry of Figure \ref{fig.SPP_quiver}. 
The green region is the fundamental region of the two-dimensional torus.
The periodic quiver compactly encodes both the quiver diagram $Q$ as an abstract graph and the superpotential $W$.}
\label{fig.SPPPQ}
\end{figure}

In general it is a highly non-trivial problem to identify the pair $(Q,W)$ for a Calabi-Yau manifold $X$.
Fortunately, for a non-compact toric Calabi-Yau threefold, there is a systematic algorithm to obtain such a pair, starting with the toric diagram $\Delta$ \cite{Hanany:2005ss, Ueda:2006jn, Gulotta:2008ef, Goncharov:2011hp}. 
We will not need details of this procedure in this paper, except to note that the algorithm generates the pair $(Q,W)$ in the form of the periodic quiver, a quiver diagram realized on the torus (see Figure \ref{fig.SPPPQ} for an example). 

Given a periodic quiver, we can first forget the fact that the quiver is realized on a two-dimensional torus, and thus obtain a quiver diagram $Q$ as an abstract graph. 
The periodic quiver, however, contains more information --- for each polygonal region of the torus the arrows of the quiver diagram point in the same direction (either clockwise or counterclockwise), and the product of the bifundamental fields along the polygonal region represents a gauge-invariant superpotential term.
The superpotential $W$ is recovered as a sum of such monomial contributions
\begin{align}
W=\sum_{F: \textrm{face of the periodic quiver}} \pm \textrm{Tr} \left(\prod_{e\in F}  \Phi_e \right) \;,
\end{align}
where the product over $e\in F$ is taken along the edges of the face $F$ and the sign $\pm$ is determined by the orientations (clockwise or counter-clockwise, respectively) of the arrows.
For the example in Figure \ref{fig.SPPPQ}, the periodic quiver gives the superpotential
\begin{align}
W=\textrm{Tr}\left(-\Phi_{11} \Phi_{13} \Phi_{31}+ \Phi_{11} \Phi_{12} \Phi_{21} - \Phi_{21}\Phi_{12} \Phi_{23} \Phi_{32}+
\Phi_{32}\Phi_{23}\Phi_{31}\Phi_{13}\right) \;,
\end{align}
where we have denoted the bifundamental chiral multiplet associated to an arrow from vertex $a$ to $b$ by $\Phi_{ab}$ (in this example there are at most one arrow for any vertices $a,b$).

In the periodic quiver description, monomial terms of the superpotential $W$ are associated with the faces of the periodic quiver, the set of which we denote by $Q_2$.
In this notation, the quiver $Q=(Q_0, Q_1)$ and the superpotential $W$ combines nicely into the data $\tilde{Q}=(Q_0, Q_1, Q_2)$ of the periodic quiver.

\begin{figure}[htbp]
\centering\includegraphics[scale=0.4]{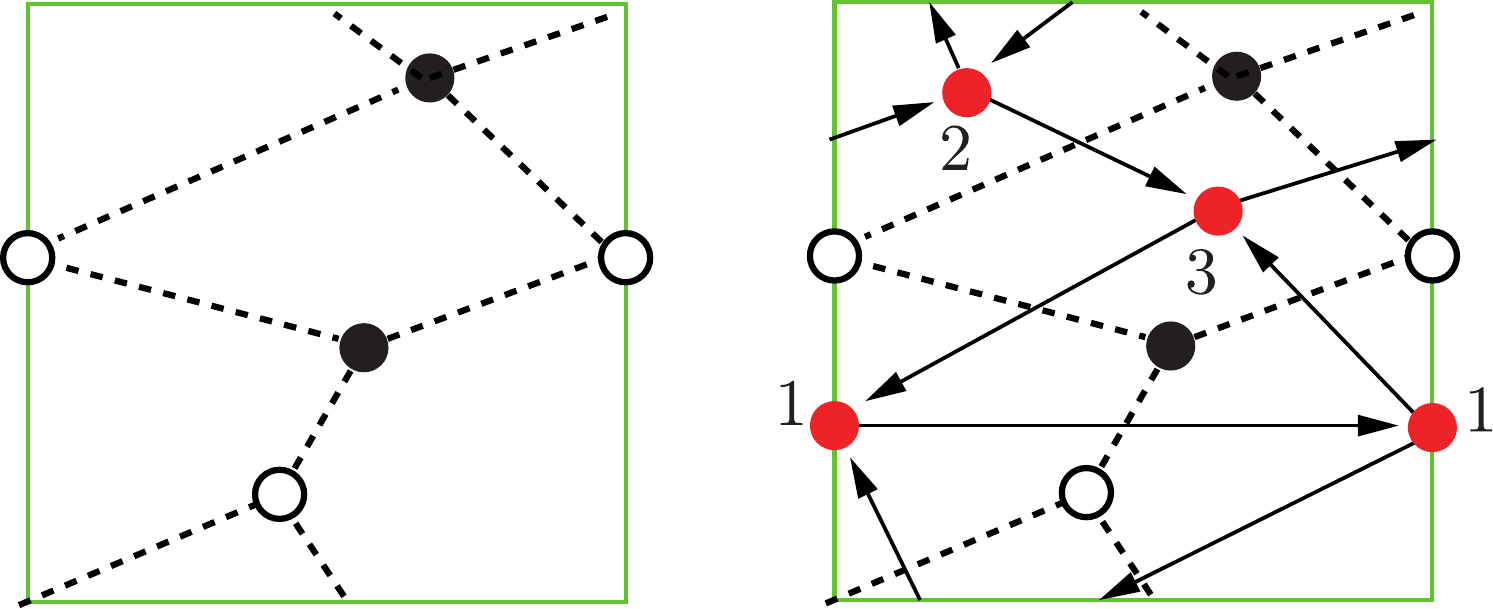}
\caption{The bipartite graph for the Suspended Pinched Point geometry of Figure \ref{fig.SPP_quiver} (shown on the left),
which is a dual graph to the periodic quiver of Figure \ref{fig.SPPPQ} (as shown on the right).
The color of a vertex of the bipartite graph is determined from the orientations of the quiver arrows surrounding it (black for counterclockwise, and white for clockwise).}
\label{fig.SPPbipartite}
\end{figure}

The dual of the periodic quiver is often represented as a bipartite graph,
i.e.\ a graph where vertices are colored with two colors (black and white) and every black vertex is connected to one single white vertex and vice versa.
The orientation of the quiver diagram canonically determines the colors of the vertices of the bipartite graph: the rule is that quiver arrows are oriented clockwise (counterclockwise) around white (black) vertices of the bipartite graph, see Figure \ref{fig.SPPbipartite} for an example. 
Such a bipartite graph in high energy theory is often called a brane tiling \cite{Hanany:2005ve, Franco:2005rj, Franco:2005sm} (see e.g.\ \cite{Kennaway:2007tq, Yamazaki:2008bt} for reviews), and has been heavily utilized in the study of supersymmetric quiver gauge theories. 
We will later show in section~\ref{sec:truncation} that the concept of the perfect matching of the bipartite graph will help us relate the truncation of the algebra to the charges of the D4-branes. 

\begin{figure}[htbp]
\centering\includegraphics[scale=0.44]{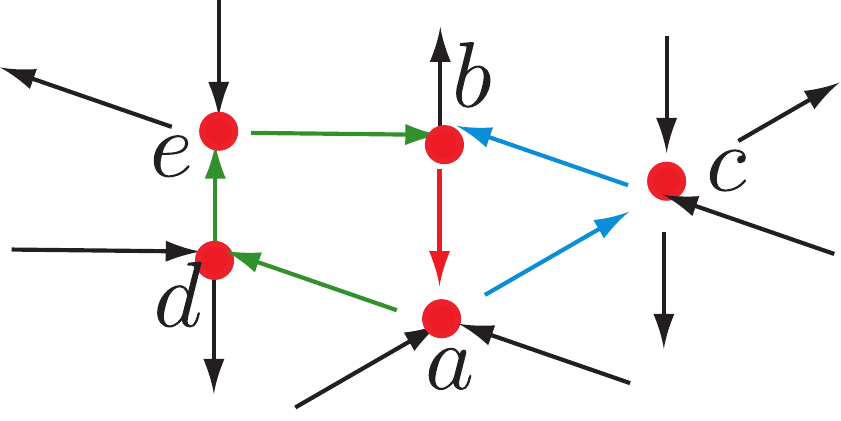}
\caption{
This figure represents a part of the periodic quiver diagram. In this example, the superpotential $W$ contains two monomial terms  $W= \textrm{Tr}(-\Phi_{ba}\Phi_{ac}\Phi_{cb}+\Phi_{ba} \Phi_{ad} \Phi_{de}\Phi_{eb})$.
The F-term relation $\partial W/\partial \Phi_{ba}=-\Phi_{ac}\Phi_{cb}+\Phi_{ad} \Phi_{de}\Phi_{eb}=0$ for the field $\Phi_{ba}$ is represented by the fact that the two different paths $a\to c\to b$ and $a\to d\to e\to b$ starting from $a$ ending at $b$ represents two F-term equivalent fields (i.e.\ same element in the chiral ring).}
\label{fig.SPPFterm}
\end{figure}

The periodic-quiver representation of the superpotential makes it easy to read off the F-term relations (see Figure \ref{fig.SPPFterm}): two paths on the periodic quiver with the same starting point and endpoint are F-term equivalent.
This will be useful when we discuss global symmetries of the quiver quantum mechanics.

\medskip

\subsection{Crystal as a Lift of Periodic Quiver}\label{subsec.crystal}

Let us next construct the BPS crystal. 
For this purpose, consider a new quiver diagram $\mathfrak{Q}$ obtained by uplifting the periodic quiver diagram to the universal cover of the two-dimensional torus (namely the two-dimensional plane).
Each vertex $\mathfrak{a}$ on the resulting quiver is still labelled (colored) by $a\in Q_0$. 
Note that as before we will use the symbols $a,b,\dots$ for the vertices of the original quiver diagram $Q$ (and hence of the periodic quiver diagram), while we use the symbols $\mathfrak{a}, \mathfrak{b}, \dots$ for vertices of the quiver $\mathfrak{Q}$ on the universal cover.

Let us choose a particular vertex $a_0\in Q_0$ as the ``initial color", 
and choose a vertex in $\mathfrak{Q}$ on the universal cover that has this color to be the origin $\mathfrak{o} \in \mathfrak{Q}$.\footnote{
This choice corresponds to the choice of framing, and represents the effect of the non-compact D6-brane filling the whole Calabi-Yau threefold.}
Let us then consider a set of paths starting with the origin $\mathfrak{o}$ modulo the F-term relation.
Any such path in $\mathfrak{Q}$, modulo the F-term relations (as described in Figure \ref{fig.SPPFterm}), defines an atom in the crystal.
This atom is placed at the location $\mathfrak{a}$ of the two-dimensional plane, where $\mathfrak{a}$ is the endpoint of the path. 
This defines the two-dimensional projection of the BPS crystal.

\begin{figure}[htbp]
\centering\includegraphics[scale=0.3]{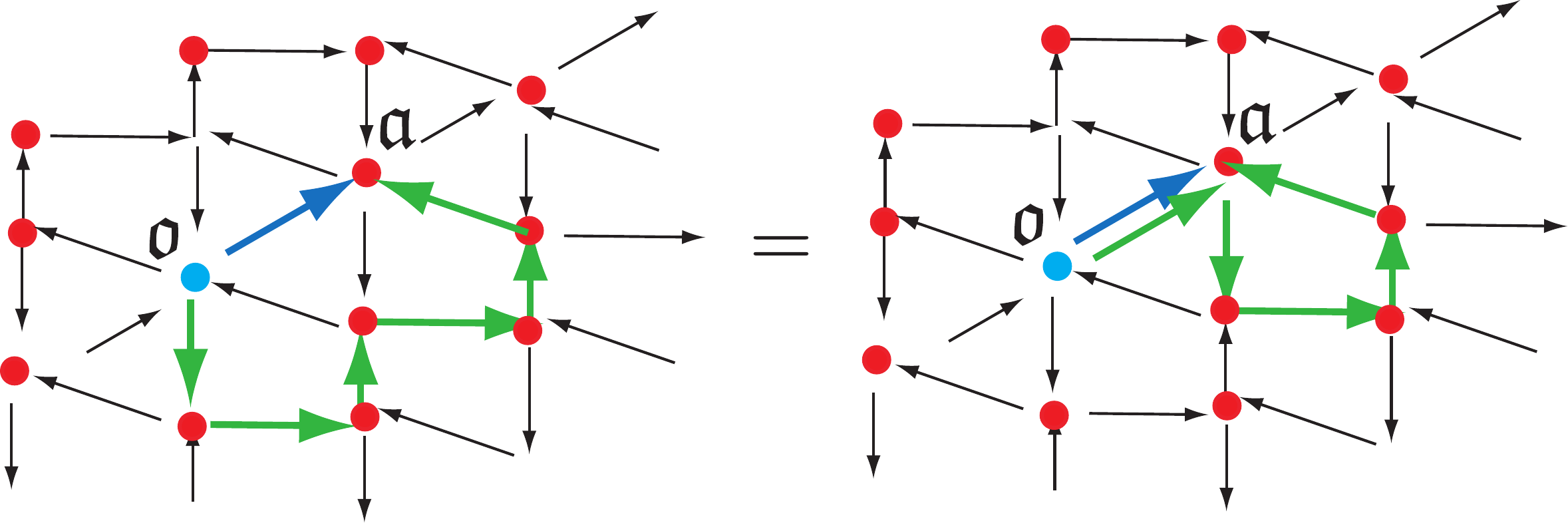}
\caption{
Module F-term relations, any path starting at the origin $\mathfrak{o}$ and ending at the position $\mathfrak{a}$ (e.g.\ the green path in the left figure) is equivalent to the shortest path $p_{\mathfrak{o}, \mathfrak{a}}$ (e.g.\ the blue arrow in both figures) times a power of the closed loop $\omega$ (e.g.\ the green loop in the right figure) along the faces of the periodic diagram.}
\label{fig.SPPpathEx}
\end{figure}

\begin{figure}[htbp]
\centering\includegraphics[scale=0.4]{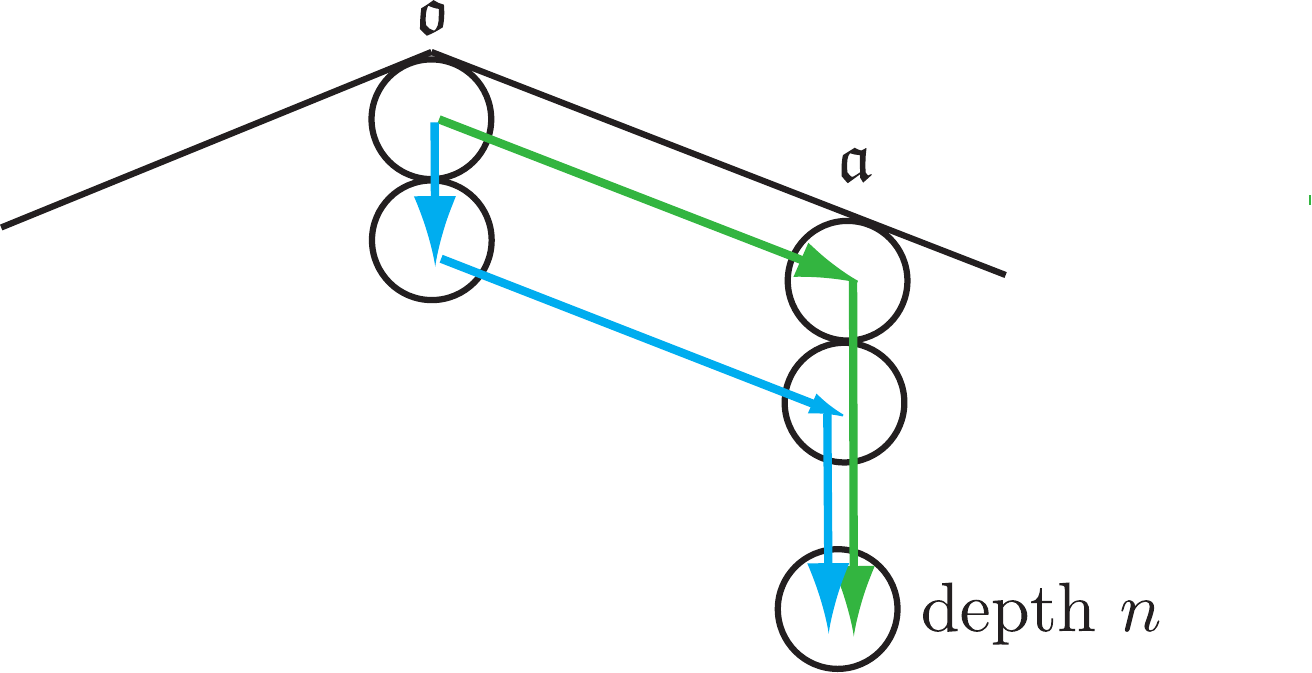}
\caption{The three-dimensional structure of the crystal configuration.
An atom in the crystal is represented by a path from the origin $\mathfrak{o} \in \mathfrak{Q}$ to $\mathfrak{a}\in \mathfrak{Q}$. 
If the path  is represented as  $p_{\mathfrak{o}, \mathfrak{a}} \,\omega^n$ modulo F-term relations, 
the corresponding atom is placed at depth $n$ at the location $\mathfrak{a}$.}
\label{fig.RuleEx}
\end{figure}

To fully describe the three-dimensional structure of the crystal, note that any path starting at the origin $\mathfrak{o}$ and ending at $\mathfrak{a}$ can be expressed in the form $p_{\mathfrak{o}, \mathfrak{a}}\, \omega^{n}$, modulo the F-term relations, 
where $p_{\mathfrak{o}, \mathfrak{a}}$ is one of the shortest paths connecting the two points $\mathfrak{o}$ and $\mathfrak{a}$, and $\omega$ represents a loop in the quiver diagram along any of the face of the periodic diagram (see Figure \ref{fig.SPPpathEx}).
The corresponding atom is then placed at depth $n$ in the crystal (see Figure \ref{fig.RuleEx}).

\begin{figure}[htbp]
\centering\includegraphics[scale=0.22]{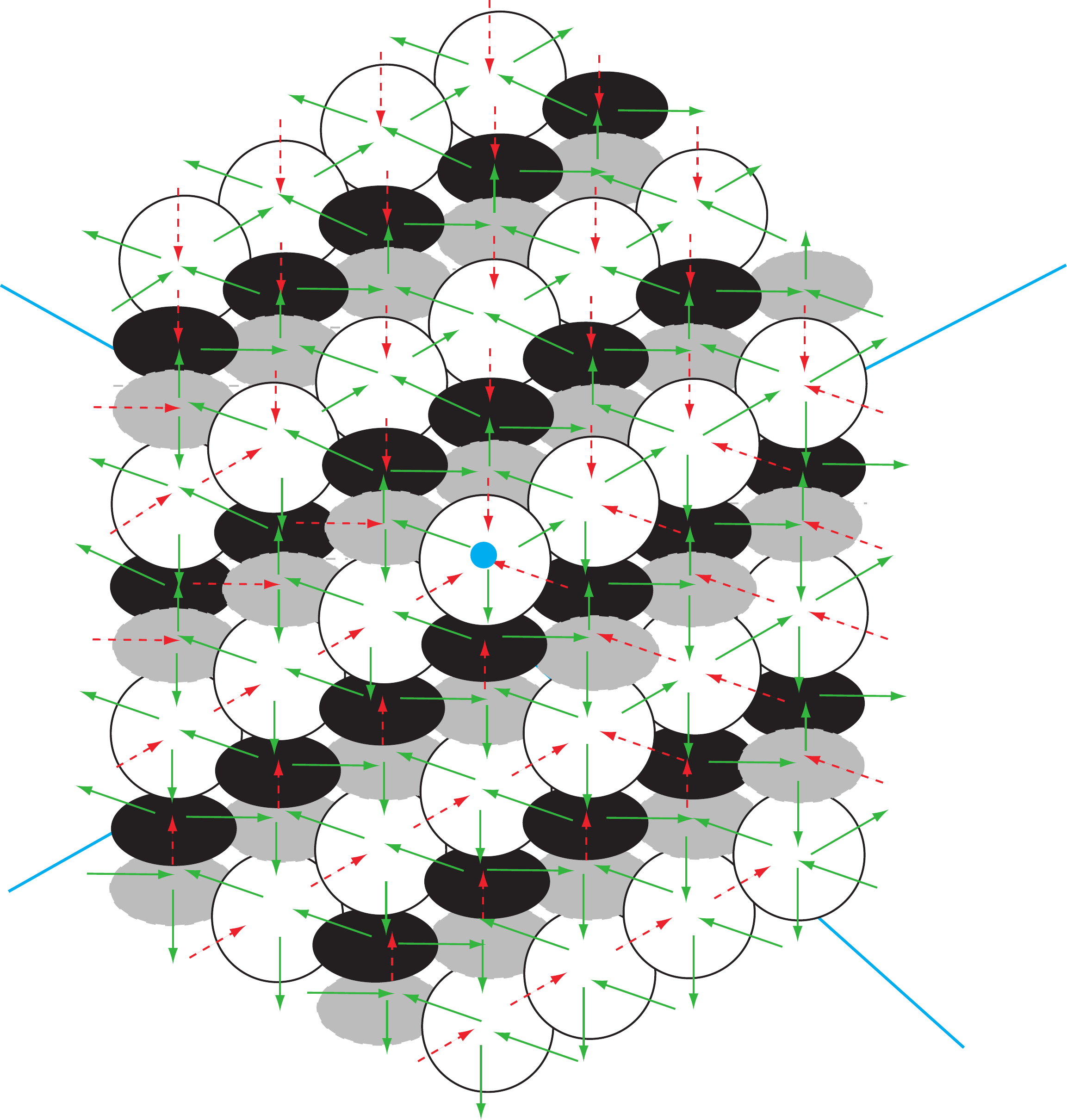}
\caption{The BPS crystal configuration for the Suspended Pinched Point singularity. 
We have chosen the vertex $1$ of Figure \ref{fig.SPP_quiver} as the origin $\mathfrak{o}$  of the crystal, whose location is shown by the blue dot in the center.}
\label{fig.SPPatom}
\end{figure}

As an example, we show in Figure \ref{fig.SPPatom} the example of the BPS crystal for the Suspended Pinched Point geometry discussed in Figure \ref{fig.SPPPQ}.

It follows from the definition that for an atom $\Box$ and an arrow $I \in Q_1$, there is a canonically-defined atom $I\cdot \Box$ in the crystal --- $I\cdot \Box$ is defined by concatenation of a path representing $\Box$ and an arrow $I$, and this definition is consistent with the identification modulo F-term relations. 
In other words, the BPS crystal naturally gives a representation of the path algebra of the quiver.

\clearpage

\medskip
\subsection{Crystal Melting and Molten Crystal}

For a given BPS crystal, we can consider a configuration of the molten crystal.

A finite set $\Kappa$ of atoms from the BPS crystal is a configuration of the molten crystal if it satisfies the following melting rule:
\begin{tcolorbox}[ams equation]\label{eq.melting_rule}
\begin{split}
&\textrm{\bf melting rule:}\\ 
&\quad \textrm{$\Box \in \Kappa$ whenever there exists an edge $I \in Q_1$ such that  $I \cdot \Box \in \Kappa$ \;. }
\end{split} 
\end{tcolorbox}
\noindent 
This is equivalent to the condition that $I\cdot \Box \notin \Kappa$ whenever $\Box \notin \Kappa$, namely the condition that the complement of $\Kappa$ is an ideal of the path algebra $A_{(Q,W)}$.
\begin{figure}[htbp]
\centering\includegraphics[scale=0.21]{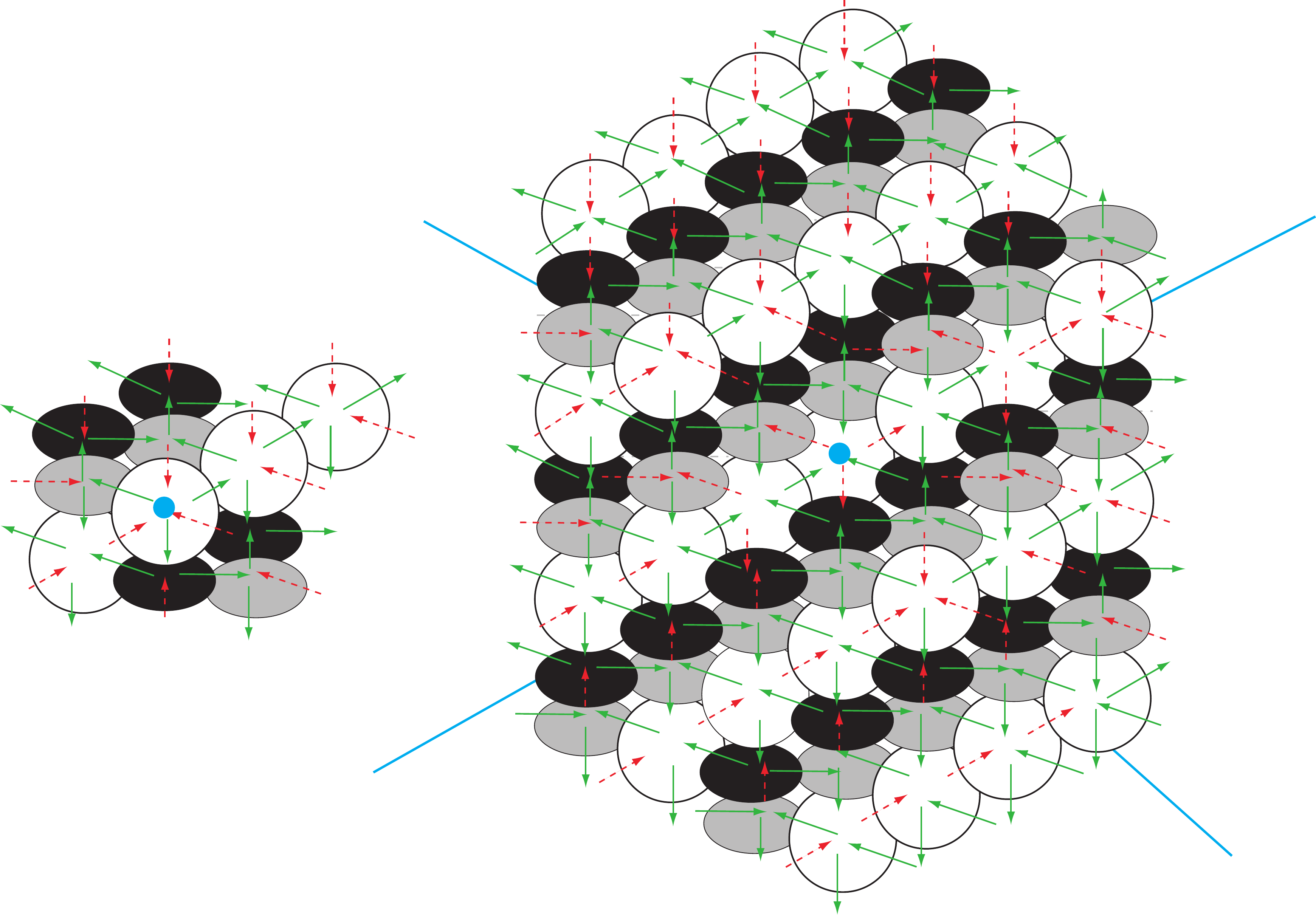}
\caption{An example of a configuration of the molten crystal (left) and the complement (right) for the crystal of 
Figure \ref{fig.SPPatom}. 
This contributes a term $q_1^4 q_2^3 q_3^2$ to the BPS partition function.}
\label{SPPmolten_rev}
\end{figure}

Since any path by definition starts at the origin $\mathfrak{o}$, it follows that the origin $\mathfrak{o}$ is always contained in $\Kappa$, unless $\Kappa$ is empty.

The molten configuration $\Kappa$ has a finite number of atoms. 
Denote the number of atoms with color $a$ as $|\Kappa(a)|$.
The statistical partition function of BPS crystal melting is then defined to be a formal power series\footnote{
More precisely we need to insert signs for this definition \cite{MR2592501,Ooguri:2008yb}.}
\begin{align}
Z(q_1,\dots, q_{|Q_0|}) =\sum_{\Kappa} \prod_{a\in Q_0} q_a^{|\Kappa(a)|} \;. 
\end{align}
The statement is that this coincides with the BPS configuration of the crystal.

The partition function has an infinite product form for the resolved conifold and more generally for toric Calabi-Yau geometries without compact $4$-cycles, as explained by M-theory \cite{Dijkgraaf:2006um, Aganagic:2009kf, Aganagic:2009cg}.
This suggests an identification of the BPS partition function as a character of some infinite-dimensional algebra. 
We will see that this is indeed the case.

\bigskip

\section{\texorpdfstring{Review: Plane Partition and Affine Yangian of $\mathfrak{gl}_1$}{Review: Plane Partition and Affine Yangian of gl(1)}}
\label{sec-Yangian-W-PP} 

As explained in Introduction, the current work is inspired by the relation between the affine Yangian of $\mathfrak{gl}_1$ and the set of plane partitions. 
We will now review the affine Yangian of $\mathfrak{gl}_1$, its relation to the $\mathcal{W}_{1+\infty}$ algebra, and its action on the set of plane partitions.  

\medskip

\subsection{\texorpdfstring{Affine Yangian of $\mathfrak{gl}_1$}{Affine Yangian of gl(1)}}

The affine Yangian of $\mathfrak{gl}_1$, which we denote by $Y(\widehat{ \mathfrak{gl}_1})$, is an infinite-dimensional associative algebra generated by the following three families of operators:
\begin{equation}
e_n\,, \qquad \psi_n\,, \qquad f_n\,, \qquad \textrm{with}\quad n\in \mathbb{Z}_{\geq 0} \;,
\end{equation}
and whose algebraic relations are
\begin{equation}\label{eq-relations-modes-gl1}
\begin{aligned}
0  & =  [\psi_n,\psi_m]  \,,\\
\sigma_3 \{\psi_n,e_m \}   & =   [\psi_{n+3},e_m] - 3  [\psi_{n+2},e_{m+1}] + 3  [\psi_{n+1},e_{m+2}]  - [\psi_{n},e_{m+3}]   \\
 & \ + \sigma_2\left( [\psi_{n+1},e_{m}] -   [\psi_{n},e_{m+1}]\right)  \,, \\
\sigma_3 \{e_n,e_m \}   &=  [e_{n+3},e_m] - 3  [e_{n+2},e_{m+1}] + 3  [e_{n+1},e_{m+2}]  - [e_{n},e_{m+3}]    \\
&  \ + \sigma_2\left( [e_{n+1},e_{m}] -   [e_{n},e_{m+1}] \right)\,, \\
- \sigma_3 \{\psi_n,f_m \}  & =   [\psi_{n+3},f_m] - 3  [\psi_{n+2},f_{m+1}] + 3  [\psi_{n+1},f_{m+2}]  - [\psi_{n},f_{m+3}]   \\
&  \ + \sigma_2 \left([\psi_{n+1},f_{m}] -  [\psi_{n},f_{m+1}] \right) \,, \\
-  \sigma_3 \{f_n, f_m \}  & =  [f_{n+3},f_m] - 3  [f_{n+2},f_{m+1}] + 3  [f_{n+1},f_{m+2}]  - [f_{n},f_{m+3}]  \\
&  \ + \sigma_2\left( [f_{n+1},f_{m}] -   [f_{n},f_{m+1}]\right)   \,, \\
\psi_{n+m}  & =  [e_n, f_m] \,. \\
\end{aligned}
\end{equation}
These relations are further supplemented by the initial conditions
\begin{equation}\label{eq-initial-gl1}
\begin{aligned}
&[\psi_0,e_m] = 0 \;, \qquad \qquad [\psi_1,e_m] = 0  \;, \qquad \qquad [\psi_2,e_m] = 2 e_m \;, \\
&[\psi_0,f_m] = 0   \;, \qquad \qquad [\psi_1,f_m] = 0 \;, \qquad \qquad [\psi_2,f_m] = - 2 f_m  \;, 
\end{aligned}
\end{equation}
and the so-called Serre relations
\begin{equation}\label{eq-Serre-modes-gl1}
\begin{aligned}
\textrm{Sym}_{(n,m, \ell)}[e_{n}\,, [e_{m}\,,e_{\ell+1} ]]&=0 \;, \\
\textrm{Sym}_{(n, m, \ell)}[f_{n}\,, [f_{m}\,,f_{\ell+1} ]]&=0 \;.\\
\end{aligned}
\end{equation}
Note that the algebra has two parameters $\sigma_2$ and $\sigma_3$ and two central elements $\psi_0$ and $\psi_1$.
\bigskip

The algebraic relations (\ref{eq-relations-modes-gl1}) can be more elegantly repackaged in terms of the following three fields
\begin{equation}\label{eq-mode-expansion-gl1}
\begin{aligned}
e(z)\equiv\sum^{\infty}_{n=0}\frac{e_n}{z^{n+1}} \,, \qquad \psi(z)\equiv 1+ \sigma_3\sum^{\infty}_{n=0}\frac{\psi_n}{z^{n+1}}\,, \qquad f(z)\equiv \sum^{\infty}_{n=0}\frac{f_n}{z^{n+1}} \;,
\end{aligned}
\end{equation}
where $z$ is sometimes called the ``spectral parameter" in this context.
In terms of $(e(z), \psi(z),f(z))$, the relations (\ref{eq-relations-modes-gl1}) can be rewritten as 
\begin{equation}\label{eq-relations-OPE-gl1}
\begin{aligned}
\psi(z)\, \psi(w)&\sim \psi(w)\,\psi(z) \;, \\
\psi(z)\, e(w) &  \sim  \varphi_3(\Delta)\, e(w)\, \psi(z)  \;, \\ 
\psi(z)\, f(w) & \sim \varphi_3^{-1}(\Delta)\, f(w)\, \psi(z) \;, \\
e(z)\, e(w) & \sim    \varphi_3(\Delta)\, e(w)\, e(z)  \;, \\
 f(z)\, f(w) &  \sim    \varphi_3^{-1}(\Delta)\, f(w)\, f(z)  \;, \\
[e(z)\,, f(w)]   &\sim- \frac{1}{\sigma_3}\, \frac{\psi(z) - \psi(w)}{z-w} \,,
\end{aligned}
\end{equation}
where here and throughout this paper $\Delta$ is defined as
\begin{equation}\label{eq-Delta-def}
\Delta\equiv z-w  \;, 
\end{equation}
and the $\varphi_3$ function is a cubic rational function defined as
\begin{equation}\label{eq-varphi3-def-gl1}
\varphi_3(z)\equiv\frac{(z+h_1)(z+h_2)(z+h_3)}{(z-h_1)(z-h_2)(z-h_3)} \;.
\end{equation}
Here the triplet parameters $(h_1,h_2,h_3)$ satisfy
\begin{equation}\label{eq-constraint-h123}
h_1+h_2+h_3=0  \;, 
\end{equation} 
and are related to the two parameters $\sigma_2$ and $\sigma_3$ introduced earlier by
\begin{equation}\label{eq-sigma23-def}
\sigma_2\equiv h_1 \, h_2+h_2\, h_3+h_3\, h_1 \qquad \textrm{and}\qquad \sigma_3 \equiv h_1 \, h_2 \, h_3 \,.
\end{equation}
(Note that the function $\varphi_3(z)$ and hence the full algebra of affine Yangian of $\mathfrak{gl}_1$ are invariant under permutation of $\{h_1,h_2,h_3\}$.)
Unless stated explicitly otherwise, the $\sim$ sign in this paper denotes equality up to $z^{n\geq 0}w^{m}$ terms and $z^{n}w^{m\geq 0}$ terms.
One can easily reproduce the relations in terms of modes (\ref{eq-relations-modes-gl1}) by expanding (\ref{eq-relations-OPE-gl1}) using (\ref{eq-mode-expansion-gl1}) (after first moving the denominator of the coefficient $\varphi(\Delta)$ or $\varphi^{-1}(\Delta)$ to the l.h.s.) and taking the $z^{-n-1}w^{-m-1}$ term.

Although the $(e(z),\psi(z),f(z))$ are not fields in a two-dimensional CFT, the relations (\ref{eq-varphi3-def-gl1}) bear some resemblance to OPE (Operator Product Expansion) relations in a two-dimensional CFT in that (1) they are written in terms of fields $(e(z),\psi(z),f(z))$ and when expanded using (\ref{eq-mode-expansion-gl1}) reproduce the algebraic relations in terms of modes; and (2) the relations in (\ref{eq-varphi3-def-gl1}) are defined up to regular terms.
Therefore throughout this paper, we will abuse the terminology and call this type of relation ``OPE relations", to distinguish them from the corresponding mode relation such as (\ref{eq-relations-modes-gl1}).

Similarly, the Serre relations (\ref{eq-Serre-modes-gl1}) can be rewritten in terms of $(e(z), f(z))$ collectively:
\begin{equation}\label{eq-Serre-gl1}
\begin{aligned}
\textrm{Sym}_{z_1,z_2,z_3} (z_2-z_3) [\,e(z_1)\,,[\,e(z_2)\,,e(z_3)]]&=0\,;\\
\textrm{Sym}_{z_1,z_2,z_3} (z_2-z_3) [f(z_1)\,,[f(z_2)\,,f(z_3)]]&=0\,.
\end{aligned}
\end{equation}
Finally, the initial conditions (\ref{eq-initial-gl1}) can be derived from the $\psi(z)\,e(w)$ and $\psi(z)\, f(w)$ OPEs in (\ref{eq-relations-OPE-gl1}), respectively, by taking the $z^{-n-1}w^{-m-1}$ with $n=-3,-2,-1$ terms of these two equations  (after first moving the denominator of the coefficient $\varphi(\Delta)$ or $\varphi^{-1}(\Delta)$ to the l.h.s.), namely, these two equations are true up to  $z^{n\geq 3}w^{m}$ terms and $z^{n}w^{m\geq 0}$ terms.
\bigskip

For the purpose of this paper, the relations in terms of fields are much more useful than those written in terms of the modes,\footnote{
Except in the discussion of the initial condition, which is necessary to define the finite part of the affine Yangian algebra, and in the computation of the vacuum module directly in terms of algebra (i.e.\ without invoking the colored crystal representations). 
The latter is important in deriving/checking Serre relations.}
for the following two reasons:
\begin{enumerate}
\item The OPE relations (\ref{eq-relations-OPE-gl1}) make manifest the $\mathcal{S}_3$ symmetry (permuting the triplet $\{h_1,h_2,h_3\}$) that is intrinsic to the algebra but is somewhat hidden in (\ref{eq-relations-modes-gl1}).
\item The action of the algebra on the representations in terms of plane partition is much more transparent in terms of the OPE relations (\ref{eq-relations-OPE-gl1}) than the mode relations (\ref{eq-relations-modes-gl1}), see later. 
\end{enumerate}

It is convenient to use the following figure to summarize the OPE relations (\ref{eq-relations-OPE-gl1}):
\begin{equation}
\begin{tikzpicture}[scale=1]
\node[state]  [draw=blue!50, very thick, fill=blue!10] (psi) at (0,0)  {$\psi$};
\node[state] [draw=blue!50, very thick, fill=blue!10] (f) at (4,0)  {$f$};
\node[state]  [draw=blue!50, very thick, fill=blue!10](e) at (-4,0)  {$e$};
\path[->] 
(psi) edge   [thin]   node [above] {$\varphi_3(\Delta)$} (e)
(psi) edge   [thin]   node [above] {$\varphi_3^{-1}(\Delta)$} (f)
(e) edge [loop above, thin] node {$\varphi_3(\Delta)$} ()
(f) edge [loop above, thin] node {$\varphi_3^{-1}(\Delta)$} ()
;
\end{tikzpicture}
\end{equation}
Finally, as already mentioned in Introduction, it is known that the affine Yangian of $\mathfrak{gl}_1$ is equivalent to the universal enveloping algebra of the $\mathcal{W}_{1+\infty}$ algebra, see \cite{SV, Maulik:2012wi, Tsymbaliuk, Tsymbaliuk:2014fvq, Prochazka:2015deb, Gaberdiel:2017dbk}.

\medskip

\subsection{Plane Partition}

A partition $\lambda$ of an integer $n$ can be characterized by a set of integers $\lambda_i$:
\begin{equation}\label{lambdap}
\textrm{partition of }n: \qquad \left\{\lambda_i \,\, \Big|\,\, \lambda_i \in
\mathbb{Z}_{\geq 0} \,, \lambda_i \geq \lambda_{i+1} \,, \sum_{i} \lambda_i=n \right\} \;.
\end{equation}
A plane partition $\Lambda$ is a three-dimensional generalization of the integer partition
\begin{equation}
\textrm{plane partition of }n: \,\, \left\{ \Lambda_{i,j}\, \Big| \,\Lambda_{i,j}\in \mathbb{Z}_{\geq 0}\,,\Lambda_{i,j} \geq \Lambda_{i+1,j} \,,\Lambda_{i,j} \geq \Lambda_{i,j+1} 
 \,, \sum_{i,j} \Lambda_{i,j}=n \right\} \,,
\end{equation}
and can be given by the stacking of three-dimensional boxes (denoted as $\square$ in this paper), which are 3D generalization of 2D Young diagrams.
The coordinates of these $\square$'s are chosen to be
\begin{equation}\label{eq-coordinates-gl1-orig}
\left(x_1(\square),x_2(\square),x_3(\square)\right) \qquad \textrm{with} \qquad x_{1,2,3}(\square)\in \mathbb{Z}_{\geq 0}\,.
\end{equation}

The generating function of plane partition counting is the MacMahon function \cite{MR1676282}
\begin{equation}\label{MacMahon} 
\begin{aligned}
 M(q)\equiv\sum_{\Lambda \in \textrm{ plane partition}} q^{|\Lambda|}&
 =  \prod^{\infty}_{k=1}\frac{1}{(1-q^k)^k}  \\
&= 1+q + 3 q^2 + 6 q^3 + 13 q^4 + 24 q^5 + 48 q^6 + \dots \;,
\end{aligned}
\end{equation}
where $|\Lambda|$ denotes the number of boxes $\square$ in the plane partition $\Lambda$. 
This partition function is also the partition function of the topological A-model on $\mathbb{C}^3$ \cite{Okounkov:2003sp}; and it is also identical to the vacuum character of $\mathcal{W}_{1+\infty}$ algebra (at general central charge $c$ and coupling $\lambda$).

\medskip

\subsection{\texorpdfstring{Action of Affine Yangian of $\mathfrak{gl}_1$ on Plane Partitions}{Action of Affine Yangian of gl(1) on Plane Partitions}}

The affine Yangian of $\mathfrak{gl}_1$ acts on the set of plane partitions (this representation is known as the MacMahon module in the literature). 
To describe this action, it is necessary to first endow plane partitions with additional structures, to accommodate the two parameters in the algebra, i.e.\ $(h_1,h_2,h_3)$ satisfying $h_1+h_2+h_3=0$.

Recall that each plane partition configuration consists of a collection of 3D-boxes, with coordinates given by (\ref{eq-coordinates-gl1-orig}).
To each box $\square$, we can associate a coordinate function
\begin{equation}\label{eq-coordinate-function-gl1}
h(\square)\equiv h_1 \, x_1(\square)+h_2 \, x_2(\square)+h_3\, x_3(\square) \,,
\end{equation}
which naturally incorporates the parameters $(h_1,h_2,h_3)$. 

The action of affine Yangian of $\mathfrak{gl}_1$ on a plane partition configuration $\Lambda$ is given by \cite{feigin2012, Tsymbaliuk:2014fvq, Prochazka:2015deb}
\begin{equation}\label{action-gl1-pp}
\begin{aligned}
\psi(z)|\Lambda \rangle & = \Psi_{\Lambda}(z)|\Lambda \rangle  \ , \\
e(z) | \Lambda \rangle & =  \sum_{ \Box \in {\rm Add}(\Lambda)}\frac{\Big[ -  \frac{1}{\sigma_3} {\rm Res}_{w = h(\Box)} \Psi_{\Lambda}(w) \Big]^{\frac{1}{2}}}{ z - h(\Box) } 
| \Lambda + \Box \rangle \ , \\
f(z) | \Lambda \rangle & =  \sum_{ \Box \in {\rm Rem}(\Lambda)}\frac{\Big[ +  \frac{1}{\sigma_3} {\rm Res}_{w = h(\Box)} \Psi_{\Lambda}(w) \Big]^{\frac{1}{2}}}{ z - h(\Box) } | \Lambda - \Box \rangle \ . 
\end{aligned}
\end{equation}
Recall the $\psi(u)$ contains all the Cartan operators of the algebra, see the first line of (\ref{eq-relations-modes-gl1}).
Each plane partition $|\Lambda\rangle$ is an eigenstate of $\psi(u)$, hence of all the Cartan modes $\psi_n$ with $n\in\mathbb{Z}_{\geq 0}$.
The eigenvalue is
\begin{equation}\label{eq-Psi-gl1}
\Psi_{\Lambda}(z)=\psi_0(z)\, \prod_{\square\in \Lambda}\, \varphi_3 (z-h(\square))  \;,
\end{equation}
where 
\begin{equation}\label{eq-vacuum-gl1}
\psi_0(z)=1+h_1h_2h_3\frac{\psi_0}{z}
\end{equation}
is the vacuum contribution.
Given a $|\Lambda\rangle$, $e(z)$ adds a $\square$ in all legitimate locations, while $f(z)$ removes  a $\square$ in all legitimate locations.\footnote{
Here ``legitimate" means that the final state $|\Lambda\pm \square\rangle$ is again a plane partition configuration. }
In summary:
\begin{equation}\label{eq-gl1action}
\square: \qquad \quad e(u):\, \textrm{creation}\,, \qquad  \psi(u):\, \textrm{charge} \,, \qquad  f(u):\, \textrm{annihilation} \,.
\end{equation}

The actions of the modes $(e_n,\psi_n,f_n)$ on the plane partition state $|\Lambda\rangle$ can be obtained from the action (\ref{action-gl1-pp}) and the mode expansion (\ref{eq-mode-expansion-gl1}).
From the action of $e_n$ one can check that the character of the vacuum module of affine Yangian of $\mathfrak{gl}_1$ reproduces the MacMahon function (\ref{MacMahon}). 
In this computation, the ($e$ part of the) Serre relations (\ref{eq-initial-gl1}) plays a crucial role, for more details see section~\ref{eq-mode-expansion-gl1}.

\bigskip

\section{BPS Quiver Yangian for General Quivers}\label{sec-quiver-Yangian} 
\label{sec:QYtopdown}

In this section let us define the BPS quiver Yangian $\mathsf{Y}_{(Q,W)}$ from a pair $(Q,W)$.\footnote{
While our interest in this paper is to those  pair $(Q,W)$ originating from toric Calabi-Yau threefolds, our definition in itself applies to more general choices of $(Q,W$). 
It is not clear, however, if the algebra acts on BPS states of some gauge/string theory in these more general situations.}
Since the pair $(Q,W)$ is obtained from a toric Calabi-Yau geometry $X$ (as we discussed in section \ref{sec:review_crystal}), the algebra $\mathsf{Y}_{(Q,W)}$ in itself can be associated with the geometry $X$.

In general, for the same Calabi-Yau manifold $X$, there exist multiple quiver gauge theories $(Q,W)$ which are dual to one another.
In these situations the quiver gauge theories are believed to be related by a sequence of Seiberg dualities (quiver mutations) \cite{Beasley:2001zp}, and we conjecture that the resulting algebras $\mathsf{Y}_{(Q,W)}$ are all isomorphic.
We will see concrete examples of this phenomenon in sections \ref{sec:glmn} and \ref{sec.P1P1}: the relevant isomorphisms are already known in the mathematical literature for the examples in  \ref{sec:glmn}, but not for those in \ref{sec.P1P1}. It would be interesting to explore this point further.
 
In this section we provide a top-down definition of the algebra.
Let us mention, however, that later in section \ref {sec:bootstrap_general} we will provide bottom-up justifications of the algebra. 
Indeed, as we will see in section \ref{sec:bootstrap_general}, under some reasonable ansatz, the condition that this algebra acts on the configurations of molten crystal can completely fix the algebra. 
In this sense our algebra and its representation on the BPS crystal are intimately connected.

\medskip
\subsection{Parameters}\label{sec:parameters}

To define the BPS quiver Yangian $\mathsf{Y}_{(Q,W)}$, we first consider a set of charge assignments $h_I$ for each arrow $I\in Q_1$. 
We impose the condition that this charge assignment is compatible with the superpotential $W$. 
In other words, the charges $h_I$ can be regarded as charges under a global symmetry of the quiver quantum mechanics.
The superpotential $W$ will enter into the definition of the algebra $\mathsf{Y}_{(Q,W)}$ through this charge-assignment constraint only.

In the periodic quiver diagram, a monomial term in the superpotential is represented by a closed loop.
This means that the constraint on the parameters $h_I$ can be written as\footnote{
Note that all arrows are in the same direction in the smallest loops.}
\begin{tcolorbox}[ams equation]\label{eq-loop-constraint-toric}
\textrm{\bf loop constraint: } \quad \sum_{I\in L} h_I=0 \;,
\end{tcolorbox}
\noindent where $L$ is an arbitrary loop in the periodic quiver. 
We will hereafter call this condition the \emph{loop constraint}, and the parameters satisfying these conditions as \emph{coordinate parameters}.
In section \ref{sec:melting_check} we will see that this constraint is instrumental in ensuring the consistency of the crystal-melting representation of the algebra.

We can count the number of coordinate parameters to be 
\begin{align}
N_h=\textrm{\# (edges of the quiver)}  - (\textrm{\# (monomial terms in the superpotential)}-1)  \;.
\end{align}
Here we have subtracted one from the superpotential constraints, since any bifundamental field appears exactly twice in the superpotential (this follows since each edge belongs to two neighboring faces in the periodic quiver) and thus one of the constraints is redundant. 
Since each monomial term in the superpotential corresponds to a polygonal region of the periodic quiver, one can also write this as
\begin{align}
N_h=\textrm{\# (edges of the periodic quiver)}  - (\textrm{\# (faces of the periodic quiver)}-1)  \;.
\end{align}
Since the periodic quiver is written on the two-dimensional torus and has Euler character zero, one can rewrite this as 
\begin{align}\label{eq.N_h}
N_h & =\textrm{\# (vertices of the periodic quiver)}  +1  \nonumber\\
& =\textrm{\# (gauge groups of the quiver)}  +1 \;. 
\end{align}
For a toric Calabi-Yau threefold this number (i.e.\ $\#=N_h-1$) is known to be the same as the area of the toric diagram $\Delta$, where the normalization of the area is chosen such that the minimal lattice triangle spanned by the three lattice points $(0,0)$, $(1,0)$, and $(0,1)$ has area $1$.
One can then use Pick's theorem to rewrite this as
\begin{align}\label{eq.N_h_2}
N_h & =E+2I-1  \;,
\end{align}
where $E$ (resp.\ $I$) is the numbers of external (resp.\ internal) lattice points of the toric diagram $\Delta$.
We will use $\{h_I\}$, with $I=1,\cdots, |Q_1|$, to denote the set of charges associated to the edges of the quiver; and we use $\{\mathsf{h}_A\}$, with $A=1,\cdots, N_h$, to denote these $N_h$ independent parameters that characterize the algebra.

\medskip

\subsection{Generators and Relations}

The algebra is generated by a triplet of fields $(e^{(a)}(u), \psi^{(a)}(u),  f^{(a)}(u))$ for each quiver vertex $a\in Q_0$:
\begin{equation}\label{eq-action-toric}
\sqbox{$a$}: \qquad e^{(a)}(u):\, \textrm{creation}\,, \qquad  \psi^{(a)}(u):\, \textrm{charge} \,, \qquad  f^{(a)}(u):\, \textrm{annihilation} \,.
\end{equation}
Generically, they have the mode expansion:\footnote{
For the fermionic generators in the NS sector, it might be more natural to expand the $e^{(a)}(z)$ and $f^{(a)}(z)$ generators in terms of half-integer modes. 
This will not by relevant within the current paper, but can be determined once we know the map between the quiver Yangians and the $\mathcal{W}$ algebras.
}
\begin{equation}\label{eq-mode-expansion-toric}
\begin{aligned}
e^{(a)}(z)\equiv\sum^{+\infty}_{n=0}\frac{e^{(a)}_n}{z^{n+1}} \,, \qquad \psi^{(a)}(z)\equiv \sum^{+\infty}_{n=-\infty}\frac{\psi^{(a)}_n}{z^{n+1}}\,, \qquad f^{(a)}(z)\equiv \sum^{+\infty}_{n=0}\frac{f^{(a)}_n}{z^{n+1}} \;,
\end{aligned}
\end{equation}
and contain infinitely many generators $e_n^{(a)}, \psi_n^{(a)},  f_n^{(a)}$. 
As we will show later in section~\ref{sec:generalno4cycle}, for Calabi-Yau threefolds without compact $4$-cycles, $\psi^{(a)}_{n< -1}=0$ and  $\psi^{(a)}_{-1}=1$.
 
We express the $\mathbb{Z}_2$-grading  (i.e.\ the Bose/Fermi statistics) of the generators $e_n^{(a)}, f_n^{(a)}$  of the generators) to be
\begin{tcolorbox}[ams equation]\label{eq.Z2_grading}
\textrm{\bf grading rule: }\quad 
|a|=\begin{cases}
0 & (\exists I\in Q_1 \quad \textrm{such that} \quad s(I)=t(I)=a ) \;,\\
1 & (\textrm{otherwise}) \;,
\end{cases}
\end{tcolorbox}
\noindent with $|a|=0$ ($|a|=1$) for bosonic (fermionic) generators. 
The operators $\psi_n^{(a)}$ are Cartan and hence are set to be even.

\subsubsection{Relations in Terms of Fields}

The generators satisfy the OPE relations
\begin{tcolorbox}[ams align]\label{eq-OPE-toric}
\begin{aligned}
\psi^{(a)}(z)\, \psi^{(b)}(w)&=   \psi^{(b)}(w)\, \psi^{(a)}(z)  \;,\\
\psi^{(a)}(z)\, e^{(b)}(w)&\simeq  \varphi^{b\Rightarrow a}(\Delta)\, e^{(b)}(w)\, \psi^{(a)}(z)  \;, \\
e^{(a)}(z)\, e^{(b)}(w)&\sim  (-1)^{|a||b|}  \varphi^{b\Rightarrow a}(\Delta) \, e^{(b)}(w)\, e^{(a)}(z)  \;, \\
\psi^{(a)}(z)\, f^{(b)}(w)&\simeq   \varphi^{b\Rightarrow a}(\Delta)^{-1} \, f^{(b)}(w)\,\psi^{(a)}(z) \;,\\
f^{(a)}(z)\, f^{(b)}(w)&\sim  (-1)^{|a||b|} \varphi^{b\Rightarrow a}(\Delta)^{-1}\,  f^{(b)}(w)\, f^{(a)}(z)   \;,\\
\left[e^{(a)}(z),f^{(b)}(w) \right\} &\sim -  \delta^{a,b} \frac{\psi^{(a)}(z)-\psi^{(b)}(w)}{z-w}  \;,
\end{aligned}
\end{tcolorbox}
\noindent where throughout this paper ``$\simeq$" means equality up to $z^n w^{m\geq 0}$ terms,  ``$\sim$" means equality up to $z^{n\geq 0} w^{m}$ and $z^{n} w^{m\geq 0}$ terms, and finally
\begin{equation}
\Delta\equiv z-w \;.
\end{equation}
The bracket $[e^{(a)}(z),f^{(b)}(w)\}$ represents the commutator in the superalgebra sense. 
Namely, it is an anti-commutator $\{ e^{(a)}(z),f^{(b)}(w)\}$ when both $a$ and $b$ are odd, and is a commutator $[e^{(a)}(z),f^{(b)}(w)]$ otherwise.

The function $\varphi^{a\Rightarrow b}(z)$, which we call the ``bond factor" since roughly speaking it describes the ``bonding" between atoms of color $a$ and atoms of color $b$, is defined to be\begingroup%
\setcounter{savefootnote}{\value{footnote}}%
\setcounter{footnote}{0}%
\renewcommand{\thefootnote}{\fnsymbol{footnote}}%
\footnote[1]{Note added for version 3: In general the bond factor \eqref{eq-charge-atob}  could have a sign factor $\textrm{sign($a$,$b$)}\in \{\pm\}$. 
The choice of these signs is not unique and in this paper we ignore this sign ambiguity;  one choice for example is  $\textrm{sign($a$,$b$)}=(-1)^{|a\rightarrow b| (1+|b\rightarrow a|)}$ used in \cite{Galakhov:2021vbo}\label{footnote11}.
This subtlety is relevant for chiral quivers that possess pairs of vertices $(a,b)$ whose total number of arrows between them is odd. 
An alternative is to not to have any sign factors in 
\eqref{eq-charge-atob}, but to modify the r.h.s.\ of \eqref{eq.varphi_sym} below to $(-1)^{|a\rightarrow b|+|b\rightarrow a|}$, and fix the ordering between $a$ and $b$ when plugging into \eqref{eq-OPE-toric}.}
\setcounter{footnote}{\value{savefootnote}}
\endgroup
\begin{tcolorbox}[ams equation]\label{eq-charge-atob}
\varphi^{a\Rightarrow b} (u)\equiv \frac{\prod_{I\in \{b\rightarrow a\}}(u+h_{I})}{\prod_{I\in \{a\rightarrow b\}}(u-h_{I})} \;,
\end{tcolorbox}
\noindent 
where $\{a\rightarrow b\}$ denotes the set of edges from vertex $a$ to vertex $b$.
When there is no arrow between vertex $a$ and vertex $b$ in the quiver (denoted as $a \centernot{\longleftrightarrow} b$), the bond factor is trivial:
\begin{tcolorbox}[ams equation]\label{eq-charge-atob-trivial}
a \centernot{\longleftrightarrow} b\, :\qquad \varphi^{a\Rightarrow b} (u)= \varphi^{b\Rightarrow a} (u)\equiv 1 \;;
\end{tcolorbox}
\noindent and the corresponding operators from $(e^{(a)}(z),\psi^{(a)}(z),f^{(a)}(z))$ commute  --- or anti-commute when the relevant  sign $(-1)^{|a||b|}$ is $-1$ --- with those from $(e^{(b)}(w),\psi^{(b)}(w),f^{(b)}(w))$.
The bond factor satisfies the reflection property
\begin{align}\label{eq.varphi_sym}
\varphi^{a\Rightarrow b} (u) \, \varphi^{b\Rightarrow a} (-u)=1 \;,
\end{align}
which is needed for the consistency of the OPE relations.
The relations (\ref{eq-OPE-toric}) (except for the relation between $e$ and $f$) are summarized in the following graph:\footnote{
Note that to reduce clutter, in the graph (\ref{eq-OPE-graph}) we have omitted the additional statistics factors in (\ref{eq-OPE-toric}), i.e.\ $(-1)^{|a|}$ for the $e^{(a)}(z) e^{(a)}(w)$ and $f^{(a)}(z)f^{(a)}(w)$ relations, $(-1)^{|b|}$ for the $e^{(b)}(z) e^{(b)}(w)$ and $f^{(b)}(z)f^{(b)}(w)$ relations, and  $(-1)^{|a||b|}$ for the $e^{(a)}(z) e^{(b)}(w)$ and $f^{(a)}(z)f^{(b)}(w)$ relations.}
\begin{equation}\label{eq-OPE-graph}
\begin{tikzpicture}[scale=1]
\node[state]  [draw=blue!50, very thick, fill=blue!10] (psia) at (0,0)  {$\psi^{(a)}$};
\node[state] [draw=blue!50, very thick, fill=blue!10] (fa) at (4,0)  {$f^{(a)}$};
\node[state]  [draw=blue!50, very thick, fill=blue!10](ea) at (-4,0)  {$e^{(a)}$};
\node[state]  [draw=blue!50, very thick, fill=blue!10] (psib) at (0,-4)  {$\psi^{(b)}$};
\node[state] [draw=blue!50, very thick, fill=blue!10] (fb) at (4,-4)  {$f^{(b)}$};
\node[state]  [draw=blue!50, very thick, fill=blue!10](eb) at (-4,-4)  {$e^{(b)}$};
\path[->] 
(psia) edge   [thin]   node [above] {$\varphi^{a\Rightarrow a}$} (ea)
(psia) edge   [thin]   node [above] {$1/\varphi^{a\Rightarrow a}$} (fa)
(ea) edge [loop above, thin] node {$\varphi^{a\Rightarrow a}$} ()
(fa) edge [loop above, thin] node {$1/\varphi^{a\Rightarrow a}$} ()
(psib) edge   [thin]   node [above] {$\varphi^{b\Rightarrow b}$} (eb)
(psib) edge   [thin]   node [above] {$1/\varphi^{b\Rightarrow b}$} (fb)
(eb) edge [loop below, thin] node {$\varphi^{b\Rightarrow b}$} ()
(fb) edge [loop below, thin] node {$1/\varphi^{b\Rightarrow b}$} ()
(psia) edge   [thin]   node [above right, xshift=-18pt, yshift=15pt] {$\varphi^{a\Rightarrow b}$} (eb)
(psib) edge   [thin]   node [below left, yshift=-5pt, xshift=-9pt] {$\varphi^{b\Rightarrow a}$} (ea)
(psia) edge   [thin]   node [below left, yshift=0pt, xshift=-7pt] {$1/\varphi^{a\Rightarrow b}$} (fb)
(psib) edge   [thin]   node [below right, xshift=7pt] {$1/\varphi^{b\Rightarrow a}$} (fa)
(ea) edge   [thin]   node [left] {$\varphi^{b\Rightarrow a}$} (eb)
(fa) edge   [thin]   node [right] {$1/\varphi^{b\Rightarrow a}$} (fb)
;
\end{tikzpicture}
\end{equation}

We emphasize that the bond factor $\varphi^{a\Rightarrow b}(u)$ (\ref{eq-charge-atob}) should be treated as a ``formal" rational function.
Namely, all the factors in its numerator and denominator, one pair (i.e.\ one in the numerator and one in the denominator) from each arrow in the quiver, need to be kept even when the charges $h_I$ take some special values such that some factors of the numerator and the denominator cancel each other.
The reason is that the algebra can also be expressed in terms of modes $(e^{(a)}_n,\psi^{(a)}_n, f^{(a)}_n)$, using the mode expansions (\ref{eq-mode-expansion-toric}), and it is important that we keep all factors in $\varphi^{a\Rightarrow b}(u)$, in order to reproduce the correct algebraic relations in terms of modes.

\subsubsection{Relations in Terms of Modes}

With the mode expansions of the fields in (\ref{eq-mode-expansion-toric}), it is straightforward to expand the OPE relations (\ref{eq-OPE-toric}) and write down the corresponding relations in terms of modes.
 
The first and the last equations in (\ref{eq-mode-expansion-toric}) do not involve the bond factor $\varphi^{b\Rightarrow a}(z-w)$ and are easy to translate into the mode relations:
\begin{equation}
\left[\psi^{(a)}_n \, , \, \psi^{(b)}_m\right]=0 \qquad \textrm{and}\qquad \left[e^{(a)}_n\, , \, f^{(b)}_m \right\}=\delta^{a,b}\,\psi^{(a)}_{n+m} \;.
\end{equation}

All the remaining ones involve the bond factor $\varphi^{b\Rightarrow a}(z-w)$ (see definition (\ref{eq-charge-atob})), whose numerator and denominator can be rewritten as
\begin{equation}\label{eq-charge-function-expand}
\begin{aligned}
\prod_{I\in \{a\rightarrow b\}}(z-w+h_{I})&=\sum^{|a\rightarrow b|}_{k=0} \sigma^{a\rightarrow b}_{|a\rightarrow b|-k}\, (z-w)^{k} \,,\\
\prod_{I\in \{b\rightarrow a\}}(z-w-h_{I})&=\sum^{|b\rightarrow a|}_{k=0}(-1)^{|b\rightarrow a|-k} \, \sigma^{b\rightarrow a}_{|b\rightarrow a|-k}\,  (z-w)^{k} \,,\\
\end{aligned}
\end{equation}
where $|a\rightarrow b|$ denotes the number of arrows from $a$ to $b$ in the quiver diagram, and $\sigma^{a\rightarrow b}_k$ denotes the $k^{\textrm{th}}$ elementary symmetric sum of the set $\{h_I\}$ with $I\in \{a\rightarrow b\}$. 

Now take the $\psi^{(a)} \, e^{(b)}$ OPE for example. 
Moving the denominator of $\varphi^{b\Rightarrow a}(z-w)$ to the l.h.s.\ of the equation, and using the expansion (\ref{eq-charge-function-expand}), one can rewrite the $\psi^{(a)} \, e^{(b)}$ OPE relation in terms of quiver data $\{h_I\}$:
\begin{equation}\label{eq-psie-OPE-1}
\begin{aligned}
&\sum^{|b\rightarrow a|}_{k=0}(-1)^{|b\rightarrow a|-k} \, \sigma^{b\rightarrow a}_{|b\rightarrow a|-k}\,  (z-w)^{k}\, \psi^{(a)}(z)\, e^{(b)}(w)\simeq \sum^{|a\rightarrow b|}_{k=0} \sigma^{a\rightarrow b}_{|a\rightarrow b|-k}\, (z-w)^{k} \,  e^{(b)}(w)\,\psi^{(a)}(z) \;.
\end{aligned}
\end{equation}
Plugging in the mode expansions of $\psi^{(a)}(z)$ and $e^{(b)}(w)$ from (\ref{eq-mode-expansion-toric}), expanding the $(z-w)^k$ in (\ref{eq-psie-OPE-1}), and extracting the terms of order $z^{-n-1}w^{-m-1}$ with $n\in \mathbb{Z}$ and $m\in\mathbb{Z}_{\geq 0}$, we have the mode relation:
\begin{equation}\label{eq-mode-relation-psi-e}
\sum^{|b\rightarrow a|}_{k=0}(-1)^{|b\rightarrow a|-k} \, \sigma^{b\rightarrow a}_{|b\rightarrow a|-k}\,  [\psi^{(a)}_n\, e^{(b)}_m]_k =\sum^{|a\rightarrow b|}_{k=0} \sigma^{a\rightarrow b}_{|a\rightarrow b|-k}\, [ e^{(b)}_m\, \psi^{(a)}_n]^k \,,
\end{equation}
for $n\in \mathbb{Z}$ and $m\in\mathbb{Z}_{\geq 0}$, where we have defined the shorthand
\begin{equation}\label{eq-ABn}
\begin{aligned}
\left[A_n\, B_m\right]_k& \equiv \sum^{k}_{j=0} (-1)^j\,\tbinom{k}{j} \,A_{n+k-j}\, B_{m+j}\,, \qquad \\
\left[B_m\,A_n\right]^k &\equiv \sum^{k}_{j=0} (-1)^j\,\tbinom{k}{j} \,B_{m+j}\, A_{n+k-j}\,.
\end{aligned}
\end{equation}

Here we can see that it is important to keep all factors in $\varphi^{b\Rightarrow a}(z-w)$, even when the charges $h_I$ take special values such that some factors in the numerator and denominator cancel each other. 
Ultimately what is important is the expansions (\ref{eq-charge-function-expand}) of the numerator and the denominator separately, which in particular control the mode shifting in the mode relation (\ref{eq-mode-relation-psi-e}).

Repeating this exercise for the remaining equations in (\ref{eq-OPE-toric}), we have their corresponding relations in terms of the modes: 
\begin{tcolorbox}[ams align]\label{eq-OPE-modes-toric}
\begin{aligned}
\left[\psi^{(a)}_n \, , \, \psi^{(b)}_m\right]&=0 \;,\\
\sum^{|b\rightarrow a|}_{k=0}(-1)^{|b\rightarrow a|-k} \, \sigma^{b\rightarrow a}_{|b\rightarrow a|-k}\,  [\psi^{(a)}_n\, e^{(b)}_m]_k &=\sum^{|a\rightarrow b|}_{k=0} \sigma^{a\rightarrow b}_{|a\rightarrow b|-k}\, [ e^{(b)}_m\, \psi^{(a)}_n]^k  \;,\\
 \sum^{|b\rightarrow a|}_{k=0}(-1)^{|b\rightarrow a|-k} \, \sigma^{b\rightarrow a}_{|b\rightarrow a|-k}\,  [e^{(a)}_n\, e^{(b)}_m]_k &=(-1)^{|a||b|}\sum^{|a\rightarrow b|}_{k=0} \sigma^{a\rightarrow b}_{|a\rightarrow b|-k}\, [ e^{(b)}_m\, e^{(a)}_n]^k  \;,\\
 \sum^{|a\rightarrow b|}_{k=0} \sigma^{a\rightarrow b}_{|a\rightarrow b|-k}\, [\psi^{(a)}_n\, f^{(b)}_m]_k &= \sum^{|b\rightarrow a|}_{k=0}(-1)^{|b\rightarrow a|-k} \, \sigma^{b\rightarrow a}_{|b\rightarrow a|-k}\,  [ f^{(b)}_m\, \psi^{(a)}_n]^k  \;,\\
\sum^{|a\rightarrow b|}_{k=0} \sigma^{a\rightarrow b}_{|a\rightarrow b|-k}\, [f^{(a)}_n\, f^{(b)}_m]_k &=(-1)^{|a||b|} \sum^{|b\rightarrow a|}_{k=0}(-1)^{|b\rightarrow a|-k} \, \sigma^{b\rightarrow a}_{|b\rightarrow a|-k}\,  [ f^{(b)}_m\, f^{(a)}_n]^k  ,\\
\left[e^{(a)}_n\, , \, f^{(b)}_m \right\}&=\delta^{a,b}\,\psi^{(a)}_{n+m} \;,
\end{aligned}
\end{tcolorbox}
\noindent where for $\psi^{(a)}_n$ modes, $n\in \mathbb{Z}$, and for $e^{(a)}_n$ and $f^{(a)}_n$ modes, $n\in\mathbb{Z}_{\geq 0}$.
When the set of charges $h_I$ with $I\in\{a\rightarrow b\}$ is identical to the set $h_I$ with $I\in \{b\rightarrow a\}$, we have $\sigma^{a\rightarrow b}_k=\sigma^{b\rightarrow a}_k$ for all $k$.
In this case, the equations in (\ref{eq-OPE-modes-toric}) can all be expressed in terms of commutators and anti-commutators.

\subsection{Some Properties of the Algebra}
\label{sec:shuffling}

\subsubsection{Grading and Filtration}

As a vector space, the algebra $\mathsf{Y}_{(Q,W)}$ has a triangular decomposition
\begin{align}\label{eq_triangular}
\mathsf{Y}_{(Q,W)}= \mathsf{Y}_{(Q,W)}^{+} \oplus \mathsf{B}_{(Q,W)} \oplus \mathsf{Y}_{(Q,W)}^{-}  \;,
\end{align}
where $\mathsf{Y}_{(Q,W)}^{+}$ ($\mathsf{Y}_{(Q,W)}^{-}$) are generated by the $e^{(a)}_n$'s ($f^{(a)}_n$'s), and $\mathsf{B}_{(Q,W)}$, which we call the Bethe subalgebra, is generated by the $\psi^{(a)}_n$'s.

First of all, we have an $\mathbb{Z}_2$ transformation 
\begin{align}
e^{(a)}(z) \leftrightarrow f^{(a)}(z)\;,
\quad 
\psi^{(a)}(z) \leftrightarrow \psi^{(a)}(z)^{-1} \;,
\end{align}
which exchanges $\mathsf{Y}_{(Q,W)}^{+}$ and $\mathsf{Y}_{(Q,W)}^{-}$ while preserving $\mathsf{B}_{(Q,W)}$.
\bigskip

The algebra has some more structures in addition to the $\mathbb{Z}_2$ grading just introduced.
First, for each vertex $a\in Q_0$ we can define an associated $\mathbb{Z}$ grading $\mathrm{deg}_a$ (termed ``grading by color $a$'' or ``mode grading") by 
\begin{align}
\mathrm{deg}_a (e^{(b)}_n) = \delta_{a,b} \;, \quad
\mathrm{deg}_a (\psi^{(b)}_n)= 0 \;, \quad
\mathrm{deg}_a (f^{(b)}_n) = - \delta_{a,b} \;.
\end{align}
Second, the algebraic relations (\ref{eq-OPE-toric}) with 
(\ref{eq-charge-atob}) have a rescaling symmetry for the parameters $h_I$, the spectral parameter $u$, and the generators:\footnote{
The scaling behaviors of $\psi^{(a)}(u)$ is determined by the consideration that in some examples (i.e.\ for  Calabi-Yau threefolds without compact $4$-cycles), we are allowed to fix $\psi^{(a)}_{-1}=1$.
(For other cases, even if we do not fix any $\psi^{(a)}_n$ mode, we are still allowed to choose the same scaling behavior for $\psi^{(a)}(u)$.)
This then gives $e^{(a)}(u)f^{(b)}(v)\rightarrow \alpha^{-1} e^{(a)}(u)f^{(b)}(v)$, following from the $e-f$ relation in  (\ref{eq-OPE-toric}). 
The most natural choice (and without loss of generality) is then the one given in (\ref{eq-rescaling-field}).} 
\begin{equation}\label{eq-rescaling-field}
\begin{aligned}
&h_I\rightarrow \alpha\, h_I\,, \quad u\rightarrow \alpha\, u\,, \quad \\
&e^{(a)}(u)\rightarrow \alpha^{-\frac{1}{2}}\, e^{(a)}(u)\,, \quad f^{(a)}(u)\rightarrow \alpha^{-\frac{1}{2}}\, f^{(a)}(u)\,,\quad \psi^{(a)}(u)\rightarrow  \psi^{(a)}(u)\,.
\end{aligned}
\end{equation}
In terms of the mode generators, (\ref{eq-rescaling-field}) is
\begin{equation}\label{eq-rescaling-mode}
\begin{aligned}
&h_I\rightarrow \alpha\, h_I \,, \quad 
e^{(a)}_n\rightarrow \alpha^{n+\frac{1}{2}}\, e^{(a)}_n\,, \quad f^{(a)}_n\rightarrow \alpha^{n+\frac{1}{2}}\, f^{(a)}_n\,,\quad \psi^{(a)}_n\rightarrow \alpha^{n+1}\, \psi^{(a)}_n\,,
\end{aligned}
\end{equation}
due to the mode expansion (\ref{eq-mode-expansion-toric}).
The rescaling symmetry (\ref{eq-rescaling-mode}) defines the grading
\begin{align}\label{level_grading}
\mathrm{deg}_{\rm level} (e^{(b)}_n) = 
\mathrm{deg}_{\rm level} (f^{(b)}_n) = n+\tfrac{1}{2} \;, \quad
\mathrm{deg}_{\rm level} (\psi^{(b)}_n)= n+1 \;,
\end{align}
together with $\mathrm{deg}_{\rm level}(h_I)=1$.
We can also regard this as a filtration (termed ``level filtration" or ``spin filtration") on the algebra when we assign zero degree to $h_I$, while keeping the assignments on mode generators \eqref{level_grading}.

\subsubsection{Spectral Shift}

One can shift the spectral parameter $z$ by an overall constant.
This linearly mixes the generators, and generates an automorphism of the algebra.
More explicitly, in terms of the mode expansions introduced in \eqref{eq-mode-expansion-toric},
one obtains under the shift $z\to z-\varepsilon$ a new set of
 modes $e'_l, \psi'_l, f'_l$:
\begin{align}\label{spectralshift}
&e'_l=\sum_{k=0}^l \binom{l}{k} \varepsilon^k e_{l-k}  \;, \quad
f'_l=\sum_{k=0}^l \binom{l}{k} \varepsilon^k f_{l-k} \;,  \quad
\psi'_l=\sum_{k=0}^l \binom{l}{k} \varepsilon^k \psi_{l-k} \quad (l=0,1, \dots)\;,\nonumber\\
&\psi'_{-l-1}=\sum_{k=l}^{\infty} \binom{k}{l} (-\varepsilon)^{k-l} \psi_{-k-1} \quad (l=0,1, \dots, )\;.
\end{align} 
Namely, since the mode expansion \eqref{eq-mode-expansion-toric} is in powers of $z^{-1}$, the shift $z\rightarrow z-\varepsilon$ mixes the generators $e^{(a)}_n$ only with $e^{(a)}_m$ with $m < n$, and similarly for $f^{(a)}_n$ and $\psi^{(a)}_n$.
The last equation involves an infinite sum and should be regarded as a formal sum.
This equation is trivialized to $\psi'_{-1}=\psi_{-1}$ for the toric Calabi-Yau threefold geometries without compact $4$-cycles,
where we have $\psi_{n< -1}=0$.

\subsubsection{Gauge-symmetry Shift}

As we discussed above, the parameters $\{h_I\}$ 
can be regarded as global-symmetry assignments of the algebra.
We have therefore imposed the loop constraints \eqref{eq-loop-constraint-toric}. 

One notices, however, that some of these symmetries are actually gauge symmetries.
Indeed, if we mix the global symmetry with a gauge symmetry associated with a particular vertex $a$, then the parameters $h_I$ are shifted as 
\begin{align}\label{eq.h_shift}
h_I \to h'_I=h_I + \varepsilon_a  \, \textrm{sign}_a(I)  \;,
\end{align}
where 
\begin{align}\label{eq.sign_def}
\textrm{sign}_a(I) \equiv \begin{cases}
+1 & \qquad (s(I)=a \;, \quad t(I)\ne a) \;, \\
-1 &\qquad (s(I)\ne a \;, \quad t(I)= a)\;, \\
0 &\qquad (\textrm{otherwise})\;,
\end{cases}
\end{align}
and $\varepsilon_a$ parametrizes the mixing between global symmetries and the $a^{\textrm{th}}$ gauge symmetry.
This shift is consistent with the loop constraint \eqref{eq-loop-constraint-toric}, which is expected since the superpotential is gauge-invariant.

What happens to the algebra under this shift?
The parameters $h_I$ enter into the algebra only through the function \eqref{eq-charge-atob}, which transforms as
\begin{align}
\varphi^{a\Rightarrow b} (u) \to  \varphi^{a\Rightarrow b}{}' (u)= \frac{\prod_{I\in \{b\rightarrow a\}}\left(u+h_{I}+\varepsilon_a  \, \textrm{sign}_a(I) \right)}{\prod_{I\in \{a\rightarrow b\}}\left(u-h_{I}-\varepsilon_a  \, \textrm{sign}_a(I) \right)} \;.
\end{align}
In other words, this amounts to constant shifts of the spectral parameter for various locations, i.e.\  $u\rightarrow u +\varepsilon_a$ for $(e^{(a)}(u),\psi^{(a)}(u), f^{(a)}(u))$ at vertex $a$.
From (\ref{spectralshift}), one concludes that the shift \eqref{eq.h_shift} mixes the generators $\psi^{(a)}_n$ only with $\psi^{(a)}_m$ with $m < n$, and similarly for $e^{(a)}_n$ and $f^{(a)}_n$.
Since automorphism merely reshuffles the generators by linear combinations, one can regard the shift \eqref{eq.h_shift} as a gauge symmetry.

Instead of modding out by the gauge shift \eqref{eq.h_shift}, we can impose gauge-fixing conditions.
One possible choice, which we adopt in this paper, is to impose the vertex constraint
\begin{tcolorbox}[ams equation]\label{eq-vertex-constraint-toric}
\textrm{\bf vertex constraint: } \quad
\sum_{I\in a} \textrm{sign}_a(I) \, h_I=0
\end{tcolorbox}
\noindent for each vertex $a$.
Note that the number of independent constraints is given by the number of vertices minus one, since the quiver quantum mechanics has only bifundamental/adjoint matters and hence the overall $U(1)$ gauge symmetry decouples.

How many parameters are there once we impose both the loop and the vertex constraints? 
Since the number of parameters with the loop constraints is given as $|Q_0|+1$ \eqref{eq.N_h}, and since we have $|Q_0|-1$ vertex constraints, there are \textit{two remaining parameters}. 
We can identify these two parameters as the coordinate parameters of the toric Calabi-Yau threefold --- a toric Calabi-Yau threefold has three $U(1)$ isometries, one of which can be identified with the R-symmetry of the supersymmetric quiver quantum mechanics, leaving behind two $U(1)$ symmetries.\footnote{
The two parameters can be regarded as 
an element of the first cohomology of the exact sequence
\begin{align}
0 \rightarrow \mathbb{C}^{Q_0} \rightarrow \mathbb{C}^{Q_1} \rightarrow \mathbb{C}^{Q_2}  \rightarrow 0 \;,
\end{align}
which can be viewed as a cohomology cochain complex for the periodic quiver, see e.g.\ section 2.3 of \cite{MR2215138}.}
Readers familiar with four-dimensional $\mathcal{N}=1$ quiver quantum gauge theories will recognize the two $U(1)$ symmetries as the so-called mesonic (non-R) global symmetries.\footnote{
In this context one also has the so-called $U(1)$ baryonic global symmetries, which, however, are not present in our context. 
The difference arises since in four-dimensional quiver gauge theories one often considers $SU(N)$ gauge groups at the nodes of the quiver, while here one considers $U(N)$ gauge groups.}

\subsection{Serre Relations}\label{sec:Serre}

For the examples in section \ref{sec:generalno4cycle}, the BPS algebra $\mathsf{Y}_{(Q,W)}$ is related to the affine Yangian of $\mathfrak{gl}_{n}$ and more generally $\mathfrak{gl}_{m|n}$.
More precisely,  while the quiver Yangians $\mathsf{Y}_{(Q,W)}$ are themselves different from affine Yangians,
we can add a set of new relations, which are traditionally called the Serre relations, to define a  {\it reduced quiver Yangian algebra}
\begin{align}
\underline{\mathsf{Y}}_{(Q,W)} = \mathsf{Y}_{(Q,W)}/ \textrm{(Serre relations)} \;,
\end{align}
and it is this algebra $\underline{\mathsf{Y}}_{(Q,W)}$ which coincides with $Y(\widehat{\mathfrak{gl}_{n}})$ or $Y(\widehat{\mathfrak{gl}_{m|n}})$. 
We will discuss explicit examples of the Serre relations in section \ref{sec:generalno4cycle}.

We will find in section \ref{sec:bootstrap_general} that the algebra $\mathsf{Y}_{(Q,W)}$ acts on the configurations of molten crystal. 
On the other hand, the reduced algebra $\underline{\mathsf{Y}}_{(Q,W)}$ also acts on the same configurations of the molten crystal. 
Namely, the extra Serre relation are also satisfied for the representations $\phi: \mathsf{Y}_{(Q,W)}\to \textrm{End}(V)$ discussed in this paper:
\begin{equation}\label{eq.Serre}
\begin{tikzcd}
\mathsf{Y}_{(Q,W)} \arrow[rd, "\phi"] \arrow[d, "\pi"]\\ 
\underline{\mathsf{Y}}_{(Q,W)}  \arrow[r, "\pi_*{\phi}"]
&\textrm{End}(V)  \;.
\end{tikzcd}
\end{equation}

We have checked that for Calabi-Yau threefolds without compact $4$-cycles, the Serre relations are precisely the conditions --- in addition to the algebraic relations (\ref{eq-OPE-modes-toric}) --- that ensure that the vacuum characters of the \textit{reduced quiver Yangian algebra} $\underline{\mathsf{Y}}_{(Q,W)}$ reproduce the correct generating functions of the corresponding crystals. 
(Namely, without the Serre relations, the vacuum character of the algebra $\mathsf{Y}_{(Q,W)}$ given by  (\ref{eq-OPE-modes-toric}) would give more states that the crystal counting.)

For a general toric Calabi-Yau threefold there seems to be no known counterpart of the affine Yangian $Y(\widehat{\mathfrak{gl}_{m|n}})$, and hence one needs to find the appropriate Serre relations such that \eqref{eq.Serre} holds.
More precisely, one wishes to find a maximum set of relations such that $\underline{\mathsf{Y}}_{(Q,W)}$ is still non-trivial and \eqref{eq.Serre} holds.
In other words, the Serre relation can be obtained by demanding that the \textit{reduced quiver Yangian algebra} $\underline{\mathsf{Y}}_{(Q,W)}$ reproduce the correct generating functions of the corresponding crystals. 
We also note that a possible alternative approach is to take advantage of the invariant bilinear pairing (Shapovalov form, see e.g.\ \cite{Prochazka:2015deb} for the case of $Y(\widehat{\mathfrak{gl}_1})$): one can define the Serre relations to be the generators for the radical for the invariant pairing, so that the Shapovalov form is non-degenerate in the reduced quiver Yangian $\underline{\mathsf{Y}}_{(Q,W)}$.
We leave detailed exploration of this for future work

\section{\texorpdfstring{Bootstrapping Affine Yangian of $\mathfrak{gl}_1$ from Plane Partitions}{Bootstrapping Affine Yangian of gl(1) from Plane Partitions}}
\label{sec:bootstrap_gl1}

In section \ref{sec:bootstrap_general} we  will discuss the representation of the BPS quiver Yangian $\mathsf{Y}_{(Q,W)}$ and motivate the definition of the algebra. 
As a preparation of the discussion, let us first discuss the case of the $\mathbb{C}^3$ and the associated algebra, the affine Yangian of $\mathfrak{gl}_1$.

Historically, the affine Yangian of $\mathfrak{gl}_1$ was constructed first and then plane partitions were found to be one of its representations. 
The review in section \ref{sec-Yangian-W-PP} followed this logic. 
However, suppose we do not know about the affine Yangian of $\mathfrak{gl}_1$, but rather want to construct an algebra that
acts transitively on the set of plane partitions.  
Within a certain ansatz, we would find that this algebra is precisely the affine Yangian of $\mathfrak{gl}_1$.

In this section, we will reconstruct the affine Yangian of $\mathfrak{gl}_1$ purely from its action on the set of plane partitions. 
Although the algebra itself is known, the goal of this section is to develop a procedure that can be generalized in the next section to construct algebras that act on the colored crystals, which describe the BPS states of type IIA string on arbitrary toric Calabi-Yau threefolds. 

The plane partition configuration as the familiar 3D box stacking can be regarded as a particular example of the crystal melting model introduced in section \ref{sec:review_crystal}. The quiver diagram for $\mathbb{C}^3$ is
\begin{equation}
\begin{tikzpicture}[scale=1]
\node[state]  [regular polygon, regular polygon sides=4, draw=blue!50, very thick, fill=blue!10] (a1) at (0,0)  {$1$};
\path[->] 
(a1) edge [in=90, out=150, loop, thin, above left] node {$h_3$} ()
(a1) edge [in=210, out=270, loop, thin, below left] node {$h_1$} ()
(a1) edge [in=330, out=30, loop, thin, right] node {$h_2$} ()
;
\end{tikzpicture}
\end{equation}
and the periodic quiver for the $\mathbb{C}^3$-theory is a triangular graph on the two-dimensional torus.
Its uplift $\mathfrak{Q}$ to the universal cover is the triangular lattice, which is the dual graph of the hexagonal tiling describing the plane partitions (in other words, the hexagonal tiling is the brane tiling graph).

\subsection{Ansatz}

The algebra consists of three families of operators
\begin{equation}\label{eq-mode-expansion-gl1_bootstrap}
\begin{aligned}
e(z)\equiv\sum^{\infty}_{n=0}\frac{e_n}{z^{n+1}} \,, \qquad \psi(z)\equiv 1+ \sigma_3\sum^{\infty}_{n=0}\frac{\psi_n}{z^{n+1}}\,, \qquad f(z)\equiv \sum^{\infty}_{n=0}\frac{f_n}{z^{n+1}} \;,
\end{aligned}
\end{equation}
where $z$ is the ``spectral parameter" and $\sigma_3$ is a parameter to be defined later.\footnote{
Note that the general mode expansion (\ref{eq-mode-expansion-toric}) specializes for this case with $\psi_{<-1}=0$, as in all quiver Yangians for Calabi-Yau threefolds with no compact $4$-cycles. Moreover, we have rescaled the modes $\psi_n$ in (\ref{eq-mode-expansion-toric}) in order to match the convention for the mode expansions of the affine Yangian of $\mathfrak{gl}_1$ in the literature. } 

The action on the plane partitions is chosen such that
\begin{enumerate}
\item
Each plane partition $\Lambda$ is an eigenstate of the Cartan generators $\psi(u)$, which means that $\Lambda$ is an eigenstate of all the zero modes $\psi_n$ with $n\in\mathbb{Z}_{\geq 0}$.
\item
Given a plane partition $\Lambda$, the action of $e(u)$ on it adds a box $\square$ at all possible positions (where a box $\square$ can be legitimately added).\footnote{
Here by ``legitimate" we mean that after a box $\square$ is added, the resulting configuration $\Lambda+\square$ should again be a plane partition.}
\item
Similarly, the action of $f(u)$ on a plane partition $\Lambda$ removes a box $\square$ from all possible positions (where a box $\square$ can be legitimately removed).
\end{enumerate}
These actions can be summarized as
\begin{equation}\label{eq-action-gl1}
\square: \qquad \quad e(u):\, \textrm{creation}\,, \qquad  \psi(u):\, \textrm{charge} \,, \qquad  f(u):\, \textrm{annihilation} \,.
\end{equation}

An ansatz  for the action of the algebra on plane partitions that satisfies the three conditions above is
\begin{equation}\label{eq-ansatz-action-gl1}
\begin{aligned}
\psi(z)|\Lambda \rangle & = \Psi_{\Lambda}(z)|\Lambda \rangle  \ , \\
e(z) | \Lambda \rangle & =  \sum_{ \Box \in {\rm Add}(\Lambda)}\frac{\Big[ -  \frac{1}{\sigma_3} {\rm Res}_{w = h(\Box)} \Psi_{\Lambda}(w) \Big]^{\frac{1}{2}}}{ z - h(\Box) } 
| \Lambda + \Box \rangle \ , \\
f(z) | \Lambda \rangle & =  \sum_{ \Box \in {\rm Rem}(\Lambda)}\frac{\Big[   \frac{1}{\sigma_3} {\rm Res}_{w = h(\Box)} \Psi_{\Lambda}(w) \Big]^{\frac{1}{2}}}{ z - h(\Box) } | \Lambda - \Box \rangle \ , 
\end{aligned}
\end{equation}
where $\Psi_{\Lambda}(z)$ is the eigenvalue of $\Lambda$.
The ansatz (\ref{eq-ansatz-action-gl1}) is the only assumption for our construction of the algebra without the Serre relations.\footnote{To obtain the Serre relations, we need the additional assumption that the Serre relations help reduce the vacuum character of the final algebra to the MacMahon function (see later).}
\bigskip

Given this ansatz, the goal of the bootstrap is to 
\begin{enumerate}
\item\label{step-fixing-hfunction-gl1} Determine the structure of poles $h(\square)$ in the $e(z)$ and $f(z)$ action part of the ansatz (\ref{eq-ansatz-action-gl1}). 
The criterion is that by applying the creation operator $e(z)$ iteratively on the vacuum $|\emptyset\rangle$, i.e.\ the plane partition configuration with no box present, one can generate all plane partitions. 
In the other way around, applying the annihilation operator $f(z)$ repeatedly on any plane partition $|\Lambda\rangle$ would eventually reduce it to the vacuum $|\emptyset\rangle$.

\item\label{step-fixing-Psi-gl1} Determine the charge function $\Psi_{\Lambda}(z)$ for arbitrary plane partition $\Lambda$.
The criterion is that the pole structures of the actions of $e(z)$ and $f(z)$ in (\ref{eq-ansatz-action-gl1}) should be encoded in the function $\Psi_{\Lambda}(z)$. 
Namely, for a given plane partition $\Lambda$, all the poles of its charge function $\Psi_{\Lambda}(z)$ correspond to either a location where a $\square$ can be added to $\Lambda$ or the location of an existing $\square$ (in $\Lambda$) that can be removed by applying $f(z)$ once.

\item\label{step-fixing-relations} 
Find all relations between the three families of operators (\ref{eq-mode-expansion-gl1_bootstrap}) that are automatically satisfied when acting on an arbitrary $\Lambda$, given the ansatz (\ref{eq-ansatz-action-gl1}) and the charge function $\Psi_{\Lambda}(z)$ determined in step-\ref{step-fixing-Psi-gl1}.

\item\label{step-Serre}
The relations found in step-\ref{step-fixing-relations} define the algebra without the Serre relations. 
Further demanding that the vacuum characters of the final algebra reproduces the generating function of the plane partitions (with trivial asymptotics), i.e.\ the MacMahon function, reproduces the Serre relations of affine Yangian of $\mathfrak{gl}_1$.

\end{enumerate}

\subsection{Analysis}

Let us now start with the step-\ref{step-fixing-hfunction-gl1}. 
A box $\square$ in the plane partition is labelled by the coordinate
\begin{equation}\label{eq-coordinates-gl1}
(x_1(\square),x_2(\square),x_3(\square)) \qquad \textrm{with} \qquad x_{1,2,3}(\square)=0,1,2,\dots,\infty \;.
\end{equation}
To each $\square$, we can associate a coordinate function
\begin{equation}\label{eq-coordinate-function-gl1-new}
h(\square)\equiv h_1 \, x_1(\square)+h_2 \, x_2(\square)+h_3\, x_3(\square) \;,
\end{equation}
where $h_i$ with $i=1,2,3$ are formal variables for now, whose role is just to translate the coordinate-triplet $(x_1(\square),x_2(\square),x_3(\square))$ to one number $h(\square)$, so that one can directly relate a $\square$ to the poles of $\Psi_{\Lambda}(z)$.
As we will see, the $h_i$ might not be mutually independent, and their relations will be determined by the criterion in step-\ref{step-fixing-Psi-gl1}.
\bigskip

Now we move on to step-\ref{step-fixing-Psi-gl1}, fixing the charge function $\Psi_{\Lambda}(z)$ for an arbitrary plane partition $\Lambda$.
Given that a plane partition consists of a set of $\square$'s, each with its coordinate function $h(\square)$, the most natural ansatz for $\Psi_{\Lambda}(z)$ is\footnote{
It will soon be clear why the other natural guess where all the contribution $\psi_{\square}(z)$ are summed over (instead of multiplied together) fails the criterion that all poles of $\Lambda$ should correspond to either a location where a $\square$ can be added or a location where a $\square$ can be removed.}
\begin{equation}
\Psi_{\Lambda}(z)=\psi_0(z) \prod_{\square\in\Lambda} \psi_{\square}(z) \;,
\end{equation}
where $\psi_0(z)$ is the contribution from the vacuum, i.e.\ before any $\square$ is added, and $\psi_{\square}(z)$ is the contribution of an individual $\square$.
Therefore we only need to fix the functions $\psi_0(z)$ and $\psi_{\square}(z)$. 
The main constraint is that for any $\Lambda$, all poles of $\Lambda$ should correspond to either a location where a $\square$ can be added or a location where a $\square$ can be removed.

\subsubsection{\texorpdfstring{Vacuum $\longrightarrow$ Level-$1$}{Vacuum -> Level-1}}

Let us start with the vacuum contribution $\psi_{0}(z)$.
Starting with the vacuum state $|\Lambda\rangle=|\emptyset\rangle$, the action of $e(z)$ should create the first $\square$ at the corner, with coordinates and $h(\square)$ given by
\begin{equation}\label{eq-states-level1-gl1}
\textrm{level-}1:\quad \square_0:\qquad x_{1}(\square)=x_2(\square)=x_3(\square)=0 \qquad \Longrightarrow \qquad h(\square)=0 \;.
\end{equation}
Since this is the very first $\square$ that can be added in the plane partition, we call it level-1 box:
\begin{equation}
\begin{tikzpicture}[scale=0.8]
\node[state]  [regular polygon, regular polygon sides=4, draw=blue!50, very thick, fill=blue!10] (a1) at (0,0)  {$1$};
\end{tikzpicture}
\end{equation}
Here the box is labelled by $1$ since we have only one vertex in this example. 
(We will encounter more general situation in the next section.)
The charge function for the vacuum $\Psi_{\Lambda}(z)=\psi_0(z)$ should have one and only one pole,\footnote{
The reason is that there is only one possible position for the first $\square$ to be added, i.e.\ only one pole for $e(z)$; and there is no $\square$ to be removed, i.e.\ no pole for $f(z)$.} at 
\begin{equation}
\textrm{adding-pole of }\square: \quad z^*=h(\square)=0 \;.
\end{equation}
Furthermore, $\psi_0(z)$'s residue at $z=0$ should be non-zero --- otherwise by the ansatz (\ref{eq-ansatz-action-gl1}) the action of $e(z)$ on vacuum would annihilate the vacuum instead of creating the first $\square$.
The simplest solution is 
\begin{equation}\label{eq-psi0-gl1}
\psi_0(z)=\frac{z+C}{z} \;,
\end{equation}
where $C\neq 0$ will be fixed later.
Finally, since there is no $\square$ to be removed, the box-removing operator $f(z)$ should annihilate the vacuum $|\emptyset\rangle$. 
(This is consistent with the fact that vacuum charge function $\psi_0(z)$ in (\ref{eq-psi0-gl1}) only has one pole, which we have already seen to be the adding-pole for the level-$1$ $\square$ in (\ref{eq-states-level1-gl1}).)
In summary, the actions of $(e(z),\psi(z),f(z))$ on the vacuum $|\emptyset\rangle$ are:
\begin{equation}\label{action-on-vacuum-gl1}
\textrm{level-}0:\quad \psi(z)|\emptyset\rangle=\psi_0(z)|\emptyset\rangle=\frac{z+C}{z}|\emptyset\rangle \,, \qquad e(z)|\emptyset\rangle =\frac{\#}{z}|\square\rangle \,, \qquad f(z)|\emptyset\rangle=0 \;.
\end{equation}

\subsubsection{\texorpdfstring{Level-$1$ $\longrightarrow$ Level-$2$}{Level-1 -> Level-2}}

To fix $\psi_{\square}(z)$, first consider the initial state $|\Lambda\rangle =|\square\rangle$, where $|\square\rangle$ denotes the configuration where only the first $\square$ at the corner is present. 
The next $\square$ to be added can be placed in three possible positions:
\begin{equation}\label{eq-states-level2-gl1}
\textrm{level-}2:\quad\begin{cases}
\begin{aligned}
&\square_1: \qquad (x_1,x_2,x_3)=(1,0,0) \qquad \Longrightarrow \qquad h(\square)=h_1 \;,\\
&\square_2: \qquad (x_1,x_2,x_3)=(0,1,0) \qquad \Longrightarrow \qquad h(\square)=h_2 \;,\\
&\square_3: \qquad (x_1,x_2,x_3)=(0,0,1)\qquad 
\Longrightarrow \qquad h(\square)=h_3  \;,
\end{aligned}
\end{cases}
\end{equation}
shown as the three blue boxes below
\begin{equation}
\begin{aligned}
&
\begin{tikzpicture}[scale=0.8]
\node[state]  [regular polygon, regular polygon sides=4, draw=black!50, very thick, fill=black!10] (a1) at (0,0)  {$1$};
\node[state]  [regular polygon, regular polygon sides=4, draw=blue!50, very thick, fill=blue!10] (a21) at (-1,-1.73205)  {$1$};
\node[state]  [regular polygon, regular polygon sides=4, draw=blue!50, very thick, fill=blue!10] (a22) at (2,0)  {$1$};
\node[state]  [regular polygon, regular polygon sides=4, draw=blue!50, very thick, fill=blue!10] (a23) at (-1,1.73205)  {$1$};
\path[->] 
(a1) edge   [thick, red]   node [right] {$h_1$} (a21)
(a1) edge   [thick, red]   node [above] {$h_2$} (a22)
(a1) edge   [thick, red]   node [left] {$h_3$} (a23)
;
\end{tikzpicture}
\end{aligned}
\end{equation}
This means that the function $\Psi_{\Lambda}(z)$ for the initial state $|\Lambda\rangle=|\square\rangle$ needs to contain these three poles $h_i$ with $i=1,2,3$.

In addition, $\Psi_{\Lambda}(z)$ for the initial state $|\Lambda\rangle=|\square\rangle$ should contain a pole at $z^*=0$, corresponding to the pole for $f(z)$ to remove this $\square$ to reduce it to vacuum:
\begin{equation}\label{eq-rempole-level1-gl1}
\textrm{removing-pole of $\square$}: \quad z^*=h(\square)=0  \;.
\end{equation}
This pole is already accounted for by the pole in $\psi_0(z)$ in (\ref{eq-psi0-gl1}). 
Namely, the pole in $\psi_0(z)$ corresponds to both the creating-pole of $e(z)$ when acting on $|\emptyset\rangle$ and the removing pole of $f(z)$ when acting on $|\square\rangle$.
Indeed, this is a general feature for $\psi_{\Lambda}(z)$ of all $\Lambda$ --- namely, a creating-pole for $e(z)$ acting on $\Lambda$ and generating a particular $\square$ is also the same pole for the (removing) action of $f(z)$ when acting on the configuration $|\Lambda+\square\rangle$ and removing this same $\square$.

Therefore, the three poles that correspond to the three $\square$'s in (\ref{eq-states-level2-gl1}) must all come from the function $\psi_{\square}(z)$ when $\square$ is the level-$1$ box in (\ref{eq-states-level1-gl1}):
\begin{equation}\label{eq-psi-pole-gl1}
\psi_{\square}(z)=\frac{N(z)}{(z-h_1)(z-h_2)(z-h_3)} \qquad \textrm{for} \qquad h(\square)=0  \;,
\end{equation}
where $N(z)$ is the numerator to be fixed momentarily.
Let us define $\psi_{\square_0}(z)\equiv\frac{N(z)}{(z-h_1)(z-h_2)(z-h_3)}$ for later use.
In summary, the charge function for $|\Lambda\rangle=|\square\rangle$ is 
\begin{equation}\label{eq-charge-level1-gl1}
\Psi_{\Lambda}(z)=\psi_0(z)\psi_{\square_0}(z)  \;,
\end{equation}
which has three adding-poles at
\begin{equation}\label{eq-addpole-level1-gl1}
\textrm{adding-pole of }\square_i: \quad z^*=h(\square)=h_i \qquad \textrm{for} \quad i=1,2,3\,,
\end{equation}
in addition to the removing-pole given in (\ref{eq-rempole-level1-gl1}).
The actions of $(e(z),\psi(z),f(z))$ on the level-$1$ state $|\square\rangle$ are:
\begin{equation}
\textrm{level-}1:\quad
\begin{cases}
\begin{aligned}
\psi(z)|\square\rangle&=\psi_0(z)\psi_{\square_0}(z)|\square\rangle=\frac{z+C}{z}\cdot\frac{N(z)}{(z-h_1)(z-h_2)(z-h_3)}|\square\rangle   \,, \\ 
e(z)|\square\rangle &=\sum_{i=1,2,3}\frac{\#}{z-h_i}|\square\square_i\rangle \,,  \\ 
 f(z)|\square\rangle &=\frac{\#}{z}|\emptyset\rangle   \;,
\end{aligned}
\end{cases}
\end{equation}
where again $\#$ denotes various numerical constants to be fixed later systematically.
\bigskip

\subsubsection{\texorpdfstring{Level-$2$ $\longrightarrow$ Level-$3$}{Level-2 -> Level-3}}
\label{sec:l2tol3-gl1}

We have just seen that to fix the denominator of the charge 
function for the state at level-$1$, we need to consider the creation of the three level-$2$ $\square$'s in (\ref{eq-states-level2-gl1}). 
By the same logic, to fix the denominator of the charge function for the level-$2$ states, we need to consider the creation of the level-$3$ $\square$'s.

There are $6$ $\square$'s at level-$3$, at position
\begin{equation}\label{eq-states-level3-gl1}
\textrm{level-}3:\quad\begin{cases}
\begin{aligned}
&\square_{(2,0,0)}: \qquad (x_1,x_2,x_3)=(2,0,0) \qquad \Longrightarrow \qquad h(\square)=2h_1  \;,\\
&\square_{(0,2,0)}: \qquad (x_1,x_2,x_3)=(0,2,0) \qquad \Longrightarrow \qquad h(\square)=2h_2 \;,\\
&\square_{(0,0,2)}: \qquad (x_1,x_2,x_3)=(0,0,2) \qquad \Longrightarrow \qquad h(\square)=2h_3 \;,\\
&\square_{(0,1,1)}: \qquad (x_1,x_2,x_3)=(0,1,1) \qquad \Longrightarrow \qquad h(\square)=h_2+h_3 \;,\\
&\square_{(1,0,1)}: \qquad (x_1,x_2,x_3)=(1,0,1) \qquad \Longrightarrow \qquad h(\square)=h_1+h_3 \;,\\
&\square_{(1,1,0)}: \qquad (x_1,x_2,x_3)=(1,1,0) \qquad \Longrightarrow \qquad h(\square)=h_1+h_2 \;,
\end{aligned}
\end{cases}
\end{equation}
shown as the $6$ red boxes below
\begin{equation}\label{fig-gl1-level3}
\begin{aligned}
&
\begin{tikzpicture}[scale=0.8]
\node[state]  [regular polygon, regular polygon sides=4, draw=black!50, very thick, fill=black!10] (a1) at (0,0)  {$1$};
\node[state]  [regular polygon, regular polygon sides=4, draw=blue!50, very thick, fill=blue!10] (a21) at (-1,-1.73205)  {$1$};
\node[state]  [regular polygon, regular polygon sides=4, draw=red!50, very thick, fill=red!10] (a31) at (-2,-3.4641)  {$1$};
\node[state]  [regular polygon, regular polygon sides=4, draw=blue!50, very thick, fill=blue!10] (a22) at (2,0)  {$1$};
\node[state]  [regular polygon, regular polygon sides=4, draw=red!50, very thick, fill=red!10] (a32) at (4,0)  {$1$};
\node[state]  [regular polygon, regular polygon sides=4, draw=blue!50, very thick, fill=blue!10] (a23) at (-1,1.73205)  {$1$};
\node[state]  [regular polygon, regular polygon sides=4, draw=red!50, very thick, fill=red!10] (a33) at (-2,3.4641)  {$1$};
\node[state]  [regular polygon, regular polygon sides=4, draw=red!50, very thick, fill=red!10] (a312) at (1,-1.73205)  {$1$};
\node[state]  [regular polygon, regular polygon sides=4, draw=red!50, very thick, fill=red!10] (a323) at (1,1.73205)  {$1$};
\node[state]  [regular polygon, regular polygon sides=4, draw=red!50, very thick, fill=red!10] (a331) at (-2,0)  {$1$};
\path[->] 
(a1) edge   [thick, red]   node [right] {$h_1$} (a21)
(a21) edge   [thick, red]   node [right] {$h_1$} (a31)
(a23) edge   [thick, red]   node [left] {$h_1$} (a331)
(a22) edge   [thick, red]   node [right] {$h_1$} (a312)
(a1) edge   [thick, red]   node [above] {$h_2$} (a22)
(a22) edge   [thick, red]   node [above] {$h_2$} (a32)
(a21) edge   [thick, red]   node [below] {$h_2$} (a312)
(a23) edge   [thick, red]   node [above] {$h_2$} (a323)
(a1) edge   [thick, red]   node [left] {$h_3$} (a23)
(a23) edge   [thick, red]   node [left] {$h_3$} (a33)
(a21) edge   [thick, red]   node [left] {$h_3$} (a331)
(a22) edge   [thick, red]   node [right] {$h_3$} (a323)
;
\end{tikzpicture}
\end{aligned}
\end{equation}
To create them, the charge function of the level-$2$ states, i.e.\ the states that contain one level-$1$ $\square$ and one or more level-$2$ $\square$'s, must contain these poles.

Let us first consider the first three $\square$'s in (\ref{eq-states-level3-gl1}).
Take the first one for example.
To create the $\square$ at the position $(x_1,x_2,x_3)=(2,0,0)$, there must be at least two existing $\square$'s sitting at $(x_1,x_2,x_3)=(0,0,0)$ and $(x_1,x_2,x_3)=(1,0,0)$.
Namely, the charge function of the (minimal) initial state\footnote{
It is possible to have more complicated initial state on which one can add these three boxes. 
But for the purpose of fixing the charge function for the level-$2$ state, it is enough --- and easier --- to consider the minimal state.} has to be $\Psi_{\Lambda}(z)=\psi_0(z)\psi_{\square_0}(z)\psi_{\square_1}(z)$, which must contain a pole at $z^{*}=h(\square_{(2,0,0)})=2h_1$.
Considering all the first three states in (\ref{eq-states-level3-gl1}), we see that the charge function of a (minimal) initial state on which one of the first three states in (\ref{eq-states-level3-gl1}) can be added is 
\begin{equation}\label{eq-charge-level2-1-gl1}
\Psi_{\Lambda}(z)=\psi_0(z)\psi_{\square_0}(z)\psi_{\square_i}(z)\qquad \textrm{for} \quad i=1,2,3\,,
\end{equation}
which must contain a pole at 
\begin{equation}\label{eq-pole-123}
z^{*}=2h_i \qquad \textrm{for} \quad i=1,2,3\,.
\end{equation}
Recall that the poles from the first two factors $\psi_0(z)$ and $\psi_{\square_0}(z)$ of (\ref{eq-charge-level2-1-gl1}) contain poles at $0$ and $h_{1,2,3}$, given by (\ref{eq-rempole-level1-gl1}) and  (\ref{eq-addpole-level1-gl1}). 
Therefore the adding-pole $z^{*}=2 h_i$ must come from the 3rd factor of (\ref{eq-charge-level2-1-gl1}). 

Similarly, for any of the last three $\square$'s to be added, we need the (minimal) initial state to have at least one level-1 $\square$ sitting at $(x_1,x_2,x_3)=(0,0,0)$ (given by (\ref{eq-states-level1-gl1}) and two of the three level-$2$ $
\square$'s in (\ref{eq-states-level1-gl1}), with charge function
\begin{equation}\label{eq-charge-level2-2-gl1}
\Psi_{\Lambda}(z)=\psi_0(z)\psi_{\square_0}(z)\psi_{\square_i}(z)\psi_{\square_j}(z)\qquad \textrm{for} \quad i,j=1,2,3 \quad \textrm{and}\quad i\neq j\,.
\end{equation}
which must contain the adding-pole of 
\begin{equation}\label{eq-pole-456} 
z^{*}=h_i+h_j \qquad \textrm{for} \quad i,j=1,2,3 \quad \textrm{and}\quad i\neq j\,.
\end{equation}

It is then easy to see that the following choice satisfies these two constraints (\ref{eq-pole-123}) and (\ref{eq-pole-456}):
\begin{equation}
\psi_{\square_i}(u)=\psi_{\square_0}(u-h_i) \;,
\end{equation}
where $\psi_{\square_0}(u)$ is given by (\ref{eq-psi-pole-gl1}).
It is more instructive to rewrite it as
\begin{equation}\label{eq-psi-general-gl1}
\psi_{\square}(u)=\psi_{\square_0}\left(u-h(\square)\right) \;.
\end{equation}

Having obtained (\ref{eq-psi-general-gl1}), we are now ready to fix the numerator in (\ref{eq-psi-pole-gl1}).
Consider a plane partition $|\Lambda\rangle$ that contains the level-$1$ $\square$ and one of the three level-$2$ $\square$'s in (\ref{eq-states-level2-gl1}).
The presence of the level-$2$ $\square$ means that the level-$1$ $\square$ can no longer to removed. 
This means that the removing-pole $z^{*}=h(\square)=0$ must be canceled by a factor in the numerator of the charge function of the level-$2$ $\square$.
Namely, $N(z-h_i)$ must contain a factor of $z$ for any $i=1,2,3$.
This constraint fixes the minimal $N(z)$ to be 
\begin{equation}
N(z)=(z+h_1)(z+h_2)(z+h_3)  \;.
\end{equation}
\smallskip

We now have the most important function in the construction of the algebra acting on the set of plane partitions:
\begin{equation}
\varphi_3(z)\equiv\frac{(z+h_1)(z+h_2)(z+h_3)}{(z-h_1)(z-h_2)(z-h_3)}=\psi_{\square}(z) \qquad \textrm{for} \qquad h(\square)=0  \;.
\end{equation}
By (\ref{eq-psi-general-gl1}), we see that the each of the level-$2$ $\square$'s contribute a factor of $\varphi_3$ function, with argument shifted by $h(\square)$.
\bigskip

Before we move on, we need to check whether the three parameters $h_{1,2,3}$ are mutually independent.
Compare the minimal initial state (\ref{eq-charge-level2-1-gl1}) (in order to add one of the first three $\square$'s in (\ref{eq-states-level3-gl1})) and the minimal initial state (\ref{eq-charge-level2-2-gl1}) (in order to add one of the last three $\square$'s in (\ref{eq-states-level3-gl1})).
For example, if one starts with the initial state (\ref{eq-charge-level2-1-gl1}), one can only add a $\square$ at $h(\square)=2h_i$, but not the $\square$ at $h(\square)=h_i+h_j$ with $j\neq i$.

But this fact has to be implemented automatically by the pole structure of the charge function (\ref{eq-charge-level2-1-gl1}). 
Without loss of generality, consider $i=1$, for which the charge function (\ref{eq-charge-level2-1-gl1}) is explicitly
\begin{equation}\label{charge-12}
\begin{aligned}
\Psi_{\Lambda}(z)&=\psi_0(z)\psi_{\square_0}(z)\psi_{\square_1}(z)\\
&=\frac{z+C}{z}\cdot \frac{(z+h_1)(z+h_2)(z+h_3)}{(z-h_1)(z-h_2)(z-h_3)} \cdot \frac{z(z+h_2-h_1)(z+h_3-h_1)}{(z-2h_1)(z-h_2-h_1)(z-h_3-h_1)}  \;.
\end{aligned}
\end{equation}
Before we proceed to consider the adding poles, first note that the factor of $z$ in the numerator of $\psi_{\square_1}(z)$ cancels the factor of $z$ in the denominator of $\psi_0(z)$, which used to be the removing pole of the first $\square$. 
This guarantees that in the presence of the second $\square$, the first one cannot be removed anymore.

Now we resume the consideration of the adding poles.
Note the three poles in the $\psi_{\square_1}(z)$.
The first one allows $e(z)$ to add a level-$3$ $\square$ at $(x_1,x_2,x_3)=(2,0,0)$, with $h(\square)=2h_1$, shown by the red box below
\begin{equation}\label{eq.allowed-PP}
\begin{aligned}
&
\begin{tikzpicture}[scale=0.8]
\node[state]  [regular polygon, regular polygon sides=4, draw=black!50, very thick, fill=black!10] (a1) at (0,0)  {$1$};
\node[state]  [regular polygon, regular polygon sides=4, regular polygon, regular polygon sides=4, draw=blue!50, very thick, fill=blue!10] (a21) at (-1,-1.73205)  {$1$};
\node[state]  [regular polygon, regular polygon sides=4, draw=red!50, very thick, fill=red!10] (a31) at (-2,-3.4641)  {$1$};
\path[->] 
(a1) edge   [thick, red]   node [right] {$h_1$} (a21)
(a21) edge   [thick, red]   node [right] {$h_1$} (a31)
;
\end{tikzpicture}
\end{aligned}
\end{equation}

\begin{figure}[htbp]
\centering\includegraphics[scale=0.45]{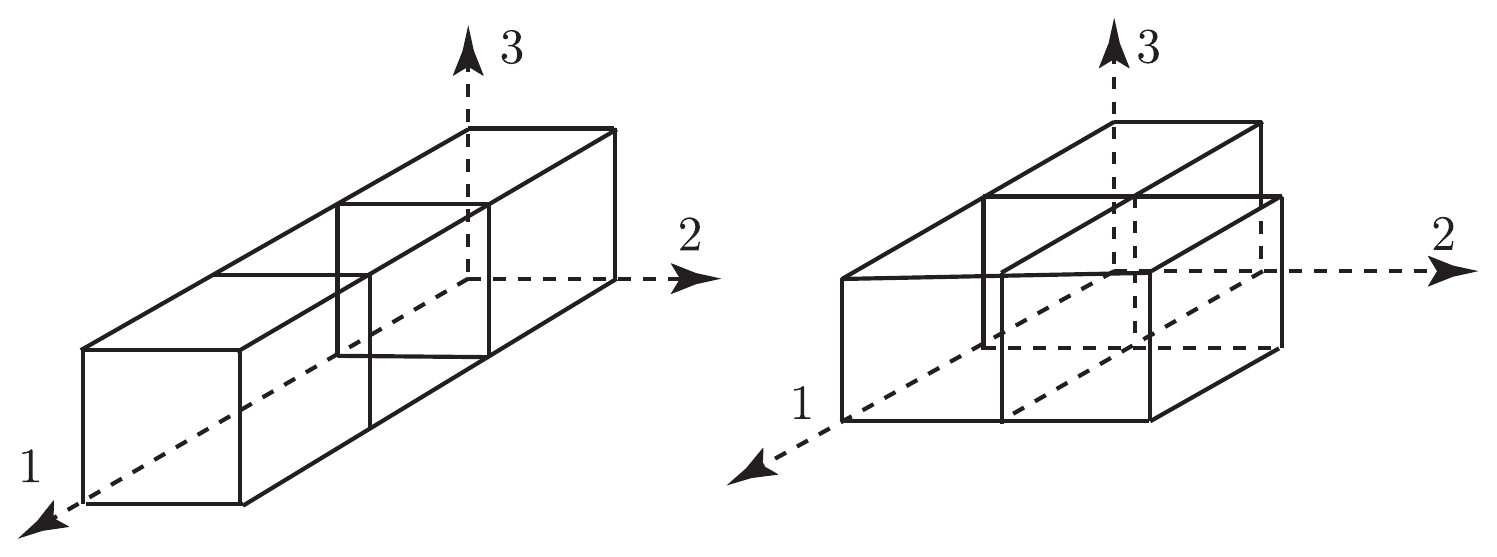}
\caption{The configuration on the left (depicting \eqref{eq.allowed-PP}) is a legitimate plane partition. 
By contrast the configuration on the right (depicting the left graph in \eqref{eq.n_allowed-PP}) violates the melting rule and is not a plane partition.}
\label{fig.bad}
\end{figure}

The other two poles, however, correspond to two $\square$'s (with $h(\square)=h_1+h_j$ for $j=2,3$) that are not allowed to be added now, shown by the two red boxes below: 
\begin{equation}\label{eq.n_allowed-PP}
\begin{tikzpicture}[scale=0.8]
\node[state]  [regular polygon, regular polygon sides=4, draw=black!50, very thick, fill=black!10] (a1) at (0,0)  {$1$};
\node[state]  [regular polygon, regular polygon sides=4, regular polygon, regular polygon sides=4, draw=blue!50, very thick, fill=blue!10] (a21) at (-1,-1.73205)  {$1$};
\node[state]  [regular polygon, regular polygon sides=4, draw=red!50, very thick, fill=red!10] (a312) at (1,-1.73205)  {$1$};
\node[state]  [regular polygon, regular polygon sides=4, draw=black!50, very thick, fill=black!10] (b1) at (6,0)  {$1$};
\node[state]  [regular polygon, regular polygon sides=4, regular polygon, regular polygon sides=4, draw=blue!50, very thick, fill=blue!10] (b21) at (5,-1.73205)  {$1$};
\node[state]  [regular polygon, regular polygon sides=4, draw=red!50, very thick, fill=red!10] (b331) at (4,0)  {$1$};
\path[->] 
(a1) edge   [thick, red]   node [right] {$h_1$} (a21)
(a21) edge   [thick, red]   node [below] {$h_2$} (a312)
(b1) edge   [thick, red]   node [right] {$h_1$} (b21)
(b21) edge   [thick, red]   node [left] {$h_3$} (b331)
;
\end{tikzpicture}
\end{equation}
Also see Figure \ref{fig.bad} for a comparison between a legitimate configuration (\ref{eq.allowed-PP}) and an illegitimate one (the left one in (\ref{eq.n_allowed-PP})) in their plane partition presentations.
This means that these two poles have to be canceled by factors in the numerator of $\psi_{\square_0}(z)$, which gives the constraint
\begin{equation}\label{eq-constraint-h123-1-gl1}
h_1+h_2+h_3=0 \;.
\end{equation}
which is precisely the loop constraint (\ref{eq-loop-constraint-toric}) for the case of $\mathbb{C}^3$:
\begin{equation}
\begin{aligned}
&
\begin{tikzpicture}[scale=0.8]
\node[state]  [regular polygon, regular polygon sides=4, draw=black!50, very thick, fill=black!10] (a1) at (0,0)  {$1$};
\node[state]  [regular polygon, regular polygon sides=4, regular polygon, regular polygon sides=4, draw=blue!50, very thick, fill=blue!10] (a21) at (-1,-1.73205)  {$1$};
\node[state]  [regular polygon, regular polygon sides=4, draw=red!50, very thick, fill=red!10] (a31) at (-2,-3.4641)  {$1$};
\node[state]  [regular polygon, regular polygon sides=4, draw=blue!50, very thick, fill=blue!10] (a22) at (2,0)  {$1$};
\node[state]  [regular polygon, regular polygon sides=4, draw=red!50, very thick, fill=red!10] (a32) at (4,0)  {$1$};
\node[state]  [regular polygon, regular polygon sides=4, draw=blue!50, very thick, fill=blue!10] (a23) at (-1,1.73205)  {$1$};
\node[state]  [regular polygon, regular polygon sides=4, draw=red!50, very thick, fill=red!10] (a33) at (-2,3.4641)  {$1$};
\node[state]  [regular polygon, regular polygon sides=4, draw=red!50, very thick, fill=red!10] (a312) at (1,-1.73205)  {$1$};
\node[state]  [regular polygon, regular polygon sides=4, draw=red!50, very thick, fill=red!10] (a323) at (1,1.73205)  {$1$};
\node[state]  [regular polygon, regular polygon sides=4, draw=red!50, very thick, fill=red!10] (a331) at (-2,0)  {$1$};
\path[->] 
(a1) edge   [thick, red]   node [right] {$h_1$} (a21)
(a21) edge   [thick, red]   node [right] {$h_1$} (a31)
(a23) edge   [thick, red]   node [left] {$h_1$} (a331)
(a22) edge   [thick, red]   node [right] {$h_1$} (a312)
(a323) edge   [thick, blue]   node [right] {$h_1$} (a1)
(a1) edge   [thick, red]   node [above] {$h_2$} (a22)
(a22) edge   [thick, red]   node [above] {$h_2$} (a32)
(a21) edge   [thick, red]   node [below] {$h_2$} (a312)
(a23) edge   [thick, red]   node [above] {$h_2$} (a323)
(a331) edge   [thick, blue]   node [above] {$h_2$} (a1)
(a1) edge   [thick, red]   node [left] {$h_3$} (a23)
(a23) edge   [thick, red]   node [left] {$h_3$} (a33)
(a21) edge   [thick, red]   node [left] {$h_3$} (a331)
(a22) edge   [thick, red]   node [right] {$h_3$} (a323)
(a312) edge   [thick, blue]   node [right] {$h_3$} (a1)
;
\end{tikzpicture}
\end{aligned}
\end{equation}
\smallskip 

From the plane partition presentation, we can see that for any of the last three $\square$'s in (\ref{eq-states-level3-gl1}) to be added, the two level-2 $\square$'s that have arrows pointing it in (\ref{fig-gl1-level3}) have to be already present. 
(This is a manifestation of the general ``melting rule" (\ref{eq.melting_rule}).)
For example, for the $\square$ with $(x_1,x_2,x_3)=(1,1,0)$ to be added (with adding pole $z^*=h(\square)=-h_3$), both the $\square$ with $(x_1,x_2,x_3)=(1,0,0)$ and the one with $(x_1,x_2,x_3)=(0,1,0)$ need to be already present.
Again, this need to be automatically implemented by the pole structure of the charge function $\Psi_{\Lambda}(z)$:
\begin{equation}
\begin{aligned}
\Psi_{\Lambda}(z)&=\psi_0(z)\psi_{\square_0}(z)\psi_{\square_1}(z)\psi_{\square_2}(z)\\
&=\frac{z+C}{z}\cdot \frac{(z+h_1)(z+h_2)(z+h_3)}{(z-h_1)(z-h_2)(z-h_3)} \cdot \frac{z(z+h_2-h_1)(z+h_3-h_1)}{(z-2h_1)(z+h_3)(z+h_2)}\\
& \qquad \quad\,\,\, \qquad\qquad \qquad \qquad\qquad \qquad \cdot \frac{(z+h_1-h_2)z(z+h_3-h_2)}{(z+h_3)(z-2h_2)(z+h_1)} \;,
\end{aligned}
\end{equation}
where we have used the constraint (\ref{eq-constraint-h123-1-gl1}).
One can see that  different from the charge function (\ref{charge-12}), the required pole $z^{*}=-h_3$ appears twice: one in $\psi_{\square_1}(z)$ and one in $\psi_{\square_2}(z)$. 
While one of them is canceled by the factor in the numerator of $\psi_{\square_0}(z)$, we are left with exactly one to account for the pole to add the $\square$ at $(x_1,x_2,x_3)=(1,1,0)$.

This example demonstrates how the ``melting rule" (\ref{eq.melting_rule}) is ensured by the pole structures of the charge functions $\Psi_{\Lambda}(z)$ and the loop constraint (\ref{eq-constraint-h123-1-gl1}), for the case of $\mathbb{C}^3$.    
As we will see in the next section, the same holds for general toric Calabi-Yau threefolds.

\subsubsection{General Levels}

One can now repeat the argument above and try to generate all possible plane partition configurations iteratively.
It is straightforward to see that the result (\ref{eq-psi-general-gl1}) applies to all $\square$'s in a plane partition.
Namely, each $\square$ in a plane partition $\Lambda$ contributes a factor of $\varphi_3$ function, with argument shifted by $h(\square)$:
\begin{equation}
\psi_{\square}(z)=\varphi_3(z-h(\square))  \;.
\end{equation}
The full charge function of the plane partition $\Lambda$ is then
\begin{tcolorbox}[ams align]\label{eq-charge-Lambda-gl1}
\begin{aligned}
\Psi_{\Lambda}(z)&=\psi_0(z)\, \prod_{\square\in \Lambda}\, \varphi_3 (z-h(\square))  \;, \\
\textrm{where}\quad \psi_0(z)&\equiv\frac{z+\sigma_3 \psi_0}{z}  \;,\\
\quad \varphi_3(z)&\equiv\frac{(z+h_1)(z+h_2)(z+h_3)}{(z-h_1)(z-h_2)(z-h_3)} \quad\textrm{with}\quad h_1+h_2+h_3=0  \;.\\
\end{aligned}
\end{tcolorbox}
\bigskip
Here we have taken advantage of this opportunity to determine the constant $C$ inside $\psi_0(z)$ to be $\sigma_3 \psi_0$, 
so that the mode expansion of $\Psi_{\Lambda}(z)$ has the same form as in \eqref{eq-mode-expansion-gl1_bootstrap}.

We now revisit the ansatz for the action of $(e(z),\psi(z),f(z))$ on arbitrary plane partitions. 
With the assignment of the charge function (\ref{eq-charge-Lambda-gl1}) for an arbitrary plane partition $\Lambda$, one can check that indeed all the poles of $\Psi_{\Lambda}(z)$ belong to one of the following two classes:
\begin{itemize}
\item The pole is equal to the coordinate function of a $\square$ that is on the ``surface" of $\Lambda$. 
This pole is  a removing-poles for the action of $f(z)$.
\item The pole is equal to the coordinate function of a $\square$ that can be added to $\Lambda$, which means that the pole is related to one of the removing-pole by a shift of $h_i$ with $i=1,2,3$ depending on the direction the $\square$ is relative to $\Lambda$. 
This pole is an adding pole for the action of $e(z)$.
\end{itemize}
Namely, the charge assignment (\ref{eq-charge-Lambda-gl1}) enables the ansatz (\ref{eq-ansatz-action-gl1}) to define the action of $(e(z),\psi(z), f(z))$ on the set of plane partitions, where the adding and removing of $\square$'s are implemented automatically due to the pole structure of the charge function (\ref{eq-charge-Lambda-gl1}).
In particular, one can apply the action of $e(z)$ in (\ref{eq-ansatz-action-gl1}) (without worrying about the coefficient given by the residue for now\footnote{
As long as the coefficients are non-zero --- we will come back to this issue later in section \ref{sec:truncation}.}) repeatedly starting from the vacuum $|\emptyset\rangle$ and generate all possible plane partition configurations. 
The ``melting rule" (\ref{eq.melting_rule}) is implemented automatically by the pole structures of the charge function  $\Psi_{\Lambda}(z)$.
\medskip

The action of the generators $(e(z),\psi(z), f(z))$ on arbitrary plane partitions (\ref{eq-ansatz-action-gl1}) together with the charge function prescription (\ref{eq-charge-Lambda-gl1}) then allows us to determine the algebraic relations.
Since we will explain the procedure of determining the algebraic relations of quiver Yangian based on the algebra's action on sets of colored crystals in detail later in section \ref{sec:ActiontoAlgebra}, and since the affine Yangian of $\mathfrak{gl}_1$ is the simplest special case,  we would not explain the details of the procedure for affine Yangian of $\mathfrak{gl}_1$ here.
One can just follow the procedure in section \ref{sec:ActiontoAlgebra}, but setting all the colors to be the same and use the charge function prescription (\ref{eq-charge-Lambda-gl1}). 
One can show that the resulting algebraic relation is given by (\ref{eq-relations-OPE-gl1}).

\subsection{Serre Relation and State Counting }
\label{sec:serre-gl1}
The algebraic relations (\ref{eq-relations-OPE-gl1}) 
we just fixed have equivalent descriptions in terms of modes, given in (\ref{eq-relations-modes-gl1}).
We have already seen that the representation of the affine Yangian of $\mathfrak{gl}_1$ can be described by set of plane partitions.
In particular, for the vacuum representation that we focus in this paper, the states in the representation are given by various plane partition configurations with trivial asymptotics, namely,  the box stacking starts from the empty room.\footnote{
A generic plane partition representation is labeled by a triplet of Young diagrams $(\lambda_1,\lambda_2,\lambda_3)$, which are the asymptotics along the $(x_1,x_2,x_3)$ directions. To bootstrap the algebra, it is enough to consider the vacuum representation, which has trivial asymptotics along all three directions.} 

On the other hand, the states in a representation 
should also be expressible in terms of the creating operators $e_n$ acting on the ground state of the representation. 
In particular, for the vacuum module, all the states should have the form 
\begin{equation}\label{states-e}
\prod^N_{i=1} e_{n_i} |\emptyset \rangle\,,
\end{equation} 
where $N$ equals the number of boxes in the corresponding plane partition configurations. 
The description by (\ref{states-e}) should match the plane partition configuration, up to suitable change of basis.
In particular, the counting of states in terms of  (\ref{states-e}) should reproduce the counting of plane partitions, which for the vacuum module should reproduce the MacMahon function (\ref{MacMahon}).

When counting the states of the form (\ref{states-e}), one need to use the algebraic relations (\ref{eq-relations-OPE-gl1}) to reduce the number of states.
However, starting from $N=3$, we see that the relations (\ref{eq-relations-OPE-gl1}) are not enough to reproduce the MacMahon function (\ref{MacMahon}).
Only after imposing the Serre relations (\ref{eq-Serre-modes-gl1}) can one obtain the correct counting. 
Let us now do this state counting starting from $N=0$, i.e.\ the vacuum. 
\subsubsection{Vacuum}
There is one state at $N=0$, the vacuum:
\begin{equation}
\textrm{vacuum}: \qquad |\emptyset\rangle\,.
\end{equation}
\subsubsection{One Box}
There is only one plane partition configuration with one box. 
To reproduce this counting, we consider all states $e_n|\emptyset\rangle$ with $n\in \mathbb{Z}_{\geq 0}$. 
The action of $e(z)$ on vacuum, given in (\ref{action-on-vacuum-gl1}), together with the mode expansion (\ref{eq-mode-expansion-gl1}), means 
\begin{equation}\label{en-on-vacuum}
e_n |\emptyset\rangle =0 \qquad \textrm{for}\quad n\geq 1\,.
\end{equation}
Namely, there is only one state for $N=1$:
\begin{equation}
1 \textrm{ box}: \qquad e_0 |\emptyset \rangle 
\end{equation}
The condition (\ref{en-on-vacuum}) can be thought of as part of the definition of the vacuum module, in terms of $e_n$ modes.
\subsubsection{Two Boxes}
There are three plane partition configurations with two boxes, see (\ref{MacMahon}).
To reproduce this counting, we start from the single state $e_0 |\emptyset \rangle$ for one box, and apply another $e_n$, and obtain 
\begin{equation}
e_n\,e_0 |\emptyset \rangle\,, \qquad \textrm{with} \quad n\in \mathbb{Z}_{\geq 0}\,.
\end{equation}

However, they are not all independent.
Using the $e-e$ algebraic relations in  (\ref{eq-relations-modes-gl1}), and also the condition (\ref{en-on-vacuum}), one can see that there are only three independent states. Without loss of generality, let us choose them to be $e_n\,e_0 |\emptyset \rangle$ with $n=0,1,2$; and all $e_n\,e_0 |\emptyset \rangle$ with $n\geq 3$ can be expressed as linear combinations of these three states.
Namely, the three states at $N=2$ are
\begin{equation}\label{state-level-3}
2 \textrm{ boxes}: \qquad e_0\,e_0 |\emptyset \rangle\,,\quad e_1\,e_0 |\emptyset \rangle \,, \quad e_2\,e_0 |\emptyset \rangle \,.
\end{equation}

\subsubsection{Three Boxes}
Now we come to the first level where the Serre relations are required in order for the counting to work right. 
First, before imposing the quadratic relations  (\ref{eq-relations-modes-gl1}), the states with three boxes can be obtained by applying $e_n$ with $n\in \mathbb{Z}_{\geq 0}$ on the three states (\ref{state-level-3}).
Then after imposing the quadratic relations  (\ref{eq-relations-modes-gl1}) and the condition (\ref{en-on-vacuum}) on the vacuum, there are $12$ independent states:
\begin{equation}\label{state-level-3-preS}
\begin{aligned}
&e_n\, e_0\,e_0 |\emptyset \rangle\,, \qquad\textrm{with}\quad 0\leq n\leq 2 \,;\\
&e_n\, e_1\,e_0 |\emptyset \rangle\,, \qquad \textrm{with}\quad 0\leq n\leq 3 \,;\\
&e_n\, e_2\,e_0 |\emptyset \rangle\,, \qquad \textrm{with}\quad 0\leq n\leq 4\,;
\end{aligned}
\end{equation}

However, we know from the MacMahon function (\ref{MacMahon}) (or from a direct counting by hand) that there are only $6$ plane partition configurations with three boxes. 
This means that we need $12-6=6$ (independent) additional relations in order to bring down the count to $6$.  
It is straightforward to check that one can choose
\begin{equation}
(n,m,\ell)=(0,0,0)\,, \quad (0,0,1)\,, \quad (0,0,2)\,,\quad (0,1,2)\,, \quad (0,2,2)\,, \quad (0,2,3)\,,
\end{equation}
for the indices in (the first line of) the Serre relation (\ref{eq-Serre-modes-gl1}) to give $6$ independent constraints on (\ref{state-level-3-preS}) in order to cut down the number of independent states to $6$.
These $6$ independent states at $N=3$ can be chosen to be
\begin{equation}\label{states-level-3-gl1}
\begin{aligned}
3 \textrm{ boxes}: \qquad &e_0\, e_0\,e_0 |\emptyset \rangle\,,\quad e_1\, e_0\,e_0 |\emptyset \rangle \,, \quad e_2\,e_0\, e_0 |\emptyset \rangle\,, \quad \\
&e_1\, e_1\,e_0 |\emptyset \rangle\,,\quad e_2\, e_1\,e_0 |\emptyset \rangle \,, \quad e_3\,e_1\, e_0 |\emptyset \rangle\,. 
\end{aligned}
\end{equation}

\subsubsection{Four Boxes and Beyond}
One can then check iteratively that the Serre relations (\ref{eq-Serre-modes-gl1}), though cubic, are enough to bring down the number of states at all $N$ to the correct counting giving by the MacMahon function (\ref{MacMahon}).\footnote{
It would be nice to have a general proof for arbitrary $N$.}
For example, at $N=4$, there should be $13$ plane partition configuration with $4$ boxes, from (\ref{MacMahon}) or a direct counting by hand.

Applying $e_n$ with $n\in \mathbb{Z}_{\geq 0}$ on the $6$ states at $N=3$ in (\ref{states-level-3-gl1}), imposing the quadratic relations  (\ref{eq-relations-modes-gl1}) and the condition (\ref{en-on-vacuum}) on the vacuum, we got $27$ states:
\begin{equation}\label{state-level-4-preS}
\begin{aligned}
&e_n\, e_0\,e_0\,e_0 |\emptyset \rangle\,, \qquad\textrm{with}\quad 0\leq n\leq 2\,;\\
&e_n\, e_1\,e_0\,e_0 |\emptyset \rangle\,, \qquad \textrm{with}\quad 0\leq n\leq 3 \,;\\
&e_n\, e_1\,e_1\,e_0 |\emptyset \rangle\,, \qquad \textrm{with}\quad 0\leq n\leq 3 \,;\\
&e_n\, e_2\,e_0\,e_0 |\emptyset \rangle\,, \qquad \textrm{with}\quad 0\leq n\leq 4 \,;\\
&e_n\, e_2\,e_1\,e_0 |\emptyset \rangle\,, \qquad \textrm{with}\quad 0\leq n\leq 4 \,;\\
&e_n\, e_3\,e_1\,e_0 |\emptyset \rangle\,, \qquad \textrm{with}\quad 0\leq n\leq 5\,,
\end{aligned}
\end{equation}
before imposing the Serre relations   (\ref{eq-Serre-modes-gl1}). 
One can then show that there are $14$ independent Serre relations applicable at this level, and bring the number of independent states to $13$:
\begin{equation}
\begin{aligned}
4 \textrm{ boxes}: \quad 
&e_1\, e_0\, e_0\,e_0 |\emptyset \rangle\,,
\quad e_1\, e_1\, e_0\,e_0 |\emptyset \rangle \,, 
\quad e_1\, e_1\, e_1\,e_0 |\emptyset \rangle \,,
\quad e_1\, e_2\, e_0\,e_0 |\emptyset \rangle \,
\quad e_1\, e_2\, e_1\,e_0 |\emptyset \rangle \,, \\
& e_2\,e_0\, e_0\,e_0 |\emptyset \rangle\,, 
\quad e_2\,e_1\, e_0\,e_0 |\emptyset \rangle\,, 
\quad e_2\,e_2\, e_0\,e_0 |\emptyset \rangle\,,
\quad e_2\,e_2\, e_1\,e_0 |\emptyset \rangle\,,\\
&e_3\, e_1\,e_0 \,e_0|\emptyset \rangle\,,
\quad e_3\, e_1\,e_1\,e_0 |\emptyset \rangle \,, 
\quad e_3\,e_2\, e_1 \,e_0|\emptyset \rangle\,,
\quad e_4\,e_2\, e_0 \,e_0|\emptyset \rangle\,.
\end{aligned}
\end{equation}
The computation for $N\geq 5$ is similar and can be automized in Mathematica.
\bigskip

\subsection{Summary}

Finally, recall that the affine Yangian of $\mathfrak{gl}_1$ has
initial conditions (\ref{eq-initial-gl1}) to supplement the quadratic relations (\ref{eq-relations-OPE-gl1}). 
They cannot be determined by the action on plane partition representations, unlike the OPE and the Serre relations. 
Instead, they are chosen to 
be consistent with the quadratic relations (\ref{eq-relations-OPE-gl1}).
As we explained earlier, they can be simply determined by demanding that the $\psi(z)\,e(w)$ and $\psi(z)\, f(w)$ OPEs in (\ref{eq-relations-OPE-gl1}) are true up to  $z^{n\geq 3}w^{m}$ terms, not just up to $z^{n\geq 0}w^{m}$ terms.

Let us summarize the relations for our algebra: 
\begin{tcolorbox}[ams align]
&\textrm{OPE:}\quad\begin{cases}\begin{aligned}
&\psi(z)\psi(w)\sim \psi(w)\psi(z) \;,\\
&\begin{aligned}
\psi(z)\, e(w) &  \sim  \varphi_3(\Delta)\, e(w)\, \psi(z)  \;,\\ 
\psi(z)\, f(w) & \sim \varphi_3^{-1}(\Delta)\, f(w)\, \psi(z)  \;,
\end{aligned}
\qquad
\begin{aligned}
e(z)\, e(w) & \sim    \varphi_3(\Delta)\, e(w)\, e(z) \;, \\
 f(z)\, f(w) &  \sim    \varphi_3^{-1}(\Delta)\, f(w)\, f(z)  \;,
\end{aligned}\\
&[e(z)\,, f(w)]   \sim  - \frac{1}{\sigma_3}\, \frac{\psi(z) - \psi(w)}{z-w}  \;,
\end{aligned}
\end{cases}\label{box-OPE-gl1}
\\
&\textrm{Initial:}\quad\begin{cases}
\begin{aligned}
&\begin{aligned}
&[\psi_0,e_n] = 0 \;,\\
&[\psi_0,f_n] = 0 \;,
\end{aligned}\qquad \qquad
\begin{aligned}
&[\psi_1,e_n] = 0 \;,\\
&[\psi_1,f_n] = 0 \;,
\end{aligned}\qquad \qquad
\begin{aligned}
&[\psi_2,e_n] = 2 e_n \;,\\
&[\psi_2,f_n] = - 2 f_n  \;,
\end{aligned}
\end{aligned}
\end{cases}\label{box-initial-gl1}
\\
&\textrm{Serre}:\quad\begin{cases}\begin{aligned}
\textrm{Sym}_{z_1,z_2,z_3} (z_2-z_3) [\,e(z_1)\,,[\,e(z_2)\,,e(z_3)]]&=0\,;\\
\textrm{Sym}_{z_1,z_2,z_3} (z_2-z_3) [f(z_1)\,,[f(z_2)\,,f(z_3)]]&=0\,.
\end{aligned}
\end{cases}\label{box-serre-gl1}
\end{tcolorbox}
\noindent

\section{Bootstrapping General BPS Algebras}\label{sec:bootstrap_general}

In this section we generalize the discussion of the previous section for $\mathbb{C}^3$ to the case of an arbitrary toric Calabi-Yau threefold.
It takes the following four steps:
\begin{equation}
\textrm{crystal}\,\, \longrightarrow\,\, 
\textrm{ansatz for action}\,\, \longrightarrow\,\, \textrm{charge function} \,\, \longrightarrow\,\,\textrm{algebra} \,\, \longrightarrow \,\, \textrm{action} \;.
\end{equation}
\begin{enumerate}
\item Fix the ansatz for the action of the algebra (to be determined) on the set of colored crystals (in section~\ref{sec:ansatz-toric}).

The ansatz contains three layers of information: 
(1) the pole structure of the action, which guarantees that applying the creation operators of the algebra iteratively on the vacuum can create the entire set of colored crystals; 
(2) the moduli of coefficients of the action, which fixes the algebra (apart from the signs); 
and (3) the signs in front of these coefficients. 
Both (1) and (2) are determined by the ``charge function" of the colored crystals; whereas (3) needs to be fixed after the algebra (including signs in it) is fixed.
\item Fix the charge function from the quiver data (in section~\ref{sec:hfunction}-\ref{sec:melting_check}), by demanding that part (1) of the ansatz, i.e.\ the pole structure of the action (for building up the crystal iteratively),  is automatically realized by the poles of the charge function. 
The result of the charge function also determines the part (2) of the ansatz.
\item Fix the algebra from the quiver data and the ansatz of the action (part (1) and (2)) (in section~\ref{sec:ActiontoAlgebra}).   
The former controls the statistics (i.e.\ bosonic or fermionic) of the operators, which manifest as various signs in the algebraic relations; whereas the latter controls the magnitudes in the algebraic relations.

\item Fix the part (3) of the action, i.e.\ signs in front of the coefficients of the action, from the statistics of the algebra (in section~\ref{sec:signs}).
\end{enumerate}

\subsection{Ansatz for Representation} 
\label{sec:ansatz-toric}

A molten configuration of the BPS crystal, which we have reviewed in section \ref{sec:review_crystal}, is a generalization of the plane partition in two ways:
\begin{enumerate}
\item The atoms in the crystal can be of multiple colors labelled by a quiver vertex $a\in Q_1$. 
We will use $\sqbox{$a$}$ to label an atom of color $a$, generalizing $\square$ for a 3D-box in the plane partition $\Lambda$.
\item The geometric crystal structure is given by the quiver diagram, following the rules outlined in section 2. 
It is still periodic since it corresponds to a tiling of a torus, but it does not need to have the hexagonal symmetry of the (rhombus tilings of the) plane partitions. 
\end{enumerate}
We use the letter $\Kappa$ (in text mode) to label a colored crystal configuration.
The plane partition can be viewed as the simplest colored crystal, with only one color and the most symmetric shape. 
\bigskip

As reviewed earlier in section~\ref{subsec.crystal}, 
for the construction of the colored crystal we need to choose an atom as the origin of the crystal.
Without loss of generality, we will choose the atom at the origin to be of color $a=1$.\footnote{
It is easy to generalize to representations with superpositions of colored crystals with the atom at the origin $\mathfrak{o}$ having colors other than $a=1$, see section~\ref{sec:vacuum-level1}. 
However, the algebra obtained from such more general representations (i.e.\ tensored representations of crystals starting with different $\sqbox{$a$}$) via the bootstrap procedure would be the same as the one obtained using the crystal starting with $\sqbox{$1$}$.} 
It corresponds to the level-$1$ box $\sqbox{$1$}$ in (\ref{eq-states-level1-gl1}).

In the $\mathbb{C}^3$ case, where there is only one type of atom, the algebra has a triplet of fields, i.e.\ family of generators, $(e(z),\psi(z),f(z))$, see (\ref{eq-action-gl1}), acting on all the atoms in the crystal (or equivalently, all the $\square$'s in the plane partition).
For a generic toric Calabi-Yau whose corresponding crystal has $|Q_0|$ colors, we need $|Q_0|$ triplets of fields labelled by $a \in Q_0$, each acting on the atoms of the corresponding color as in \eqref{eq-action-toric}; they have the mode expansion as in \eqref{eq-mode-expansion-toric}.

Now we write down the ansatz for the action of the fields (\ref{eq-action-toric}) on an arbitrary crystal configuration $|\Kappa\rangle$, as a natural generalization of the ansatz (\ref{eq-ansatz-action-gl1}) for the action of the affine Yangian of  $\mathfrak{gl}_1$ on the set of plane partitions:
\begin{equation}\label{eq-ansatz-action-toric}
\begin{aligned}
\psi^{(a)}(z)|\Kappa\rangle&= \Psi_{\Kappa}^{(a)}(z)|\Kappa\rangle \;,\\
e^{(a)}(z)|\Kappa\rangle &=\sum_{\sqbox{$a$} \,\in \,\textrm{Add}(\Kappa)} 
 \frac{E^{(a)}(\Kappa\rightarrow \Kappa+\sqbox{$a$})}{z-h(\sqbox{$a$})}|\Kappa+\sqbox{$a$}\rangle \;,\\
f^{(a)}(z)|\Kappa\rangle &=\sum_{\sqbox{$a$}\, \in\, \textrm{Rem}(\Kappa)}
\frac{F^{(a)}(\Kappa\rightarrow \Kappa-\sqbox{$a$})}{z-h(\sqbox{$a$})}|\Kappa-\sqbox{$a$}\rangle \;,\\
\end{aligned}
\end{equation}
for $a=1,\dots,|Q_0|$, where 
\begin{equation}\label{eq-EF-epsilon}
\begin{aligned}
E^{(a)}(\Kappa\rightarrow \Kappa+\sqbox{$a$})&\equiv\epsilon(\Kappa\rightarrow \Kappa+\sqbox{$a$})\sqrt{p^{(a)}\textrm{Res}_{u=h(\sqbox{$a$})}\Psi^{(a)}_{\Kappa}(u)}
\\
F^{(a)}(\Kappa\rightarrow \Kappa-\sqbox{$a$})&\equiv\epsilon(\Kappa\rightarrow \Kappa-\sqbox{$a$})\sqrt{q^{(a)}\textrm{Res}_{u=h(\sqbox{$a$})}\Psi^{(a)}_{\Kappa}(u)}\,,
\end{aligned}
\end{equation}
with
\begin{equation} 
\epsilon(\Kappa\rightarrow \Kappa+\sqbox{$a$})=\pm  \qquad \textrm{and}\qquad \epsilon(\Kappa\rightarrow \Kappa-\sqbox{$a$})=\pm\,.
\end{equation}
Here $\sqbox{$a$} \in\textrm{Add}(\Kappa)$ means that we consider an atom of color $a$ which can be added to the crystal $\Kappa$ (a similar comment applies to $\sqbox{$a$}\in\textrm{Rem}(\Kappa)$).

Before we proceed, let us explain the reason behind the ansatz (\ref{eq-ansatz-action-toric}).
First of all, as a natural generalization of the action (\ref{eq-ansatz-action-gl1}) of the affine Yangian of $\mathfrak{gl}_1$ on the set of plane partitions, each colored crystal state $|\Kappa\rangle$ is an eigenstate of the zero modes $\psi^{(a)}(z)$.
The operator $e^{(a)}(z)$  adds an atom with color $a$ to $|\Kappa\rangle$ at all allowed places, whereas $f^{(a)}(z)$  removes an atom with color $a$ from $|\Kappa\rangle$ at all allowed places.

The important part of the ansatz is that the actions of $e^{(a)}(z)$ and $f^{(a)}(z)$ are determined by the $\psi^{(a)}(z)$ eigenfunction of the initial state $|\Kappa\rangle$, i.e.\ $\Psi^{(a)}_{\Kappa}(z)$, which is called ``charge function" here.
In particular, the position of the atom $\sqbox{$a$}$ to be added to $|\Kappa\rangle$ by $e^{(a)}(z)$ or removed from $|\Kappa\rangle$ by $f^{(a)}(z)$ is given by the poles of the charge function $\Psi^{(a)}_{\Kappa}(z)$ of $|\Kappa\rangle$.

The coefficients of the action, i.e.\ $E^{(a)}(\Kappa\rightarrow \Kappa+\sqbox{$a$})$ and $F^{(a)}(\Kappa\rightarrow \Kappa-\sqbox{$a$})$, are then given by the residue of the 
charge function at that particular pole.
The square roots in the coefficients in (\ref{eq-EF-epsilon}) are due to the natural generalization of (\ref{eq-ansatz-action-gl1}). 
The factors $p^{(a)}$ and $q^{(a)}$ are constants allowed by the ansatz, i.e.\ they do not affect the pole structure of the $e^{(a)}(z)$ and $f^{(a)}(z)$ actions.
As we will see later, only their product $p^{(a)}q^{(a)}$ matters; hence without loss of generality we can set $q^{(a)}=1$.
Later in section~\ref{sec:ef-relations} we will show that $p^{a}$ is given by
\begin{equation}
p^{(a)}= \varphi^{a\Rightarrow a}(0)=(-1)^{\#(\textrm{self-loops of $a$})}\,,
\end{equation}
and is related to the bosonic/fermionic nature of the operators $e^{(a)}(z)$ and $f^{(a)}(z)$ via:
\begin{equation}\label{eq-p-epsilon}
p^{(a)}=-(-1)^{|a|}\,.
\end{equation}
This explains the grading rule that we stated earlier in (\ref{eq.Z2_grading}).\footnote{
Quivers in which vertices have $2n$, with $n\in \mathbb{N}$, number of self-loops do not  seem to exist, when we consider toric Calabi-Yau threefolds.}
We postpone the proof of (\ref{eq-p-epsilon}) to section~\ref{sec:ef-relations}.
For the affine Yangian of $\mathfrak{gl}_1$, $p^{(1)}=(-1)^3=-1$, which explains the sign difference inside the square roots of the $e(z)$'s and $f(z)$'s actions in (\ref{eq-ansatz-action-gl1}).

Finally, the $\pm$ signs in front of the square roots, i.e.\ $\epsilon(\Kappa\rightarrow \Kappa+\sqbox{$a$})$ and $\epsilon(\Kappa\rightarrow \Kappa-\sqbox{$a$})$ in (\ref{eq-EF-epsilon}), depend both on the initial crystal state $\Kappa$ and on the atom $\sqbox{$a$}$ (to be added or removed).
As explained earlier, they need to be chosen so as to reproduce the statistics of the algebra, which will be fixed by the quiver data.
\bigskip

Although a crystal $\Kappa$ consists of atoms of possibly multiple colors, each triplet $(e^{(a)},\psi^{(a)},f^{(a)})$ has its own charge function $\Psi^{(a)}_{\Kappa}(z)$ and only acts upon the atoms of its own color $a$.
However, we emphasize that this does not mean that each charge function $\Psi^{(a)}_{\Kappa}(z)$ only receive contribution from atoms of color $a$ --- it is just that the action of $(e^{(a)},\psi^{(a)},f^{(a)})$ is only controlled by the charge function $\Psi^{(a)}_{\Kappa}(z)$ of color $a$. 
\bigskip

With the ansatz (\ref{eq-ansatz-action-toric}), we are now ready to construct the algebra that realizes (\ref{eq-ansatz-action-toric}). 
Similar to the case of the affine Yangian of $\mathfrak{gl}_1$, the procedure is the following.
\begin{enumerate}
\item\label{step-fixing-hfunction-toric} Determine the structure of the poles $h(\sqbox{$a$})$ in the $e^{(a)}(z)$ and $f^{(a)}(z)$ action part of the ansatz (\ref{eq-ansatz-action-toric}). 
The criterion is that by applying all the creation operators $e^{(a)}(z)$, with $a=1,\dots,|Q_0|$, iteratively on the vacuum $|\emptyset\rangle$, i.e.\ the crystal with no atom present yet, one can generate all possible crystal configurations of this type. 
In a similar manner, applying all the annihilation operators $f^{(a)}(z)$, with $a=1,\dots,|Q_0|$, iteratively on any crystal $|\Kappa\rangle$ would eventually reduce it to the vacuum $|\emptyset\rangle$.

\item\label{step-fixing-Psi-toric} Determine the charge function $\Psi^{(a)}_{\Kappa}(z)$ for an arbitrary crystal $\Kappa$ and for all the colors $a=1,\dots, |Q_0|$.
The criterion is that the pole structures of the actions of $e^{(a)}(z)$ and $f^{(a)}(z)$ in (\ref{eq-ansatz-action-toric}) should be encoded in the $\Psi^{(a)}_{\Kappa}(z)$. 
Namely, for a given crystal $\Kappa$, and for any color $a$, all the poles of the charge function $\Psi^{(a)}_{\Kappa}(z)$ of color $a$ correspond to either a location where an $\sqbox{$a$}$ can be added to $\Kappa$ or the location of an existing $\sqbox{$a$}$ in $\Kappa$ that can be removed.

\item\label{step-fixing-relations-new} Find all relations between the three families of operators (\ref{eq-mode-expansion-toric}) that are automatically satisfied when acting on an arbitrary crystal $\Kappa$, given the ansatz (\ref{eq-ansatz-action-toric}) and the charge function $\Psi^{(a)}_{\Kappa}(z)$ determined in step-\ref{step-fixing-Psi-toric}.
\end{enumerate}
The relations found in step-\ref{step-fixing-relations-new} then define the BPS algebra $\mathsf{Y}_{(Q,W)}$, without the Serre relations. 
The Serre relations that are needed to define the (reduced) quiver Yangian algebra $\underline{\mathsf{Y}}_{(Q,W)}$ are to be determined by demanding that the vacuum character of $\underline{\mathsf{Y}}_{(Q,W)}$ reproduces the generating function of the corresponding colored crystal.
In this paper, we will treat the Serre relations for each $(Q,W)$ separately.

Let us note that the strategy above already suggests the $\mathbb{Z}_2$-grading (Bose or Fermi statistics) of the 
generators. 
Suppose that none of the arrows $I\in Q_1$ satisfies $s(I)=t(I)=a$, namely none of the arrows starts and ends at the same vertex $a$. 
When we apply $e^{(a)}(z)$ multiplet times as $e^{(a)}(z_1) e^{(a)}(z_2) \dots |\Kappa\rangle$, this then necessarily vanishes after a finite number of $e^{(a)}$'s. 
This is because for any finite $\Kappa$ we can add only a finite number of atoms of color $a$.
In this case we expect $e^{(a)}(z)$ to have Fermi statistics.
A similar argument suggests Fermi statistics for $f^{(a)}(z)$. 
The remaining generator $\psi^{(a)}(z)$ are even, since they are Cartan generators, under which each crystal state $|\Kappa\rangle$ is an eigenvector; consistent with this claim,
 we will also later find that $\psi^{(a)}(z)$ is obtained from the commutators between $e^{(a)}(z)$ and $f^{(a)}(z)$, see e.g.\ \eqref{eq-OPE-toric}.
This suggests the $\mathbb{Z}_2$-grading as in \eqref{eq.Z2_grading}.

\subsection{Fixing Coordinate Function}
\label{sec:hfunction}

We first need a coordinate system generalizing (\ref{eq-coordinates-gl1}).
For a crystal of generic shape, it is no longer natural to assign each atom a 3D coordinate.
Instead, an atom $\sqbox{$a$}$ can be (non-uniquely) characterized by a path in the periodic quiver $\mathfrak{Q}$ staring from the origin $\mathfrak{o}$. 
Let us denote this path as 
\begin{equation}\label{eq-coordinates-toric}
\sqbox{$a$}: \qquad \textrm{path}[\mathfrak{o}\rightarrow \sqbox{$a$}\,]  \;.
\end{equation} 
Note that there are infinitely many such paths for each $\sqbox{$a$}$, due to the presence of loops in the periodic quiver.

For each color $a$, we would like to define a coordinate function that is adapted to the coordinate system (\ref{eq-coordinates-toric}), generalizing the coordinate function (\ref{eq-coordinate-function-gl1-new}). The most natural way would be to associate a charge $h_I$ to each edge $I$ in the quiver diagram, where $I\in \{a\rightarrow b\}$ for two vertices $a$ and $b$ (which are possibly identical). We then define the coordinate function for $\sqbox{$a$}$ to be the sum of all the charges along the path $[\mathfrak{o}\rightarrow \sqbox{$a$}]$:
\begin{tcolorbox}[ams equation]\label{eq-coordinate-function-toric}
h(\sqbox{$a$})\equiv\sum_{I\, \in\,  \textrm{path}[\mathfrak{o}\rightarrow \sqbox{$a$}\,]} h_I  \;.
\end{tcolorbox}

Recall that in the case of plane partitions, the coordinate function for an atom $\square$ is the way to translate the position of the $\square$ to the pole of the charge function $\Psi_{\Lambda}(z)$. 
We need the same for the colored crystal. 
Therefore, although for a given $\sqbox{$a$}$, the path $[\mathfrak{o}\rightarrow \sqbox{$a$}]$ is not unique, we need its coordinate function to be uniquely defined, in order to associate it to the poles of $\Psi^{(a)}_{\Kappa}(z)$.
This requires that the sum over charges on the edges around any loop has to vanish, which is precisely the loop constraint \eqref{eq-loop-constraint-toric}.
This condition is the generalization of (\ref{eq-constraint-h123-1-gl1}) for plane partitions.

\subsection{Fixing Charge Function}
\label{sec:chargefunction}

We are now ready to fix the charge function $\Psi^{(a)}_{\Kappa}(z)$ for an arbitrary colored crystal $\Kappa$ and any color $a$.

\subsubsection{Ansatz}

Generically, the charge function of $\Psi^{(a)}_{\Kappa}(z)$ can receive contributions from all the atoms in the crystal configuration $\Kappa$. 
Generalizing the result for $\mathbb{C}^3$ in (\ref{eq-Psi-gl1}), we write down the ansatz for the charge function $\Psi^{(a)}_{\Kappa}(z)$
\begin{tcolorbox}[ams equation]\label{eq-Psi-ansatz-toric}
\Psi^{(a)}_{\Kappa}(u)=\psi^{(a)}_{0}(z)\prod_{b\in Q_0} \prod_{\sqbox{$b$}\in \Kappa} \varphi^{b\Rightarrow a}(u-h(\sqbox{$b$}))  \;,
\end{tcolorbox}
\noindent where $\psi^{(a)}_{0}(z)$ is the vacuum contribution,  and we have grouped the atoms in $\Kappa$ by their colors, with the color label $b$ running over all vertices in the quiver diagram, including the color $a$ itself.
For each color $a$, each atom of color $b$ contributes a factor of $\varphi^{b\Rightarrow a}$ function, with argument shifted by the coordinate function of that atom $h(\sqbox{b})$, given by (\ref{eq-coordinate-function-toric}) with the charges subject to the loop constraint (\ref{eq-loop-constraint-toric}).

Given the ansatz for the charge function (\ref{eq-Psi-ansatz-toric}), the goal is to determine the bond factor $\varphi^{b\Rightarrow a}(z)$ (so called because it describes the ``bonding" between atoms of color $a$ and those of color $b$).  We use the ansatz for the algebra's action (\ref{eq-ansatz-action-toric}) on crystals $|\Kappa\rangle$, following the procedure outline in section \ref{sec:ansatz-toric}. 
As in the case of $\mathbb{C}^3$, we first consider how to grow the first few layers (or levels) of the crystal by applying $e^{(a)}(z)$ (for all $a$) starting from the vacuum.

In particular, the poles for the charge function at level-$n$ are fixed by considering adding atoms at level-$(n+1)$, since they control the (creation) action of the operator $e^{(a)}(z)$.
Similarly, the numerators for the charge function at level-$n$ are fixed by demanding that they should cancel relevant poles in the charge function at the level $n-1$, since the presence of these level-$n$ atoms prevents the level-$(n-1)$ atoms from being removed by the operators $f^{(a)}(z)$.
The whole computation is facilitated by the fact that to the charge function $\Psi^{(a)}_{\Kappa}(u)$, atoms of the same color (e.g.\ $b$) contribute the same factor $\varphi^{b\Rightarrow a}$, with only the argument shifted by the coordinate function of the atoms, see ansatz (\ref{eq-Psi-ansatz-toric}).

\subsubsection{\texorpdfstring{Vacuum $\longrightarrow$ Level-$1$}{Vacuum -> Level-1}}\label{sec:vacuum-level1}

The vacuum contribution to the charge function $\Psi_{\Kappa}^{(a)}(z)$ determines the creation of the first atom in the crystal. 
Since the first atom sits at the origin, namely, its coordinate function $h(\sqbox{$a$})=0$, the vacuum contribution $\psi^{(a)}_0(z)$ is a straightforward generalization of (\ref{eq-psi0-gl1}):
\begin{tcolorbox}[ams equation]\label{eq-psi0-a}
\psi^{(a)}_0(z)=1+\frac{C^{(a)}}{z}\,,
\end{tcolorbox}
\noindent where $\{C^{(a)}\}$ is the set of $|Q_0|$ numerical constants that label the representation.
The action of $(e^{(a)}(z), \psi^{(a)}(z),f^{(a)}(z))$ on the vacuum is
\begin{equation}
e^{(a)}(z)|\emptyset\rangle=\frac{\pm \sqrt{p^{(a)}\, C^{(a)}}}{z}|\sqbox{a}\rangle\,, \qquad \psi^{(a)}(z)|\emptyset\rangle= \left(1+\frac{C^{(a)}}{z}\right)|\emptyset\rangle\,,\qquad f^{(a)}(z)|\emptyset\rangle=0\,.
\end{equation}
When acting on the vacuum, $e^{(a)}(z)$ creates an atom $\sqbox{$a$}$ at the origin if $C^{(a)}\neq 0$.

In general, one can allow arbitrary $\{C^{(a)}\}$.
The vacuum representation, labeled by $\{C^{(a)}\}$, would consist of tensored representations in which each irreducible representation, labeled by $C^{(a)}\neq 0$ with $a\in Q_0$, consists of crystal states whose leading atom (at the origin) has color $a$.
However, the algebra obtained from such more general representations (i.e.\ tensored representations of crystals starting with different $\sqbox{$a$}$) via the bootstrap procedure would be the same as the one obtained using the crystal starting with $\sqbox{$a$}$ with a particular color $a\in Q_0$.
Therefore it is enough to consider the irreducible representation, where only one $C^{(a)}$ is nonzero.\footnote{
In this paper we only focus on the vacuum representation, which is enough to bootstrap the algebra. } 

Without loss of generality, we assume that the first atom in the crystal has color $a=1$, namely
\begin{equation}
C^{(a)}=C \,\delta^{a,1}\,,
\end{equation}
therefore the vacuum contribution to the charge function is
\begin{equation}
\psi^{(a)}_0(z)=( \psi_0(z))^{\delta_{a,1}}=\begin{cases}
\psi_0(z)\qquad &(a=1)\\
1\qquad &(\textrm{otherwise})
\end{cases}\qquad \textrm{with} \quad \psi_0(z)=1+\frac{C}{z} \;,
\end{equation}
with $C\neq 0$ a constant to be fixed later. 
Therefore, the charge function for the vacuum $|\Kappa\rangle=|\emptyset\rangle$, for any color $a$, is
\begin{equation}\label{eq-vacuum-toric}
\Psi^{(a)}_{\Kappa}(z)=( \psi_0(z))^{\delta_{a,1}}= \left(1+\frac{C}{z}\right)^{\delta_{a,1}}\,,
\end{equation} 
whose pole corresponds to the adding pole for $e^{(a)}(z)$ at level-$1$:
\begin{equation}\label{eq-adding-pole-1-1-toric}
\textrm{level}-1:\qquad \textrm{adding-pole } z^*=h(\sqbox{1})=0  \;.
\end{equation}
The resulting state at the level-$1$ is denoted by
\begin{equation}
\begin{tikzpicture}[scale=1]
\node[state]  [regular polygon, regular polygon sides=4, draw=blue!50, very thick, fill=blue!10] (a1) at (0,0)  {$1$};
\end{tikzpicture}
\end{equation}
In summary, the action of $(e^{(a)}(z), \psi^{(a)}(z),f^{(a)}(z))$ on the vacuum is
\begin{equation}
e^{(a)}(z)|\emptyset\rangle=\delta_{a,1}\frac{\pm \sqrt{p^{(a)}\,C}}{z}|\sqbox{1}\rangle\,, \qquad \psi^{(a)}(z)|\emptyset\rangle= \left(1+\frac{C}{z}\right)^{\delta_{a,1}}|\emptyset\rangle\,,\qquad f^{(a)}(z)|\emptyset\rangle=0\,.
\end{equation}

\subsubsection{\texorpdfstring{Level-$1$ $\longrightarrow$ Level-$2$}{Level-1 -> Level-2}}

The level-$1$ atom is unique, and has coordinate function
\begin{equation}
h(\sqbox{1})=0
\end{equation}
(see (\ref{eq-adding-pole-1-1-toric})) and its color $a$ charge function, for any $a\in Q_0$, is
\begin{equation}\label{eq-Psi-a-level-1}
\Psi^{(a)}_{\sqbox{1}}(z)=( \psi_0(z))^{\delta_{a,1}}\,\varphi^{1\Rightarrow a}(z) \;.
\end{equation}
We need to fix $\varphi^{1\Rightarrow a}(z)$.

As in the case of $\mathbb{C}^3$, the poles of the charge function at level-$1$ is fixed by considering adding the level-$2$ atoms. 
In the quiver diagram, consider the arrows that emit from the vertex $b=1$, the vertices these arrows end at correspond to the atoms to be added at the level-$2$. 
In order for the creation operators $e^{(a)}(z)$ for these colors to be able to create these atoms, the factors $\varphi^{1\Rightarrow a}(z)$ in the charge function (\ref{eq-Psi-a-level-1}) has to contain the pole $\frac{1}{z-h_{I}}$, where $h_I$ is the charge associated to the arrow $1\rightarrow a$. 
(Note that there can be multiple arrows going from $1$ to $a$, then for each arrow there is an independent $h_I$.)
If a vertex $a$ is not connected by any arrow starting from $1$, it doesn't contribute any pole factor.
Namely, the factor $\varphi^{1\Rightarrow a}(z)$ for all $a$ contains:
\begin{equation}\label{eq-pole-1-toric}
\varphi^{1\Rightarrow a}(z)\supset 
\begin{cases}
\begin{aligned}
\frac{1}{\prod_{I \in \{1\rightarrow a \}}(z-h_I)} \qquad\qquad &(1\rightarrow a) \;, \\
1\qquad\qquad &(1\centernot{\rightarrow}a) \;,
\end{aligned}
\end{cases}
\end{equation}
where $a\rightarrow b$ denotes the case when there is an arrow from $a$ to $b$, whereas $a\centernot{\rightarrow} b$ indicates otherwise.
Define the numerator and denominator of the factor $\varphi^{b\Rightarrow a}(z)$
\begin{equation}
\varphi^{b\Rightarrow a}(z)\equiv \frac{N^{b\Rightarrow a}(z)}{D^{b\Rightarrow a}(z)} \;.
\end{equation}
The minimal solution for the denominator $D^{1\Rightarrow a}(z)$ is
\begin{equation}
D^{1\Rightarrow a}(z) = \prod_{I \in \{1\rightarrow a \}}(z-h_I)  \;.
\end{equation}
\smallskip

The pole structure (\ref{eq-pole-1-toric}) is for the color-$a$ charge function of the leading atom which has color $b=1$.
However, the argument in deriving (\ref{eq-pole-1-toric}) applies to all colors $b$, given that we could have chosen the leading atom in the crystal to be of any color.
Thus we have
\begin{equation}
D^{b\Rightarrow a}(z) = \prod_{I \in \{b\rightarrow a \}}(z-h_I) 
\end{equation}
for any two colors $a$ and $b$.
Note that this notation allows us to express the contribution from atoms of color $b$ to the color-$a$ charge function uniformly, irrespective of whether there is an arrow from $b$ to $a$ in the quiver $Q$ or not.

Therefore, each color $b$ contributes a factor $\varphi^{b\Rightarrow a}(z)$ to the color-$a$ charge function.
The charge functions of the level-$1$ atom are thus
\begin{equation}\label{eq-Psi-level-1-1-toric}
\begin{aligned}
&\Psi^{(a)}_{\Kappa}(z)= \psi^{(a)}_0(z)\, \varphi^{1\Rightarrow a}(z) \,,
\\
&\textrm{with}\qquad \psi_0^{(a)}(z)=(1+\frac{C}{z})^{\delta_{a,1}}\;, \qquad \varphi^{1\Rightarrow a}(z)=\frac{N^{1\Rightarrow a}(z)}{\prod_{I \in \{1\rightarrow a\}} (z-h_I) } \;.
\end{aligned}
\end{equation}
We could have chosen another color $b\neq 1$ as the color for the leading atom of the crystal; in this case the charge function of the level-1 atom would be
\begin{align}\label{eq-Psi-level-1-a-toric}
\begin{split}
&\Psi^{(a)}_{\Kappa}(z)= \psi^{(a)}_0(z)\, \varphi^{b\Rightarrow a}(z)\\ 
&\textrm{with}\qquad \psi_0^{(a)}(z)=(1+\frac{C}{z})^{\delta_{a,b}}\;, \qquad \varphi^{b\Rightarrow a}(z)=\frac{N^{b\Rightarrow a}(z)}{\prod_{I \in \{b\rightarrow a\}} (z-h_I) } \;.
\end{split}
\end{align}
To determine the numerator $N^{1\Rightarrow a}(z)$, or more generally $N^{b\Rightarrow a}(z)$, we need to move to the next level.

The level-$1$ charge function (\ref{eq-Psi-a-level-1}) has poles at
\begin{equation}
\begin{aligned}
\textrm{removing-pole}: \qquad &z^*=0  \;,\\
 \textrm{adding-pole}: \qquad &z^*=h_I \quad \textrm{with}\quad I \in \bigcup_{a\in Q_0} \{1 \rightarrow a\}  \;.
 \end{aligned}
\end{equation}
The action of $(e^{(a)},\psi^{(a)},f^{(a)}(z))$ on the level-1 state $|\Kappa\rangle=|\sqbox{$1$}\rangle$ is then 
\begin{equation}
\begin{aligned}
\psi^{(a)}(z)|\Kappa\rangle &=\psi^{(a)}_0(z)\,  \varphi^{1\Rightarrow a}(z)|\Kappa\rangle \;,\\
e^{(a)}(z)|\Kappa\rangle &=\sum_{a\in \textrm{n}[1\rightarrow]} \sum_{i}\,\frac{\#}{z-h(\sqbox{$a$}_i)}|\Kappa+\sqbox{$a$}_i\rangle  \;,\\
f^{(a)}(z)|\Kappa\rangle &=\delta_{a,1}\frac{\#}{z}|\emptyset\rangle \;,
\end{aligned}
\end{equation}
where by $\textrm{n}[1\rightarrow]$ we mean the set of vertices connected to by an arrow from $1$ in the periodic quiver $\mathfrak{Q}$, $i$ distinguishes different such vertices, and $h(\sqbox{$a$}_i)$ measures the distance of the added atom to the origin: 
\begin{equation}
h(\sqbox{$a$}_i)=h_I \qquad \textrm{with}\qquad I:\sqbox{1}\rightarrow \sqbox{$a$}_i \;.
\end{equation}

\subsubsection{\texorpdfstring{Level-$2$}{Level-2}}

Let us consider the state with one level-$1$ atom of color $b=1$ and one level-$2$ atom of color $c=2$, for which the arrow from $1\rightarrow 2$ has to exist in the quiver diagram.
The charge function of this state is
\begin{equation}\label{eq-Psi-level-2-1-toric}
\Psi^{(a)}_{\Kappa}(z)=\left[\psi^{(a)}_0(z)\varphi^{1\Rightarrow a}(z) \right]
\left[\varphi^{2\Rightarrow a} (z-h(\sqbox{2}))\right]  \;,
\end{equation}
where the first bracket contains contributions from the vacuum and the level-$1$ atom and hence is identical to (\ref{eq-Psi-level-1-a-toric}), and the second bracket is the contribution from the one level-$2$ atom $\sqbox{$2$}$ that we are considering, with
\begin{equation}
\varphi^{2\Rightarrow a} (z)=\frac{N^{2\Rightarrow a}(z)}{\prod_{I \in \{2\rightarrow a\}} (z-h_I) }  \;.
\end{equation}

Compare the charge function (\ref{eq-Psi-level-2-1-toric}) to the level-$1$ charge function (\ref{eq-Psi-level-1-1-toric}).
Since the presence of the level-$2$ atom prevents the level-$1$ atom from being removed by the $f^{(1)}(z)$ operator, we need a numerator factor in the second bracket of (\ref{eq-Psi-level-2-1-toric}) when $a=1$, i.e.\ the numerator of $\varphi^{2\Rightarrow 1}(z)$ has to contain the factor that cancels the $z^*=0$ removing pole in $\psi^{(a=1)}_0(z)$.
In addition, since this needs to happen for any atom of color $c=2$ at the level-$2$, we have
\begin{equation}
N^{2 \Rightarrow 1}(z) = \prod_{I\in\{1\rightarrow 2 \}} (z+h_I)  \;.
\end{equation}
Take 
\begin{equation}\label{pic-level-12-toric}
\begin{tikzpicture}[scale=1]
\node[state]  [regular polygon, regular polygon sides=4, draw=blue!50, very thick, fill=blue!10] (a1) at (0,0)  {$1$};
\node[state]  [regular polygon, regular polygon sides=4, draw=blue!50, very thick, fill=blue!10] (a2r) at (2,0)  {$2$};
\node[state]  [regular polygon, regular polygon sides=4, draw=blue!50, very thick, fill=blue!10] (a2l) at (-2,0)  {$2$};
\path[->] 
(a1) edge   [thick, red]   node [above] {$h_1$} (a2r)
(a1) edge   [thick, red]   node [above] {$h_3$} (a2l)
;
\end{tikzpicture}
\end{equation} 
for example, which indicate the relevant part in the periodic quiver  $\mathfrak{Q}$,  we have
\begin{equation}
N^{2 \Rightarrow 1}(z) =(z+h_1)(z+h_3)  \;.
\end{equation}
As with the denominator, this argument works for any two colors $a$ and $b$, therefore we have
\begin{equation}
N^{b \Rightarrow a}(z) = \prod_{I\in\{a\rightarrow b \}} (z+h_I) \;.
\end{equation}
\bigskip

To summarize, the contribution to the color-$b$ charge function from atoms of color $a$ is given by the bond factor \eqref{eq-charge-atob}.
An atom of color $b$ at a position with coordinate function $h(\sqbox{$b$})$ contributes to the color-$a$ charge function by:
\begin{equation}
\varphi^{b\Rightarrow a}(u-h(\sqbox{$b$})) \;.
\end{equation}
Taking the contributions from all the atoms in the crystal --- all colors including color $a$, we get the color-$a$ charge function: 
\begin{equation}\label{eq-Psi-ansatz}
\Psi^{(a)}_{\Kappa}(u)=\psi^{(a)}_0(u)\prod_{b\in Q_0} \prod_{\sqbox{$b$}\in \Kappa} \varphi^{b\Rightarrow a}(u-h(\sqbox{$b$})) \;,
\end{equation}
where the vacuum contributes only to the charge function of color $a=1$: 
\begin{equation}\label{eq-psi0-ansatz}
\psi^{(a)}_0(u)=\left(1+\frac{C}{z}\right)^{\delta_{a,1}}\,.
\end{equation}

\subsection{Melting Rule in General and Loop Constraint}\label{sec:melting_check}


We can now verify in general that the representation given by (\ref{eq-ansatz-action-toric}) and (\ref{eq-EF-epsilon}), with charge function $\Psi^{(a)}_{\Kappa}(u)$ defined in (\ref{eq-Psi-ansatz-toric}) and bond factors $\varphi^{b\Rightarrow a}(u)$ defined by (\ref{eq-charge-atob}), 
realizes the melting rule of \eqref{eq.melting_rule}.
In particular, we will show that the loop constraint (\ref{eq-loop-constraint-toric}), which is a generalization of the constraint (\ref{eq-constraint-h123-1-gl1}) that comes from the box-stacking rules for the plane partitions, 
is necessary to ensure the general melting rule  \eqref{eq.melting_rule}.
The crux of the argument is the same as in the low-level examples above.

Suppose that we have an atom $\sqbox{$a$}\in \Kappa$ inside a molten crystal configuration $\Kappa$.
Suppose moreover there exists another atom $\sqbox{$b$}$ which is connected to the atom $\sqbox{$a$}$ by an arrow $I: \sqbox{$a$} \to \sqbox{$b$}$ and which moreover is not in the crystal configuration $\Kappa$. 
We can now try to create atom $\sqbox{$b$}$ by applying the generator $e^{(b)}(z)$ to the state $|\Kappa\rangle$.

One might expect that it is always possible to create atom $\sqbox{$b$}$.
However, the melting rule says that this is not possible if there exists another atom $\sqbox{$c$}$ which is not in the crystal $\Kappa$ and is connected to the atom $\sqbox{$b$}$ by an arrow $J:\sqbox{$c$} \to \sqbox{$b$}$, shown below:
\begin{equation}\label{not_melting_rule-2}
\begin{tikzpicture}[scale=1]
\node[state]  [regular polygon, regular polygon sides=4, draw=blue!50, very thick, fill=blue!10] (a) at (0,0)  {$a$};
\node[state]  [regular polygon, regular polygon sides=4, draw=blue!50, very thick, fill=blue!10] (b) at (2,0)  {$b$};
\node[state]  [regular polygon, regular polygon sides=4, draw=red!50, very thick, fill=red!10] (c) at (4,-1) {$c$};
\path[->] 
(a) edge   [thick, red]   node [above, pos=0.2] {$h_I$} (b)
(c) edge   [thick, red]   node [above] {$h_J$} (b)
;
\draw (1,-1)--(1,2);
\draw[->] (1,1)--(0,1);
\node[above] at (0,1) {crystal $\Kappa$};
\end{tikzpicture}
\end{equation}
This obstruction should be reflected in the charge function $\Psi^{(b)}_{\Kappa}(u)$.

In order to see this, let us first project the crystal down to the periodic quiver $\mathfrak{Q}$. (Recall that an atom $\sqbox{$a$}$ of the crystal is specified by a pair $(\mathfrak{a},n)$, where $\mathfrak{a}$ is a vertex of the periodic quiver and the non-negative integer $n$ specifies the depth along the crystal (see Figure \ref{fig.RuleEx}).)
In the periodic quiver one can find a vertex $\mathfrak{d}$ such that (1) there exists an arrow $K: \mathfrak{b} \to \mathfrak{d}$ and (2) there exists a path $\mathfrak{d} \to \mathfrak{a}$, shown below:
\begin{equation}\label{not_melting_rule-3}
\begin{tikzpicture}[scale=1]
\node[state]  [regular polygon, regular polygon sides=4, draw] (2a) at (5,0)  {$\mathfrak{a}$};
\node[state]  [regular polygon, regular polygon sides=4, draw] (2b) at (7,0)  {$\mathfrak{b}$};
\node[state]  [regular polygon, regular polygon sides=4, draw] (2c) at (9,-1) {$\mathfrak{c}$};
\node[state]  [regular polygon, regular polygon sides=4, draw] (2d) at (6,-2) {$\mathfrak{d}$};
\path[->] 
(2a) edge   [thick, red]   node [above] {$h_I$} (2b)
(2c) edge   [thick, red]   node [above] {$h_J$} (2b)
(2b) edge   [thick, red]   node [left] {$h_K$} (2d)
(2d) edge   [thick, red, decorate, decoration=snake]   node [left] {$h_{\mathfrak{d}\to \mathfrak{a}}$} (2a)
;
\end{tikzpicture}
\end{equation}
In other words, the arrows $I: \mathfrak{a} \to \mathfrak{b}$ and $K: \mathfrak{b} \to \mathfrak{d}$ belong to the same polygonal region of the periodic quiver.
In figure (\ref{not_melting_rule-3}) we have used the wiggly line between $\mathfrak{d}$ and $\mathfrak{a}$ to emphasize that this is in general not a single arrow, but a path consisting of several arrows. 
We have denoted the weight for this path by $h_{\mathfrak{d}\to \mathfrak{a}}$. 

Let us now uplift this picture to the three-dimensional crystal.
Since $\sqbox{$b$}$ is not contained in the crystal $\Kappa$, the melting rule says that another atom $\sqbox{$d$}=K\cdot \sqbox{$b$}$ is also not in the crystal $\Kappa$.
However, since there exists a path from $\sqbox{$d$}$ to $\sqbox{$a$}$, the melting rule also suggests that there exists another atom $\sqbox{$\tilde{d}$}$ in the same position $\mathfrak{d}$ of the periodic quiver such that $\sqbox{$\tilde{d}$}$ belongs to the crystal configuration $\Kappa$, shown by
\begin{equation}\label{not_melting_rule-1}
\begin{tikzpicture}[scale=1]
\node[state]  [regular polygon, regular polygon sides=4, draw=blue!50, very thick, fill=blue!10] (a) at (0,0)  {$a$};
\node[state]  [regular polygon, regular polygon sides=4, draw=blue!50, very thick, fill=blue!10] (b) at (2,0)  {$b$};
\node[state]  [regular polygon, regular polygon sides=4, draw=red!50, very thick, fill=red!10] (c) at (4,-1) {$c$};
\node[state]  [regular polygon, regular polygon sides=4, draw=blue!50, very thick, fill=blue!10] (td) at (0,-2) {$\tilde{d}$};
\node[state]  [regular polygon, regular polygon sides=4, draw=red!50, very thick, fill=red!10] (d) at (2,-2) {$d$};
\path[->] 
(a) edge   [thick, red]   node [above, pos=0.2] {$h_I$} (b)
(c) edge   [thick, red]   node [above] {$h_J$} (b)
(b) edge   [thick, red]   node [left] {$h_K$} (d)
(d) edge   [thick, red, dotted]   node [below] {} (td)
(td) edge   [thick, red, decorate, decoration=snake]   node [left] {$h_{\mathfrak{d}\to \mathfrak{a}}$} (a)
;
\draw (1,-3)--(1,2);
\draw[->] (1,1)--(0,1);
\node[above] at (0,1) {crystal $\Kappa$};
\end{tikzpicture}
\end{equation}
Assume that the depth of the atom $\sqbox{$d$}$ is $n$, i.e.\ $\sqbox{$d$}=(\mathfrak{d},n)$.
We can choose the atom $\sqbox{$\tilde{d}$}$ to have the maximal depth, which is $n-1$, i.e.\ $\sqbox{$\tilde{d}$}=(\mathfrak{d}, n-1)$.
We emphasize that $\sqbox{$\tilde{d}$}=(\mathfrak{d}, n-1)$ and $\sqbox{d$$}=(\mathfrak{d}, n)$ are in different depths inside the crystal.
\smallskip

Let us now consider the charge function $\Psi_{\Kappa}^{(b)}(u)$.
Since the crystal configuration $\Kappa$ contains the atom $\sqbox{$a$}$, the charge function $\Psi_{\Kappa}^{(b)}(u)$ has a factor $\varphi^{a\Rightarrow b}(u-h(\sqbox{$a$}))$, which contains the factor $(u-h_I-h(\sqbox{$a$}))^{-1}=(u-h(\sqbox{$b$}))^{-1}$, as expected. 
However, $\Psi_{\Kappa}^{(b)}(u)$  also contains the factor $\varphi^{d\Rightarrow b}(u-h(\sqbox{$\tilde{d}$}))$ from the atom $\sqbox{$\tilde{d}$}$, which contains a factor $(u+h_K-h(\sqbox{$\tilde{d}$}))$. 
Since $h(\sqbox{$d$})-h_K=h(\sqbox{$b$})$, this factor can cancel the pole $(u-h(\sqbox{$b$}))^{-1}$ if 
\begin{equation}\label{eq-h-d-dtilde}
h(\sqbox{$\tilde{d}$})=h(\sqbox{$d$})\,.
\end{equation}
On the other hand, since the difference between $\sqbox{$d$}$ and $\sqbox{$\tilde{d}$}$ is only a loop around the periodic quiver, we have
\begin{equation}
h(\sqbox{$d$})-h(\sqbox{$\tilde{d}$})=\sum_{I\in \textrm{loop }L} h_I\,,
\end{equation}
where $L$ is the loop in the periodic quiver that characterizes the difference between  $\sqbox{$d$}$ and $\sqbox{$\tilde{d}$}$. 
For (\ref{eq-h-d-dtilde}) to hold in general, we need to impose the loop constraint \eqref{eq-loop-constraint-toric} for all loops in the period quiver.\footnote{
One small caveat to our argument occurs when the atom $\sqbox{$d$}$ is on the surface of the crystal ($n=1$ in our previous notation), such that we cannot find $\sqbox{$\tilde{d}$}$. 
For example, the atom $\sqbox{$a$}$ in itself can be an origin $\mathfrak{o}$ of the crystal. 
However, one can check that this is not possible in the crystal with $\sqbox{$c$}\notin |\Kappa \rangle$ as in \eqref{not_melting_rule-2}.
}

A careful reader might have noticed that we have not used explicitly the condition that the atom  $\sqbox{$c$}$  is not contained in the crystal. 
This condition is needed because, had $\sqbox{$c$}$ been in the crystal, it would have contributed to the pole of the charge function $\Psi_{\Kappa}^{(b)}(u)$, just like the atom $\sqbox{$a$}$ does.
More generally, each atom that is inside the $\Kappa$ and has an arrow pointing to the atom $\sqbox{$b$}$ contributes an adding-pole $z^*=h(\sqbox{$b$})$. 
Let us call them $\sqbox{$a$}_i$.
By the previous argument, for each of them there exists at least one atom deeper inside the crystal $\Kappa$ and playing the role of $\sqbox{$\tilde{d}$}$ that can cancel this adding pole. 
However, these deeper atoms  $\sqbox{$\tilde{d}$}_j$ are shared among the atoms $\sqbox{$a$}_i$ leading to $\sqbox{$b$}$; on the other hand, each atom  $\sqbox{$a$}_i$ can have more than one deeper atom  $\sqbox{$\tilde{d}$}_j$ to cancel the adding-pole $z^*=h(\sqbox{$b$})$.
The end result is that if all the atoms that have arrows pointing to $\sqbox{$b$}$ are already inside the $\Kappa$, there is precisely one adding pole $z^*=h(\sqbox{$b$})$ left after all the cancellation from the deeper atoms $\sqbox{$\tilde{d}$}_j$.
Therefore, when there is an atom $\sqbox{$c$}$ that has an arrow pointing to $\sqbox{$b$}$ yet is not inside $\Kappa$, the cancellation is such that there is no adding pole at $z^*=h(\sqbox{$b$})$. 
\smallskip

In section~\ref{sec:l2tol3-gl1}, we gave an example of this melting rule for the case of affine Yangian of $\mathfrak{gl}_1$.
See the left figure of (\ref{eq.n_allowed-PP}) for the quiver and the right figure of Fig.~\ref{fig.bad} for the plane partition.
In this example, the atoms $\sqbox{$a$}$,  $\sqbox{$b$}$, and $\sqbox{$c$}$ have coordinate
\begin{equation}
\sqbox{$a$}: (x_1,x_2,x_3)=(1,0,0)\,,\quad \sqbox{$b$}: (x_1,x_2,x_3)=(1,1,0)\,,\quad \sqbox{$c$}: (x_1,x_2,x_3)=(0,1,0)\,,
\end{equation}
where $\sqbox{$c$}$ is the missing box in the right figure of Fig.~\ref{fig.bad}.
The atoms $\sqbox{$d$}$ and $\sqbox{$\tilde{d}$}$ 
have coordinate
\begin{equation}
\sqbox{$d$}: (x_1,x_2,x_3)=(1,1,1)\,,\qquad \sqbox{$\tilde{d}$}: (x_1,x_2,x_3)=(0,0,0)\,;
\end{equation}
namely, $\sqbox{$d$}$ (not shown in Fig.~\ref{fig.bad}) should be right on top of the box $\sqbox{$b$}$ and $\sqbox{$\tilde{d}$}$ is the box at the origin.
\bigskip

Finally, for later convenience, let us summarize the action of the algebra on any colored crystal state $|\Kappa\rangle$:
\begin{tcolorbox}[ams align]\label{eq.efpsi}
\begin{aligned}
&\textrm{action}: \quad \begin{cases}
\begin{aligned}
\psi^{(a)}(z)|\Kappa\rangle&= \Psi_{\Kappa}^{(a)}(z)|\Kappa\rangle \;,\\
e^{(a)}(z)|\Kappa\rangle &=\sum_{\sqbox{$a$} \,\in \,\textrm{Add}(\Kappa)} 
 \frac{\pm\sqrt{p^{(a)}\textrm{Res}_{u=h(\sqbox{$a$})}\Psi^{(a)}_{\Kappa}(u)}}{z-h(\sqbox{$a$})}|\Kappa+\sqbox{$a$}\rangle \;,\\
f^{(a)}(z)|\Kappa\rangle &=\sum_{\sqbox{$a$}\, \in\, \textrm{Rem}(\Kappa)}
\frac{\pm\sqrt{q^{(a)}\textrm{Res}_{u=h(\sqbox{$a$})}\Psi^{(a)}_{\Kappa}(u)}}{z-h(\sqbox{$a$})}|\Kappa-\sqbox{$a$}\rangle \;,\\
\end{aligned}
\end{cases}
\\
\\
&\textrm{with}:\qquad \Psi^{(a)}_{\Kappa}(u)\equiv \psi_0(u)^{\delta_{a,1}} \prod_{b\in Q_0} \prod_{\sqbox{$b$}\in \Kappa} \varphi^{b\Rightarrow a}(u-h(\sqbox{$b$})) \;,\\
&\qquad \quad\qquad \psi_0(z)\equiv 1+\frac{C}{z} \;,\\
&\qquad \quad\qquad\varphi^{b\Rightarrow a} (u)\equiv \frac{\prod_{I\in \{a\rightarrow b\}}(u+h_{I})}{\prod_{I\in \{b\rightarrow a\}}(u-h_{I})} \;,\\
&\qquad \quad\qquad h(\sqbox{$a$})\equiv\sum_{I\, \in\,  \textrm{path}[\mathfrak{o}\rightarrow \sqbox{$a$}\,]} h_I \,, \\
&\textrm{subject to}:\qquad 
\sum_{I\in \textrm{loop }L} h_I= 
0  \;,
\end{aligned}
\end{tcolorbox}
\noindent where $q^{(a)}=1$, $p^{(a)}\equiv \varphi^{a\Rightarrow a}(0)=\pm 1$ and is related to the statistics of the operators $e^{(a)}(z)$ and $f^{(a)}(z)$.
The $\pm$ signs in the coefficients of the $e^{(a)}(z)$ and $f^{(a)}(z)$ actions depend on both the initial state $\Kappa$ and the atom $\sqbox{$a$}$, and will only be fixed after we determine the statistics of the algebra. 

The numerical constant $C$ in the vacuum contribution to the charge function should be considered as a parameter that defines the quiver Yangian algebra. 
As we will see later, it enters the eigenvalues of the zero modes $\psi^{(a)}_0$ on crystal state $|\Kappa\rangle$, together with charges $\{h_I\}$ on the quiver. 
In particular, when the quiver satisfies certain conditions under which the quiver Yangian has central terms, $C$ is directly related to the (leading) central term. 
Since this discussion depends crucially on the quiver, in particular, whether the corresponding Calabi-Yau threefold has compact $4$-cycles, we will discuss the two classes in section~\ref{sec:generalno4cycle} (without compact $4$-cycles) and section~\ref{sec:general4cycle} (with compact $4$-cycles), respectively.

\subsection{From Action on Colored Crystals to Relations of Quiver Yangian}
\label{sec:ActiontoAlgebra}

In section~\ref{sec:QYtopdown}, we summarized the relations of the quiver Yangian (see (\ref{eq-OPE-toric})) before deriving them.
The goal of the current section is to derive the algebra (\ref{eq-OPE-toric}) from its action on the set of colored crystals.
We will show that, starting from our ansatz for the action of the algebra on the set of colored crystals $|\Kappa\rangle$ given in (\ref{eq.efpsi}), one can derive the relations (\ref{eq-OPE-toric}) by demanding that the set of colored crystals $|\Kappa\rangle$ furnishes a representation of the quiver Yangian.

\subsubsection{\texorpdfstring{$\psi-\psi$ Relations}{psi-psi Relations}}

First of all, since any crystal state $|\Kappa\rangle$ is an eigenstate of all $\psi^{(a)}(z)$ (see the first equation in (\ref{eq.efpsi})), we have
\begin{equation}
\psi^{(a)}(z)\, \psi^{(b)}(w)\, |\Kappa\rangle = \Psi^{(a)}_{\Kappa} (z)\,  \Psi^{(b)}_{\Kappa} (w)\,|\Kappa\rangle 
=\psi^{(b)}(w)\, \psi^{(a)}(z)\, |\Kappa\rangle\,.
\end{equation}
Since this is true for any $ |\Kappa\rangle$, we have
\begin{equation}
\psi^{(a)}(z)\, \psi^{(b)}(w)= \psi^{(b)}(w)\, \psi^{(a)}(z) \;,
\end{equation}
as shown in the first equation in (\ref{eq-OPE-toric}).

\subsubsection{\texorpdfstring{$\psi-e$ and $\psi-f$ Relations}{psi-e and psi-f Relations}}
\label{sec:psi-e-f-relations}

To derive the $\psi-e$ relation in  (\ref{eq-OPE-toric}), apply first  $e^{(b)}(w)$ and then $\psi^{(a)}(z)$ on an arbitrary crystal state $|\Kappa\rangle$, and use the actions of $\psi^{(a)}(z)$ and $e^{(b)}(w)$ in (\ref{eq-ansatz-action-toric}):
\begin{equation}\label{eq-psi-e-K}
\begin{aligned}
\psi^{(a)}(z) \, e^{(b)}(w) \, |\Kappa\rangle=\sum_{\sqbox{$b$} \,\in \,\textrm{Add}(\Kappa)} 
 \frac{ \Psi^{(a)}_{\Kappa+\sqbox{$b$}} (z)\, E^{(b)}(\Kappa\rightarrow \Kappa+\sqbox{$b$})}{w-h(\sqbox{$b$})}\, |\Kappa+\sqbox{$b$}\rangle \,.
\end{aligned}
\end{equation}
Reversing the order of $\psi^{(a)}(z)$ and $e^{(b)}(w)$, we have
\begin{equation}\label{eq-e-psi-K}
\begin{aligned}
e^{(b)}(w) \,\psi^{(a)}(z) \,  |\Kappa\rangle=\sum_{\sqbox{$b$} \,\in \,\textrm{Add}(\Kappa)} 
 \frac{ \Psi^{(a)}_{\Kappa} (z)\, E^{(b)}(\Kappa\rightarrow \Kappa+\sqbox{$b$})}{w-h(\sqbox{$b$})}\, |\Kappa+\sqbox{$b$}\rangle \,.
\end{aligned}
\end{equation}

Now compare the coefficients in (\ref{eq-psi-e-K}) and (\ref{eq-e-psi-K}): for each final state $ |\Kappa+\sqbox{$b$}\rangle$, the ratio between the coefficient in (\ref{eq-psi-e-K}) and the one in (\ref{eq-e-psi-K}) is
\begin{equation}
\frac{\Psi^{(a)}_{\Kappa+\sqbox{$b$}} (z)}{ \Psi^{(a)}_{\Kappa} (z)} 
=\varphi^{b\Rightarrow a} (z-h(\sqbox{$b$}))\,,
\end{equation}
which can be written as $\varphi^{b\Rightarrow a}(z-w)$ in (\ref{eq-psi-e-K}) and (\ref{eq-e-psi-K}) since in both equations $w\rightarrow h(\sqbox{$b$})$ for each final state $ |\Kappa+\sqbox{$b$}\rangle$, and for the $\psi-e$ relation we only care about the singular terms $\sim w^{-m-1}$ with $m\in\mathbb{Z}_{\geq 0}$.
Since this is true for any $ |\Kappa\rangle$, we have
\begin{equation}\label{eq-psi-e-fromK}
\psi^{(a)}(z)\, e^{(b)}(w)\simeq \varphi^{b\Rightarrow a}(z-w)\, e^{(b)}(w)\, \psi^{(a)}(z) \;,
\end{equation}
as shown in the second equation of (\ref{eq-OPE-toric}).

Similarly, to derive the $\psi-f$ relations, we consider
\begin{equation}\label{eq-psi-f-K}
\begin{aligned}
\psi^{(a)}(z) \, f^{(b)}(w) \, |\Kappa\rangle&=\sum_{\sqbox{$b$} \,\in \,\textrm{Rem}(\Kappa)} 
 \frac{ \Psi^{(a)}_{\Kappa-\sqbox{$b$}} (z)\, F^{(b)}(\Kappa\rightarrow \Kappa-\sqbox{$b$})}{w-h(\sqbox{$b$})}\, |\Kappa-\sqbox{$b$}\rangle \,,\\
 f^{(b)}(w) \,\psi^{(a)}(z) \,  |\Kappa\rangle&=\sum_{\sqbox{$b$} \,\in \,\textrm{Rem}(\Kappa)} 
 \frac{ \Psi^{(a)}_{\Kappa} (z)\, F^{(b)}(\Kappa\rightarrow \Kappa-\sqbox{$b$})}{w-h(\sqbox{$b$})}\, |\Kappa-\sqbox{$b$}\rangle \,,
\end{aligned}
\end{equation}
and compute the ratio between the two coefficients as
\begin{equation}
\frac{\Psi^{(a)}_{\Kappa-\sqbox{$b$}} (z)}{ \Psi^{(a)}_{\Kappa} (z)} 
=\left(\varphi^{b\Rightarrow a} (z-h(\sqbox{$b$}))\right)^{-1}\,,
\end{equation}
which gives 
\begin{equation}\label{eq-psi-f-fromK}
\psi^{(a)}(z)\, f^{(b)}(w)\simeq \varphi^{b\Rightarrow a}(z-w)^{-1}\, f^{(b)}(w)\, \psi^{(a)}(z) \;,
\end{equation}
as shown by the fourth equation of (\ref{eq-OPE-toric}).

\subsubsection{\texorpdfstring{$e-f$ Relations and Statistics of $e$ and $f$ Operators}{e-f Relations Statistics of e and f Operators}}
\label{sec:ef-relations}

Next, let us consider the $e-f$ relation. 
This would also fix the bosonic/fermionic nature of the operators $e^{(a)}(z)$ and $f^{(a)}(z)$, namely, we will prove the relation (\ref{eq-p-epsilon}).

To derive the $e-f$ relation, consider applying
\begin{equation}
\mathcal{O}^{(a,b)}(z, w)\equiv e^{(a)}(z)\, f^{(b)}(w)+\epsilon^{ab} f^{(b)}(w)\,e^{(a)}(z)
\end{equation}
on an arbitrary initial state $|\Kappa\rangle$, where $\epsilon^{ab}$ is a shorthand for $-(-1)^{|a||b|}$, which characterizes the (mutual) statistics of the operators $e^{(a)}(z)$ and $f^{(b)}(w)$ and will be determined (in terms of the self bond factor $\varphi^{a\Rightarrow a}$) shortly.

First, on an arbitrary initial state $|\Kappa\rangle$, we have
\begin{equation}\label{eq-e-f-K}
\begin{aligned}
&e^{(a)}(z) \, f^{(b)}(w) \, |\Kappa\rangle\\
&= \sum_{\sqbox{$b$} \,\in \,\textrm{Rem}(\Kappa)}   
\sum_{\sqbox{$a$} \,\in \,\textrm{Add}(\Kappa-\sqbox{$b$})}
\frac{ E^{(a)}(\Kappa-\sqbox{$b$}\rightarrow\Kappa-\sqbox{$b$}+\sqbox{$a$} )}{z-h(\sqbox{$a$})}
\cdot \frac{ F^{(b)}(\Kappa\rightarrow \Kappa -\sqbox{$b$})}{w-h(\sqbox{$b$})}
\,|\Kappa-\sqbox{$b$}+\sqbox{$a$}\rangle \,,
\end{aligned}
\end{equation}
and
\begin{equation}\label{eq-f-e-K}
\begin{aligned}
&f^{(b)}(w) \, e^{(a)}(z) \, |\Kappa\rangle\\
&= \sum_{\sqbox{$a$} \,\in \,\textrm{Add}(\Kappa)}   
\sum_{\sqbox{$b$} \,\in \,\textrm{Rem}(\Kappa+\sqbox{$a$})}
\frac{F^{(b)}(\Kappa +\sqbox{$a$}\rightarrow \Kappa+\sqbox{$a$}-\sqbox{$b$})}{w-h(\sqbox{$b$})}
\cdot \frac{ E^{(a)}(\Kappa \rightarrow \Kappa+\sqbox{$a$})}{z-h(\sqbox{$a$})}
\,|\Kappa+\sqbox{$a$}-\sqbox{$b$}\rangle \,.
\end{aligned}
\end{equation}
Generically there are three scenarios
\begin{enumerate}
\item $a=b$, and the atom $\sqbox{$a$}$ removed by $f^{(a)}(w)$ coincides with the atom $\sqbox{$a$}$ added by $e^{(a)}(z)$. 
\item $a=b$, the atom $\sqbox{$a$}$ removed by $f^{(a)}(w)$ is different from the atom $\sqbox{$a$}'$ added by $e^{(a)}(z)$. 
\item $a\neq b$, which implies that the atom $\sqbox{$b$}$ removed by $f^{(b)}(w)$ is different from the atom $\sqbox{$a$}$ added by $e^{(a)}(z)$. 
\end{enumerate}

\subsubsubsection{Scenario (1)}

Let us first consider the scenario (1) and (2), in which $a=b$.
Namely we consider the operator
\begin{equation}
\mathcal{O}^{(a)}(z, w)\equiv e^{(a)}(z)\, f^{(a)}(w)-(-1)^{|a|} f^{(a)}(w)\, e^{(a)}(z) \;,
\end{equation}
where the factor $-(-1)^{|a|}$ is to be fixed in terms of the self bond factor $\varphi^{a\Rightarrow a}(u)$.
For the scenario (1), where the atom $\sqbox{$a$}$ removed by $f^{(a)}(w)$ coincides with the atom $\sqbox{$a$}$ added by $e^{(a)}(z)$, we have
\begin{equation}\label{eq-e-f-K-1}
\begin{aligned}
&e^{(a)}(z) \, f^{(a)}(w) \, |\Kappa\rangle\ni \sum_{\sqbox{$a$} \,\in \,\textrm{Rem}(\Kappa)}   
\frac{ E^{(a)}(\Kappa-\sqbox{$a$}\rightarrow\Kappa)}{z-h(\sqbox{$a$})}
\cdot \frac{ F^{(a)}(\Kappa\rightarrow \Kappa -\sqbox{$a$})}{w-h(\sqbox{$a$})}
\,|\Kappa\rangle \,,\\
&f^{(a)}(w) \, e^{(a)}(z) \, |\Kappa\rangle\ni \sum_{\sqbox{$a$} \,\in \,\textrm{Add}(\Kappa)}   
\frac{F^{(a)}(\Kappa +\sqbox{$a$}\rightarrow \Kappa)}{w-h(\sqbox{$a$})}
\cdot \frac{ E^{(a)}(\Kappa \rightarrow \Kappa+\sqbox{$a$})}{z-h(\sqbox{$a$})}\,|\Kappa\rangle \,.
\end{aligned}
\end{equation}
Consider the first equation in (\ref{eq-e-f-K-1}).
For each atom $\sqbox{$a$}$ to be removed, the coefficient is
\begin{equation}\label{eq-EF-K}
\begin{aligned}
&E^{(a)}(\Kappa-\sqbox{$a$}\rightarrow \Kappa)\, F^{(a)}(\Kappa\rightarrow \Kappa-\sqbox{$a$})\\
&=\epsilon(\Kappa-\sqbox{$a$}\rightarrow \Kappa)\cdot \epsilon(\Kappa\rightarrow \Kappa-\sqbox{$a$})
\sqrt{p^{(a)}\textrm{Res}_{u=h(\sqbox{$a$})}\Psi^{(a)}_{\Kappa-\sqbox{$a$}}(u)} \sqrt{q^{(a)}\textrm{Res}_{u=h(\sqbox{$a$})}\Psi^{(a)}_{\Kappa}(u)}\\
&=\left[\epsilon(\Kappa-\sqbox{$a$}\rightarrow \Kappa)\cdot \epsilon(\Kappa\rightarrow \Kappa-\sqbox{$a$})\sqrt{p^{(a)}q^{(a)}\varphi^{a\Rightarrow a}(0)}\right]
\textrm{Res}_{u=h(\sqbox{$a$})}\Psi^{(a)}_{\Kappa}(u) \;,
\end{aligned}
\end{equation}
where we have used that $\varphi^{a\Rightarrow a}(0)^{-1}=\varphi^{a\Rightarrow a}(0)=\pm 1$ in the last step.
If the constant $p^{(a)}$ and $q^{(a)}$ satisfy
\begin{equation}\label{eq-pq-a}
p^{(a)}q^{(a)}\varphi^{a\Rightarrow a}(0)=1\,,
\end{equation}
and if we further demand
\begin{equation}\label{eq-epsilon-1}
\epsilon(\Kappa-\sqbox{$a$}\rightarrow \Kappa)\cdot \epsilon(\Kappa\rightarrow \Kappa-\sqbox{$a$})=1 \;,
\end{equation}
we have
\begin{equation}
E^{(a)}(\Kappa-\sqbox{$a$}\rightarrow \Kappa)\, F^{(a)}(\Kappa\rightarrow \Kappa-\sqbox{$a$})=\textrm{Res}_{u=h(\sqbox{$a$})}\Psi^{(a)}_{\Kappa}(u) \;,
\end{equation}
which implies that the coefficient for the second equation of (\ref{eq-e-f-K-1}) is
\begin{equation}
\begin{aligned}
F^{(a)}(\Kappa+\sqbox{$a$}\rightarrow \Kappa)\, E^{(a)}(\Kappa\rightarrow \Kappa+\sqbox{$a$})&=\textrm{Res}_{u=h(\sqbox{$a$})}\Psi^{(a)}_{\Kappa+\sqbox{$a$}}(u)\\
&=\textrm{Res}_{u=h(\sqbox{$a$})}\Psi^{(a)}_{\Kappa}(u) \cdot \varphi^{a\Rightarrow a}(0)\,.
\end{aligned}
\end{equation}

If further the statistics factor $-(-1)^{|a|}$ is related to the $\varphi^{a\Rightarrow a}(0)$ by
\begin{equation}\label{eq-statistics-a}
-(-1)^{|a|}\, \varphi^{a\Rightarrow a}(0)=1\,,
\end{equation}
we have
\begin{equation}\label{eq-effe-K}
\begin{aligned}
&e^{(a)}(z) \, f^{(a)}(w) \, |\Kappa\rangle -(-1)^{|a|}f^{(a)}(w) \, e^{(a)}(z) \, |\Kappa\rangle\\
&\ni\left(\sum_{\sqbox{$a$}=\textrm{Rem}(\Kappa)} +\sum_{\sqbox{$a$}=\textrm{Add}(\Kappa)}\right)\frac{ \textrm{Res}_{u=h(\sqbox{$a$})}\Psi^{(a)}_{\Kappa}(u)}{(z-h(\sqbox{$a$}))(w-h(\sqbox{$a$}))}|\Kappa\rangle\sim -\frac{\Psi^{(a)}_{\Kappa}(z)-\Psi^{(a)}_{\Kappa}(w)}{z-w}|\Kappa\rangle\,,
\end{aligned}
\end{equation}
when the atom $\sqbox{$a$}$ removed by $f^{(a)}(w)$ coincides with the atom $\sqbox{$a$}$ added by $e^{(a)}(z)$, 
where in the last step we have used 
\begin{equation}
\sum_{p=\textrm{pole}(\Psi_{\Kappa}^{(a)}(u))} \frac{\textrm{Res}_{u=p}\Psi^{(a)}_{\Kappa}(u)}{(z-p)(w-p)}
\sim -\frac{\Psi^{(a)}_{\Kappa}(z)-\Psi^{(a)}_{\Kappa}(w)}{z-w}\,.
\end{equation}

\subsubsubsection{Scenario (2)}

For the scenario (2), i.e.\ when the atom $\sqbox{$a$}$ removed by $f^{(a)}(w)$ is different from the atom $\sqbox{$a$}'$ added by $e^{(a)}(z)$, we have
\begin{equation}
\begin{aligned}
&e^{(a)}(z) \, f^{(a)}(w) \, |\Kappa\rangle\\
&= \sum_{\sqbox{$a$} \,\in \,\textrm{Rem}(\Kappa)}   
\sum_{\sqbox{$a$}' \,\in \,\textrm{Add}(\Kappa-\sqbox{$a$})}
\frac{ E^{(a)}(\Kappa-\sqbox{$a$}\rightarrow\Kappa-\sqbox{$a$}+\sqbox{$a$}' )}{z-h(\sqbox{$a$}')}
\cdot \frac{ F^{(a)}(\Kappa\rightarrow \Kappa -\sqbox{$a$})}{w-h(\sqbox{$a$})}
\,|\Kappa-\sqbox{$a$}+\sqbox{$a$}'\rangle \,,
\end{aligned}
\end{equation}
and
\begin{equation}
\begin{aligned}
&f^{(a)}(w) \, e^{(a)}(z) \, |\Kappa\rangle\\
&= \sum_{\sqbox{$a$}' \,\in \,\textrm{Add}(\Kappa)}   
\sum_{\sqbox{$a$} \,\in \,\textrm{Rem}(\Kappa+\sqbox{$a$}')}
\frac{F^{(a)}(\Kappa +\sqbox{$a$}'\rightarrow \Kappa+\sqbox{$a$}'-\sqbox{$a$})}{w-h(\sqbox{$a$})}
\cdot \frac{ E^{(a)}(\Kappa \rightarrow \Kappa+\sqbox{$a$}')}{z-h(\sqbox{$a$}')}
\,|\Kappa+\sqbox{$a$}'-\sqbox{$a$}\rangle \,,
\end{aligned}
\end{equation}
where we have focused on the generic situation where the addition of the atom $\sqbox{$a$}'$ by $e^{(a)}(z)$ and the removal of the atom $\sqbox{$a$}$ by $f^{(a)}(w)$ do not depend on each other.\footnote{
We have checked that the non-generic situations give the same result.
} 
The ratio of the two coefficients, for the same final state $|\Kappa+\sqbox{$a$}'-\sqbox{$a$}\rangle$, is
\begin{equation}
\begin{aligned}
&\frac{E^{(a)}(\Kappa-\sqbox{$a$}\rightarrow \Kappa-\sqbox{$a$}+\sqbox{$a$}')\cdot F^{(a)}(\Kappa\rightarrow \Kappa-\sqbox{$a$})}{F^{(a)}(\Kappa+\sqbox{$a$}'\rightarrow \Kappa+\sqbox{$a$}'-\sqbox{$a$})\cdot E^{(a)}(\Kappa\rightarrow \Kappa+\sqbox{$a$}')}\\
&=\frac{\epsilon(\Kappa-\sqbox{$a$}\rightarrow \Kappa-\sqbox{$a$}+\sqbox{$a$}')\cdot \epsilon(\Kappa\rightarrow \Kappa-\sqbox{$a$})}{\epsilon(\Kappa+\sqbox{$a$}'\rightarrow \Kappa+\sqbox{$a$}'-\sqbox{$a$})\cdot \epsilon(\Kappa\rightarrow \Kappa+\sqbox{$a$}')}\sqrt{\frac{\textrm{Res}_{u=h(\sqbox{$a$}')}\Psi^{(a)}_{\Kappa-\sqbox{$a$}}(u)}{\textrm{Res}_{u=h(\sqbox{$a$}')}\Psi^{(a)}_{\Kappa}(u)}\cdot \frac{\textrm{Res}_{u=h(\sqbox{$a$})}\Psi^{(a)}_{\Kappa}(u)}{\textrm{Res}_{u=h(\sqbox{$a$})}\Psi^{(a)}_{\Kappa+\sqbox{$a$}'}(u)}} \\
&=\frac{\epsilon(\Kappa-\sqbox{$a$}\rightarrow \Kappa-\sqbox{$a$}+\sqbox{$a$}')\cdot \epsilon(\Kappa\rightarrow \Kappa-\sqbox{$a$})}{\epsilon(\Kappa+\sqbox{$a$}'\rightarrow \Kappa+\sqbox{$a$}'-\sqbox{$a$})\cdot \epsilon(\Kappa\rightarrow \Kappa+\sqbox{$a$}')}=\pm 1\,,
\end{aligned}
\end{equation}
where we have used the reflection property of the bond factor (\ref{eq.varphi_sym}) to reduce the square root factor to $1$.

If we demand that this ratio is related to the statistics factor of $a$ by
\begin{equation}\label{eq-epsilon-2}
\frac{\epsilon(\Kappa-\sqbox{$a$}\rightarrow \Kappa-\sqbox{$a$}+\sqbox{$a$}')\cdot \epsilon(\Kappa\rightarrow \Kappa-\sqbox{$a$})}{\epsilon(\Kappa+\sqbox{$a$}'\rightarrow \Kappa+\sqbox{$a$}'-\sqbox{$a$})\cdot \epsilon(\Kappa\rightarrow \Kappa+\sqbox{$a$}')}=(-1)^{|a|} \;,
\end{equation}
the process in which the atom $\sqbox{$a$}$ removed by $f^{(a)}(w)$ is independent from the atom $\sqbox{$a$}'$ added by $e^{(a)}(z)$ would not contribute to the action of $e^{(a)}(z)\, f^{(a)}(w)-(-1)^{|a|} f^{(a)}(w)\, e^{(a)}(z)$ on $|\Kappa\rangle$.
Namely, the $\ni$ in (\ref{eq-effe-K}) would become an $=$ sign:
\begin{equation}\label{eq-effe-K-final}
\begin{aligned}
&e^{(a)}(z) \, f^{(a)}(w) \, |\Kappa\rangle -(-1)^{|a|}f^{(a)}(w) \, e^{(a)}(z) \, |\Kappa\rangle\\
&=\left(\sum_{\sqbox{$a$}=\textrm{Rem}(\Kappa)} +\sum_{\sqbox{$a$}=\textrm{Add}(\Kappa)}\right)\frac{ \textrm{Res}_{u=h(\sqbox{$a$})}\Psi^{(a)}_{\Kappa}(u)}{(z-h(\sqbox{$a$}))(w-h(\sqbox{$a$}))}|\Kappa\rangle \sim -\frac{\Psi^{(a)}_{\Kappa}(z)-\Psi^{(a)}_{\Kappa}(w)}{z-w}|\Kappa\rangle\,.
\end{aligned}
\end{equation}
Since (\ref{eq-effe-K-final}) is true for arbitrary $|\Kappa\rangle$, we have the relation of the operators:
\begin{equation}\label{eq-ef-a}
e^{(a)}(z) \, f^{(a)}(w)  -(-1)^{|a|}f^{(a)}(w) \, e^{(a)}(z) \sim -  \frac{\psi^{(a)}(z)-\psi^{(a)}(w)}{z-w}   \;.
\end{equation}
We will explain in section~\ref{sec:signs} how to determine the $\epsilon$ function such that (\ref{eq-epsilon-2}) holds.
\subsubsubsection{Statistics of Generators from Crystal}

Before we proceed to the scenario (3), let us first determine the statistics of the generators $\{\psi^{(a)}(z), e^{(a)}(z),f^{(a)}(z)\}$.
The result was already summarized in (\ref{eq.Z2_grading}).

First of all, given the action of $\psi^{(a)}(z)$ on the crystal state $|\Kappa\rangle$ (see \eqref{eq.efpsi}), we see that all the $\psi^{(a)}$ generators are bosonic, since the eigenvalues $\Psi^{(a)}_{\Kappa}(u)$ all commute with each other.

For the operators $e^{(a)}(z)$ and $f^{(a)}(z)$, we have just seen that in order to have (\ref{eq-ef-a}), we need the condition (\ref{eq-statistics-a}), namely
\begin{tcolorbox}[ams equation]\label{eq-statistics-afinal}
(-1)^{|a|}=-\varphi^{a\Rightarrow a}(0) \;,
\end{tcolorbox}
\noindent
where we have used the fact that $\varphi^{a\Rightarrow a}(0)=\pm 1$.
Since $\varphi^{a \Rightarrow a}(0)=(-1)^{\#(\textrm{self-loops of $a$})}$, we conclude that the $e^{(a)}(z)$ and $f^{(a)}(z)$ operators are bosonic (i.e.\ $|a|=0$) when there are odd number of self-loops for the vertex $a$ in the quiver, and fermionic (i.e.\ $|a=1|$) otherwise.
Since quivers in which vertices have 
even positive number of self-loops do not  seem to exist for toric Calabi-Yau threefolds, this proves the grading rule (\ref{eq.Z2_grading}).

In particular, for a vertex $a$ that has no self-loop in the quiver, the corresponding  $e^{(a)}(z)$ and $f^{(a)}(z)$ operators are fermionic.
This is consistent with the intuition that when there is no self-loop for $a$, $e^{(a)}(z)e^{(a)}(w)|\emptyset\rangle=0$, even when we choose the atom at the origin to have color $a$, signaling the fermionic nature of the creation operator $e^{(a)}(z)$ (and therefore for its corresponding annihilation operator).
This is also consistent with the conclusion drawn from the vacuum characters of known examples.
\bigskip

Finally, by the condition (\ref{eq-pq-a}), the constants $p^{(a)}$ and $q^{(a)}$ are also related to the statistics of  $e^{(a)}(z)$ and $f^{(a)}(z)$ operators:
\begin{equation}
p^{(a)}\,q^{(a)}=\varphi^{a\Rightarrow a}(0)=(-1)^{\#(\textrm{self-loops of $a$})}=-(-1)^{|a|}\,.
\end{equation}
We are free to set $q^{(a)}=1$, and have (\ref{eq-p-epsilon}).

\subsubsubsection{Scenario (3)}

Now let us resume with the scenario (3), where $a\neq b$, and consider the generic situation where the addition of $\sqbox{$a$}$ and the removal of $\sqbox{$b$}$ are independent. 
The computation is similar to scenario (2), where the atom $\sqbox{$a$}$ removed by $f^{(a)}(w)$ is independent from the atom $\sqbox{$a$}'$ added by $e^{(a)}(z)$.

Compare the two processes (\ref{eq-e-f-K}) and  (\ref{eq-f-e-K}). 
The ratio between the coefficients at the two sides (for the same final state) is 
\begin{equation}\label{eq-ratio-ee}
\begin{aligned}
&\frac{E^{(a)}(\Kappa-\sqbox{$b$}\rightarrow \Kappa-\sqbox{$b$}+\sqbox{$a$})\cdot F^{(b)}(\Kappa\rightarrow \Kappa-\sqbox{$b$})}{F^{(a)}(\Kappa+\sqbox{$a$}\rightarrow \Kappa+\sqbox{$a$}-\sqbox{$b$})\cdot E^{(a)}(\Kappa\rightarrow \Kappa+\sqbox{$a$})}\\
&=\frac{\epsilon(\Kappa-\sqbox{$b$}\rightarrow \Kappa-\sqbox{$b$}+\sqbox{$a$})\cdot\epsilon(\Kappa\rightarrow \Kappa-\sqbox{$b$})}{\epsilon(\Kappa+\sqbox{$a$}\rightarrow \Kappa+\sqbox{$a$}-\sqbox{$b$})\cdot \epsilon(\Kappa\rightarrow \Kappa+\sqbox{$a$})}\sqrt{\frac{\textrm{Res}_{u=h(\sqbox{$a$})}\Psi^{(a)}_{\Kappa-\sqbox{$b$}}(u)}{\textrm{Res}_{u=h(\sqbox{$a$})}\Psi^{(a)}_{\Kappa}(u)}\cdot \frac{\textrm{Res}_{u=h(\sqbox{$b$})}\Psi^{(b)}_{\Kappa}(u)}{\textrm{Res}_{u=h(\sqbox{$b$})}\Psi^{(b)}_{\Kappa+\sqbox{$a$}}(u)}} \\
&=\frac{\epsilon(\Kappa-\sqbox{$b$}\rightarrow \Kappa-\sqbox{$b$}+\sqbox{$a$})\cdot\epsilon(\Kappa\rightarrow \Kappa-\sqbox{$b$})}{\epsilon(\Kappa+\sqbox{$a$}\rightarrow \Kappa+\sqbox{$a$}-\sqbox{$b$})\cdot \epsilon(\Kappa\rightarrow \Kappa+\sqbox{$a$})}=\pm 1 \;,
\end{aligned}
\end{equation}
where again we have used the reflection property of the bond factor (\ref{eq.varphi_sym}) to reduce the square root factor to $1$.

Demanding 
\begin{equation}\label{eq-epsilon-3}
\frac{\epsilon(\Kappa-\sqbox{$b$}\rightarrow \Kappa-\sqbox{$b$}+\sqbox{$a$})\cdot\epsilon(\Kappa\rightarrow \Kappa-\sqbox{$b$})}{\epsilon(\Kappa+\sqbox{$a$}\rightarrow \Kappa+\sqbox{$a$}-\sqbox{$b$})\cdot \epsilon(\Kappa\rightarrow \Kappa+\sqbox{$a$})}=(-1)^{|a||b|} \;,
\end{equation}
we have
\begin{equation}
a\neq b: \qquad e^{(a)}(z) \, f^{(b)}(w) \, |\Kappa\rangle -(-1)^{|a||b|}f^{(b)}(w) \, e^{(a)}(z) \, |\Kappa\rangle=0  \;,
\end{equation}
for an arbitrary state $|\Kappa\rangle$, which gives
\begin{equation}\label{eq-ef-aneqb}
a\neq b: \qquad e^{(a)}(z) \, f^{(b)}(w)  -(-1)^{|a||b|}f^{(b)}(w) \, e^{(a)}(z) \sim 0  \;.
\end{equation} 
Together with the result for $a= b$ in (\ref{eq-ef-a}), we have
\begin{equation}\label{eq-ef-ab-final}
e^{(a)}(z) \, f^{(b)}(w)  -(-1)^{|a||b|}f^{(b)}(w) \, e^{(a)}(z) \sim -  \delta^{a,b}\,\frac{\psi^{(a)}(z)-\psi^{(b)}(w)}{z-w} \,,
\end{equation}
given in the last equation of (\ref{eq-OPE-toric}).
\bigskip

Finally, we emphasize that to reach (\ref{eq-ef-ab-final}), we have demanded the $\epsilon$ function to satisfy various constraints, i.e.\ (\ref{eq-epsilon-1}), (\ref{eq-epsilon-2}), and (\ref{eq-epsilon-3}).
We will solve these constraints, together with two more coming from the $e-e$ and $f-f$ relations, after we fix the algebra.

\subsubsection{\texorpdfstring{$e-e$ and $f-f$ Relations}{e-e and f-f Relations}}

The computation for the $e-e$ and $f-f$ relations are similar to the ones for the scenario (2) and (3) of the $e-f$ relation.
First, use the action of $e^{(a)}$ on arbitrary $|\Kappa\rangle$, given in (\ref{eq.efpsi}), we have
\begin{equation}\label{eq-e-e-K}
\begin{aligned}
&e^{(a)}(z) \, e^{(b)}(w) \, |\Kappa\rangle\\
&= \sum_{\sqbox{$b$} \,\in \,\textrm{Add}(\Kappa)}   
\sum_{\sqbox{$a$} \,\in \,\textrm{Add}(\Kappa+\sqbox{$b$})}
\frac{  E^{(a)}(\Kappa+\sqbox{$b$}\rightarrow \Kappa+\sqbox{$b$}+\sqbox{$a$})}{z-h(\sqbox{$a$})}
\cdot \frac{  E^{(b)}(\Kappa\rightarrow \Kappa+\sqbox{$b$})}{w-h(\sqbox{$b$})}
\,|\Kappa+\sqbox{$b$}+\sqbox{$a$}\rangle \,,
\end{aligned}
\end{equation}
and
\begin{equation}\label{eq-e-e-K-2}
\begin{aligned}
&e^{(b)}(w) \, e^{(a)}(z) \, |\Kappa\rangle\\
&= \sum_{\sqbox{$a$} \,\in \,\textrm{Add}(\Kappa)}   
\sum_{\sqbox{$b$} \,\in \,\textrm{Add}(\Kappa+\sqbox{$a$})}
\frac{E^{(a)}(\Kappa+\sqbox{$a$}\rightarrow \Kappa+\sqbox{$a$}+\sqbox{$b$})}{w-h(\sqbox{$b$})}
\cdot \frac{ E^{(a)}(\Kappa\rightarrow \Kappa+\sqbox{$a$})}{z-h(\sqbox{$a$})}
\,|\Kappa+\sqbox{$a$}+\sqbox{$b$}\rangle \,.
\end{aligned}
\end{equation}
Consider the generic situation\footnote{
The non-generic situation where the adding of $\sqbox{$a$}$ requires the adding of $\sqbox{$b$}$ first need to be discussed within concrete examples. 
We have checked that the result remains the same.} where the creation of the atom $\sqbox{$a$}$ by $e^{(a)}(z)$ and the creation of the atom $\sqbox{$b$}$ by $e^{(b)}(w)$ do not depend on each other. 
In such cases, the ratio between the coefficients in (\ref{eq-e-e-K}) and (\ref{eq-e-e-K-2}) is
\begin{equation}
\begin{aligned}
&\frac{E^{(a)}(\Kappa+\sqbox{$b$}\rightarrow \Kappa+\sqbox{$b$}+\sqbox{$a$})\cdot E^{(b)}(\Kappa\rightarrow \Kappa+\sqbox{$b$})}{E^{(a)}(\Kappa+\sqbox{$a$}\rightarrow \Kappa+\sqbox{$a$}+\sqbox{$b$})\cdot E^{(a)}(\Kappa\rightarrow \Kappa+\sqbox{$a$})}\\
&=\frac{\epsilon(\Kappa+\sqbox{$b$}\rightarrow \Kappa+\sqbox{$b$}+\sqbox{$a$})\cdot\epsilon(\Kappa\rightarrow \Kappa+\sqbox{$b$})}{\epsilon(\Kappa+\sqbox{$a$}\rightarrow \Kappa+\sqbox{$a$}+\sqbox{$b$})\cdot \epsilon(\Kappa\rightarrow \Kappa+\sqbox{$a$})}\sqrt{\frac{\textrm{Res}_{u=h(\sqbox{$a$})}\Psi^{(a)}_{\Kappa+\sqbox{$b$}}(u)}{\textrm{Res}_{u=h(\sqbox{$a$})}\Psi^{(a)}_{\Kappa}(u)}\cdot \frac{\textrm{Res}_{u=h(\sqbox{$b$})}\Psi^{(b)}_{\Kappa}(u)}{\textrm{Res}_{u=h(\sqbox{$b$})}\Psi^{(b)}_{\Kappa+\sqbox{$a$}}(u)}} \;.
\end{aligned}
\end{equation}
The square root factor gives
\begin{equation}
\begin{aligned}
&\sqrt{\varphi^{b\Rightarrow a}(h(\sqbox{$a$})-h(\sqbox{$b$}))\cdot \frac{1}{\varphi^{a\Rightarrow b}(h(\sqbox{$b$})-h(\sqbox{$a$}))}}=
\varphi^{b\Rightarrow a}(h(\sqbox{$a$})-h(\sqbox{$b$}))\sim  \varphi^{b\Rightarrow a}(z-w) \,,
\end{aligned}
\end{equation}
where in the first step we have used the reflection property of the bond factor (\ref{eq.varphi_sym}), and in the second step we have used the fact that in (\ref{eq-e-e-K}) and (\ref{eq-e-e-K-2}), $z\rightarrow h(\sqbox{$a$})$ and $w\rightarrow h(\sqbox{$b$})$ and we only care about terms $\sim z^{-n-1}w^{-m-1}$ with $n,m\in\mathbb{Z}_{\geq 0}$. 
For each $\Kappa$, $\sqbox{$a$}$, and $\sqbox{$b$}$, the $\epsilon$ factors should be chosen such that 
\begin{equation}\label{eq-epsilon-4}
\frac{\epsilon(\Kappa+\sqbox{$b$}\rightarrow \Kappa+\sqbox{$b$}+\sqbox{$a$})\cdot\epsilon(\Kappa\rightarrow \Kappa+\sqbox{$b$})}{\epsilon(\Kappa+\sqbox{$a$}\rightarrow \Kappa+\sqbox{$a$}+\sqbox{$b$})\cdot \epsilon(\Kappa\rightarrow \Kappa+\sqbox{$a$})}=(-1)^{|a||b|}\,.
\end{equation}
Namely, the sign should be $-$ when both $e^{(a)}$ and $e^{(b)}$ are fermions, and $+$ otherwise. 
In summary we have
\begin{equation}
e^{(a)}(z)\, e^{(b)}(w)\sim (-1)^{|a| \cdot|b|}\varphi^{b\Rightarrow a}(z-w)\, e^{(b)}(w)\, e^{(a)}(z)\,,
\end{equation}
as shown in the third equation in (\ref{eq-OPE-toric}).

Finally, a parallel derivation gives
\begin{equation}
f^{(a)}(z)\, f^{(b)}(w)\sim (-1)^{|a| \cdot|b|}\varphi^{b\Rightarrow a}(z-w)^{-1}\, f^{(b)}(w)\, f^{(a)}(z)\,,
\end{equation}
as shown in the fifth equation of (\ref{eq-OPE-toric}).
The constraint on $\epsilon$ needed for the $f-f$ relation is   
\begin{equation}\label{eq-epsilon-5}
\begin{aligned}
&\frac{\epsilon(\Kappa-\sqbox{$b$}\rightarrow \Kappa-\sqbox{$b$}-\sqbox{$a$})\cdot\epsilon(\Kappa\rightarrow \Kappa-\sqbox{$b$})}{\epsilon(\Kappa-\sqbox{$a$}\rightarrow \Kappa-\sqbox{$a$}-\sqbox{$b$})\cdot \epsilon(\Kappa\rightarrow \Kappa-\sqbox{$a$})}=(-1)^{|a||b|}\;.
\end{aligned}
\end{equation}

\subsection{\texorpdfstring{Prescription for Choice of $\epsilon$}{Prescription for Choice of epsilon}}
\label{sec:signs}

In deriving the algebraic relations (\ref{eq-OPE-toric}) from the ansatz of the action (\ref{eq-ansatz-action-toric}), we have demanded the five conditions on the $\epsilon$ signs, namely (\ref{eq-epsilon-1}), (\ref{eq-epsilon-2}), (\ref{eq-epsilon-3}), (\ref{eq-epsilon-4}), and (\ref{eq-epsilon-5}).
Now we need to show that there always exists an assignment for $\epsilon$ such that this set of five equations hold.

First of all, these five equations are not all independent. 
First, the condition (\ref{eq-epsilon-2}) is merely a specialization of (\ref{eq-epsilon-3}).
Second, since all $\epsilon=\pm 1$, (\ref{eq-epsilon-1}) can be rewritten into the reciprocity condition 
\begin{equation}\label{eq-epsilon-RC}
\epsilon(\Kappa\rightarrow \Kappa +\sqbox{$a$})=\epsilon(\Kappa +\sqbox{$a$}\rightarrow \Kappa) \;,
\end{equation}
using which we can show that of the three equations (\ref{eq-epsilon-3}), (\ref{eq-epsilon-4}), and (\ref{eq-epsilon-5}), only one is independent. 
For example, we can use (\ref{eq-epsilon-RC}) to reverse the directions of various processes in (\ref{eq-epsilon-3}) and (\ref{eq-epsilon-5}), thus bringing both of them into the form of (\ref{eq-epsilon-4}), with the new initial state being $|\Kappa-\sqbox{$b$}\rangle$ for (\ref{eq-epsilon-3}) and $|\Kappa-\sqbox{$a$}-\sqbox{$b$}\rangle$ for (\ref{eq-epsilon-5}). 
Therefore we only need to impose (\ref{eq-epsilon-RC}) and (\ref{eq-epsilon-4}), the latter of which we repeat here:
\begin{equation}\label{eq-epsilons}
\begin{aligned}
&\frac{\epsilon(\Kappa+\sqbox{$b$}\rightarrow \Kappa+\sqbox{$b$}+\sqbox{$a$})\cdot\epsilon(\Kappa\rightarrow \Kappa+\sqbox{$b$})}{\epsilon(\Kappa+\sqbox{$a$}\rightarrow \Kappa+\sqbox{$a$}+\sqbox{$b$})\cdot \epsilon(\Kappa\rightarrow \Kappa+\sqbox{$a$})}=(-1)^{|a||b|} \;.\\
\end{aligned}
\end{equation}

Given the reciprocity condition (\ref{eq-epsilon-RC}), we can simply assign the value of the $\epsilon$ function for each adding process $\Kappa \rightarrow \Kappa +\sqbox{$a$}$ iteratively, starting from the vacuum $\Kappa=\emptyset$. 
The value for the $\epsilon$ function for a removing process is taken to be identical to the one for the corresponding adding process, due to (\ref{eq-epsilon-RC}).
In this iterative assigning process, one only needs to observe (\ref{eq-epsilons}), which is a condition that is associated to the ``faces" of the adding diagram.
But since we are starting from the atom at the origin and adding atoms according to a (two-dimensional) periodic quiver, the condition (\ref{eq-epsilons}) is very easy to satisfy.
For example, we can choose the $\epsilon$ for the first few processes to be $+$, and for the new adding processes switch the sign whenever demanded by (\ref{eq-epsilons}), and since this is an iterative process, the sign assignment demanded by (\ref{eq-epsilons}) is always pushed to the outskirt of the adding diagram, such that we are always free to assign $\epsilon$ to whatever value that satisfy (\ref{eq-epsilons}). 
Thus we conclude that one can always fix a prescription of the $\epsilon$ function in the ansatz (\ref{eq-ansatz-action-toric}) such that by this action, the set of colored crystals furnishes a representation of the quiver Yangian algebra, whose algebraic relations are bootstrapped from the (\ref{eq-ansatz-action-toric}) and the quiver data.

\section{Truncations of Quiver Yangians and D4-branes}\label{sec:truncation}

\subsection{Truncations of Quiver Yangians}

The representation we constructed in the previous section is generically a cyclic module of the algebra, since we can arrive at any molten crystal configuration starting with the empty room (i.e.\ the vacuum) and applying a finite number of creation operators $e^{(a)}(u)$. 
Conversely, starting from any molten crystal configuration we can arrive at the vacuum by appropriately applying a finite number of annihilation operators $f^{(a)}(u)$.
This ensures that the representation is irreducible.

As explained in section \ref{sec:parameters}, the algebra associated to the quiver $(Q,W)$ has $|Q_1|$ coordinate parameters $\{h_I\}$, corresponding to the $|Q_1|$ edges of the quiver diagram. 
After the loop constraint (\ref{eq-loop-constraint-toric}) is imposed, they reduce to $|Q_0|+1=E+2I-1$ independent parameters $\{\mathsf{h}_A\}$.
In this section we will show that the representation can become reducible when the coordinate parameters $\{h_I \}$ (or more precisely $\{\mathsf{h}_A\}$) take certain fine-tuned values, causing the residue of the charge function $\textrm{Res}_{u=h(\sqbox{$a$})} \Psi_{\Kappa}^{(a)}(u)$ in \eqref{eq.efpsi} to vanish for some atom $\sqbox{$a$}$. 
In this case, it is impossible to add this atom  (and hence all subsequent ones) to the crystal, and consequently this stops the growth of the crystal for the part beyond this atom. 
The representation is then no longer irreducible.

Since we have motivated the definition of the algebra $\mathsf{Y}_{(Q,W)}$ by its action on the crystal, it is natural to translate the truncation of the growth of the crystal into a truncation of the algebra.
Namely, when the coordinate parameters take certain fine-tuned values, the algebra develops an ideal, quotienting out which gives the truncation of the algebra.  
The representation that is reducible with respect to the original algebra becomes irreducible in the truncated algebra.

Now we will show that the special values for $\{h_I\}$ that characterize the truncation of the algebra are defined by certain linear equations with integer coefficients $\vec{N}$. 
We denote the corresponding truncated algebra by $\mathsf{Y}^{\vec{N}}_{(Q,W)}$.\footnote{
For earlier discussions on truncations of the affine Yangian of $\mathfrak{gl}_1$, see \cite{Fukuda:2015ura, Prochazka:2015deb, Gaiotto:2017euk, Harada:2018bkb}. 
For truncations of $\mathcal{W}$ algebras that are related to some of the quiver Yangians constructed in this paper, see \cite{Prochazka:2017qum, Rapcak:2019wzw}.}
Moreover we find that the (linear combination of) integers $\vec{N}$ corresponds to the number of D4-branes.\bigskip

Suppose the growth of the crystal stops at an atom $\sqbox{$a$}$ of color $a$.
Let us express its coordinate function as
\begin{equation}\label{eq-truncation-coordinate-function}
h(\sqbox{$a$})= \sum_{I\, \in\,  \textrm{path}[\mathfrak{o}\rightarrow \sqbox{$a$}\,]} h_I  =\sum_{I\in Q_1} N_I \, h_I 
=\sum^{|Q_0|+1}_{A=1} \mathsf{N}_A \, \mathsf{h}_A \,,
\end{equation}
where $N_{I}\in \mathbb{Z}_{\geq 0}$ and $\mathsf{N}_A\in \mathbb{Z}$; in the last step we have used the loop constraint (\ref{eq-loop-constraint-toric}) to reduce the parameters $\{h_I\}$ to the independent ones $\{\mathsf{h}_A\}$.
Note that since the edges have fixed positions on the quiver, on each path from $\mathfrak{o}$ to $\sqbox{$a$}$, only certain $h_I$ appear. 
As a result, the non-negative integers $N_I$ do not take arbitrary values in $\mathbb{Z}_{\ge 0}$, and $\mathsf{N}_A$ do not take arbitrary values in $\mathbb{Z}$.

When we add this atom to the initial state $\Kappa$, the numerical coefficient is
\begin{equation}\label{eq-truncation-residue}
\left(\textrm{Res}_{u=h(\sqbox{$a$})}\, \Psi^{(a)}_{\Kappa}(u)\right)^{\frac{1}{2}} = \left(\textrm{Res}_{u=h(\sqbox{$a$})} \, \psi^{(a)}_0(u)\, \psi^{(a)}_{\Kappa} (u)\right)^{\frac{1}{2}} \,,
\end{equation} 
where in the last step we have extracted out the vacuum part of the charge function
\begin{equation}
\psi_0^{(a)}(u)=\left(1+\frac{C}{u}\right)^{\delta_{a,1}}\,.
\end{equation}
(Recall that we label the color of the atom at the origin of the crystal to be $a=1$, and the vacuum only contributes to the charge function of color $a=1$.)

Only the atoms on the surface of $\Kappa$ contribute to the residue in (\ref{eq-truncation-residue}).
In particular, to add the atom $\sqbox{$a$}$ , we need to consider all the paths from the origin $\mathfrak{o}$ to the atom $\sqbox{$a$}$: for the atom $\sqbox{$a$}$ to be added, all the atoms right before the atom $\sqbox{$a$}$ in these paths need to be already present. 
However, none of the contributions from these penultimate atoms on the surface of $\Kappa$ contain all the information of $\{N_I\}$, since it is the difference between their coordinate functions and the coordinate function $h(\sqbox{$a$})$ in (\ref{eq-truncation-coordinate-function}) that enters the residue (\ref{eq-truncation-residue}). 
Instead, when $a=1$, the contribution from the vacuum part of the charge function $\psi_0(u)$ contains all $\{N_I\}$:
\begin{equation}
\textrm{Res}_{u=h(\sqbox{$a$})} \psi_0^{(a)}(u)\sim \sum_{I\in Q_1} N_I \, h_I+C \qquad \textrm{for}\quad a=1\,,
\end{equation}
where $\sim$ means that we only take into account of the numerator in $\psi_0(u)$ here.

The condition for the algebra to truncate at the level $\{N_I\}$, i.e.\ for the growth of the crystal to stop beyond the atom $\sqbox{$a$}$ with $a=1$ and at the position defined in (\ref{eq-truncation-coordinate-function}), is that 
\begin{tcolorbox}[ams equation]\label{eq-truncation-condition}
\sum_{I\in Q_1} N_I \, h_I+ C=0\,.
\end{tcolorbox}
\noindent One can use these non-negative integers $\{N_I\}$ to label the truncation of the algebra.

As we have already described, only $|Q_0|+1$ out of the 
$|Q_1|$ non-negative integers $\{N_I\}$ are independent.
This is realized by the fact that, due to the loop constraint (\ref{eq-loop-constraint-toric}), the truncation condition (\ref{eq-truncation-condition}) is invariant under the shift
\begin{equation}
N_I \rightarrow N_I +n \qquad \forall I \in L  \;,
\end{equation}
where $L$ is any loop in the periodic quiver. We can use these shifts to 
obtain $|Q_0|+1$ non-negative integers, whose linear combinations map to the $|Q_0|+1$ integers $\mathsf{N}_A$.

\subsection{Multiple Truncations and Rational Algebras}

An important motivation to consider truncations of the algebra is to obtain ``rational" versions of the algebra, namely the quiver Yangian analogue of rational $\mathcal{W}$ algebras, which has only finitely many irreducible representations. 

The truncation condition (\ref{eq-truncation-condition}) is one condition on the charge parameters $h_I$, imposed by the fact that the growth of the crystal stops at one particular atom, labeled by $\{N_I\}$.
It is possible for the truncation of the growth of the crystal to happen at multiple locations, each characterized by integers $\{N_{i,I}\}$, where $i$ labels the different obstructing atoms. 
The truncation condition (\ref{eq-truncation-condition}) is then enhanced to
\begin{equation}
\sum_{I\in Q_1} N_{i,I} \, h_I+ C=0\qquad i=1,\dots,T\,.
\end{equation}
where $T$ is the number of ``obstructions".
\bigskip

Consider the simplest quiver Yangian: the affine Yangian of $\mathfrak{gl}_1$.
Let us demand that the growth of the crystal, in this case the plane partition where all atoms have the same color $a=1$, stops at an atom with position $(x_1,x_2,x_3)=(N_1,N_2,N_3)$. Correspondingly the parameter $\{h_i\}$ with $i=1,2,3$ must satisfy
\begin{equation}\label{eq-truncation-gl1}
\sum^3_{i=1}\, N_i\, h_i +(h_1h_2 h_3 \psi_0)=0
\end{equation}
up to the loop constraint $h_1+h_2+h_3=0$ and the scaling freedom $(h_i,\psi_0)\rightarrow (\alpha h_i, \alpha^{-2}\psi_0)$.

Now suppose the growth of the plane partition happens at two positions:\footnote{
For earlier discussions on double truncations of the affine Yangian of $\mathfrak{gl}_1$, see \cite{Fukuda:2015ura, Prochazka:2015deb, Harada:2018bkb}.}
\begin{equation}\label{eq-two-obstructions-gl1}
(N_1,N_2,N_3)=(0,0,N) \qquad \textrm{and}\qquad (N_1,N_2,N_3)=(k, k+1,0)\,.
\end{equation}
The first atom effectively truncates the plane partition along the $x_3$ direction with the $x_3=N$ plane, which ensures that no box can be added with $x_3\geq N$.
The second atom acts as a ``pit" on the $x_1-x_2$ plane, at position $(x_1,x_2)=(k, k+1)$, which means that no box can be added with $x_1\geq k$ and $x_2\geq k+1$.

Recall that the representation of the affine Yangian of $\mathfrak{gl}_1$ is labeled by the three Young diagrams $(\lambda_1,\lambda_2,\lambda_3)$ as the asymptotic along the $(x_1,x_2,x_3)$ directions.
When both of these obstructing atoms are present, the three Young diagrams $(\lambda_1,\lambda_2,\lambda_3)$ cannot take arbitrary values anymore. 
First of all, the presence of the cutoff along $x_3=N$ means there is no non-trivial asymptotic along the $x_3$ direction, namely $\lambda_3=\emptyset$.
Moreover, it also means that the heights of both $\lambda_1$ and $\lambda_2$ cannot exceed $N$.
Lastly, the presence of the ``pit" at $(x_1,x_2)=(k, k+1)$ means that the width of $\lambda_1$ cannot exceed $k$ and that of 
$\lambda_2$ cannot exceed $k+1$.
Therefore, there are only finitely many representations, suggesting that the corresponding algebra is rational. 

One can check this by direct computation of the $\{h_i\}$ parameter of the affine Yangian of $\mathfrak{gl}_1$. 
Solving the double truncation condition (\ref{eq-truncation-gl1}) with the $\{N_i\}$ taking the two triplets in (\ref{eq-two-obstructions-gl1}), we get
\begin{equation}\label{eq-h123-Nk}
h_1 =  -\sqrt{\frac{N+k+1}{N+k}} \ , \quad h_2 =  \sqrt{\frac{N+k}{N+k+1}} \ , \quad h_3 = \frac{1}{\sqrt{(N+k)(N+k+1)}} \ ,
\end{equation}
together with 
\begin{equation}
\psi_0=N\,,
\end{equation}
up to the scaling freedom $(h_i,\psi_0)\rightarrow (\alpha \,h_i, \alpha^{-2}\,\psi_0)$.\footnote{
To compare with the literature, here we are using the mode expansion (\ref{eq-mode-expansion-gl1_bootstrap}) adopted in \cite{Prochazka:2015deb, Gaberdiel:2017dbk}, instead of  (\ref{eq-mode-expansion-no4}), which is universal for all quiver Yangians of Calabi-Yau threefolds without compact $4$-cycles. 
Had we adopted the convention (\ref{eq-mode-expansion-no4}), the solution for $\psi_0$ would have been $\psi_{0}=-\frac{N}{\sqrt{(N+k)(N+k+1)}}$ (while the solutions for $h_i$ remain unchanged), and up to the scaling freedom $(h_i,\psi_0)\rightarrow (\alpha\, h_i, \alpha\,\psi_0)$.} 
This is precisely the values of $\{h_{i}\}$ and $\psi_0$ obtained by a direct translation between the affine Yangian of $\mathfrak{gl}_1$ and the $\mathcal{W}_{N,k}$ algebra in \cite{Gaberdiel:2017dbk}, where $N,k$ are both positive integers (or one of the $\mathcal{S}_3$ image of the ``triality symmetry" of the $\mathcal{W}_{N,k}$ algebra \cite{Gaberdiel:2012ku}).

Note that to obtain a rational algebra, only the first condition in (\ref{eq-two-obstructions-gl1}) is necessary: one can relax the second condition by choosing
\begin{equation}\label{eq-N123-2-2}
(N_1,N_2,N_3)=(k, k+m, 0)
\end{equation}
with $m\in \mathbb{N}$.
The coupled equation (\ref{eq-truncation-gl1}) whose second one having coefficient in the form (\ref{eq-N123-2-2}) can be brought back to the one with coefficients in the forms of (\ref{eq-two-obstructions-gl1}) but with $k$ non-integer:
\begin{equation}\label{eq-kprime}
 (k, k+m, 0) \qquad \longrightarrow \qquad 
 (k',k'+1,0) \quad \textrm{with}\quad k'=\frac{k+N}{m}-N
\end{equation}
This corresponds to a rational $\mathcal{W}_{N,k}$ algebra with $k$ no longer an integer, but a rational number of the form $\frac{k+N}{m}-N$ with $N, k, m\in\mathbb{N}$.
(Moreover, one can check that the $(N,k')$ pair from (\ref{eq-kprime}) is not an triality image from an integer pair $(M,k)$.\footnote{
The two generators of the $\mathcal{S}_3$ symmetry (so-called ``triality symmetry") are $(N,k)\rightarrow (N,-2N-k-1)$ and $(N,k)\rightarrow (\frac{N}{N+k}, \frac{1-N}{N+k})$ \cite{Gaberdiel:2012ku}.})
These are precisely the admissible (non-integer) levels for the rational $\mathcal{W}_{N,k}$ algebra when $p\equiv N+k$ and $p'\equiv k+N+m$ are coprime \cite{MathieuWalton90}.\footnote{
The map of the parameters $(N, k, m)$ to those of the non-unitary $\mathcal{W}_{N}(p, p')$ minimal model (with $\textrm{gcd}(p, p')=1$) in \cite{MathieuWalton90} is $p=N+k, p'=k+N+m$.}
We have just obtained them by an easily-visualizable truncation of the affine Yangian of $\mathfrak{gl}_1$.
\bigskip

This can be generalized to all quiver Yangian algebras of this paper.
Namely, for each quiver Yangian algebra, one can study its multiple truncations and use them to obtain the ``rational" version of the algebra, whose number of irreducible representations becomes finite.
The procedure is actually easier than the one for the corresponding $\mathcal{W}$ algebra.
The rational $\mathcal{W}$ algebras usually belong to a family of generically irrational $\mathcal{W}$ algebras; when the parameters of the family take specific values, enough null vectors arise and the algebra becomes rational. 
Locating such rational points requires an analysis of the null vector structure and needs to be done case by case for each family. 
In contrast, the truncations of the quiver Yangian algebras follow a universal mechanism which is easy to visualize and to classify.
One can use the truncation of the quiver Yangian algebras to find new rational $\mathcal{W}$ algebras.

\subsection{Relation with D4-branes}

We now claim that these non-negative integers correspond to the number of D4 branes wrapping the $4$-cycles (divisors) in the Calabi-Yau threefold. 
These $4$-cycles can be either compact or non-compact. 
In short, adding D4-branes corresponds to truncating the algebra.

In order to see this, we first need to see the effect of the D4-branes to the supersymmetric quiver quantum mechanics (see \cite{Franco:2006es, Imamura:2008fd} for discussions
in the context of brane tilings\footnote{
Since we are discussing quiver quantum mechanics and not four-dimensional quiver gauge theories, we need to dimensionally-reduce the setup. 
For example, a D3-brane probing the toric Calabi-Yau threefold is turned into a D0-brane probe in our context.}).
Let us first recall that in the absence of the D4-brane we have an effective D0-brane quantum mechanics.
This D0-brane probes the geometry of the toric Calabi-Yau threefold, and hence the vacuum moduli space of the quiver quantum mechanics (when the gauge group is Abelian) reproduces the geometry of the toric Calabi-Yau threefold.

When the D4-brane wraps a non-compact $4$-cycle we have a non-dynamical gauge symmetry on it, whereas for a compact $4$-cycle a dynamical gauge symmetry appears. 
In either case, from the viewpoint of the D0-brane quantum mechanics the D4-brane looks like a flavor brane.

The divisors in question are regions of the $(p,q)$ 5-brane webs. 
This is also in one-to-one correspondence with a lattice point of the toric diagram. 
Since we have denoted the number of external (internal) lattice points by $E$ ($I$), we have $E$ non-compact ($I$ compact) D4-brane divisors.

When we include D4-branes, we need to include strings connecting D0-brane to the D4-brane, which gives a pair of the quark chiral multiplet $q$ and the anti-quark chiral multiplet $\tilde{q}$.
They couple to one of the bifundamental fields $\Phi_I$ of the D0-brane quiver quantum mechanics, with superpotential
\begin{align}
W= \tilde{q} \, \Phi_I \, q \;.
\end{align}

Which bifundamental field do we get? 
To answer this, it is useful to take T-duality twice, so that both the D0-brane and the D4-brane are turned into D2-branes \cite{Imamura:2008fd}.
We then have a brane configuration consisting of D2-branes and an NS5-brane (see \cite{Yamazaki:2008bt} for a detailed analysis), which gives a physical realization of the brane tilings and the periodic quiver.

Let us consider the situation where the flavor D4-brane (which is now a flavor D2-brane) is associated with a non-compact region corresponding to the corner external vertex of the toric diagram. 
One then finds that the D2-brane in the D2/NS5 brane configuration is sandwiched between two asymptotic NS5-brane cylinders,
which are related by string duality to two asymptotic lines of $(p,q)$-webs surrounding D4-brane region.
The string at the intersection of the two NS5-branes gives rise to a bifundamental chiral multiplet, which  can be identified with the bifundamental field $\Phi_I$ in question.

When we include the D4-brane, the bifundamental chiral multiplet $\Phi_I$ will in general have a VEV (Vacuum Expectation Value), and this gives masses to the quarks. 
This means that the probe D0-brane and the flavor D4-brane are separate. 
In order to identify the D4-brane divisor, one therefore needs to probe the locus where the VEV of the chiral multiplet vanishes:
$\Phi_I=0$. Since we have one complex equation, we could expect a divisor.

\begin{figure}[htbp]
\centering\includegraphics[scale=0.22]{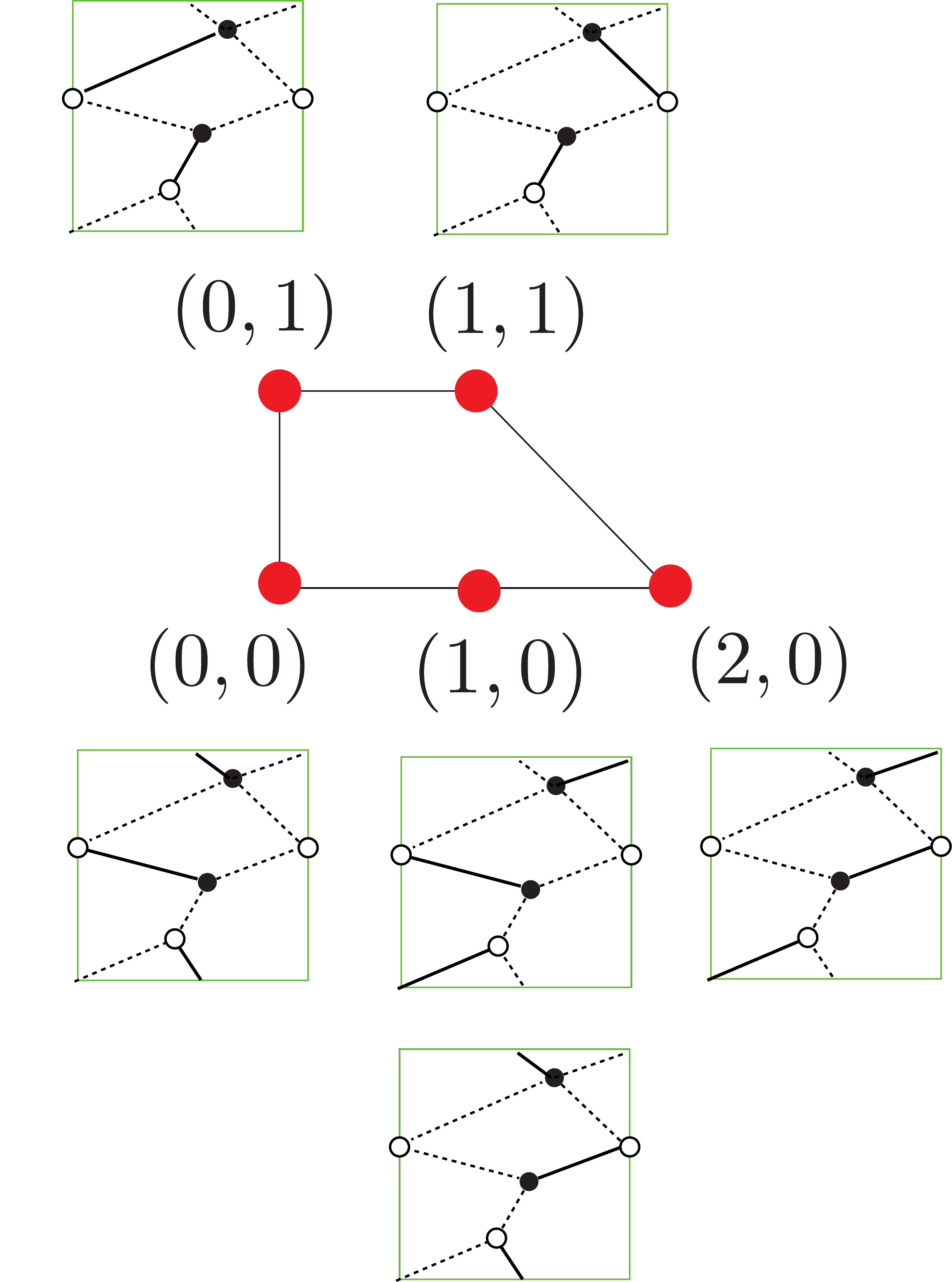}
\caption{The perfect matchings for the Suspended Pinched Point geometry of Figure \ref{fig.SPP_quiver},
whose bipartite graph was shown in Figure \ref{fig.SPPbipartite}.
There are six perfect matchings, each of which is associated with one of the five lattice points of the toric diagram (with multiplicity two for the lattice point $(1,0)$); this association is determined by the height function, as explained in the main text.}
\label{fig.SPPPM}
\end{figure}

While $\Phi_I=0$ is a legitimate equation, one needs to remember that we need to take into account the F-term relations arising from the derivatives of the superpotential. 
One systematic approach is to solve the F-term equations first, and then impose the condition $\Phi_I=0$. 
This process is helped greatly by the fact that the F-term equations can be solved by a set of fields $\tilde{\Phi}_p$ associated with perfect matchings $p$ of the dimer model \cite{Franco:2005rj}.
Here a perfect matching refers to a subset of the edges of the bipartite graph such that any vertex of the bipartite graph is contained in exactly one edge (see Figure \ref{fig.SPPPM}).
Since the periodic quiver is the dual of the bipartite graph, this means that a perfect matching can be regarded as a subset of $Q_1$, the set of arrows of the quiver.
The relation between $\Phi_I$ and $\tilde{\Phi}_p$ can now be stated as
\begin{align}
\Phi_I =\prod_{p \ni I} \tilde{\Phi}_p \;.   
\end{align}
This means that the divisor $\{ \Phi_I=0 \}$ can now be regarded as the union of the submanifolds $\{ \tilde{\Phi}_p=0 \}$.

Now, for each perfect matching we can associate a lattice point of the toric diagram (in general this can either
be on the boundary or inside of the toric diagram), see Figure \ref{fig.SPPPM} for an example.
This is determined by the so-called height function introduced in the dimer model literature --- one chooses one of the perfect matchings as a reference matching,
and when this is superimposed with another perfect matching we have a set of closed paths on the torus. The total winding numbers of the resulting 
paths, labeled by two integers corresponding to winding in $\alpha$ and $\beta$ cycles of the two-dimensional torus, 
determine the corresponding lattice point of the toric diagram. (One can show that the resulting toric diagram is independent
of the choice of the reference perfect matching, up to a $\mathrm{GL}(2, \mathbb{Z})$-transformation on the toric diagram.)

In general multiple perfect matchings can be associated with the same lattice point.
In this paper we consider the case of D4-branes associated with a corner lattice point of the toric diagram.
In this case, it is known that there is a unique perfect matching $p$ corresponding to the lattice point (cf.\ \cite{Hanany:2005ss}),
and one can show \cite[Theorem 2]{Imamura:2008fd} that the D4-brane divisor can be identified with the locus $\{ \tilde{\Phi}_p=0 \}$ associated with that perfect matching $p$.
In this locus $\{ \tilde{\Phi}_p=0 \}$, we set all the bifundamental fields belonging to the perfect matching to zero. 
Such a truncation for the BPS crystal melting model was discussed previously in \cite{Nishinaka:2013mba} (see also \cite{Nishinaka:2010qk, Nishinaka:2011sv}). 

For the present purpose of identifying the number of D4-branes, when we consider a D4-brane wrapping the divisor $\{ \tilde{\Phi}_p=0 \}$, we impose the condition
\begin{align}
h_I= 0 \quad \textrm{when} \quad I\in p \;.
\end{align}
This leaves a restricted set of parameters, which we regard as the parameter space needed for the truncation with D4-branes.
\begin{align}\label{sum_Np}
\sum_p N_p  \left(\sum_{I\in p} h_I \right) + C=0 \;.
\end{align}

Since we consider divisors associated with perfect matchings for the corner lattice points of the toric diagram, 
this will be specified by $E_{\rm corner}$, the number of such lattice points.
This should be compared with the set of  $|Q_0|+1=E+2I-1$ integers $\mathsf{N}_A$ associated with truncations of the algebra.
Note, however, not all the possible sets of integers $\mathsf{N}_A$ are realized in the 
quiver diagram, and hence the actual possible truncations are much more limited.
We will discuss many examples in section \ref{sec:generalno4cycle} and section \ref{sec:general4cycle},
and find that the truncations of the algebra are always labelled by a set of two integers,
at least for all the examples studied in this paper. This is actually smaller 
than the number $E_{\rm corner}$ of independent D4-brane charges associated with corner perfect matchings.
Moreover, we in general have many more complex submanifolds described by non-corner perfect matchings.
This suggests that there exist more general representations of the quiver Yangians than studied in this paper.
We leave a detailed discussion of these subtleties for future work.

\section{\texorpdfstring{Examples: Calabi-Yau Threefolds without Compact $4$-cycles}{Examples: Calabi-Yau Threefolds without Compact 4-cycles}}
\label{sec:generalno4cycle}

In the next two sections we apply the algorithm outlined in section 4 on various toric Calabi-Yau threefolds.
For each example, we will construct explicitly its associated algebra, and define its truncations. 
For some examples, we will also discuss special points in the parameter space where our algebra reduces to known affine Yangian algebras. 

This section deals with toric Calabi-Yau threefolds without compact $4$-cycles and section~\ref{sec:general4cycle} will study those with compact $4$-cycles.

\subsection{\texorpdfstring{Simplification when No Compact $4$-Cycles are Present}{Simplification when No Compact 4-Cycles are Present}}
\label{sec:generalno4cycle_special}

\subsubsection{Mode Expansion}

For Calabi-Yau threefolds without compact $4$-cycles, the corresponding quiver has the property that the number of arrows from $a$ to $b$ is the same as the number of arrows from $b$ to $a$:
\begin{equation}
|a\rightarrow b|=|b\rightarrow a|\,.
\end{equation}
As a result, the bond factor $\varphi^{a\Rightarrow b}(u)$, defined in (\ref{eq-charge-atob}), become homogeneous rational functions. 
Therefore the eigenvalue of the charge function $\Psi^{(a)}_{\Kappa}(u)$, which is a product of all $\varphi^{b\Rightarrow a}(u-h(\sqbox{$b$}))$ together with the possible vacuum contribution (see definition (\ref{eq-Psi-ansatz-toric})) for any crystal state $\Kappa$, is also a homogeneous rational function, which has the expansion 
\begin{equation}\label{eq-Psi-expansion-no4}
 \Psi^{(a)}_{\Kappa}(u)=1+\sum^{+\infty}_{n=0}\frac{\Psi^{(a)}_{n}(\Kappa)}{u^{n+1}}\,.
\end{equation}

Since the expansion (\ref{eq-Psi-expansion-no4}) is true for any $\Kappa$ and $a$, the operator $\psi^{(a)}(u)$ has the same expansion. 
Namely, for Calabi-Yau threefolds without compact $4$-cycles, the general mode expansion (\ref{eq-mode-expansion-toric}) specializes to
\begin{equation}\label{eq-mode-expansion-no4}
\begin{aligned}
e^{(a)}(z)\equiv\sum^{+\infty}_{n=0}\frac{e^{(a)}_n}{z^{n+1}} \,, \qquad \psi^{(a)}(z)\equiv 1+ \sum^{+\infty}_{n=0}\frac{\psi^{(a)}_n}{z^{n+1}}\,, \qquad f^{(a)}(z)\equiv \sum^{+\infty}_{n=0}\frac{f^{(a)}_n}{z^{n+1}} \;.
\end{aligned}
\end{equation}
Accordingly, in the algebraic relations in terms of modes (\ref{eq-OPE-modes-toric}), the $\psi^{(a)}_n$, $e^{(a)}_n$, and $f^{(a)}_n$ modes all have $n\in\mathbb{Z}_{\geq 0}$.
Finally, the mode expansion of $\psi^{(a)}(z)$ in (\ref{eq-mode-expansion-no4}) also gives 
\begin{equation}\label{eq-leading-psi}
\psi^{(a)}_{-1}=1\,,
\end{equation}
for all $a$.
\subsubsection{Initial Conditions}


As explained just now, for Calabi-Yau threefolds without compact $4$-cycles, the mode relations (\ref{eq-OPE-modes-toric}) are all from terms of order $z^{-n-1}w^{-m-1}$, with $n,m\in\mathbb{Z}_{\geq 0}$, in the corresponding OPE relation (\ref{eq-OPE-toric}). 
In particular, for the $\psi  e$ and $\psi f$ relations, the mode relations are from terms of  order $z^{-n-1}w^{-m-1}$, with $n, m\in\mathbb{Z}_{\geq 0}$.

One can also supplement these $\psi e$ and $\psi f$ relations with ``initial conditions" that come from terms of order $z^{\ell-1}w^{-m-1}$ with $\ell=1, \dots, |a\rightarrow b|$ and $m \in \mathbb{Z}_{\ge 0}$.
Note that these additional initial conditions are allowed by the algebraic relations (\ref{eq-OPE-toric}) with the mode expansions (\ref{eq-mode-expansion-no4}), and are consistent with the algebra's action (\ref{eq.efpsi}) on colored crystals $\Kappa$.
For Calabi-Yau threefolds without compact $4$-cycles, these initial conditions contain the finite part of the affine Yangian.
One can derive from these initial conditions the central elements of the algebra. 

Let us again take the $\psi^{(a)} \, e^{(b)}$ OPE for example. 
Plugging the mode expansions of $\psi^{(a)}(z)$ and $e^{(b)}(w)$ from (\ref{eq-mode-expansion-no4}) into (\ref{eq-psie-OPE-1}) and extracting the terms of order $z^{\ell-1}w^{-m-1}$ with $\ell=1, \dots, |a\rightarrow b|$ and $m \in \mathbb{Z}_{\geq 0}$, and using (\ref{eq-leading-psi}), we have the mode relation:
\begin{equation}\label{eq-initial-toric}
\begin{aligned}
&\sum^{|b\rightarrow a|}_{k=0}(-1)^{|b\rightarrow a|-k} \, \sigma^{b\rightarrow a}_{|b\rightarrow a|-k}\,  [\psi^{(a)}_{-\ell}\, e^{(b)}_m]_k
=\sum^{|a\rightarrow b|}_{k=0} \sigma^{a\rightarrow b}_{|a\rightarrow b|-k}\, [ e^{(b)}_m\, \psi^{(a)}_{-\ell}]^k \;, 
\end{aligned}
\end{equation}
where the notations for $ [\psi^{(a)}_{-\ell}\, e^{(b)}_m]_k$ and $ [ e^{(b)}_m\, \psi^{(a)}_{-\ell}]^k$ were defined in (\ref{eq-ABn}), and 
$\psi^{(a)}_{n<-1}=0$ due to the mode expansion (\ref{eq-mode-expansion-no4}).
Similarly, the $\psi f$ initial condition is
\begin{equation}\label{eq-initial-f-toric}
\begin{aligned}
&\sum^{|a\rightarrow b|}_{k=0} \sigma^{a\rightarrow b}_{|a\rightarrow b|-k}\,  [\psi^{(a)}_{-\ell}\, f^{(b)}_m]_k
=\sum^{|b\rightarrow a|}_{k=0}(-1)^{|b\rightarrow a|-k} \, \sigma^{b\rightarrow a}_{|b\rightarrow a|-k}\, [ f^{(b)}_m\, \psi^{(a)}_{-\ell}]^k  \;,
\end{aligned}
\end{equation}
for $\ell=1, \dots, |a\rightarrow b|$ and $m \in \mathbb{Z}_{\geq 0}$.

One can impose the initial conditions (\ref{eq-initial-toric}) and (\ref{eq-initial-f-toric}) to supplement the mode relations (\ref{eq-OPE-modes-toric}) with $n, m\in\mathbb{Z}_{\geq 0}$.
Note that since $\psi^{(a)}_{n<-1}=0$, the relations (\ref{eq-initial-toric}) and (\ref{eq-initial-f-toric}) are non-empty only for $\ell \leq |a\rightarrow b|$.
 For example, let us first consider the case with $\ell=|a\rightarrow b|= |b\rightarrow a|$, which gives
\begin{equation}\label{eq-psi-e-f-initial-0}
\begin{aligned}
\left[\psi^{(a)}_0\,,\,e^{(b)}_m\right] &=\left(\sigma^{a\rightarrow b}_{1}+\sigma^{b\rightarrow a}_{1}\right)\,e^{(b)}_{m}=\left(\sum_{I\in \{a\rightarrow b\}}h_I+\sum_{I\in \{b\rightarrow a\}}h_I\right) \,e^{(b)}_{m} \;,\\
\left[\psi^{(a)}_0\,,\,f^{(b)}_m\right] &=-\left(\sigma^{a\rightarrow b}_{1}+\sigma^{b\rightarrow a}_{1}\right)\,f^{(b)}_{m}=-\left(\sum_{I\in \{a\rightarrow b\}}h_I+\sum_{I\in \{b\rightarrow a\}}h_I\right) \,f^{(b)}_{m} \;,
\end{aligned}
\end{equation}
where we have used $\psi^{(a)}_{-1}=1$.
Next, for $|a\rightarrow b|= |b\rightarrow a|\geq 2$, consider $\ell=|a\rightarrow b|-1$, which gives
\begin{equation}\label{eq-psi-e-f-initial-1}
\begin{aligned}
\left[\psi^{(a)}_1\,,\,e^{(b)}_m\right] &=\left(\sigma^{a\rightarrow b}_{2}-\sigma^{b\rightarrow a}_{2}\right)\,e^{(b)}_{m}
+\left(\sigma^{b\rightarrow a}_{1}\, \psi^{(a)}_0\, e^{(b)}_{m}+\sigma^{a\rightarrow b}_{1}\, e^{(b)}_{m}\, \psi^{(a)}_0\right) \;,\\
\left[\psi^{(a)}_1\,,\,f^{(b)}_m\right] &=-\left(\sigma^{a\rightarrow b}_{2}-\sigma^{b\rightarrow a}_{2}\right)\,f^{(b)}_{m}
-\left(\sigma^{a\rightarrow b}_{1}\, \psi^{(a)}_0\, f^{(b)}_{m}+\sigma^{b\rightarrow a}_{1}\, f^{(b)}_{m}\, \psi^{(a)}_0\right) \;.
\end{aligned}
\end{equation}
The initial conditions with $\psi^{(a)}_{\ell \geq 2}$, if exist, can be derived similarly from the general formulae (\ref{eq-initial-toric}) and (\ref{eq-initial-f-toric}).
Since details of these initial conditions depend on the quiver data $\{a\rightarrow b\}$, we will discuss them further when we consider concrete examples.

\subsubsection{Central Element of the Algebra}

From the initial condition, one can construct various central terms (if exist) of the algebra.
For a given vertex $b$, define
\begin{equation}\label{eq-Sigma-def}
\Sigma_{b}\equiv\sum_{a\in Q_0} \left(\sum_{I\in \{a\rightarrow b\}}h_I+\sum_{I\in \{b\rightarrow a\}}h_I\right) = \sum_{I\in b} h_I  \;,
\end{equation}
where the sum in the last term runs over all charges both incoming and outgoing (without signs) from the vertex $b$. 
The combination 
\begin{tcolorbox}[ams equation]\label{eq-central-generic}
\psi_0\equiv \sum_{a\in Q_0} \psi_0^{(a)}
\end{tcolorbox}
\noindent obeys
\begin{equation}\label{eq-psi-e-f-initial}
[\psi_0\,,\, e^{(b)}_k]=\Sigma_{b} \cdot  e^{(b)}_k\ \qquad \textrm{and} \qquad  [\psi_0\,,\, f^{(b)}_k]=-\Sigma_{b} \cdot f^{(b)}_k\,.
\end{equation}

If the following condition is satisfied:
\begin{tcolorbox}[ams equation]\label{eq-central-condition}
\textrm{\bf central condition}:\qquad 
\sum_{I\in a} h_I=0 \qquad \textrm{for} \quad \forall\, a\,,
\end{tcolorbox}
\noindent then the combination (\ref{eq-central-generic}) is a central term of the algebra. 
(Note the difference between the vertex constraint (\ref{eq-vertex-constraint-toric}) and the central condition (\ref{eq-central-condition})).
For the Calabi-Yau threefolds without compact $4$-cycles, this condition is always guaranteed by the loop constraint (\ref{eq-loop-constraint-toric}).
The central term $\psi_0$ defined in (\ref{eq-central-generic}) is thus the universal central term in the quiver Yangians for all the Calabi-Yau threefolds without compact $4$-cycles.
There could be other central terms, depending on specifics of each quiver diagram. 
We will define these additional central terms when we consider specific examples later.

\subsubsection{\texorpdfstring{Identification between Universal Central Term $\psi_0$ and Vacuum Charge $C$}{Identification between Universal Central Term psi0 and Vacuum Charge C}}

Now we show that in quiver Yangians for Calabi-Yau threefolds without compact $4$-cycles, the numerical constant $C$ in the vacuum contribution to the charge function can be identified as the universal central term $\psi_0$ defined in (\ref{eq-central-generic}).

It is enough to consider the state where only the first atom $\sqbox{1}$ is present, i.e.\ $|\Kappa\rangle=|\sqbox{1}\rangle $. (Recall that we have assumed that the atom at the origin of the crystal is labelled by color $1$.)
From the ansatz (\ref{eq-Psi-ansatz}) with (\ref{eq-psi0-ansatz}), this state has a charge function $\Psi^{(a)}_{\Kappa}(u)$ for each color $a$:
\begin{equation}
\Psi^{(a)}_{\Kappa}(u)= \left(1+\frac{C}{z}\right)^{\delta_{a,1}} \, \varphi^{1\Rightarrow a}(u) \;.
\end{equation}
We can now expand this charge function to obtain its charges $\psi^{(a)}_n$ using the expansion (\ref{eq-mode-expansion-toric}).
In particular, we are interested in the leading charge $\psi^{(a)}_0$, which satisfies
\begin{equation}\label{eq-psi0-C}
 \psi^{(a)}_0 =\delta_{a,1}\, C+ \sum_{I\in \{1\rightarrow a \}} h_I + \sum_{I \in \{a\rightarrow 1 \}} h_I\,.
\end{equation}
Summing (\ref{eq-psi0-C}) over all atoms $a$, and recalling the definition of the generic central term in (\ref{eq-central-generic}) and that of $\Sigma_a$ in (\ref{eq-Sigma-def}), we have
\begin{equation}\label{eq-psi0-C-1}
\psi_0 =C+\Sigma_1\,. 
\end{equation}

Now we can impose the central condition $\Sigma_1=0$ (\ref{eq-central-condition}), which has two consequences for (\ref{eq-psi0-C-1}). 
First, $\psi_0$ is central, due to (\ref{eq-psi-e-f-initial}).
Second, 
\begin{equation}\label{eq-C-psi0}
C=\psi_0\,.
\end{equation} 
It is also straightforward to check that one can obtain (\ref{eq-C-psi0}) if we start with an arbitrary state $|\Kappa \rangle$.
The analogue of (\ref{eq-psi0-C-1}) for an arbitrary crystal state $|\Kappa \rangle$ is
\begin{equation}
\psi_0=C+\sum_{\sqbox{$a$}\in \Kappa}  \Sigma_{a}  \;,
\end{equation} 
where each atom $\sqbox{$a$}$ in the crystal $|\Kappa\rangle$ contributes a term $\Sigma_a$, where $a$ is the color of the atom $\sqbox{$a$}$. 
Due to the central condition (\ref{eq-central-condition}), all $\Sigma_a=0$, and we have (\ref{eq-C-psi0}) for any $|\Kappa \rangle$.
The identification (\ref{eq-C-psi0}) is a natural generalization of the $\mathfrak{gl}_1$ case (\ref{eq-vacuum-gl1}).

\subsection{\texorpdfstring{Quiver Yangians for $(\mathbb{C}^2/\mathbb{Z}_n) \times \mathbb{C}$ and Affine Yangian of $\mathfrak{gl}_n$}{Quiver Yangians for (C2/Zn)x C and Affine Yangian of gl(n)}}
\label{sec:An}

We start with the toric Calabi-Yau threefold $(\mathbb{C}^2/\mathbb{Z}_n)\times \mathbb{C}$. 
The quiver algebra has $n+1$ parameters.
If we impose the $n-1$ vertex constraints (\ref{eq-vertex-constraint-toric}), we can reduce the number of parameters to $2$, which are the two coordinate parameters. 
We find that the reduced quiver Yangian in this sub-parameter space is the affine Yangian of $\mathfrak{gl}_n$ constructed in \cite{MR2199856,MR2323534}, which are rational limits of quantum toroidal algebra of $\mathfrak{gl}_n$ constructed in \cite{MR1324698} (see also \cite{Feigin:2013fga}).

Let us study the cases of $n=1$, $n=2$, and $n\geq 3$ in turn.

\subsubsection{\texorpdfstring{$\mathbb{C}^3$ and Affine Yangian of $\mathfrak{gl}_1$}{C3 and Affine Yangian of gl(1)}}
\label{sec:gl1-8211}

\subsubsubsection{Quiver Yangian for $\mathbb{C}^3$}

For $\mathbb{C}^3$, the toric diagram and its dual graph are
\begin{equation}\label{fig-toric-C3}
\begin{tikzpicture} 
\filldraw [red] (0,0) circle (2pt); 
\filldraw [red] (0,1) circle (2pt); 
\filldraw [red] (1,0) circle (2pt); 
\node at (-0.5,-0.5) {(0,0)}; 
\node at (-0.5,1.5) {(0,1)}; 
\node at (1.5,-0.5) {(1,0)}; 
\draw (0,0) -- (0,1); 
\draw (0,0) -- (1,0); 
\draw (1,0) -- (0,1); 
\end{tikzpicture}
\qquad \qquad \qquad
\begin{tikzpicture} 
\draw[->] (0,0) -- (-1,0); 
\draw[->] (0,0) -- (0,-1); 
\draw[->] (0,0) -- (1,1); 
\node at (-1.5,0) {3}; 
\node at (0,-1.5) {1}; 
\node at (1.5,1.5) {2}; 
\end{tikzpicture}
\end{equation}
Its associated quiver diagram is
\begin{equation}\label{fig-quiver-C3}
\begin{tikzpicture}[scale=1]
\node[state]  [regular polygon, regular polygon sides=4, draw=blue!50, very thick, fill=blue!10] (a1) at (0,0)  {$1$};
\path[->] 
(a1) edge [in=90, out=150, loop, thin, above left] node {$(X_3,\, h_3)$} ()
(a1) edge [in=210, out=270, loop, thin, below left] node {$(X_1,\, h_1)$} ()
(a1) edge [in=330, out=30, loop, thin, right] node {$(X_2, \, h_2)$} ()
;
\end{tikzpicture}
\end{equation}
where we have labelled the three adjoints $X_{1,2,3}$, together with their three charges $h_{1,2,3}$. 
The super-potential is 
\begin{equation}
W=\textrm{Tr}[-X_1\, X_2\, X_3 +X_1\, X_3\, X_2] \;.
\end{equation}
Since in the quiver the vertex $1$ has a self-loop, it is bosonic:
$|1|=0$.

\begin{figure}[htbp]
\begin{minipage}{0.5\linewidth}
\centering
\begin{tikzpicture}[scale=1]
\filldraw[mygreen] (-1,1.73205)--(1,1.73205)-- (0,0)--(-2,0) -- cycle; 
\node[state]  [regular polygon, regular polygon sides=4, draw=blue!50, very thick, fill=blue!10] (a1) at (0,0)  {$1$};
\node[state]  [regular polygon, regular polygon sides=4, regular polygon, regular polygon sides=4, draw=blue!50, very thick, fill=blue!10] (a21) at (-1,-1.73205)  {$1$};
\node[state]  [regular polygon, regular polygon sides=4, draw=blue!50, very thick, fill=blue!10] (a22) at (2,0)  {$1$};
\node[state]  [regular polygon, regular polygon sides=4, draw=blue!50, very thick, fill=blue!10] (a23) at (-1,1.73205)  {$1$};
\node[state]  [regular polygon, regular polygon sides=4, draw=blue!50, very thick, fill=blue!10] (a312) at (1,-1.73205)  {$1$};
\node[state]  [regular polygon, regular polygon sides=4, draw=blue!50, very thick, fill=blue!10] (a323) at (1,1.73205)  {$1$};
\node[state]  [regular polygon, regular polygon sides=4, draw=blue!50, very thick, fill=blue!10] (a331) at (-2,0)  {$1$};
\path[->] 
(a1) edge   [thick, red]   node [left] {$h_1$} (a21)
(a23) edge   [thick, red]   node [left] {$h_1$} (a331)
(a22) edge   [thick, red]   node [right] {$h_1$} (a312)
(a323) edge   [thick, red]   node [left] {$h_1$} (a1)
(a1) edge   [thick, red]   node [above] {$h_2$} (a22)
(a21) edge   [thick, red]   node [below] {$h_2$} (a312)
(a23) edge   [thick, red]   node [above] {$h_2$} (a323)
(a331) edge   [thick, red]   node [above] {$h_2$} (a1)
(a1) edge   [thick, red]   node [left] {$h_3$} (a23)
(a21) edge   [thick, red]   node [left] {$h_3$} (a331)
(a22) edge   [thick, red]   node [right] {$h_3$} (a323)
(a312) edge   [thick, red]   node [right] {$h_3$} (a1)
;
\end{tikzpicture}

\end{minipage}
\hfill
\begin{minipage}{0.5\linewidth}
\centering
\begin{tikzpicture}[scale=0.57]
\filldraw[mygreen] (-3,0) -- (0,0) --(0,3) -- (-3,3)-- cycle; 
\node[state]  [regular polygon, regular polygon sides=4, draw=blue!50, very thick, fill=blue!10] (a1) at (0,0)  {$1$};
\node[state]  [regular polygon, regular polygon sides=4, draw=blue!50, very thick, fill=blue!10] (a21) at (3,0)  {$1$};
\node[state]  [regular polygon, regular polygon sides=4, draw=blue!50, very thick, fill=blue!10] (a22) at (-3,0)  {$1$};
\node[state]  [regular polygon, regular polygon sides=4, draw=blue!50, very thick, fill=blue!10] (a41) at (0,3)  {$1$};
\node[state]  [regular polygon, regular polygon sides=4, draw=blue!50, very thick, fill=blue!10] (a42) at (0,-3)  {$1$};
\node[state]  [regular polygon, regular polygon sides=4, draw=blue!50, very thick, fill=blue!10] (a31) at (3,3)  {$1$};
\node[state]  [regular polygon, regular polygon sides=4, draw=blue!50, very thick, fill=blue!10] (a32) at (3,-3)  {$1$};
\node[state]  [regular polygon, regular polygon sides=4, draw=blue!50, very thick, fill=blue!10] (a34) at (-3,-3)  {$1$};
\node[state]  [regular polygon, regular polygon sides=4, draw=blue!50, very thick, fill=blue!10] (a33) at (-3,3)  {$1$};
\path[->] 
(a1) edge   [thick, red]   node [above] {$h_2$} (a21)
(a22) edge   [thick, red]   node [above] {$h_2$} (a1)
(a31) edge   [thick, red]   node [right] {$h_1$} (a21)
(a21) edge   [thick, red]   node [right] {$h_1$} (a32)
(a33) edge   [thick, red]   node [left] {$h_1$} (a22)
(a22) edge   [thick, red]   node [left] {$h_1$} (a34)
(a41) edge   [thick, red]   node [above] {$h_2$} (a31)
(a33) edge   [thick, red]   node [above] {$h_2$} (a41)
(a34) edge   [thick, red]   node [above] {$h_2$} (a42)
(a42) edge   [thick, red]   node [above] {$h_2$} (a32)
(a41) edge   [thick, red]   node [right] {$h_1$} (a1)
(a1) edge   [thick, red]   node [right] {$h_1$} (a42)
(a1) edge   [thick, red]   node [right] {$h_3$} (a33)
(a32) edge   [thick, red]   node [right] {$h_3$} (a1)
(a21) edge   [thick, red]   node [right] {$h_3$} (a41)
(a42) edge   [thick, red]   node [right] {$h_3$} (a22)
;
\end{tikzpicture}
\end{minipage}
\caption{Two ways to draw the periodic quiver for $\mathbb{C}^3$. 
The left one emphasizes its connection to the projection of the plane partitions and the triality symmetry of the three directions, whereas the right one is for later comparison with the periodic quiver for $(\mathbb{C}^{2}/\mathbb{Z}_n)\times\mathbb{C}$ and generalized conifolds.}
\label{fig-periodic-quiver-C3-big}
\end{figure}
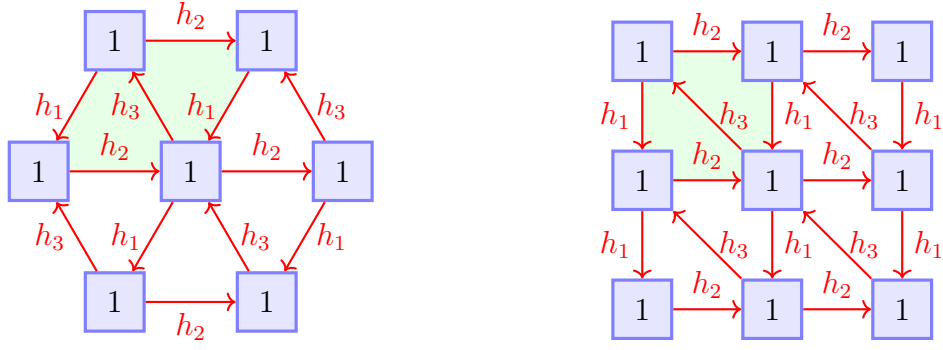

The periodic quiver is
\begin{equation}\label{fig-periodic-quiver-C3}
\begin{aligned}
&
\begin{tikzpicture}[scale=1]
\filldraw[mygreen] (-1,-1.73205)--(1,-1.73205)  -- (2,0)--(0,0)--cycle; 
\node[state]  [regular polygon, regular polygon sides=4, draw=blue!50, very thick, fill=blue!10] (a1) at (0,0)  {$1$};
\node[state]  [regular polygon, regular polygon sides=4, regular polygon, regular polygon sides=4, draw=blue!50, very thick, fill=blue!10] (a21) at (-1,-1.73205)  {$1$};
\node[state]  [regular polygon, regular polygon sides=4, draw=blue!50, very thick, fill=blue!10] (a22) at (2,0)  {$1$};
\node[state]  [regular polygon, regular polygon sides=4, draw=blue!50, very thick, fill=blue!10] (a312) at (1,-1.73205)  {$1$};
\path[->] 
(a1) edge   [thick, red]   node [left] {$h_1$} (a21)
(a22) edge   [thick, red]   node [right] {$h_1$} (a312)
(a1) edge   [thick, red]   node [above] {$h_2$} (a22)
(a21) edge   [thick, red]   node [below] {$h_2$} (a312)
(a312) edge   [thick, red]   node [right] {$h_3$} (a1)
;
\end{tikzpicture}
\end{aligned}
\end{equation}
where the fundamental region of the torus is shown as a shaded region.
The map to the crystal configuration is easier to visualize from a bigger domain, shown in the left of Figure \ref{fig-periodic-quiver-C3-big}. In the right of Figure \ref{fig-periodic-quiver-C3-big}, we have redrawn this period quiver in a slightly different shape, for the later comparison with periodic quivers for $(\mathbb{C}^{2}/\mathbb{Z}_n)\times\mathbb{C}$ and generalized conifolds.

The loop constraint (\ref{eq-loop-constraint-toric}) gives
\begin{equation}\label{eq-loop-constraint-C3}
h_1+h_2+h_3=0 \,.
\end{equation}
Therefore we have two coordinate parameters, corresponding to the two equivariant parameters $(\epsilon_{1}, \epsilon_2)$.
Note that the central condition (\ref{eq-central-condition}) is guaranteed by the loop constraint (\ref{eq-loop-constraint-C3}).

\subsubsubsection{Affine Yangian of $\mathfrak{gl}_1$}

Note that in this case the vertex constraint (\ref{eq-vertex-constraint-toric}) also gives (\ref{eq-loop-constraint-C3}).
Therefore the minimal number of parameters we can have is two, corresponding to the $U(1)^2$ toric isometries.

There is only one bond factor:
\begin{equation}\label{eq-algebraM-C3}
\varphi^{1\Rightarrow 1}(u)=\varphi_3(u)=\frac{(u+h_1)(u+h_2)(u+h_3)}{(u-h_1)(u-h_2)(u-h_3)} \;.
\end{equation}
Plugging this into the general formulae for the OPE relations (\ref{eq-OPE-toric}) and the initial conditions (\ref{eq-initial-toric}) and (\ref{eq-initial-f-toric}), and supplementing them with Serre relations, we have the full list of algebra relations of the affine Yangian of $\mathfrak{gl}_1$:
\begin{tcolorbox}[ams align]
&\textrm{OPE:}\quad\begin{cases}\begin{aligned}
&\psi(z)\psi(w)\sim \psi(w)\psi(z) \;,\\
&\begin{aligned}
\psi(z)\, e(w) &  \sim  \varphi_3(\Delta)\, e(w)\, \psi(z)  \;,\\ 
\psi(z)\, f(w) & \sim \varphi_3^{-1}(\Delta)\, f(w)\, \psi(z)  \;,
\end{aligned}
\qquad
\begin{aligned}
e(z)\, e(w) & \sim    \varphi_3(\Delta)\, e(w)\, e(z)  \;,\\
 f(z)\, f(w) &  \sim    \varphi_3^{-1}(\Delta)\, f(w)\, f(z)  \;,
\end{aligned}\\
&[e(z)\,, f(w)]   \sim  - \, \frac{\psi(z) - \psi(w)}{z-w}  \;,
\end{aligned}
\end{cases}\label{box-OPE-gl1-our}
\\
&\textrm{Initial:}\quad\begin{cases}
\begin{aligned}
&\begin{aligned}
&[\psi_0,e_m] = 0 \;,\\
&[\psi_0,f_m] = 0 \;,
\end{aligned}\qquad \qquad
\begin{aligned}
&[\psi_1,e_m] = 0 \;,\\
&[\psi_1,f_m] = 0 \;,
\end{aligned}\qquad \qquad
\begin{aligned}
&[\psi_2,e_m] = 2 \, \sigma_3\, e_m \;,\\
&[\psi_2,f_m] = - 2  \,\sigma_3\, f_m \;,
\end{aligned}
\end{aligned}
\end{cases}\label{box-initial-gl1-our}
\\
&\textrm{Serre}:\quad\begin{cases}\begin{aligned}
&\textrm{Sym}_{z_1,z_2,z_3}\, (z_2-z_3)\,\left[ e(z_1)\,, \left[ e(z_2)\,, e(z_3)\right] \right]\sim 0  \;,\\
&\textrm{Sym}_{z_1,z_2,z_3}\, (z_2-z_3)\left[ f(z_1)\,, \left[ f(z_2)\,,f(z_3)\right]\right] \sim 0  \;.
\end{aligned}
\end{cases}\label{box-serre-gl1-our}
\end{tcolorbox}
\noindent where $\sigma_3\equiv h_1 h_2 h_3$. 
It is straightforward to write down the relation in terms of modes, following (\ref{eq-OPE-modes-toric}). 

In the $e f$ relation in (\ref{box-OPE-gl1-our}), note its difference from (\ref{box-initial-gl1}) in the factor of $\frac{1}{\sigma_3}$. 
This is due to the different convention in our mode expansion of $\psi(u)$ in (\ref{eq-mode-expansion-no4}) --- which is the universal for all quiver Yangian of Calabi-Yau threefolds without $4$-cycle --- from the one (\ref{eq-mode-expansion-gl1}) in the literature. 
(This difference also manifests itself in the two initial conditions involving $\psi_2$.)

In the derivation of the initial conditions (\ref{box-initial-gl1-our}), we have used $|a\rightarrow a|=3$, and setting $\ell=3,2,1$ in the general formulae (\ref{eq-initial-toric}) and (\ref{eq-initial-f-toric}) gives the initial conditions involving $\psi_{0,1,2}$, respectively, and we have also used $\sigma_1\equiv h_1+h_2+h_3=0$.
We see that there are two central terms, $\psi_0$ and $\psi_1$.

\subsubsubsection{Truncation}

For the affine Yangian of $\mathfrak{gl}_1$, the vacuum charge $C$ is identical to the leading central term $\psi_0$.
Applying the general truncation condition (\ref{eq-truncation-condition}) on the quiver (\ref{fig-periodic-quiver-C3}), we have the truncation condition
\begin{tcolorbox}[ams equation]\label{eq-truncation-condition-gl1}
N_1\, h_1+N_2 \, h_2 + N_3\, h_3+\psi_0=0  \;,
\end{tcolorbox}
\noindent which is invariant under the shift
\begin{equation}
N_i \rightarrow N_i+n
\end{equation}
due to the loop constraint (\ref{eq-loop-constraint-C3}).

\begin{figure}[htbp]
\centering\includegraphics[scale=0.3]{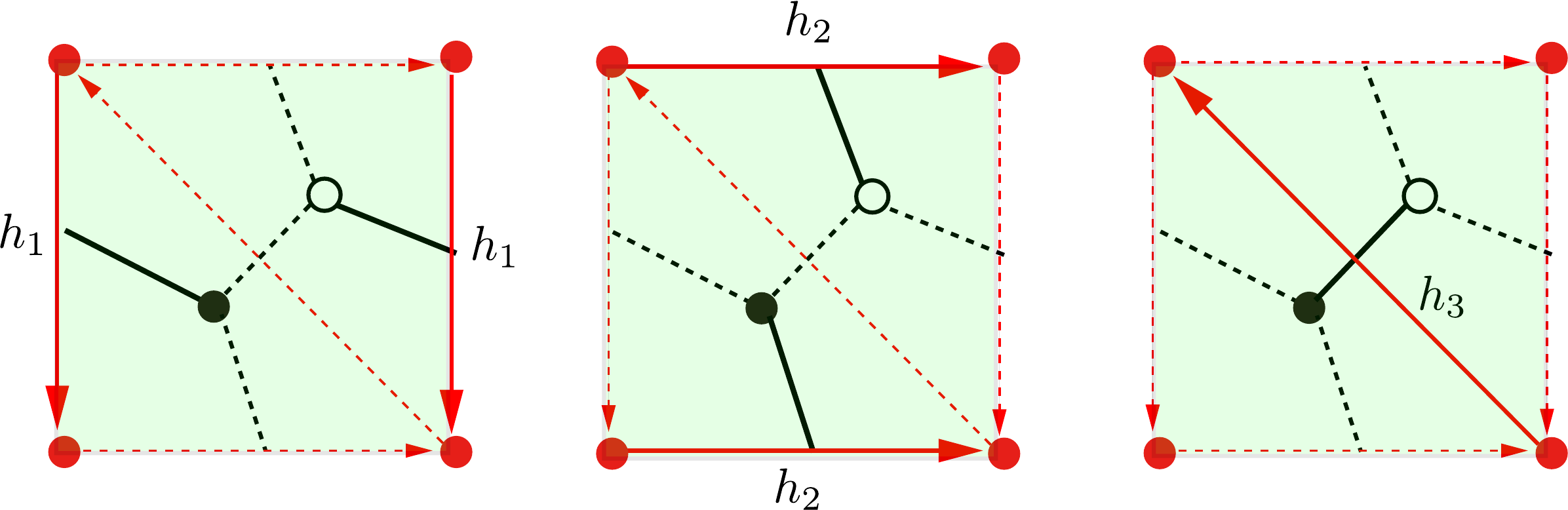}
\caption{The three perfect matchings for the $\mathbb{C}^3$ geometry.
Each of these perfect matchings corresponds to one of the non-compact regions of the $(p,q)$-web,
and to one of the parameters $h_1, h_2, h_3$.}
\label{fig.C3PM}
\end{figure}

There are three non-compact divisors where the D4-branes wrap, which are related by the permutation ($\mathcal{S}_3$) symmetry.
These correspond to the three perfect matchings of the dimer model (see Figure \ref{fig.C3PM}), 
each of which corresponds to one of the triple $\{h_1, h_2, h_3\}$,
thus giving rise to the same condition \eqref{eq-truncation-condition-gl1}.
This means that the $N_i$'s can indeed be regarded as the number of D4-branes in that region.
We therefore obtain the algebra $\mathsf{Y}^{N_1, N_2, N_3}_{(Q,W)}$ and $\underline{\mathsf{Y}}^{N_1, N_2, N_3}_{(Q,W)}$. 
The latter was studied previously in \cite{bershtein2015plane, Litvinov:2016mgi, Gaiotto:2017euk, Prochazka:2017qum}.
Note that due to the loop constraint (\ref{eq-loop-constraint-C3}) one can simultaneously shift all the $N_i$'s by the same amount, hence leaving two non-negative integers.

\begin{figure}[htbp]
\centering\includegraphics[scale=0.3]{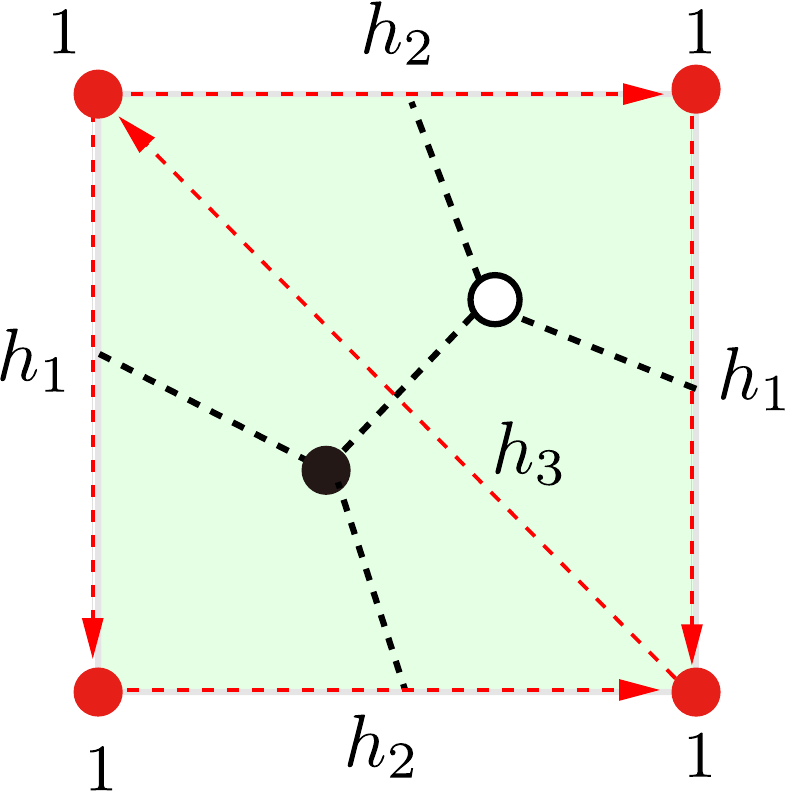}
\caption{The bipartite graph and its dual, the periodic quiver, for the $\mathbb{C}^3$ geometry.}
\label{fig.C3bipartite}
\end{figure}

\subsubsection{\texorpdfstring{$(\mathbb{C}^2/\mathbb{Z}_2) \times \mathbb{C}$ and Affine Yangian of $\mathfrak{gl}_2$}{(C2/Z2)xC and Affine Yangian of gl(2)}}

\subsubsubsection{Quiver Yangian for $(\mathbb{C}^2/\mathbb{Z}_2)\times \mathbb{C}$}

For $\mathbb{C}^2/\mathbb{Z}_2$$\times$$\mathbb{C}$, the toric diagram and its dual graph are
\begin{equation}\label{fig-toric-Z2-1}
\begin{tikzpicture} 
\filldraw [red] (0,0) circle (2pt); 
\filldraw [red] (0,1) circle (2pt); 
\filldraw [red] (0,2) circle (2pt); 
\filldraw [red] (1,0) circle (2pt); 
\node at (-.5,-.5) {(0,0)}; 
\node at (-.5,1) {(0,1)}; 
\node at (-.5,2.5) {(0,2)}; 
\node at (1.5,-0.5) {(1,0)}; 
\draw (0,0) -- (0,2); 
\draw (0,0) -- (1,0); 
\draw (1,0) -- (0,2); 
\end{tikzpicture}
\qquad \qquad \qquad
\begin{tikzpicture} 
\draw[->] (0,0) -- (-1,0); 
\draw[->] (0,0) -- (0,-1); 
\draw (0,0) -- (1,1); 
\draw[->] (1,1) -- (-1,1); 
\draw[->] (1,1) -- (3,2); 
\node at (-1.5,0) {3}; 
\node at (-1.5,1) {$\hat{3}$}; 
\node at (0,-1.5) {1}; 
\node at (3.5,2.5) {$\hat{1}$}; 
\end{tikzpicture}
\end{equation}
Its associated quiver diagram is the $A_2$-quiver
\begin{equation}\label{figure-quiver-Z2}
\begin{tikzpicture}[scale=1]
\node[state]  [regular polygon, regular polygon sides=4, draw=blue!50, very thick, fill=blue!10] (a1) at (-2,0)  {$1$};
\node[state]  [regular polygon, regular polygon sides=4, draw=blue!50, very thick, fill=blue!10] (a2) at (2,0)  {$2$};
\path[->] 
(a1) edge [in=150, out=210, loop, thin, left] node {$(C_1, \gamma_1)$} ()
(a2) edge [in=330, out=30, loop, thin, right] node {$(C_2,\gamma_2)$} ();
\path[->>] 
(a1) edge   [thin, bend left]  node [above] {$(A_1,\alpha_{1}), \,\, (B_2,\beta_{2})$} (a2) 
(a2) edge   [thin, bend left]  node [below]{$(B_1,\beta_1), \,\, (A_2,\alpha_{2})$} (a1);
\end{tikzpicture}
\end{equation}
with super-potential 
\begin{equation}
W=\textrm{Tr}[-C_1\, A_1\, B_1+C_1 \, B_2\, A_2 -C_2\, A_2\, B_2+C_2 \, B_1\, A_1]  \;.
\end{equation}
Both vertices are bosonic:
\begin{equation}\label{eq-gl2-boson-1}
|a|=0 \,,\qquad a=1,2\,,
\end{equation}
since there is a self-loop for each of them in the quiver (\ref{figure-quiver-Z2}).

\begin{figure}[htbp]
\begin{minipage}{0.5\linewidth}
\centering
\begin{tikzpicture}[scale=1]
\filldraw[mygreen] (1,1.73205)-- (-2,0)--(-1,-1.73205)--(2,0) -- cycle; 
\node[state]  [regular polygon, regular polygon sides=4, draw=blue!50, very thick, fill=blue!10] (a1) at (0,0)  {$1$};
\node[state]  [regular polygon, regular polygon sides=4, regular polygon, regular polygon sides=4, draw=blue!50, very thick, fill=blue!10] (a21) at (-1,-1.73205)  {$2$};
\node[state]  [regular polygon, regular polygon sides=4, draw=blue!50, very thick, fill=blue!10] (a22) at (2,0)  {$2$};
\node[state]  [regular polygon, regular polygon sides=4, draw=blue!50, very thick, fill=blue!10] (a23) at (-1,1.73205)  {$1$};
\node[state]  [regular polygon, regular polygon sides=4, draw=blue!50, very thick, fill=blue!10] (a312) at (1,-1.73205)  {$1$};
\node[state]  [regular polygon, regular polygon sides=4, draw=blue!50, very thick, fill=blue!10] (a323) at (1,1.73205)  {$2$};
\node[state]  [regular polygon, regular polygon sides=4, draw=blue!50, very thick, fill=blue!10] (a331) at (-2,0)  {$2$};
\path[->] 
(a1) edge   [thick, red]   node [left] {$\alpha_1$} (a21)
(a23) edge   [thick, red]   node [left] {$\alpha_1$} (a331)
(a22) edge   [thick, red]   node [right] {$\alpha_2$} (a312)
(a323) edge   [thick, red]   node [left] {$\alpha_2$} (a1)
(a1) edge   [thick, red]   node [above] {$\beta_2$} (a22)
(a21) edge   [thick, red]   node [below] {$\beta_1$} (a312)
(a23) edge   [thick, red]   node [above] {$\beta_2$} (a323)
(a331) edge   [thick, red]   node [above] {$\beta_1$} (a1)
(a1) edge   [thick, red]   node [left] {$\gamma_1$} (a23)
(a21) edge   [thick, red]   node [left] {$\gamma_2$} (a331)
(a22) edge   [thick, red]   node [right] {$\gamma_2$} (a323)
(a312) edge   [thick, red]   node [right] {$\gamma_1$} (a1)
;
\end{tikzpicture}
\end{minipage}
\hfill
\begin{minipage}{0.5\linewidth}
\centering
\begin{tikzpicture}[scale=0.57]
\filldraw[mygreen] (0,3)-- (-3,0)--(0,-3)-- (3,0) -- cycle; 
\node[state]  [regular polygon, regular polygon sides=4, draw=blue!50, very thick, fill=blue!10] (a1) at (0,0)  {$1$};
\node[state]  [regular polygon, regular polygon sides=4, draw=blue!50, very thick, fill=blue!10] (a21) at (3,0)  {$2$};
\node[state]  [regular polygon, regular polygon sides=4, draw=blue!50, very thick, fill=blue!10] (a22) at (-3,0)  {$2$};
\node[state]  [regular polygon, regular polygon sides=4, draw=blue!50, very thick, fill=blue!10] (a41) at (0,3)  {$2$};
\node[state]  [regular polygon, regular polygon sides=4, draw=blue!50, very thick, fill=blue!10] (a42) at (0,-3)  {$2$};
\node[state]  [regular polygon, regular polygon sides=4, draw=blue!50, very thick, fill=blue!10] (a31) at (3,3)  {$1$};
\node[state]  [regular polygon, regular polygon sides=4, draw=blue!50, very thick, fill=blue!10] (a32) at (3,-3)  {$1$};
\node[state]  [regular polygon, regular polygon sides=4, draw=blue!50, very thick, fill=blue!10] (a34) at (-3,-3)  {$1$};
\node[state]  [regular polygon, regular polygon sides=4, draw=blue!50, very thick, fill=blue!10] (a33) at (-3,3)  {$1$};
\path[->] 
(a1) edge   [thick, red]   node [above] {$\beta_2$} (a21)
(a22) edge   [thick, red]   node [above] {$\beta_1$} (a1)
(a31) edge   [thick, red]   node [right] {$\alpha_1$} (a21)
(a21) edge   [thick, red]   node [right] {$\alpha_2$} (a32)
(a33) edge   [thick, red]   node [left] {$\alpha_1$} (a22)
(a22) edge   [thick, red]   node [left] {$\alpha_2$} (a34)
(a41) edge   [thick, red]   node [above] {$\beta_1$} (a31)
(a33) edge   [thick, red]   node [above] {$\beta_2$} (a41)
(a34) edge   [thick, red]   node [above] {$\beta_2$} (a42)
(a42) edge   [thick, red]   node [above] {$\beta_1$} (a32)
(a41) edge   [thick, red]   node [right] {$\alpha_2$} (a1)
(a1) edge   [thick, red]   node [right] {$\alpha_1$} (a42)
(a1) edge   [thick, red]   node [right] {$\gamma_1$} (a33)
(a32) edge   [thick, red]   node [right] {$\gamma_1$} (a1)
(a21) edge   [thick, red]   node [right] {$\gamma_2$} (a41)
(a42) edge   [thick, red]   node [right] {$\gamma_2$} (a22)
;
\end{tikzpicture}
\end{minipage}
\caption{Two ways to draw the periodic quiver $(\mathbb{C}^{2}/\mathbb{Z}_2)\times\mathbb{C}$. 
The left one shows that the representation can be realized by coloring the plane partitions, whereas the right one is for later comparison with the periodic quiver for the conifold and $(\mathbb{C}^{2}/\mathbb{Z}_n)\times\mathbb{C}$. 
For clarity we have shown several copies of the fundamental region of the two-dimensional torus; one choice of the fundamental region is shown as a shaded region.}
\label{fig-periodic-quiver-gl2}
\end{figure}
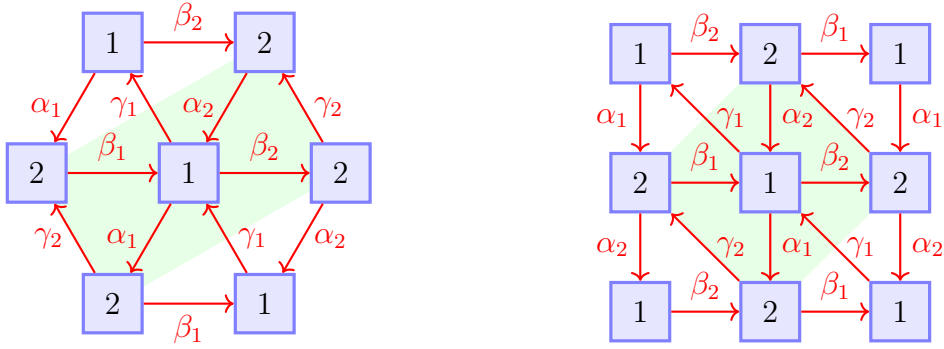

The periodic quiver is shown in Figure \ref{fig-periodic-quiver-gl2}, drawn in two slightly different ways. Comparing the left one to the left drawing in Figure~\ref{fig-periodic-quiver-C3-big}, one can see the representation of the algebra for $(\mathbb{C}^{2}/\mathbb{Z}_2)\times\mathbb{C}$ can be realized by coloring plane partitions accordingly --- the color alternates between $1$ and $2$ as one moves along the  $x_1$ or $x_2$ directions, but remains unchanged along the $x_3$ direction. 
The right drawing in Figure~\ref{fig-periodic-quiver-C3-big} is for later comparison with the conifold and $(\mathbb{C}^{2}/\mathbb{Z}_n)\times\mathbb{C}$. 

Applying the loop constraint (\ref{eq-loop-constraint-toric}) gives the constraints on the charges:
\begin{equation}\label{eq-loop-constraint-Z2}
\gamma_1=\gamma_2\equiv \gamma\,, \qquad \alpha_1+\beta_1+\gamma=0 \,,\qquad \alpha_2+\beta_2+\gamma=0  \;.
\end{equation}
Namely, there are only three independent parameters for the algebra for $\mathbb{C}^2/\mathbb{Z}_2$$\times$$\mathbb{C}$.
The central condition (\ref{eq-central-condition}) is guaranteed by the loop constraint (\ref{eq-loop-constraint-Z2}).

One can immediately read off the bond factors from the periodic quiver shown in Figure~\ref{fig-periodic-quiver-gl2} by the definition (\ref{eq-charge-atob})
\begin{equation}\label{eq-charge-function-Z2}
\varphi^{a\Rightarrow a}(u)=\frac{u+\gamma}{u-\gamma}  \qquad \textrm{and}\qquad \varphi^{a\Rightarrow a+1}(u)=\frac{(u+\alpha_{a+1})(u+\beta_{a})}{(u-\alpha_{a})(u-\beta_{a+1})} \;,
\end{equation}
where the indices are understood as mod $2$.
The resulting algebra is
\begin{tcolorbox}[ams align]\label{eq-algebra-shifted-gl2-1}
&\textrm{OPE:}\quad
\begin{cases}
\begin{aligned}
\psi^{(a)}(z)\, \psi^{(b)}(w)&\sim \psi^{(b)}(w)\, \psi^{(a)}(z) \;,\\
\psi^{(a)}(z)\, e^{(a)}(w)   &\sim \tfrac{\Delta+\gamma}{\Delta-\gamma} \, e^{(a)}(w)\, \psi^{(a)}(z) \;, \\ 
e^{(a)}(z)\, e^{(a)}(w) & \sim  \tfrac{\Delta+\gamma}{\Delta-\gamma} \, e^{(a)}(w)\, e^{(a)}(z)  \;,\\
\psi^{(a)}(z)\, f^{(a)}(w) &  \sim \tfrac{\Delta-\gamma}{\Delta+\gamma} \, f^{(a)}(w)\, \psi^{(a)}(z)  \;,\\
 f^{(a)}(z)\, f^{(a)}(w) &  \sim   \tfrac{\Delta-\gamma}{\Delta+\gamma} \, f^{(a)}(w)\, f^{(a)}(z)  \;,\\
 \psi^{(a+1)}(z)\, e^{(a)}(w)   &\sim \tfrac{(\Delta+\alpha_{a+1})(\Delta+\beta_{a})}{(\Delta-\alpha_{a})(\Delta-\beta_{a+1})}  \, e^{(a)}(w)\, \psi^{(a+1)}(z)  \;,\\ 
e^{(a+1)}(z)\, e^{(a)}(w) & \sim  \tfrac{(\Delta+\alpha_{a+1})(\Delta+\beta_{a})}{(\Delta-\alpha_{a})(\Delta-\beta_{a+1})} \, e^{(a)}(w)\, e^{(a+1)}(z)  \;,\\
\psi^{(a+1)}(z)\, f^{(a)}(w) &  \sim \tfrac{(\Delta-\alpha_{a})(\Delta-\beta_{a+1})}{(\Delta+\alpha_{a+1})(\Delta+\beta_{a})}  \, f^{(a)}(w)\, \psi^{(a+1)}(z)  \;,\\
 f^{(a+1)}(z)\, f^{(a)}(w) &  \sim \tfrac{(\Delta-\alpha_{a})(\Delta-\beta_{a+1})}{(\Delta+\alpha_{a+1})(\Delta+\beta_{a})}   \,f^{(a)}(w)\, f^{(a+1)}(z) \;,\\
[e^{(a)}(z)\,, f^{(b)}(w)]   &=  - \delta^{a,b}\, \frac{\psi^{(a)}(z) - \psi^{(a)}(w)}{z-w}  \;.
\end{aligned}
\end{cases}
\end{tcolorbox}
The initial conditions can be computed using the general formula (\ref{eq-initial-toric}). 
For $a=b$, only the equation with $\ell=2$ is non-empty, giving the initial condition on $[\psi^{(a)}_0, e^{(a)}_m]$. 
For $b=a-1$, the equation with $\ell=2$ gives the relation on $[\psi^{(a+1)}_0, e^{(a)}_m]$ whereas the one with $\ell=1$ gives $[\psi^{(a+1)}_1, e^{(a)}_m]$.
\begin{tcolorbox}[ams align]\label{eq-algebra-shifted-gl2-2}
&\textcolor{black}{\textrm{Initial:}}\quad\begin{cases}
\begin{aligned}
&\begin{aligned}
&[\psi^{(a)}_0,e^{(a)}_m] = 2\gamma\, e^{(a)}_{m} \;, \qquad\quad\, [\psi^{(a+1)}_0,e^{(a)}_m] =- 2\gamma\, e^{(a)}_{m} \;,\\
&[\psi^{(a)}_0,f^{(a)}_m] = -2\gamma\, f^{(a)}_{m} \;,\qquad [\psi^{(a+1)}_0,f^{(a)}_m] =2\gamma\,  f^{(a)}_{m} \;\\
&[\psi^{(a+1)}_1,e^{(a)}_m] =( \alpha_{a+1}\beta_{a}-\alpha_{a}\beta_{a+1})\,e^{(a)}_m \\
&\qquad  \qquad\qquad+( \alpha_{a}+\beta_{a+1})\,\psi^{(a+1)}_0\,e^{(a)}_m +(\alpha_{a+1}+\beta_{a})\, e^{(a)}_m \,\psi^{(a+1)}_0\;,\\
&[\psi^{(a+1)}_1,f^{(a)}_m] =  -(\alpha_{a+1}\beta_{a}-\alpha_{a}\beta_{a+1})\,f^{(a)}_m \\
&\qquad  \qquad\qquad-( \alpha_{a+1}+\beta_{a})\,\psi^{(a+1)}_0\,f^{(a)}_m-(\alpha_{a}+\beta_{a+1})\, f^{(a)}_m \,\psi^{(a+1)}_0\;.\\
\end{aligned}
\end{aligned}
\end{cases}
\end{tcolorbox}
\noindent One can check that $\psi_0\equiv \psi^{(1)}_0+\psi^{(2)}_0$ is the central term. 
Now one can study how to find additional relations (i.e.\ Serre relations) such that the vacuum module of the reduced quiver Yangian reproduces the generating function of the $2$-colored plane partitions. 
For simplicity, we leave the discussion of Serre relation for later (in section~\ref{sec:gl2}) 
after we imposed the vertex constraint.

\subsubsubsection{Truncation}

To study the truncation of the algebra,
consider a path from the origin $\mathfrak{o}$ (on which the atom has color $a=1$) 
to another atom $\sqbox{$1$}$ with the same color, at which the growth of the crystal stops.
The coordinate function of the second atom can be written as
\begin{equation}\label{C3Z2_charge_truncate}
h(\sqbox{$1$})=N_{\gamma}\,\gamma+N_{1}\, (\alpha_1+\alpha_2) 
+N_{2}\, (\alpha_1+\beta_1) 
+N_{3}\, (\beta_2+\alpha_2) 
+N_{4}\, (\beta_2+\beta_1) \,.
\end{equation}
Here $N_{\gamma}$ counts the number of self-loops at the vertices $1$ and $2$ (recall $\gamma_1=\gamma_2=\gamma$),
and $N_{1,2,3,4}$ counts the number of loops $1\rightarrow 2 \rightarrow 1$;
since there are two choices of arrows for both $1\rightarrow 2$ and $2 \rightarrow 1$,
one obtains $2^2=4$ different choices.

We still need to fully take into account the loop constraints (\ref{eq-loop-constraint-Z2}).
Eliminating $\beta_1$ and $\beta_2$, one obtains the truncation condition to be
\begin{equation}\label{C3Z2_charge_truncate-1}
\gamma \, (N_{\gamma}-N_2-N_3-2N_4)+(\alpha_1+\alpha_2)  \, (N_1-N_4)+(\psi^{(1)}_0+\psi^{(2)}_0)=0 \;.
\end{equation}
We find that the truncation is described by a set of two integers, namely
the two coefficients in front of $\gamma$ and $\alpha_1+\alpha_2$, respectively. 

\begin{figure}[htbp]
\centering\includegraphics[scale=0.2]{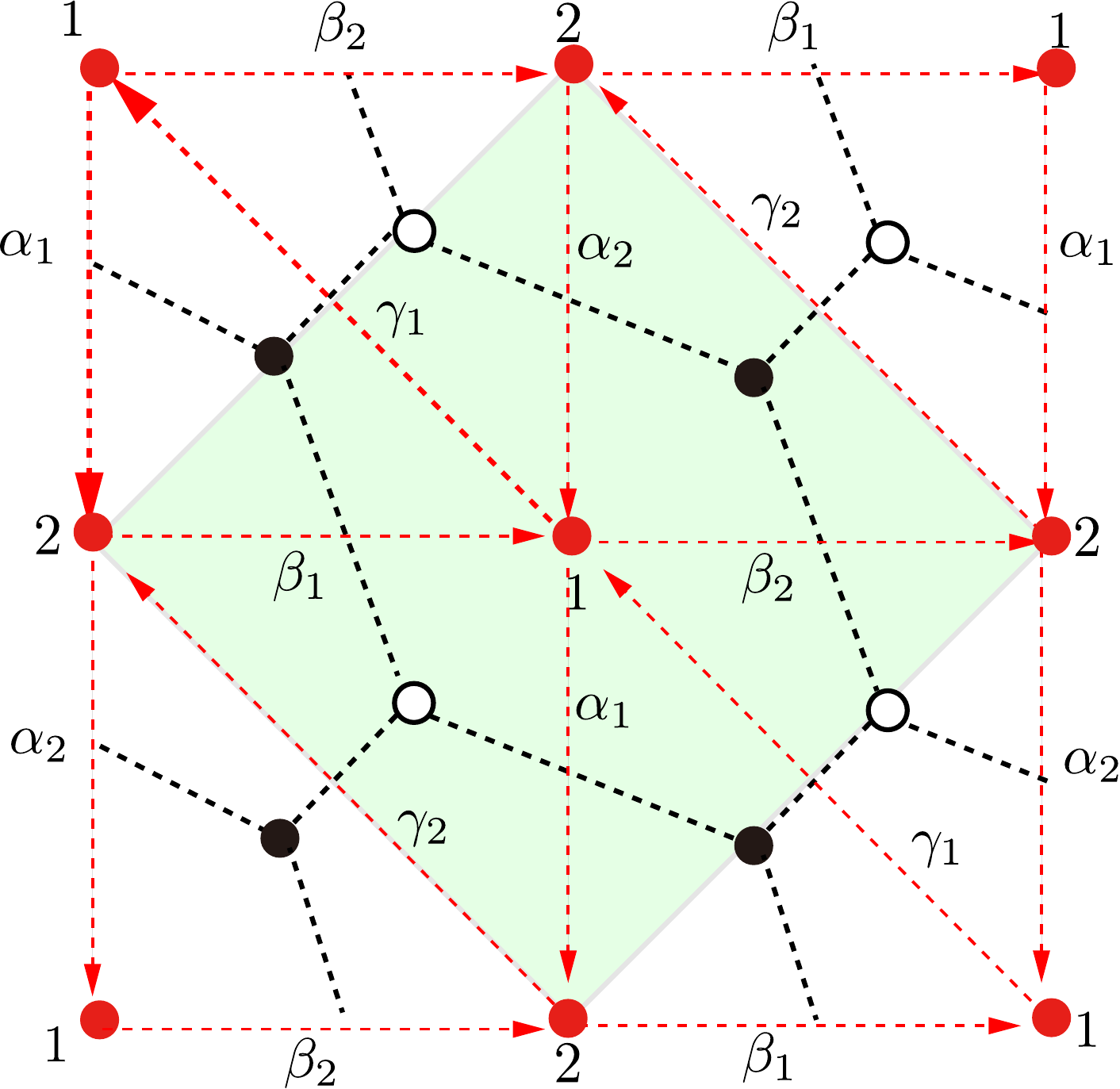}
\caption{The bipartite graph for the $(\mathbb{C}^2/\mathbb{Z}_2)\times \mathbb{C}$ geometry.}
\label{fig.C3Z2bipartite}
\end{figure}

\begin{figure}[htbp]
\centering\includegraphics[scale=0.2]{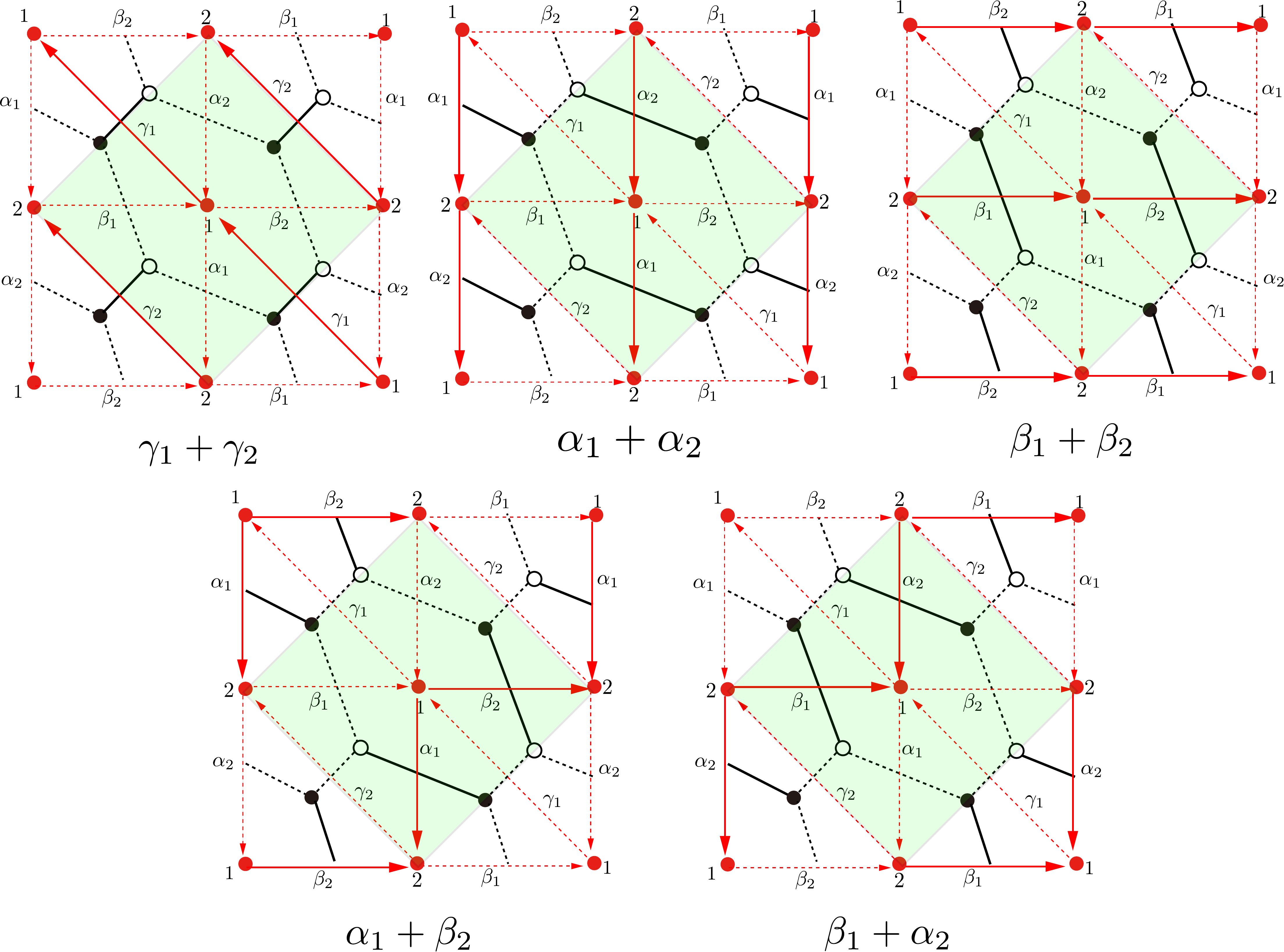}
\caption{The perfect matchings for the $(\mathbb{C}^2/\mathbb{Z}_2)\times \mathbb{C}$ geometry.
There are five perfect matchings, corresponding to the combination of parameters
$\alpha_1+\alpha_2, \beta_1+\beta_2, \gamma_1+\gamma_2, \alpha_1+\beta_2, \alpha_2+\beta_1$.}
\label{fig.C3Z2PM}
\end{figure}

We can compare this with the expectation from the perfect matchings.
The bipartite graph is shown in Figure \ref{fig.C3Z2bipartite},
and leads to five perfect matchings as shown in Figure \ref{fig.C3Z2PM}.
They correspond to five different combinations of parameters 
\begin{align}
\gamma_1+\gamma_2=2\gamma\;, \quad \alpha_1+\alpha_2\;,  \quad \beta_1+\beta_2 \;, \quad \alpha_1+\beta_2\;, \quad   \alpha_2+\beta_1 \;.
\end{align}
Of these five, only the first three correspond to the corner lattice points of the toric diagram:
\begin{align}
\gamma_1+\gamma_2=2\gamma\;, \quad \alpha_1+\alpha_2\;,  \quad \beta_1+\beta_2 =2 \gamma-( \alpha_1+\alpha_2) \;,
\end{align}
and these span almost the same combinations as in \eqref{C3Z2_charge_truncate-1} above, except that the D4-branes give even integer coefficients in front of $\gamma$.

\clearpage

\subsubsubsection{Affine Yangian of $\mathfrak{gl}_2$}
\label{sec:gl2}

In addition to the loop constraint (\ref{eq-loop-constraint-Z2}), we can also impose the vertex constraint (\ref{eq-vertex-constraint-toric}), which in this case give
\begin{equation}\label{eq-vertex-constraint-Z2}
\alpha_1+\beta_2=\alpha_2+\beta_1\,.
\end{equation}
The loop constraint (\ref{eq-loop-constraint-Z2}) and the vertex constraint (\ref{eq-vertex-constraint-Z2}) together give
\begin{equation}\label{eq-constraint-gl2}
\begin{aligned}
&\alpha_1=\alpha_2\equiv h_1  \;,\qquad \beta_1=\beta_2\equiv h_2 \;, \qquad \gamma_1=\gamma_2=\gamma\equiv h_3 \;,\\
&\textrm{and}\qquad h_1+h_2+h_3=0 \;.
\end{aligned}
\end{equation}
Namely, after imposing the vertex constraints on top of the loop constraint, we have two parameters $(h_1,h_2)$, same as in the case of the affine Yangian of $\mathfrak{gl}_1$ for $\mathbb{C}^3$.

Restricting the parameters to (\ref{eq-constraint-gl2}), the bond factors (\ref{eq-charge-function-Z2}) become:
\begin{equation}
\begin{aligned}
&\varphi^{a\Rightarrow a}(u)=\frac{u+h_3}{u-h_3} \qquad \textrm{and}\qquad\varphi^{a+1\Rightarrow a}(u)=\frac{(u+h_1)(u+h_2)}{(u-h_1)(u-h_2)} \;,
\end{aligned}
\end{equation}
The resulting algebra is given by restricting the charges in (\ref{eq-algebra-shifted-gl2-1}) as in (\ref{eq-constraint-gl2}).
The initial conditions becomes
\begin{tcolorbox}[ams align]\label{eq-algebra-shifted-gl2-3}
&\textcolor{black}{\textrm{Initial:}}\quad\begin{cases}
\begin{aligned}
&\begin{aligned}
&[\psi^{(a)}_0,e^{(a)}_m] = 2\,h_3\, e^{(a)}_{m} \;, \qquad\quad\, [\psi^{(a+1)}_0,e^{(a)}_m] =- 2\,h_3\, e^{(a)}_{m} \;,\\
&[\psi^{(a)}_0,f^{(a)}_m] = -2\,h_3\, f^{(a)}_{m} \;,\qquad [\psi^{(a+1)}_0,f^{(a)}_m] =2\, h_3\,  f^{(a)}_{m} \;,\\
&[\psi^{(a+1)}_1,e^{(a)}_m] =-h_3\,\{\psi^{(a+1)}_0\,,\,e^{(a)}_m\} \;,\\
&[\psi^{(a+1)}_1,f^{(a)}_m] = h_3\, \{\psi^{(a+1)}_0\,,\,f^{(a)}_m\}\;.
\end{aligned}
\end{aligned}
\end{cases}
\end{tcolorbox}
\noindent Finally, the relations above can be supplemented by the Serre relations
\begin{tcolorbox}[ams align]\label{eq-Serre-gl2}
&\textrm{Serre}:\quad\begin{cases}\begin{aligned}
&\textrm{Sym}_{z_1,z_2,z_3}\, \left[ e^{(a)}(z_1)\,, \left[ e^{(a)}(z_2)\,, \left[e^{(a)}(z_3)\,, e^{(a\pm1)}(w)\right]\right]\right] \sim 0  \;,\\
&\textrm{Sym}_{z_1,z_2,z_3}\, \left[ f^{(a)}(z_1)\,, \left[ f^{(a)}(z_2)\,,\left[f^{(a)}(z_3)\,,  f^{(a\pm1)}(w)\right]\right]\right] \sim 0  \;,
\end{aligned}
\end{cases}
\end{tcolorbox}
It only remains to check that after imposing the Serre relations, the character of the vacuum module of the algebra reproduces the generating function of the $2$-colored partition functions.
In the class of $(\mathbb{C}^2/\mathbb{Z}_n)\times \mathbb{C}$, which corresponds to affine Yangian of $\mathfrak{gl}_2$, the case $n=2$ is special in that the Serre relations given in the literature is quartic. 
However, as we show in Appendix, we already need additional relations at the level of three atoms; namely, we need Serre relations in the form of cubic relations. 
One can derive a pair of cubic relations using the quartic Serre relations (\ref{eq-Serre-gl2}) and the mode version of the OPE relation (\ref{eq-algebra-shifted-gl2-1}).
However, the computation is rather involved, and we leave it to future work.

\subsubsection{\texorpdfstring{$(\mathbb{C}^2/\mathbb{Z}_n)\times \mathbb{C}$ and Affine Yangian of $\mathfrak{gl}_n$}{(C2/Zn)xC and Affine Yangian of gl(n)}}
\label{sec:C2Zn}

\subsubsubsection{$(\mathbb{C}^2/\mathbb{Z}_n)\times \mathbb{C}$}
\label{sec:An-general}

For $\mathbb{C}^2/\mathbb{Z}_n$$\times$$\mathbb{C}$, the toric diagram and its dual graph are
\begin{equation}\label{fig-toric-Zn}
\begin{tikzpicture} 
\filldraw [red] (0,0) circle (2pt); 
\filldraw [red] (0,1) circle (2pt); 
\filldraw [red] (0,2) circle (2pt); 
\filldraw [red] (0,5) circle (2pt); 
\filldraw [red] (1,0) circle (2pt); 
\node at (-.5,-.5) {(0,0)}; 
\node at (-.5,1) {(0,1)}; 
\node at (-.5,2) {(0,2)}; 
\node at (-.5,5.5) {(0,n)}; 
\node at (1.5,-0.5) {(1,0)}; 
\draw (0,0) -- (0,2); 
\draw  (0,2) -- (0,5); 
\draw (0,0) -- (1,0); 
\draw (1,0) -- (0,5); 
\end{tikzpicture}
\qquad  \scalebox{0.6}{
\begin{tikzpicture} 
\draw[->] (0,0) -- (-1,0); 
\draw[->] (0,0) -- (0,-1); 
\draw (0,0) -- (1,1); 
\draw[->] (1,1) -- (-1,1); 
\draw (1,1) -- (3,2); 
\draw[->] (3,2) -- (-1,2); 
\draw (3,2) -- (6,3); 
\draw[->] (6,3) -- (-1,3); 
\draw (6,3) -- (10,4); 
\draw[->] (10,4) -- (-1,4); 
\draw[->] (10,4) -- (15,5); 
\node at (-1.5,0) {$3^{(1)}$}; 
\node at (-1.5,1) {$3^{(2)}$}; 
\node at (-1.5,4) {$3^{(n)}$}; 
\node at (0,-1.5) {1}; 
\end{tikzpicture}
}
\end{equation}

The quiver diagram for $\mathbb{C}^2/\mathbb{Z}_n$$\times$$\mathbb{C}$ is
\begin{equation}\label{fig-quiver-Zn}
\begin{tikzpicture}[scale=1]
\node[state]  [regular polygon, regular polygon sides=4, draw=blue!50, very thick, fill=blue!10] (a1) at (0,0)  {$1$};
\node[state]  [regular polygon, regular polygon sides=4, draw=blue!50, very thick, fill=blue!10] (a2) at (2,-1)  {$2$};
\node[state]  [regular polygon, regular polygon sides=4, draw=blue!50, very thick, fill=blue!10] (a3) at (2,-3)  {$3$};
\node[state]  [regular polygon, regular polygon sides=4, draw=blue!50, very thick, fill=blue!10] (an) at (-2,-1)  {$n$};
\node[state]  [regular polygon, regular polygon sides=4, draw=blue!50, very thick, fill=blue!10, label=center:$n\!\!-\!\! 1$] (adot) at (-2,-3)  {};
\path[->] 
(a1) edge [in=60, out=120, loop, thin, above] node {$\gamma_1$} ()
(a2) edge [in=330, out=30, loop, thin, right] node {$\gamma_2$} ()
(a3) edge [in=330, out=30, loop, thin, right] node {$\gamma_3$} ()
(an) edge [in=150, out=210, loop, thin, left] node {$\gamma_n$} ()
(adot) edge [in=150, out=210, loop, thin, left] node {$\gamma_{n-1}$} ()
(a1) edge   [thin, bend left]  node [above] {$\alpha_{1}$} (a2)
(a2) edge   [thin, bend left]  node [below]{$\beta_{1}$} (a1)
(a2) edge   [thin, bend left]  node [right] {$\alpha_{2}$} (a3)
(a3) edge   [thin, bend left]  node [left]{$\beta_{2}$} (a2)
(an) edge   [thin, bend left]  node [above] {$\alpha_{n}$} (a1)
(a1) edge   [thin, bend left]  node [below]{$\beta_{n}$} (an)
(adot) edge   [thin, bend left]  node [left] {$\alpha_{n-1}$} (an)
(an) edge   [thin, bend left]  node [right]{$\beta_{n-1}$} (adot)
(a3) edge   [dotted, bend left]  node [] {} (adot)
(adot) edge   [dotted, bend right]  node [] {}  (a3)
;
\end{tikzpicture}
\end{equation}
with super potential
\begin{equation}
W=\sum^n_{a=1}\textrm{Tr}[-\Phi_{a,a} \, \Phi_{a,a+1} \, \Phi_{a+1,a} + \Phi_{a,a} \, \Phi_{a, a-1}\, \Phi_{a-1,a}] \;,
\end{equation}
and the charges assignment 
\begin{equation}
\Phi_{a, a+1}: \quad \alpha_a \,, \qquad \Phi_{a+1, a}: \quad \beta_a \,, \qquad \Phi_{a,a}:\quad \gamma_a\,.
\end{equation}
We see that all vertices are bosonic:
\begin{equation}\label{eq-gl2-boson-2}
|a|=0 \,,\qquad a=1,2,\cdots,n\,,
\end{equation}
since there is a self-loop for each of them in the quiver.

\begin{figure}[htbp]
\begin{minipage}{0.5\linewidth}
\centering
\begin{tikzpicture}[scale=1]
\node[state]  [regular polygon, regular polygon sides=4, draw=blue!50, very thick, fill=blue!10] (a1) at (0,0)  {$1$};
\node[state]  [regular polygon, regular polygon sides=4, regular polygon, regular polygon sides=4, draw=blue!50, very thick, fill=blue!10] (a21) at (-1,-1.73205)  {$2$};
\node[state]  [regular polygon, regular polygon sides=4, draw=blue!50, very thick, fill=blue!10] (a22) at (2,0)  {$n$};
\node[state]  [regular polygon, regular polygon sides=4, draw=blue!50, very thick, fill=blue!10] (a23) at (-1,1.73205)  {$1$};
\node[state]  [regular polygon, regular polygon sides=4, draw=blue!50, very thick, fill=blue!10] (a312) at (1,-1.73205)  {$1$};
\node[state]  [regular polygon, regular polygon sides=4, draw=blue!50, very thick, fill=blue!10] (a323) at (1,1.73205)  {$n$};
\node[state]  [regular polygon, regular polygon sides=4, draw=blue!50, very thick, fill=blue!10] (a331) at (-2,0)  {$2$};
\path[->] 
(a1) edge   [thick, red]   node [left] {$\alpha_1$} (a21)
(a23) edge   [thick, red]   node [left] {$\alpha_1$} (a331)
(a22) edge   [thick, red]   node [right] {$\alpha_n$} (a312)
(a323) edge   [thick, red]   node [left] {$\alpha_n$} (a1)
(a1) edge   [thick, red]   node [above] {$\beta_n$} (a22)
(a21) edge   [thick, red]   node [below] {$\beta_1$} (a312)
(a23) edge   [thick, red]   node [above] {$\beta_n$} (a323)
(a331) edge   [thick, red]   node [above] {$\beta_1$} (a1)
(a1) edge   [thick, red]   node [left] {$\gamma_1$} (a23)
(a21) edge   [thick, red]   node [left] {$\gamma_2$} (a331)
(a22) edge   [thick, red]   node [right] {$\gamma_n$} (a323)
(a312) edge   [thick, red]   node [right] {$\gamma_1$} (a1)
;
\end{tikzpicture}
\end{minipage}
\hfill
\begin{minipage}{0.5\linewidth}
\centering
\begin{tikzpicture}[scale=0.6]
\node[state]  [regular polygon, regular polygon sides=4, draw=blue!50, very thick, fill=blue!10] (a1) at (0,0)  {$1$};
\node[state]  [regular polygon, regular polygon sides=4, draw=blue!50, very thick, fill=blue!10] (a21) at (3,0)  {$n$};
\node[state]  [regular polygon, regular polygon sides=4, draw=blue!50, very thick, fill=blue!10] (a22) at (-3,0)  {$2$};
\node[state]  [regular polygon, regular polygon sides=4, draw=blue!50, very thick, fill=blue!10] (a41) at (0,3)  {$n$};
\node[state]  [regular polygon, regular polygon sides=4, draw=blue!50, very thick, fill=blue!10] (a42) at (0,-3)  {$2$};
\node[state]  [regular polygon, regular polygon sides=4, draw=blue!50, very thick, fill=blue!10, label=center:$n\!\!-\!\! 1$] (a31) at (3,3)  {};
\node[state]  [regular polygon, regular polygon sides=4, draw=blue!50, very thick, fill=blue!10] (a32) at (3,-3)  {$1$};
\node[state]  [regular polygon, regular polygon sides=4, draw=blue!50, very thick, fill=blue!10] (a34) at (-3,-3)  {$3$};
\node[state]  [regular polygon, regular polygon sides=4, draw=blue!50, very thick, fill=blue!10] (a33) at (-3,3)  {$1$};
\path[->] 
(a1) edge   [thick, red]   node [above] {$\beta_n$} (a21)
(a22) edge   [thick, red]   node [above] {$\beta_1$} (a1)
(a31) edge   [thick, red]   node [right] {$\alpha_{n-1}$} (a21)
(a21) edge   [thick, red]   node [right] {$\alpha_n$} (a32)
(a33) edge   [thick, red]   node [left] {$\alpha_1$} (a22)
(a22) edge   [thick, red]   node [left] {$\alpha_2$} (a34)
(a41) edge   [thick, red]   node [above] {$\beta_{n-1}$} (a31)
(a33) edge   [thick, red]   node [above] {$\beta_n$} (a41)
(a34) edge   [thick, red]   node [above] {$\beta_2$} (a42)
(a42) edge   [thick, red]   node [above] {$\beta_1$} (a32)
(a41) edge   [thick, red]   node [right] {$\alpha_{n}$} (a1)
(a1) edge   [thick, red]   node [right] {$\alpha_1$} (a42)
(a1) edge   [thick, red]   node [right] {$\gamma_1$} (a33)
(a32) edge   [thick, red]   node [right] {$\gamma_1$} (a1)
(a21) edge   [thick, red]   node [right] {$\gamma_n$} (a41)
(a42) edge   [thick, red]   node [right] {$\gamma_2$} (a22)
;
\end{tikzpicture}
\end{minipage}
\caption{Two ways to draw the periodic quiver $(\mathbb{C}^{2}/\mathbb{Z}_n)\times\mathbb{C}$. 
The left one shows that the representation can be realized by coloring the plane partitions, whereas the right one is for later comparison with the periodic quiver for the generalized conifolds. Note that this shows only part of the periodic quiver diagram around the vertex $1$.}
\label{fig-periodic-quiver-gln}
\end{figure}
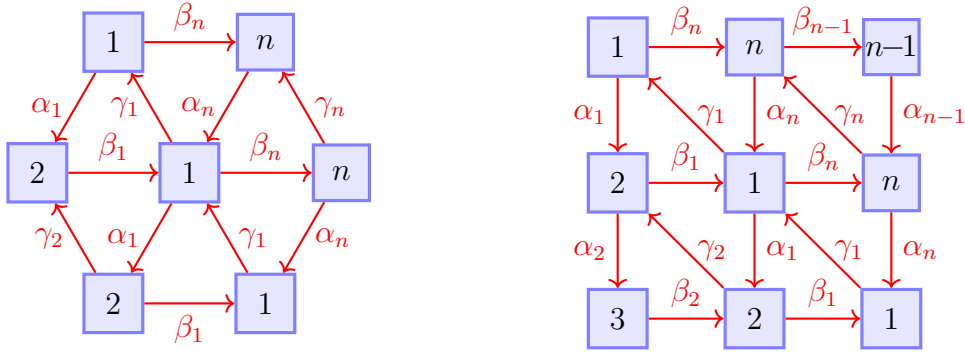

The periodic quiver for (\ref{fig-quiver-Zn}) is given in Figure \ref{fig-periodic-quiver-gln}, where we have shown only the part of the graph around the vertex $1$; the full graph is obtained by periodically extending the graph.
Comparing the left drawing in Figure \ref{fig-periodic-quiver-gln} with the left one in Figure~\ref{fig-periodic-quiver-C3-big} (i.e.\ the periodic quiver that gives the affine Yangian of $\mathfrak{gl}_1$), we see the representation of the algebra for $(\mathbb{C}^{2}/\mathbb{Z}_n)\times\mathbb{C}$ can be obtained by coloring the plane partitions by the following rules: the box at the origin has color $1$; the color increases by $1$ as one moves by one step along the positive $x_1$ direction, decreases by $1$ by each step along the positive $x_2$ direction, and remains the same along the $x_3$ direction.

Again, the loop constraint (\ref{eq-loop-constraint-toric}) gives
\begin{equation}\label{eq-loop-constraint-Zn}
\alpha_a+\beta_a+\gamma_a=0 \qquad \textrm{and} \qquad \alpha_{a}+\beta_{a}+\gamma_{a-1}=0\,,
\qquad \textrm{for}\quad a=1,2,\dots, n\,,
\end{equation}
which gives
\begin{equation}\label{eq-loop-constraint-Zn-2}
\gamma_1=\gamma_2=\cdots=\gamma_n \equiv \gamma \qquad \textrm{and} \qquad \alpha_{a}+\beta_{a}=\alpha_{a+1}+\beta_{a+1} \,, \quad \textrm{for}\quad a=1,2,\dots, n\,,
\end{equation}
which are in total $2n-1$ independent constraints on the 3n variables $(\alpha_a,\beta_a, \gamma_a)$ with $a=1,2,\dots,n$.
Namely, the algebra for $\mathbb{C}^2/\mathbb{Z}_n$$\times$$\mathbb{C}$ has $n+1$ parameters.
Again, the central condition (\ref{eq-central-condition}) is guaranteed by the loop constraint (\ref{eq-loop-constraint-Zn-2}).

One can immediately read off the bond factors from the periodic quiver in Figure~\ref{fig-periodic-quiver-gln}  by the definition (\ref{eq-charge-atob})
\begin{equation}\label{eq-charge-function-Zn}
\begin{aligned}
&\varphi^{a\Rightarrow a}(u)=\frac{u+\gamma}{u-\gamma}\,, \qquad \varphi^{a\Rightarrow a+1}(u)=\frac{u+\beta_a}{u-\alpha_a} \,, \qquad \varphi^{a\Rightarrow a-1}(u)=\frac{u+\alpha_{a-1}}{u-\beta_{a-1}} \,,
\\
&\varphi^{a\Rightarrow b}(u)=1 \,, \quad ( b\neq a\,,\,a\pm 1)\,,
\end{aligned}
\end{equation}
where the indices are understood as mod $n$.

The bond factors (\ref{eq-charge-function-Zn}) give the algebra
\begin{tcolorbox}[ams align]
&\textrm{OPE:}\quad\begin{cases}\begin{aligned}\label{eq-OPE-gln}
\psi^{(a)}(z)\, \psi^{(b)}(w)&\sim \psi^{(b)}(w)\, \psi^{(a)}(z) \;,\\
\psi^{(a)}(z)\, e^{(a)}(w)   &\sim \tfrac{\Delta+\gamma}{\Delta-\gamma} \, e^{(a)}(w)\, \psi^{(a)}(z)  \;,\\ 
e^{(a)}(z)\, e^{(a)}(w) & \sim  \tfrac{\Delta+\gamma}{\Delta-\gamma} \, e^{(a)}(w)\, e^{(a)}(z)  \;,\\
\psi^{(a)}(z)\, f^{(a)}(w) &  \sim \tfrac{\Delta-\gamma}{\Delta+\gamma} \, f^{(a)}(w)\, \psi^{(a)}(z)  \;,\\
 f^{(a)}(z)\, f^{(a)}(w) &  \sim   \tfrac{\Delta-\gamma}{\Delta+\gamma} \, f^{(a)}(w)\, f^{(a)}(z)  \;,\\
 \psi^{(a+1)}(z)\, e^{(a)}(w)   &\sim \tfrac{\Delta+\beta_a}{\Delta-\alpha_a}  \, e^{(a)}(w)\, \psi^{(a+1)}(z)  \;,\\ 
   \psi^{(a-1)}(z)\, e^{(a)}(w)   &\sim \tfrac{\Delta+\alpha_{a-1}}{\Delta-\beta_{a-1}} \, e^{(a)}(w)\, \psi^{(a-1)}(z)  \;,\\ 
e^{(a+1)}(z)\, e^{(a)}(w) & \sim  \tfrac{\Delta+\beta_a}{\Delta-\alpha_a} \, e^{(a)}(w)\, e^{(a+1)}(z)  \;,\\
\psi^{(a+1)}(z)\, f^{(a)}(w) &  \sim \tfrac{\Delta-\alpha_a}{\Delta+\beta_a} \, f^{(a)}(w)\, \psi^{(a+1)}(z)  \;,\\
\psi^{(a-1)}(z)\, f^{(a)}(w) &  \sim \tfrac{\Delta-\beta_{a-1}}{\Delta+\alpha_{a-1}}  \, f^{(a)}(w)\, \psi^{(a-1)}(z)  \;,\\
 f^{(a+1)}(z)\, f^{(a)}(w) &  \sim \tfrac{\Delta-\alpha_a}{\Delta+\beta_a} \,f^{(a)}(w)\, f^{(a+1)}(z) \;,\\
\psi^{(b)}(z)\, e^{(a)}(w)   &\sim  \, e^{(a)}(w)\, \psi^{(b)}(z)\quad\, (b\ne a, a\pm 1)   \;,\\ 
e^{(b)}(z)\, e^{(a)}(w) & \sim  \, e^{(a)}(w)\, e^{(b)}(z) \quad \,\,(b\ne a, a\pm 1)  \;,\\
\psi^{(b)}(z)\, f^{(a)}(w) &  \sim  \, f^{(a)}(w)\, \psi^{(b)}(z)  \quad (b\ne a, a\pm 1) \;,\\
 f^{(b)}(z)\, f^{(a)}(w) &  \sim   \, f^{(a)}(w)\, f^{(b)}(z)  \quad (b\ne a, a\pm 1) \;,\\
[e^{(a)}(z)\,, f^{(b)}(w)]   &\sim  - \delta^{a,b}\, \frac{\psi^{(a)}(z) - \psi^{(b)}(w)}{z-w} \ , 
\end{aligned}
\end{cases} 
\\
&\textcolor{black}{\textrm{Initial:}}\quad\begin{cases}
\begin{aligned}\label{eq-initial-shifted-gl2}
&\begin{aligned}
[\psi^{(a-1)}_0,e^{(a)}_m] &=- \gamma  \, e^{(a)}_m\;,\\
[\psi^{(a)}_0,e^{(a)}_m] &= 2\gamma   \, e^{(a)}_m\;,\\
[\psi^{(a+1)}_0,e^{(a)}_m] &=- \gamma \, e^{(a)}_m \;,\\
[\psi^{(b)}_0,e^{(a)}_m] &=0\;,
\end{aligned}\qquad 
&\begin{aligned}
[\psi^{(a-1)}_0,f^{(a)}_m] &=  \gamma  \, f^{(a)}_m\;,\\
[\psi^{(a)}_0,f^{(a)}_m] &= -2\gamma\, f^{(a)}_m\;,\\
[\psi^{(a+1)}_0,f^{(a)}_m] &=  \gamma \, f^{(a)}_m \;,\\
[\psi^{(b)}_0,f^{(a)}_m] &=0\;,\quad ( b\neq a\,,\, a\pm 1)\,,
\end{aligned}
\end{aligned}
\end{cases}
\end{tcolorbox}
\noindent where in the computation of the initial conditions, only equations with $\ell=1$ in the general formula (\ref{eq-initial-toric}) is non-empty, since all the bond factors are of order $1$. 
As a result, we only have initial conditions on $[\psi^{(a)}_0\, , \, e^{(b)}_m]$ and $[\psi^{(a)}_0\, , \, f^{(b)}_m]$.
From the initial conditions one can check that the combination $\sum^{n}_{a=1}\psi^{(a)}_0$ is indeed the central term of the algebra.

\subsubsubsection{Serre Relations}
\label{sec:serre-gl3}

We now demonstrate how to check that the vacuum module of the reduced quiver Yangian $\underline{\mathsf{Y}}_{(Q,W)}$ reproduces the generating function of the colored crystals, using the example of $(\mathbb{C}^2/\mathbb{Z}_n)\times \mathbb{C}$ with $n\geq 3$.
In particular, we will show that the Serre relations are necessary in reducing the number of the states to the correct counting.
The case of $n=1$ corresponds to affine Yangian of $\mathfrak{gl}_1$ and was explained in section~\ref{sec:serre-gl1}.
See appendix for a discussion on the case of $n=2$.

As we will explain later, the analysis is the same for all the $(\mathbb{C}^2/\mathbb{Z}_n)\times \mathbb{C}$ with $n\geq 3$, since both the OPE relations and the Serre relations have the same structures for all $n\geq 3$. (As a contrast, the case of $n=1$ and $n=2$ are special.)
For concreteness, we explain in detail the case of $n=3$.

The generating function of the $3$-colored plane partition (with the coloring scheme explained in section~\ref{sec:An-general}) is
\begin{equation}\label{eq-count-gl3}
\begin{aligned}
Z(q_1,q_2,q_3)=&\sum_{n_1,n_2,n_3}d(n_1,n_2,n_3)q^{n_1}_1\,q_2^{n_2}\,q_3^{n_3}\\
=&1+q_1+(q_1^2+q_1\, q_2+q_1\, q_3)+(q_1^3+ q_1^2\,q_2+q_1^2\,q_3+3\,q_1\,q_2\,q_3)\\
& +(q_1^4+q_1^3\,q_2+q_1^3\,q_3+q_1^2\,q_2^2+q_1^2\,q_3^2+6\,q_1^2\,q_2\,q_3+q_1\,q_2^2\,q_3+ q_1\,q_2\,q_3^2)+\dots
\end{aligned}
\end{equation}
where $d(n_1,n_2,n_3)$ counts the number of distinct configurations with $n_i$ number of $i^{\textrm{th}}$ colored atom in the crystal. 
Inside each bracket we have grouped all the terms with the same total number of atoms $N=n_1+n_2+n_3$.

Let us now reproduce the counting  (\ref{eq-count-gl3}) level by level.
As we will show presently, the first non-trivial level is $N=3$, which is the first level where one needs the Serre relations in order to reduce the counting of the vacuum module of the quiver Yangian to that of the reduced quiver Yangian in order to match the counting in (\ref{eq-count-gl3}).
The Serre relations introduced for the level $N=3$ are actually sufficient for all higher levels. 
We will demonstrate this explicitly for $N=4$.
\medskip

\noindent\underline{Vacuum.}
There is a unique vacuum:
\begin{equation}
(n_1,n_2,n_3)=(0,0,0):\qquad \qquad |\emptyset\rangle
\end{equation}
shown by the $1$ in (\ref{eq-count-gl3}).
\smallskip

\noindent\underline{One atom.}
Since the leading atom has color $a=1$ by our convention, there is only one state with one atom: $|\sqbox{1}\rangle$. 
Let us verify this in terms of $e^{(a)}_n |\emptyset\rangle$.

The action of $e^{(a)}(z)$ on vacuum is
\begin{equation}
e^{(1)}(z)|\emptyset\rangle =\frac{\#}{z} |\sqbox{1}\rangle\ \quad \textrm{and}\quad e^{(2)}(z)|\emptyset\rangle=e^{(3)}(z)|\emptyset\rangle=0\,,
\end{equation}
which, when translated into modes, gives
\begin{equation}
e^{(1)}_{0}|\emptyset\rangle=\#|\sqbox{1}\rangle\quad \textrm{and}\quad e^{(1)}_{n\geq 1}|\emptyset\rangle =e^{(2)}_{n\geq 0}|\emptyset\rangle =e^{(3)}_{n\geq 0}|\emptyset\rangle =0 \;.
\end{equation}
Namely, there is only one state at $N=1$:
\begin{equation}\label{eq-gl3-N1}
(n_1,n_2,n_3)=(1,0,0):\qquad \qquad e^{(1)}_0 |\emptyset\rangle\,.
\end{equation}
\smallskip

\noindent\underline{Two atoms.}
There are three possible configurations with two atoms: with $(n_1,n_2,n_3)=(2,0,0)$, $(1,1,0)$, and $(1,0,1)$, respectively. 
We need to reproduce this counting  in terms of $e^{(b)}_m \, e^{(a)}_n |\emptyset\rangle$.

Starting from the unique $N=1$ state (\ref{eq-gl3-N1}), the potential $N=2$ states are
\begin{equation}\label{eq-gl3-N2-preOPE}
e^{(1)}_n \, e^{(1)}_0 |\emptyset\rangle\,,\quad e^{(2)}_n \, e^{(1)}_0 |\emptyset\rangle \,,\quad e^{(3)}_n \, e^{(1)}_0 |\emptyset\rangle \qquad \textrm{with}\quad n\in \mathbb{Z}_{\geq 0}\,.
\end{equation}
Now we use the mode version of the relevant algebraic relations from (\ref{eq-OPE-gln}), namely
\begin{equation}\label{eq-mode-gl3}
\begin{aligned}
&[e^{(1)}_{n+1}, e^{(1)}_{m} ]-[e^{(1)}_n, e^{(1)}_{m+1} ]=\gamma \{e^{(1)}_{n}, e^{(1)}_{m} \}\,,\\
& [e^{(2)}_{n+1}, e^{(1)}_{m} ]-[e^{(2)}_n, e^{(1)}_{m+1} ]=\alpha_1\, e^{(2)}_{n}\,e^{(1)}_{m} +\beta_1\,e^{(1)}_{m} \, e^{(2)}_{n}\,,\\
& [e^{(3)}_{n+1}, e^{(1)}_{m} ]-[e^{(3)}_n, e^{(1)}_{m+1} ]=\beta_3\, e^{(3)}_{n}\,e^{(1)}_{m} +\alpha_3\,e^{(1)}_{m}\, e^{(3)}_{n} \,,
\end{aligned}
\end{equation}
to eliminate the set (\ref{eq-gl3-N2-preOPE}) down to only three independent states:
\begin{equation}\label{eq-gl3-N2-postOPE}
e^{(1)}_0 \, e^{(1)}_0 |\emptyset\rangle\,,\quad e^{(2)}_0 \, e^{(1)}_0 |\emptyset\rangle \,,\quad e^{(3)}_0 \, e^{(1)}_0 |\emptyset\rangle\,.
\end{equation}
They correspond to the three configurations with $(n_1,n_2,n_3)=(2,0,0)$, $(1,1,0)$, and $(1,0,1)$, respectively.
Note that for all affine Yangian of $\mathfrak{gl}_{\mathrm{n}\geq 3}$, all the $e^{(a)}_n e^{(b)}_m$ relations are step-1 relations, irrespective of the relation between $a$ and $b$. 
We see that the OPE relations are enough to reproduce the counting up to $N=2$ level, same as the case of affine Yangian of $\mathfrak{gl}_1$
\smallskip

\noindent\underline{Three atoms.}
From the generating function (\ref{eq-count-gl3}), we see that there are six different configurations with $N=3$ atoms, one with $(n_1,n_2,n_3)=(3,0,0)$, one with $(n_1,n_2,n_3)=(2,1,0)$, one with $(n_1,n_2,n_3)=(2,0,1)$, and finally three with $(n_1,n_2,n_3)=(1,1,1)$.
We need to reproduce this counting in terms of states of the form $e^{(c)}_{\ell} \, e^{(b)}_m \,e^{(a)}_{n}|\emptyset\rangle$.

Starting with the three independent states (\ref{eq-gl3-N2-postOPE}) with $N=2$, applying $e^{(a)}_n$, and finally using the relations (\ref{eq-mode-gl3}) to eliminate dependent states, we have
\begin{equation}\label{eq-gl3-preS}
\begin{aligned}
&(n_1,n_2,n_3)=(3,0,0):\quad e^{(1)}_0 \,e^{(1)}_0 \, e^{(1)}_0 |\emptyset\rangle\,, \\
&(n_1,n_2,n_3)=(2,1,0):\quad e^{(2)}_0 \,e^{(1)}_0 \, e^{(1)}_0 |\emptyset\rangle\,,\quad e^{(1)}_0 \,e^{(2)}_0 \, e^{(1)}_0 |\emptyset\rangle\,,\quad e^{(1)}_1 \,e^{(2)}_0 \, e^{(1)}_0 |\emptyset\rangle\,, \\
&(n_1,n_2,n_3)=(1,2,0):\quad e^{(2)}_0 \,e^{(2)}_0 \, e^{(1)}_0 |\emptyset\rangle\,, \\
&(n_1,n_2,n_3)=(2,0,1):\quad e^{(3)}_0 \,e^{(1)}_0 \, e^{(1)}_0 |\emptyset\rangle\,,\quad e^{(1)}_0 \,e^{(3)}_0 \, e^{(1)}_0 |\emptyset\rangle\,,\quad e^{(1)}_1 \,e^{(3)}_0 \, e^{(1)}_0 |\emptyset\rangle\,, \\
&(n_1,n_2,n_3)=(1,0,2):\quad e^{(3)}_0 \,e^{(3)}_0 \, e^{(1)}_0 |\emptyset\rangle\,, \\
&(n_1,n_2,n_3)=(1,1,1):\quad e^{(3)}_0 \,e^{(2)}_0 \, e^{(1)}_0 |\emptyset\rangle\,, \quad e^{(2)}_0 \,e^{(3)}_0 \, e^{(1)}_0 |\emptyset\rangle\,, \quad e^{(2)}_1 \,e^{(3)}_0 \, e^{(1)}_0 |\emptyset\rangle\,. \\
\end{aligned}
\end{equation}
Comparing (\ref{eq-gl3-preS}) with the counting (\ref{eq-count-gl3}) from the $3$-colored crystal at the level $N=3$, we see that the states with $(n_1,n_2,n_3)=(3,0,0)$ and $(1,1,1)$ match on the nose, whereas (\ref{eq-gl3-preS}) contains too many states for the other four classes, which all have the form  $e^{(b)}_{\ell} \, e^{(b)}_m \,e^{(a)}_{n}|\emptyset\rangle$ or $e^{(a)}_{\ell} \, e^{(b)}_m \,e^{(a)}_{n}|\emptyset\rangle$ with $b\neq a$.

We see that we need Serre relations involving terms of the form $e^{(b)}_{\ell} \, e^{(b)}_m \,e^{(a)}_{n}$ and $e^{(a)}_{\ell} \, e^{(b)}_m \,e^{(a)}_{n}$ with $b\neq a$.
The correct Serre relations (for all $n\geq 3$) turn out to be
\begin{equation}
\begin{aligned}\label{eq-serre-gl3}
&\textrm{Sym}_{z_1,z_2}\, \left[ e^{(a)}(z_1)\,, \left[ e^{(a)}(z_2)\,, e^{(a\pm1)}(w)\right]\right] \sim 0  \;,\\
\end{aligned}
\end{equation}
whose mode version is
\begin{equation}\label{eq-serre-gl3-mode}
\textrm{Sym}_{n_1,n_2} \left[ e^{(a)}_{n_1}\,, \left[ e^{(a)}_{n_2}\,, e^{(a\pm1)}_{m}\right]\right] = 0  \,.
\end{equation}
(There are also parallel Serre equations with $e$ replaced by $f$. However we do not need them for the computation of the vacuum characters.)

Using the Serre relation (\ref{eq-serre-gl3-mode}) to eliminate additional dependent states from the 2nd to 4th lines of (\ref{eq-gl3-preS}), we have
\begin{equation}\label{eq-gl3-N3-postS}
\begin{aligned}
&(n_1,n_2,n_3)=(3,0,0):\quad e^{(1)}_0 \,e^{(1)}_0 \, e^{(1)}_0 |\emptyset\rangle\,, \\
&(n_1,n_2,n_3)=(2,1,0):\quad e^{(2)}_0 \,e^{(1)}_0 \, e^{(1)}_0 |\emptyset\rangle\,, \\
&(n_1,n_2,n_3)=(2,0,1):\quad e^{(3)}_0 \,e^{(1)}_0 \, e^{(1)}_0 |\emptyset\rangle\,, \\
&(n_1,n_2,n_3)=(1,1,1):\quad e^{(3)}_0 \,e^{(2)}_0 \, e^{(1)}_0 |\emptyset\rangle\,, \quad e^{(2)}_0 \,e^{(3)}_0 \, e^{(1)}_0 |\emptyset\rangle\,, \quad e^{(2)}_1 \,e^{(3)}_0 \, e^{(1)}_0 |\emptyset\rangle\,, \\
\end{aligned}
\end{equation}
which match precisely with the counting  (\ref{eq-count-gl3}) from the $3$-colored crystal.
\smallskip

\noindent\underline{Four atoms and beyond.}
The generating function (\ref{eq-count-gl3}) indicates that there are $13$ different configurations with $N=4$ atoms, in $8$ different classes.
We need to reproduce this counting in terms of states of the form $e^{(d)}_{k} \, e^{(c)}_{\ell} \, e^{(b)}_m \,e^{(a)}_{n}|\emptyset\rangle$.

The analysis is similar to the case with three atoms. 
Applying the $e^{(a)}_{n}$ generator on the $6$ states (\ref{eq-gl3-N3-postS}) from the level $N=3$ and imposing the OPE relations (\ref{eq-mode-gl3}), we get 
\begin{equation}\label{eq-gl3-N4-preS}
\begin{aligned}
(n_1,n_2,n_3)=(4,0,0):
&\quad e^{(1)}_0 \,e^{(1)}_0 \,e^{(1)}_0 \, e^{(1)}_0 |\emptyset\rangle\,, \\
(n_1,n_2,n_3)=(3,1,0):
&\quad e^{(2)}_0 \,e^{(1)}_0 \,e^{(1)}_0 \, e^{(1)}_0 |\emptyset\rangle\,, 
\quad e^{(1)}_0 \,e^{(2)}_0 \,e^{(1)}_0 \, e^{(1)}_0 |\emptyset\rangle\,,
\quad e^{(1)}_1 \,e^{(2)}_0 \,e^{(1)}_0 \, e^{(1)}_0 |\emptyset\rangle\,,
\\
(n_1,n_2,n_3)=(2,2,0):
&\quad e^{(2)}_0 \,e^{(2)}_0 \,e^{(1)}_0 \, e^{(1)}_0 |\emptyset\rangle\,, 
\\
(n_1,n_2,n_3)=(3,0,1):
&\quad e^{(3)}_0 \,e^{(1)}_0 \,e^{(1)}_0 \, e^{(1)}_0 |\emptyset\rangle\,, 
\quad e^{(1)}_0 \,e^{(3)}_0 \,e^{(1)}_0 \, e^{(1)}_0 |\emptyset\rangle\,,
\quad e^{(1)}_1 \,e^{(3)}_0 \,e^{(1)}_0 \, e^{(1)}_0 |\emptyset\rangle\,,
\\
(n_1,n_2,n_3)=(2,0,2):
&\quad e^{(3)}_0 \,e^{(3)}_0 \,e^{(1)}_0 \, e^{(1)}_0 |\emptyset\rangle\,, 
\\
(n_1,n_2,n_3)=(2,1,1):
&\quad e^{(3)}_0 \,e^{(2)}_0 \,e^{(1)}_0 \, e^{(1)}_0 |\emptyset\rangle\,, 
\quad e^{(3)}_1 \,e^{(2)}_0 \,e^{(1)}_0 \, e^{(1)}_0 |\emptyset\rangle\,, 
\quad e^{(2)}_0 \,e^{(3)}_0 \,e^{(1)}_0 \, e^{(1)}_0 |\emptyset\rangle\,, \\
&
\quad e^{(1)}_0 \,e^{(3)}_0 \,e^{(2)}_0 \, e^{(1)}_0 |\emptyset\rangle\,, 
\quad e^{(1)}_0 \,e^{(2)}_0 \,e^{(3)}_0 \, e^{(1)}_0 |\emptyset\rangle\,, 
\quad e^{(1)}_0 \,e^{(2)}_1 \,e^{(3)}_0 \, e^{(1)}_0 |\emptyset\rangle\,, \\
(n_1,n_2,n_3)=(1,2,1):
&\quad e^{(2)}_0 \, e^{(2)}_0 \,e^{(3)}_0 \, e^{(1)}_0 |\emptyset\rangle\,, 
\quad e^{(2)}_0 \, e^{(2)}_1 \,e^{(3)}_0 \, e^{(1)}_0 |\emptyset\rangle\,,
\quad e^{(2)}_1 \, e^{(2)}_1 \,e^{(3)}_0 \, e^{(1)}_0 |\emptyset\rangle\,, \\
&\quad e^{(2)}_0 \, e^{(3)}_0 \,e^{(2)}_0 \, e^{(1)}_0 |\emptyset\rangle\,, \\
(n_1,n_2,n_3)=(1,1,2):
&\quad e^{(3)}_0 \, e^{(3)}_0 \,e^{(2)}_0 \, e^{(1)}_0 |\emptyset\rangle\,, 
\quad e^{(3)}_0 \, e^{(3)}_1 \,e^{(2)}_0 \, e^{(1)}_0 |\emptyset\rangle\,,
\quad e^{(3)}_1 \, e^{(3)}_1 \,e^{(2)}_0 \, e^{(1)}_0 |\emptyset\rangle\,, \\
&\quad e^{(3)}_0 \, e^{(2)}_0 \,e^{(3)}_0 \, e^{(1)}_0 |\emptyset\rangle\,,
\end{aligned}
\end{equation}
which contain more states than the counting (\ref{eq-count-gl3}) indicates.

However, after applying Serre relation (\ref{eq-serre-gl3-mode}), the independent states in (\ref{eq-gl3-N4-preS}) are reduced to
\begin{equation}\label{eq-gl3-N4-postS}
\begin{aligned}
(n_1,n_2,n_3)=(4,0,0):
&\quad e^{(1)}_0 \,e^{(1)}_0 \,e^{(1)}_0 \, e^{(1)}_0 |\emptyset\rangle\,, \\
(n_1,n_2,n_3)=(3,1,0):
&\quad e^{(2)}_0 \,e^{(1)}_0 \,e^{(1)}_0 \, e^{(1)}_0 |\emptyset\rangle\,, 
\\
(n_1,n_2,n_3)=(2,2,0):
&\quad e^{(2)}_0 \,e^{(2)}_0 \,e^{(1)}_0 \, e^{(1)}_0 |\emptyset\rangle\,, 
\\
(n_1,n_2,n_3)=(3,0,1):
&\quad e^{(3)}_0 \,e^{(1)}_0 \,e^{(1)}_0 \, e^{(1)}_0 |\emptyset\rangle\,, 
\\
(n_1,n_2,n_3)=(2,0,2):
&\quad e^{(3)}_0 \,e^{(3)}_0 \,e^{(1)}_0 \, e^{(1)}_0 |\emptyset\rangle\,, 
\\
(n_1,n_2,n_3)=(2,1,1):
&\quad e^{(3)}_0 \,e^{(2)}_0 \,e^{(1)}_0 \, e^{(1)}_0 |\emptyset\rangle\,, 
\quad e^{(3)}_1 \,e^{(2)}_0 \,e^{(1)}_0 \, e^{(1)}_0 |\emptyset\rangle\,, 
\quad e^{(2)}_0 \,e^{(3)}_0 \,e^{(1)}_0 \, e^{(1)}_0 |\emptyset\rangle\,, \\
&
\quad e^{(1)}_0 \,e^{(3)}_0 \,e^{(2)}_0 \, e^{(1)}_0 |\emptyset\rangle\,, 
\quad e^{(1)}_0 \,e^{(2)}_0 \,e^{(3)}_0 \, e^{(1)}_0 |\emptyset\rangle\,, 
\quad e^{(1)}_0 \,e^{(2)}_1 \,e^{(3)}_0 \, e^{(1)}_0 |\emptyset\rangle\,, \\
(n_1,n_2,n_3)=(1,2,1):
&\quad e^{(2)}_0 \, e^{(2)}_0 \,e^{(3)}_0 \, e^{(1)}_0 |\emptyset\rangle\,, \\
(n_1,n_2,n_3)=(1,1,2):
&\quad e^{(3)}_0 \, e^{(3)}_0 \,e^{(2)}_0 \, e^{(1)}_0 |\emptyset\rangle\,,
\end{aligned}
\end{equation}
which match precisely with the counting  (\ref{eq-count-gl3}) from the $3$-colored crystal. 
The higher level (with more than $5$ atoms) can be checked in this way. 

We emphasize that although we have only shown the detailed computation for $(\mathbb{C}^{2}/\mathbb{Z}_n)\times\mathbb{C}$ with $n=3$, 
the computation for all  $n\geq 3$ works in exactly the same way, and the expressions for all $n\geq 3$ have the same  structure.
The reason is that both the OPE relations (\ref{eq-OPE-gln}) and the Serre relations (\ref{eq-serre-gl3}) have exactly the same structure for all $n\geq 3$.

\subsubsubsection{Truncation}
\label{sec:truncation-gln}

For the truncation, consider a path from the origin $\mathfrak{o}$ to an atom $\sqbox{$1$}$ of color $a=1$, at which the growth of the crystal stops.
The coordinate function of this atom $\sqbox{$1$}$ is
\begin{equation}\label{eq-truncation-1}
h(\sqbox{$1$})=\sum^n_{a=1}N_{\gamma_a}\gamma_a +\sum^{n}_{a=1}N_{a} (\alpha_a+\beta_a)+N_{\alpha}\sum^{n}_{a=1}\alpha_a+N_{\beta}\sum^{n}_{a=1}\beta_a  \;,
\end{equation}
where $N_{\gamma_a}$ denotes the number of edges with $\gamma_a$ in the path, $N_{a}$ the number of segment $a\rightarrow a+1 \rightarrow a$, $N_{\alpha}$ the number of the segment $1\rightarrow 2\rightarrow \cdots \rightarrow n\rightarrow 1$, and $N_{\beta}$ the number of segment $1\rightarrow n\rightarrow \cdots \rightarrow 2\rightarrow 1$.
Using the loop constraint (\ref{eq-loop-constraint-Zn}) and (\ref{eq-loop-constraint-Zn-2}), the coordinate function can be rewritten as
\begin{equation}
h(\sqbox{$1$})=\gamma \sum^n_{a=1}\left(N_{\gamma_a}-N_{a}-N_{\beta}\right) +\left(\sum^{n}_{a=1}\alpha_a\right)(N_{\alpha}-N_{\beta})  \;,
\end{equation}
Therefore the algebra truncates when the parameter $\{\alpha_a,  \gamma\}$ satisfy
\begin{equation}\label{eq-truncation-gln}
\gamma \sum^n_{a=1}\left(N_{\gamma_a}-N_{a}-N_{\beta}\right) +\left(\sum^{n}_{a=1}\alpha_a\right)(N_{\alpha}-N_{\beta})+\sum^n_{a=1}\psi^{(a)}_0=0\,,
\end{equation}
namely the truncation can be characterized by the two integer coefficients multiplying $\gamma$ and $(\sum^{n}_{a=1}\alpha_a)$. 

We can compare this result with expectations from perfect matchings.
While the number of perfect matchings grows quickly as $n$ increases,
one can draw the bipartite graphs and perfect matchings
(as in Figures \ref{fig.C3Z2bipartite} and \ref{fig.C3Z2PM} for $n=2$),
and one finds that the perfect matchings correspond to linear combinations of the form
\begin{align}
n\gamma \;, \quad
 \genfrac{\{ }{\} }{0pt}{}{\alpha_1}{\beta_1}  + \dots + \genfrac{\{ }{\} }{0pt}{}{\alpha_n}{\beta_n}   \;,
\end{align}
where in the second term we choose either $\alpha_i$ or $\beta_i$
for each $i=1, \dots, n$. Out of these combinations
only three correspond to lattice points in the corner of the 
toric diagram (which in this case is a triangle):
\begin{align}
n\gamma \;, \quad
\alpha_1+\dots + \alpha_n \;, \quad
\beta_1+\dots + \beta_n=n \gamma-(\alpha_1+\dots + \alpha_n)  \;.
\end{align}
The span of the three again gives rise to
integer span of $n\gamma$ and $\sum_{a=1}^n \alpha_a$.
This almost matches the result above in (\ref{eq-truncation-gln}), except that the coefficient for $\gamma$ is a
multiplet of $n$, as in the case of $n=2$ before.

\subsubsubsection{Affine Yangian of $\mathfrak{gl}_{n}$}

If in addition to the loop constraint (\ref{eq-loop-constraint-Zn-2}), we impose the vertex constraint (\ref{eq-vertex-constraint-toric}), which in this case give
\begin{equation}\label{eq-vertex-constraint-Zn}
\alpha_a-\beta_{a}=\alpha_{a+1}-\beta_{a+1}\qquad \textrm{for}\quad a=1,2,\cdots, n\,,
\end{equation}
which together with the loop constraint (\ref{eq-loop-constraint-Zn-2}) give
\begin{equation}\label{eq-constraint-gln}
\begin{aligned}
&\alpha_1=\alpha_2=\cdots=\alpha_n \equiv h_1\,, \quad \beta_1=\beta_2=\cdots=\beta_n \equiv h_2\,, \quad \gamma_1=\gamma_2=\cdots=\gamma_n \equiv h_3 \,,\\
&\textrm{and}\qquad h_1+h_2+h_3=0  \;.
\end{aligned}
\end{equation}
Namely, after imposing the vertex constraints (\ref{eq-vertex-constraint-Zn}) on top of the loop constraint (\ref{eq-loop-constraint-Zn-2}), we have two parameters $(h_1,h_2)$, same as in the case of the affine Yangian of $\mathfrak{gl}_1$ for $\mathbb{C}^3$ and the affine Yangian of $\mathfrak{gl}_2$ for $\mathbb{C}^3/\mathbb{Z}_2$.

With the restriction of the $n+1$ parameters to the two parameters $(h_1,h_2)$, the bond factors in (\ref{eq-charge-function-Zn}) become
\begin{equation}
\begin{aligned}
&\varphi^{a\Rightarrow a}(u)=\frac{u+h_3}{u-h_3} \,,\qquad
 \varphi^{a\Rightarrow a+1}(u)=\frac{u+h_2}{u-h_1}\,, \qquad
  \varphi^{a\Rightarrow a-1}(u)=\frac{u+h_1}{u-h_2}  \;,
  \\
&\varphi^{a\Rightarrow b}(u)=1 \,, \quad ( b\neq a\,,\,a\pm 1)\,,
\end{aligned}
\end{equation}
which gives the algebra
\begin{tcolorbox}[ams align]
&\textrm{OPE:}\quad\begin{cases}\begin{aligned}\label{eq-OPE-gln-f}
\psi^{(a)}(z)\, \psi^{(b)}(w)&\sim \psi^{(b)}(w)\, \psi^{(a)}(z) \;,\\
\psi^{(a)}(z)\, e^{(a)}(w)   &\sim \tfrac{\Delta+h_3}{\Delta-h_3} \, e^{(a)}(w)\, \psi^{(a)}(z)  \;,\\ 
e^{(a)}(z)\, e^{(a)}(w) & \sim  \tfrac{\Delta+h_3}{\Delta-h_3} \, e^{(a)}(w)\, e^{(a)}(z) \;, \\
\psi^{(a)}(z)\, f^{(a)}(w) &  \sim \tfrac{\Delta-h_3}{\Delta+h_3} \, f^{(a)}(w)\, \psi^{(a)}(z) \;, \\
 f^{(a)}(z)\, f^{(a)}(w) &  \sim   \tfrac{\Delta-h_3}{\Delta+h_3} \, f^{(a)}(w)\, f^{(a)}(z)  \;,\\
 \psi^{(a+1)}(z)\, e^{(a)}(w)   &\sim \tfrac{\Delta+h_2}{\Delta-h_1} \, e^{(a)}(w)\, \psi^{(a+1)}(z)  \;,\\ 
   \psi^{(a-1)}(z)\, e^{(a)}(w)   &\sim \tfrac{\Delta+h_1}{\Delta-h_2} \, e^{(a)}(w)\, \psi^{(a-1)}(z) \;, \\ 
e^{(a+1)}(z)\, e^{(a)}(w) & \sim   \tfrac{\Delta+h_2}{\Delta-h_1}\, e^{(a)}(w)\, e^{(a+1)}(z)  \;,\\
\psi^{(a+1)}(z)\, f^{(a)}(w) &  \sim  \tfrac{\Delta-h_1}{\Delta+h_2}\, f^{(a)}(w)\, \psi^{(a+1)}(z)  \;,\\
\psi^{(a-1)}(z)\, f^{(a)}(w) &  \sim  \tfrac{\Delta-h_2}{\Delta+h_1}\, f^{(a)}(w)\, \psi^{(a-1)}(z)  \;,\\
 f^{(a+1)}(z)\, f^{(a)}(w) &  \sim  \tfrac{\Delta-h_1}{\Delta+h_2}\,f^{(a)}(w)\, f^{(a+1)}(z) \;,\\
\psi^{(b)}(z)\, e^{(a)}(w)   &\sim  \, e^{(a)}(w)\, \psi^{(b)}(z)\quad (b\ne a, a\pm 1)   \;,\\ 
e^{(b)}(z)\, e^{(a)}(w) & \sim  \, e^{(a)}(w)\, e^{(b)}(z) \quad (b\ne a, a\pm 1)  \;,\\
\psi^{(b)}(z)\, f^{(a)}(w) &  \sim  \, f^{(a)}(w)\, \psi^{(b)}(z)  \quad (b\ne a, a\pm 1) \;,\\
 f^{(b)}(z)\, f^{(a)}(w) &  \sim   \, f^{(a)}(w)\, f^{(b)}(z)  \quad (b\ne a, a\pm 1) \;,\\
[e^{(a)}(z)\,, f^{(b)}(w)]   &\sim  - \delta^{a,b}\, \frac{\psi^{(a)}(z) - \psi^{(a)}(w)}{z-w}  \;, 
\end{aligned}
\end{cases}
\\
&\textcolor{black}{\textrm{Initial:}}\quad\begin{cases}
\begin{aligned}
&\begin{aligned}
[\psi^{(a-1)}_0,e^{(a)}_m] &= -\, e^{(a)}_m\;,\\
[\psi^{(a)}_0,e^{(a)}_m] &=2  \, e^{(a)}_m\;,\\
[\psi^{(a+1)}_0,e^{(a)}_m] &= -\, e^{(a)}_m \;,\\
[\psi^{(b)}_0,e^{(a)}_m] &=0\;,
\end{aligned}\qquad 
&\begin{aligned}
[\psi^{(a-1)}_0,f^{(a)}_m] &=  f^{(a)}_m\;,\\
[\psi^{(a)}_0,f^{(a)}_m] &= -2 f^{(a)}_m\;,\\
[\psi^{(a+1)}_0,f^{(a)}_m] &= f^{(a)}_m \;,\\
[\psi^{(b)}_0,f^{(a)}_m] &=0\;\quad ( b\neq a\,,\, a\pm 1)\,,
\end{aligned}
\end{aligned}
\end{cases}\\
&\textrm{Serre}:\quad\begin{cases}\begin{aligned}\label{eq-serre-gl3-n}
&\textrm{Sym}_{z_1,z_2}\, \left[ e^{(a)}(z_1)\,, \left[ e^{(a)}(z_2)\,, e^{(a\pm1)}(w)\right]\right] \sim 0  \;,\\
&\textrm{Sym}_{z_1,z_2}\, \left[ f^{(a)}(z_1)\,, \left[ f^{(a)}(z_2)\,, f^{(a\pm1)}(w)\right]\right] \sim 0  \;,
\end{aligned}
\end{cases}
\end{tcolorbox}
\noindent where we have supplemented the algebra with Serre relations.
Note that the initial conditions on  $\psi^{(a)}_0$ and $e^{(b)}_m$ give the algebra of $\mathfrak{sl}_n$.

Finally, one can check that the Serre relations are needed to reproduce the generating function of the $n$-colored plane partitions by the vacuum module of the reduced quiver Yangian. 
We leave detailed discussions to Appendix.

\subsection{\texorpdfstring{Quiver Yangian for Generalized Conifolds and Affine Yangian of $\mathfrak{gl}_{m|n}$}{Quiver Yangian for Generalized Conifolds and Affine Yangian of gl(m|n)}}
\label{sec:glmn}

Let us next discuss the toric Calabi-Yau geometries described by the algebraic equation
\begin{align}
x y=z^m w^n\;,
\label{eq.gc}
\end{align} 
where $x, y, z, w$ are complex numbers and $m, n$ are non-negative integers (excluding $m=n=0$).
We can assume $m\ge n$ without loss of generality.

\begin{figure}[htbp]
\centering\includegraphics[scale=0.38]{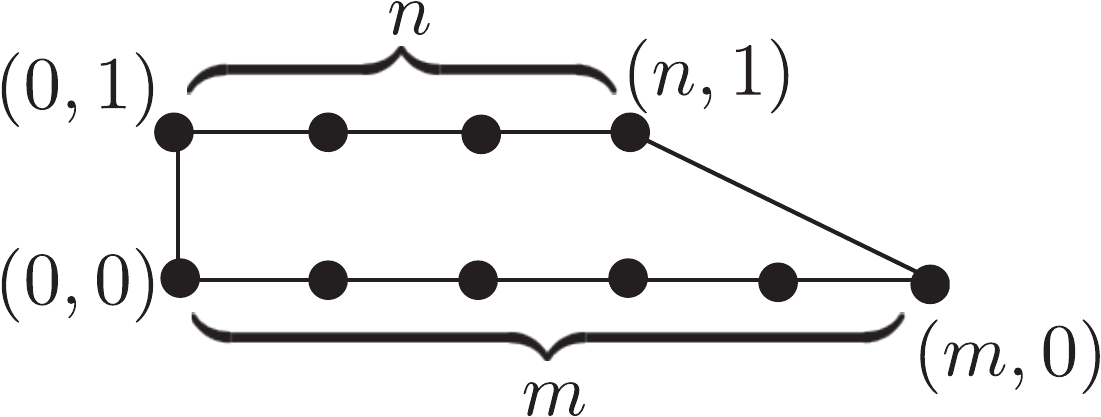}
\caption{The toric diagram for the generalized conifold geometry of \eqref{eq.gc}.}
\label{fig.gc}
\end{figure}

The geometry \eqref{eq.gc} is sometimes called the generalized conifold, and is the most general toric Calabi-Yau geometry without compact $4$-cycles (mathematically, such toric Calabi-Yau singularities are known to have small resolutions). 
The case of $n=0$ are the $A_m$ singularities $(\mathbb{C}^2/\mathbb{Z}_m) \times \mathbb{C}$ studied in section~\ref{sec:An} (which includes $\mathbb{C}^3$ as the special case $m=1, n=0$), and the case of $m=n=1$ is the conifold (see section~\ref{sec:conifold}).
The toric diagram, up to a suitable $\mathrm{SL}(2, \mathbb{Z})$-transformation, can be chosen as in Figure \ref{fig.gc}.

\subsubsection{Quivers and Superpotentials}

When discussing BPS crystals it is important to note that there are several different quiver gauge theories corresponding to the same geometry \eqref{eq.gc}; their quiver diagrams are different but they all have the same moduli space of vacua, and the module categories of their associated path algebras are derived-equivalent.

\begin{figure}[htbp]
\centering\includegraphics[scale=0.45]{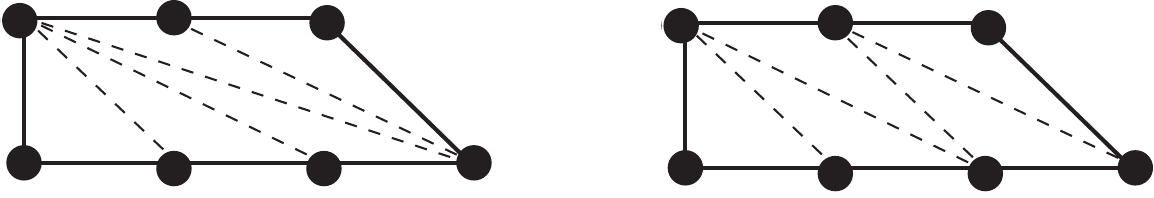}
\caption{The choice of the resolution of the singularity is encoded by the signs $\sigma$.
We here show two choices of $\sigma$, $\sigma_1=\{+1, +1,  +1,-1,  -1\}$ (left) and $\sigma_2=\{+1, +1,  -1,+1,  -1\}$ (right).}
\label{fig.gc_sgn}
\end{figure}

Geometrically, such ambiguities arise from the choice of the resolution of the singularity \eqref{eq.gc}.
This is described by a choice of the triangulation of the toric diagram, and any two such choices are related by a sequence of flop transitions.
Combinatorially, this choice is encoded by a set of
signs $\sigma$, which has $m$ $+1$'s and $n$ $-1$'s \cite{MR2999994,Nagao:2009rq} (see Figure \ref{fig.gc_sgn}): 
\begin{align}
\sigma: \quad \{ 1, 2, \dots, m+n  \}  \to \{ +1, -1 \} \quad \textrm{such that}
\quad \#( +1)= m, \quad \# (-1)=n \;. 
\label{eq.sgn}
\end{align}
For our later purposes we can regard the domain periodically as $\mathbb{Z}_{m+n}$, so that $\sigma$ is a map from $\mathbb{Z}_{m+n}$ to  $\{ +1, -1 \}$.

Given these data, we can identify the quiver diagram as follows \cite{MR2999994}:
\begin{itemize}
\item We have $m+n$ vertices $a=1, \dots, m+n$.
\item For each vertex  $a$ we have an arrow from $a$ to $a+1$, and another from $a+1$ to $a$ (the quiver is therefore non-chiral).
\item We have an arrow starting and ending at the same vertex $a$ when $\sigma_a=\sigma_{a+1}$; otherwise we do not have such an arrow. 
From the grading rule \eqref{eq.Z2_grading} one finds that in the former case the vertex $a$ is an even vertex ($|a|=0$), where in the latter case an odd vertex  ($|a|=1$).
\item There are no arrows from vertex $a$ to $b$ when $|a-b|\ge 2$, where $a$ and $b$ are considered mod $m+n$.
\end{itemize}
Here the indices $a,b,\dots $ are regarded as an element of $\mathbb{Z}_{m+n}$.

Let us describe the superpotential $W$.
For each vertex $a$, we add superpotential terms
\begin{equation}\label{eq.gc_W}
W \ni
\begin{cases}
-\sigma_a\textrm{Tr}(\Phi_{a,a} \Phi_{a,a+1} \Phi_{a+1,a}) + \sigma_a\textrm{Tr}(\Phi_{a,a} \Phi_{a,a-1} \Phi_{a-1,a}) & (\sigma_a=\sigma_{a+1}) \;, \\
\sigma_a\textrm{Tr}(\Phi_{a,a+1} \Phi_{a+1,a} \Phi_{a,a-1} \Phi_{a-1,a})  &(\sigma_a= - \sigma_{a+1}) \;,
\end{cases}
\end{equation}
where as before $\Phi_{a,b}$ denotes the bifundamental chiral multiplet corresponding to the arrow from a vertex $a$ to a vertex $b$.

\begin{figure}[htbp]
\begin{center}
\begin{minipage}{0.48\linewidth}
\centering
\begin{tikzpicture}[scale=1]
\node[state][regular polygon, regular polygon sides=4, draw=blue!50, very thick, fill=blue!10, label=center:$a$,minimum width=17mm] (a1) at (0,2.236)  {};
\node[state][regular polygon, regular polygon sides=4, draw=blue!50, very thick, fill=blue!10, label=center:$a+1$,minimum width=17mm] (a2) at (2.126,0.691)  {};
\node[state][regular polygon, regular polygon sides=4, draw=blue!50, very thick, fill=blue!10, label=center:$a-1$,minimum width=17mm] (a5) at (-2.127,0.691)  {};
\path[->] 
(a1) edge [in=60, out=120, loop, thin, above] node {$\Phi_{a, a}$} ()
(a1) edge   [thin, bend left]  node [right] {$\Phi_{a, a+1}$} (a2)
(a2) edge   [thin, bend left]  node [right, pos=0.9]{$\Phi_{a+1,a}$} (a1)
(a5) edge   [thin, bend left]  node [left] {$\Phi_{a-1,a}$} (a1)
(a1) edge   [thin, bend left]  node [left, pos=0.1]{$\Phi_{a, a-1}$} (a5) 
;
\end{tikzpicture}
\end{minipage}
\hfill
\begin{minipage}{0.48\linewidth}
\vspace{2.0cm}
\centering
\begin{tikzpicture}[scale=1]
\node[state][regular polygon, regular polygon sides=4, draw=blue!50, very thick, fill=blue!10, label=center:$a$,minimum width=17mm] (a1) at (0,2.236)  {};
\node[state][regular polygon, regular polygon sides=4, draw=blue!50, very thick, fill=blue!10, label=center:$a+1$,minimum width=17mm] (a2) at (2.126,0.691)  {};
\node[state][regular polygon, regular polygon sides=4, draw=blue!50, very thick, fill=blue!10, label=center:$a-1$,minimum width=17mm] (a5) at (-2.127,0.691)  {};
\path[->] 
(a1) edge   [thin, bend right]  node [right, pos=0.1] {$\Phi_{a, a+1}$} (a2)
(a2) edge   [thin, bend right]  node [right, pos=0.4]{$\Phi_{a+1,a}$} (a1)
(a5) edge   [thin, bend left]  node [left] {$\Phi_{a-1,a}$} (a1)
(a1) edge   [thin, bend left]  node [left, pos=0.1]{$\Phi_{a,a-1}$} (a5) 
;
\end{tikzpicture}
\end{minipage}
\end{center}
\caption{The quiver diagram around a vertex $a$. Depending on whether we have $\sigma_{a}=\sigma_{a+1}$ or $\sigma_{a}=-\sigma_{a+1}$
we require superpotential terms as in \eqref{eq.gc_W}. }
\label{fig.gc_quiver_local}
\end{figure}
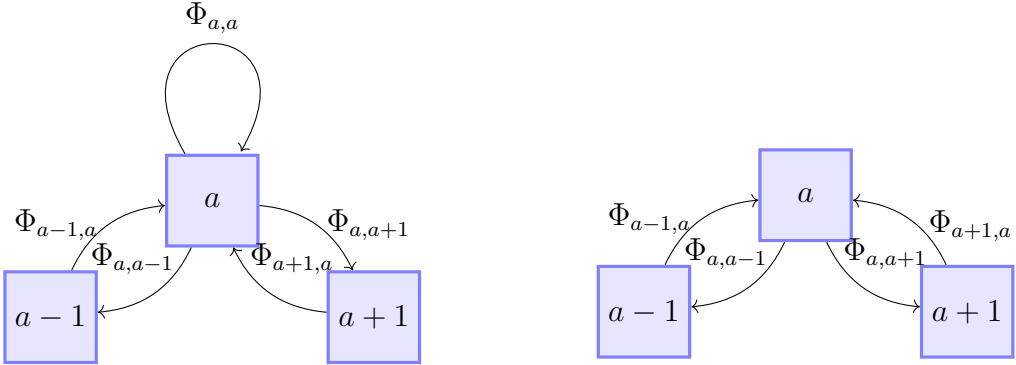

The charge assignment for the bifundamental multiplets is described in terms of $(\alpha_a, \beta_a, \gamma_a)$:
\begin{equation}
 \Phi_{a,a+1} \,\, \mapsto \,\, \alpha_a\,, \qquad \Phi_{a+1,a} \,\, \mapsto \,\, \beta_a \,, \qquad \Phi_{a,a} \,\, \mapsto \,\, \gamma_a\,.
\end{equation}
The loop constraints imposed by the superpotential terms (\ref{eq.gc_W}) have different forms for the case  with $\sigma_{a}=\sigma_{a+1}$ and $\sigma_{a}=-\sigma_{a+1}$:
\begin{equation}\label{eq-loop-constraint-glmn}
\begin{aligned}
\sigma_a=\sigma_{a+1} &: \qquad  \begin{cases}
\begin{aligned}
\alpha_{a-1}+\beta_{a-1}+\gamma_{a}&=0  \;,\\
\alpha_{a}+\beta_{a}+\gamma_{a}&=0 \;,
\end{aligned}
\end{cases}\\
\sigma_a=-\sigma_{a+1} &:\qquad \alpha_{a-1}+\beta_{a-1}+\alpha_{a}+\beta_a=0 \;.
\end{aligned}
\end{equation}
Again, one can check that the central condition (\ref{eq-central-condition}) is guaranteed by the loop constraint (\ref{eq-loop-constraint-glmn}).

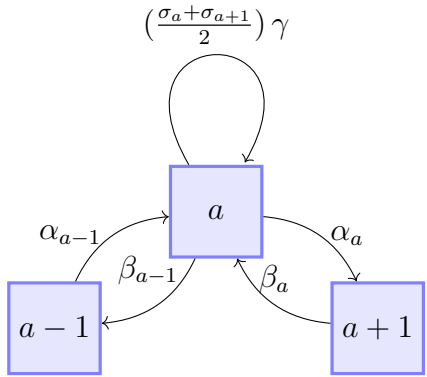
\begin{figure}[htbp]
\begin{center}
\begin{minipage}{0.48\linewidth}
\begin{tikzpicture}[scale=1]
\node[state][regular polygon, regular polygon sides=4, draw=blue!50, very thick, fill=blue!10, label=center:$a$,minimum width=17mm] (a1) at (0,2.236)  {};
\node[state][regular polygon, regular polygon sides=4, draw=blue!50, very thick, fill=blue!10, label=center:$a+1$,minimum width=17mm] (a2) at (2.126,0.691)  {};
\node[state][regular polygon, regular polygon sides=4, draw=blue!50, very thick, fill=blue!10, label=center:$a-1$,minimum width=17mm] (a5) at (-2.127,0.691)  {};
\path[->] 
(a1) edge [in=60, out=120, loop, thin, above] node {$(\frac{\sigma_a+\sigma_{a+1}}{2})\, \gamma$} ()
(a1) edge   [thin, bend left]  node [right] {$\alpha_{a} $} (a2)
(a2) edge   [thin, bend left]  node [above]{$\beta_{a} $} (a1)
(a5) edge   [thin, bend left]  node [left] {$\alpha_{a-1}$} (a1)
(a1) edge   [thin, bend left]  node [left, pos=0.1]{$\beta_{a-1}$} (a5) 
;
\end{tikzpicture}
\end{minipage}
\end{center}
\caption{The charge assignment for the bifundamental/adjoint chiral multiplets around a vertex $a$, with the constraint $\alpha_a+\beta_a=-\sigma_{a+1}\, \gamma $.}
\label{fig:gc_h}
\end{figure}

The loop constraints (\ref{eq-loop-constraint-glmn}) for the two scenarios can be rewritten in a uniformed way:
\begin{equation}\label{eq-loop-constraint-glmn-2}
\alpha_a+\beta_a=-\sigma_{a+1}\, \gamma \qquad \textrm{and} \qquad \gamma_a=\left(\frac{\sigma_a+\sigma_{a+1}}{2}\right)\, \gamma \,,
\end{equation}
for all $a$.
Imposing (\ref{eq-loop-constraint-glmn-2}) leaves us with $m+n+1$ variables, i.e.\ $\alpha_a$ for $a=1,2,\dots, m+n$ and $\gamma$.
The charge assignment is summarized in Figure \ref{fig:gc_h},  satisfying (\ref{eq-loop-constraint-glmn-2}).

\subsubsection{Periodic Quivers and Dimers}

The quiver and the superpotential described above are sufficient for the discussion of the BPS quiver Yangian.
Let us nevertheless describe the periodic quiver \cite{MR2999994}, which will be needed for the explicit construction of the BPS crystal melting,
as well as for the discussion of the truncation of the algebra later in section \ref{gc_truncation}.

Instead of directly writing down the periodic quiver, it is useful to discuss its dual graph, which is a bipartite graph known as the brane tiling.

\begin{figure}[htbp]
\centering\includegraphics[scale=0.4]{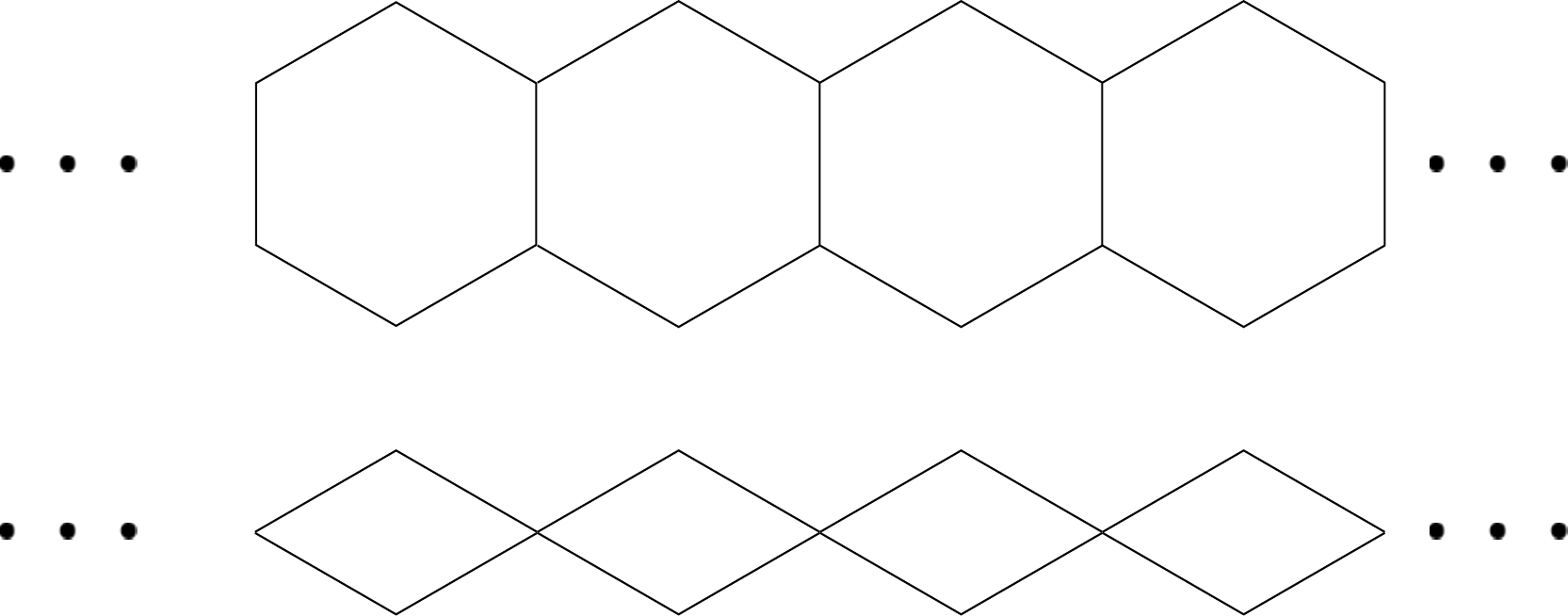}
\caption{The building blocks for the bipartite graphs for the generalized conifold geometry. 
For $a\in \mathbb{Z}_{m+n}$ one stacks the hexagons as above (squares as below) when $\sigma_{a+1}=\sigma_a$ ($\sigma_{a+1}=-\sigma_a$).}
\label{fig.gc_block}
\end{figure}

\begin{figure}[htbp]
\centering\includegraphics[scale=0.3]{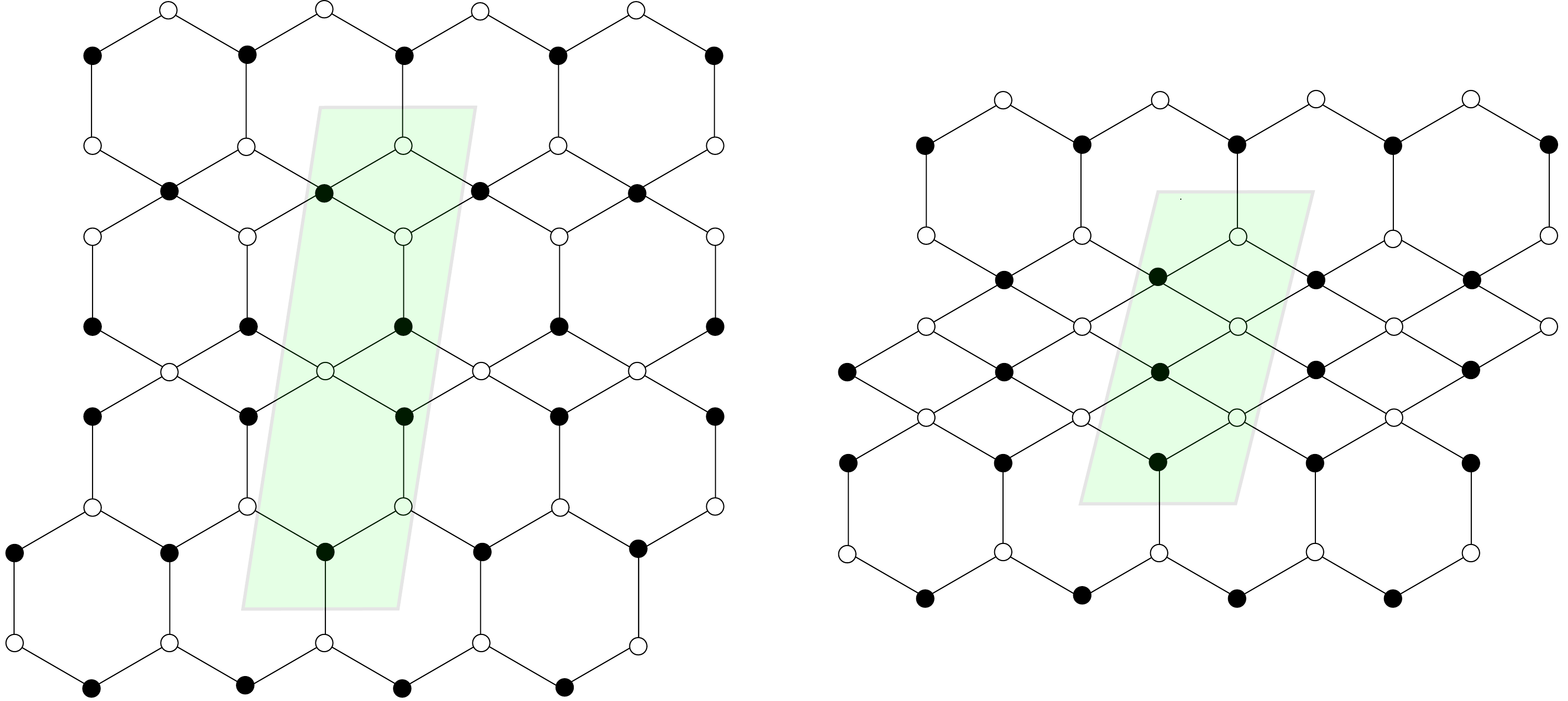}
\caption{The bipartite graphs for $m=3, n=2$, for the two sign choices $\sigma_1=\{+1, +1,  +1,-1,  -1\}$ and $\sigma_2=\{+1, +1,  -1,+1,  -1\}$. 
The shaded regions are the fundamental regions of the torus.}
\label{fig.gc_dimer}
\end{figure}

Let us again start with a choice of the signs $\sigma$.
We consider $m+n$ stacks of fundamental building blocks as shown in Figure \ref{fig.gc_block}, either hexagons or squares (with length of each edge $1$).
For $a\in \mathbb{Z}_{m+n}$ one consider hexagons (squares)  when $\sigma_{a+1}=\sigma_a$ ($\sigma_{a+1}=-\sigma_a$).
One then chooses a fundamental region such that the parallelogram representing the fundamental region is shifted by $m-n$ units. 
The examples of $m=3, n=2$ are shown in Figure \ref{fig.gc_dimer}.
We can then choose a fundamental region as in Figure \ref{fig.gc_dimer}.

\begin{figure}[htbp]
\centering\includegraphics[scale=0.3]{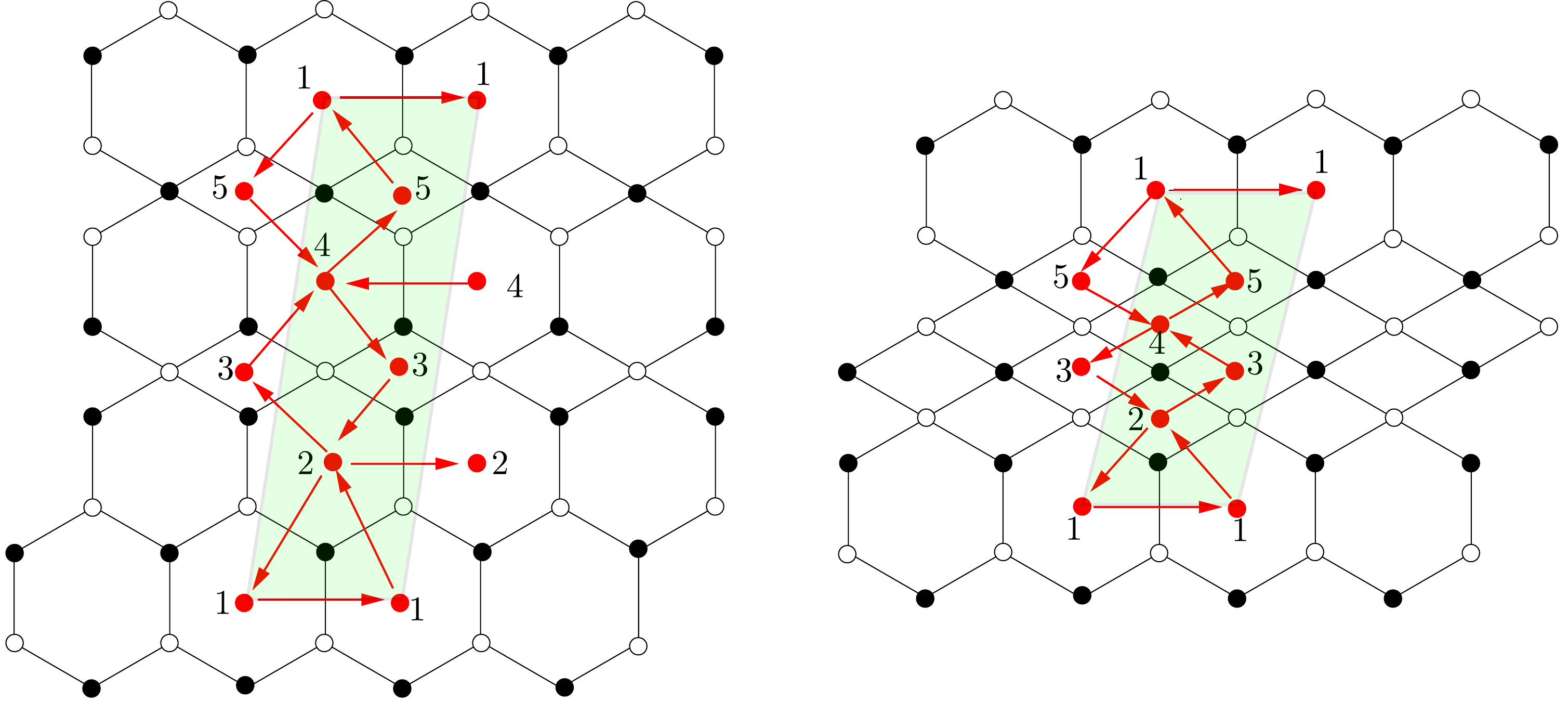}
\caption{The bipartite graphs for $m=3, n=2$, for the two sign choices $\sigma_1=\{+1, +1,  +1,-1,  -1\}$ and $\sigma_2=\{+1, +1,  -1,+1,  -1\}$. 
The region enclosed by the dotted lines is the fundamental region of the torus.}
\label{fig.gc_stack_quiver}
\end{figure}

The dual graph of the bipartite graph gives the periodic quiver, which in turn gives the quiver and the superpotential.
For the examples of Figure \ref{fig.gc_dimer}, they are shown as in Figure \ref{fig.gc_stack_quiver}.

\subsubsection{Algebra}

We can now write down the algebra.

In order to write down the OPE relations one first needs to know the Bose/Fermi statistics of the generators.
From the rules of the quiver diagrams above, the presence/absence of the arrows starting and ending on the same vertex $a$ depends on the relative sign of $\sigma_a$ and $\sigma_{a+1}$ ---
it then follows from the grading rule of \eqref{eq.Z2_grading} that the generators $e^{(a)}(z), f^{(a)}(z), \psi^{(a)}(z)$ are bosonic (even) when $\sigma_a=\sigma_{a+1}$, and fermionic (odd) when $\sigma_a=-\sigma_{a+1}$.
The OPE relations are then determined by the function
\begin{align}\label{eq-charge-function-glmn}
\varphi^{a\Rightarrow b}(u)= \frac{\prod_{I\in \{b\rightarrow a\}}(u+h_{I})}{\prod_{J\in \{a\rightarrow b\}}(u-h_{J})} \;,
\end{align}
which gives for example
\begin{align}
e^{(a)}(z) e^{(b)}(w) \sim (-1)^{|a||b|} \, \varphi^{b\Rightarrow a}(\Delta) \, e^{(b)}(w) \,  e^{(a)}(z)   \;.
\end{align} 

Let us first concentrate on the general case $m+n\ge 3$.
The special case of $m+n\le 2$ need to be considered separately: in particular, the case of $\mathbb{C}^3$ ($m=1, n= 0$) and $\mathbb{C}^2/\mathbb{Z}_2\times \mathbb{C}$ ($m=2,n=0$) have been considered in section~\ref{sec:gl1-8211} and \ref{sec:C2Zn}, and the resolved conifold ($m=n=1$) will be considered in section~\ref{sec:conifold}.
When $m+n\ge 3$, for any pair of vertices $a,b$ there is at most one arrow in the quiver from vertex $a$ to vertex $b$ (this is the case even for the cases with $a=b$):
\begin{equation}\label{eq-charge-function-glmn-1}
\begin{aligned}
&\varphi^{a\Rightarrow a+1}(u)=\frac{u+\beta_a}{u-\alpha_a}\,,   \qquad \varphi^{a\Rightarrow a-1}(u)=\frac{u+\alpha_{a-1}}{u-\beta_{a-1}}\,,\\
&\varphi^{a\Rightarrow a}(u)=\frac{u+(\sigma_a+\sigma_{a+1}) \gamma/2}{u-(\sigma_a+\sigma_{a+1}) \gamma/2} \,,
\end{aligned}
\end{equation}
and all other $\varphi^{a\Rightarrow b}$ trivial.
Now it is straightforward to write down the algebra
\begin{tcolorbox}[ams align]\label{eq-OPE-generalizedO}
&\textrm{OPE:}\quad\begin{cases}\begin{aligned}
\psi^{(a)}(z)\, \psi^{(b)}(w)&\sim \psi^{(b)}(w)\, \psi^{(a)}(z) \;,\\
\psi^{(a)}(z)\, e^{(a)}(w)   &\sim \tfrac{\Delta+(\sigma_a+\sigma_{a+1}) \gamma/2}{\Delta-(\sigma_a+\sigma_{a+1}) \gamma/2} \, e^{(a)}(w)\, \psi^{(a)}(z)  \;,\\ 
e^{(a)}(z)\, e^{(a)}(w) & \sim  (-1)^{|a|}\tfrac{\Delta+(\sigma_a+\sigma_{a+1}) \gamma/2}{\Delta-(\sigma_a+\sigma_{a+1}) \gamma/2}\, e^{(a)}(w)\, e^{(a)}(z) \;, \\
\psi^{(a)}(z)\, f^{(a)}(w) &  \sim \tfrac{\Delta-(\sigma_a+\sigma_{a+1}) \gamma/2}{\Delta+(\sigma_a+\sigma_{a+1}) \gamma/2}\, f^{(a)}(w)\, \psi^{(a)}(z) \;, \\
 f^{(a)}(z)\, f^{(a)}(w) &  \sim  (-1)^{|a|}\tfrac{\Delta-(\sigma_a+\sigma_{a+1}) \gamma/2}{\Delta+(\sigma_a+\sigma_{a+1}) \gamma/2}\, f^{(a)}(w)\, f^{(a)}(z)  \;,\\
 \psi^{(a+1)}(z)\, e^{(a)}(w)   &\sim \tfrac{\Delta+\beta_a}{\Delta-\alpha_a} \, e^{(a)}(w)\, \psi^{(a+1)}(z)  \;,\\ 
   \psi^{(a-1)}(z)\, e^{(a)}(w)   &\sim \tfrac{\Delta+\alpha_{a-1}}{\Delta-\beta_{a-1}}\, e^{(a)}(w)\, \psi^{(a-1)}(z) \;, \\ 
e^{(a+1)}(z)\, e^{(a)}(w) & \sim  (-1)^{|a||a+1|}\tfrac{\Delta+\beta_a}{\Delta-\alpha_a}\, e^{(a)}(w)\, e^{(a+1)}(z)  \;,\\
\psi^{(a+1)}(z)\, f^{(a)}(w) &  \sim \tfrac{\Delta-\alpha_a}{\Delta+\beta_a}\, f^{(a)}(w)\, \psi^{(a+1)}(z)  \;,\\
\psi^{(a-1)}(z)\, f^{(a)}(w) &  \sim   \tfrac{\Delta-\beta_{a-1}}{\Delta+\alpha_{a-1}}\, f^{(a)}(w)\, \psi^{(a-1)}(z)  \;,\\
 f^{(a+1)}(z)\, f^{(a)}(w) &  \sim (-1)^{|a||a+1|}\tfrac{\Delta-\alpha_a}{\Delta+\beta_a}\,f^{(a)}(w)\, f^{(a+1)}(z) \;,\\
\psi^{(b)}(z)\, e^{(a)}(w)   &\sim  \, e^{(a)}(w)\, \psi^{(b)}(z)\qquad\qquad (b\ne a, a\pm 1)   \;,\\ 
e^{(b)}(z)\, e^{(a)}(w) & \sim  \, (-1)^{|a||b|} e^{(a)}(w)\, e^{(b)}(z) \quad (b\ne a, a\pm 1)  \;,\\
\psi^{(b)}(z)\, f^{(a)}(w) &  \sim  \, f^{(a)}(w)\, \psi^{(b)}(z)  \qquad\qquad (b\ne a, a\pm 1) \;,\\
 f^{(b)}(z)\, f^{(a)}(w) &  \sim   \,(-1)^{|a||b|}   f^{(a)}(w)\, f^{(b)}(z)  \quad (b\ne a, a\pm 1) \;,\\
\left[e^{(a)}(z)\,, f^{(b)}(w)\right\}   &\sim  - \delta^{a,b}\, \frac{\psi^{(a)}(z) - \psi^{(b)}(w)}{z-w}  \;, 
\end{aligned}
\end{cases}
\end{tcolorbox}
\noindent together with the initial conditions
\begin{tcolorbox}[ams align]\label{eq-initial-generalizedO}
&\textcolor{black}{\textrm{Initial:}}\quad\begin{cases}
\begin{aligned}
&[\psi^{(a)}_0,e^{(a)}_m] = (\sigma_a+\sigma_{a+1})\,\gamma\, e^{(a)}_m \;,\\
&[\psi^{(a)}_0,f^{(a)}_m] =-(\sigma_a+\sigma_{a+1})\,\gamma\,f^{(a)}_m \;,\\
&[\psi^{(a+1)}_0,e^{(a)}_m] =-\sigma_{a+1}\,\gamma\,e^{(a)}_m \;,\\
&[\psi^{(a+1)}_0,f^{(a)}_m] =\sigma_{a+1}\,\gamma\,f^{(a)}_m  \;,\\
&[\psi^{(a-1)}_0,e^{(a)}_m] =-\sigma_{a}\,\gamma\,e^{(a)}_m \;,\\
&[\psi^{(a-1)}_0,f^{(a)}_m] =\sigma_{a}\,\gamma\,f^{(a)}_m  \;,\\
&[\psi^{(b)}_0,e^{(a)}_m] =[\psi^{(b)}_0,f^{(a)}_m] =0\;\quad ( b\neq a\,,\, a\pm 1)\,,
\\
\end{aligned}
\end{cases}
\end{tcolorbox}
\noindent from which one can check that the combination $\psi_0\equiv \sum^{m+n}_{a=1} \psi^{(a)}_0$ is a central term.

\subsubsection{Truncation}
\label{gc_truncation}

The truncation condition for the algebra for the generalized conifold can be derived in a similar way to the one for the algebra of $(\mathbb{C}^2/\mathbb{Z}_n)\times\mathbb{C}$, given in (\ref{eq-truncation-gln}).

Consider a path from the origin $\mathfrak{o}$ to an atom $\sqbox{$1$}$ of color $a=1$, at which the growth of the crystal stops.
The coordinate function of this atom $\sqbox{$1$}$ is
\begin{equation}\label{eq-truncation-glmn-1}
h(\sqbox{$1$})=\sum^{m+n}_{a=1}N_{\gamma_a}\gamma_a +\sum^{m+n}_{a=1}N_{a} (\alpha_a+\beta_a)+N_{\alpha}\sum^{m+n}_{a=1}\alpha_a+N_{\beta}\sum^{m+n}_{a=1}\beta_a  \;,
\end{equation}
where $N_{\gamma_a}$ denotes the number of edges with $\gamma_a$ in the path, $N_{a}$ the number of segment $a\rightarrow a+1 \rightarrow a$, $N_{\alpha}$ the number of the segment $1\rightarrow 2\rightarrow \cdots \rightarrow m+n\rightarrow 1$, and $N_{\beta}$ the number of segment $1\rightarrow m+n\rightarrow \cdots \rightarrow 2\rightarrow 1$.
This is identical to the coordinate function of the truncation atom $\sqbox{$1$}$ for $(\mathbb{C}^2/\mathbb{Z}_n)\times \mathbb{C}$ (see (\ref{eq-truncation-1})).

The difference from the case of $(\mathbb{C}^2/\mathbb{Z}_n)\times\mathbb{C}$ enters through the loop constraints (\ref{eq-loop-constraint-glmn-2}) (cf.\ (\ref{eq-loop-constraint-Zn}) and (\ref{eq-loop-constraint-Zn-2}) for $(\mathbb{C}^2/\mathbb{Z}_n)\times\mathbb{C}$).
Imposing (\ref{eq-loop-constraint-glmn-2}) reduces the coordinate function (\ref{eq-truncation-glmn-1}) to
\begin{equation}
h(\sqbox{$1$})=\gamma\sum^{m+n}_{a=1}\left(\frac{N_{\gamma_a}(\sigma_a+\sigma_{a+1})}{2}-(N_{a}+N_{\beta})\sigma_{a+1}
\right) + \left(\sum^{m+n}_{a=1}\alpha_a\right)(N_{\alpha}-N_{\beta}) \;.
\end{equation}
Therefore the algebra truncates when the parameters $\{\alpha_a,  \gamma\}$ satisfy
\begin{equation}\label{eq-truncation-glmn-final}
\gamma\sum^{m+n}_{a=1}\left(\frac{N_{\gamma_a}(\sigma_a+\sigma_{a+1})}{2}-(N_{a}+N_{\beta})\sigma_{a+1}
\right) +\left(\sum^{m+n}_{a=1}\alpha_a\right)(N_{\alpha}-N_{\beta})+\sum^{m+n}_{a=1}\psi^{(a)}_0=0\,,
\end{equation}
namely the truncation can be characterized by the two integer coefficients multiplying $\gamma$ and $(\sum^{m+n}_{a=1}\alpha_a)$.

We can compare this result with perfect matchings.
Since in general there are many ($2^{n}+2^{m}$) perfect matchings,
we will present all the details only for examples of $m=n=1$ and $m=2, n=1$ below.
We can nevertheless present here the 
linear combinations which appear in perfect matchings, for the choice of sign
$\sigma=\{ +, \dots, +, -, \dots, -\}$:
\begin{align}
(n-1) \gamma  + \sum_{a: \, \sigma_{a+1}=+} \genfrac{\{ }{\} }{0pt}{}{\alpha_a}{\beta_a}  \;,
\quad
(m-1) \gamma  + \sum_{a: \, \sigma_{a+1}=-} \genfrac{\{ }{\} }{0pt}{}{\alpha_a}{\beta_a}  \;,
\end{align}
of which four correspond to corner perfect matchings:
\begin{align}
(n-1) \gamma  + \genfrac{\{ }{\} }{0pt}{}{\sum_{a: \, \sigma_{a+1}=+}  \alpha_a}{\sum_{a: \, \sigma_{a+1}=+}  \beta_a}  \;,
\quad
(m-1) \gamma  + \genfrac{\{ }{\} }{0pt}{}{\sum_{a: \, \sigma_{a+1}=-}  \alpha_a}{\sum_{a: \, \sigma_{a+1}=-}  \beta_a}  \;,
\end{align}
One can choose a basis for the integer span of these vectors to be
\begin{align}
(n-1) \gamma  + \sum_{a: \, \sigma_{a+1}=+}  \alpha_a  \;,
\quad
(m-1) \gamma  + \sum_{a: \, \sigma_{a+1}=-}  \alpha_a  \;,
\quad 
\textrm{gcd}(n-2, m-2)  \gamma  \;.
\end{align}

\subsubsection{\texorpdfstring{Vertex Constraint and Affine Yangian of $\mathfrak{gl}_{m|n}$}{Vertex Constraint and Affine Yangian of gl(m|n)}}

The vertex constraint (\ref{eq-vertex-constraint-toric}) is
\begin{equation}\label{eq-vertex-constraint-glmn}
\alpha_{a-1}-\beta_{a-1}=\alpha_a-\beta_{a}\,
\end{equation}
for any vertex $a$. 
Imposing the vertex constraint (\ref{eq-vertex-constraint-glmn}) on top of the loop constraints (\ref{eq-loop-constraint-glmn-2}), we have
\begin{equation}\label{eq-constraint-glmn-v}
\begin{aligned}
\sigma_a=\sigma_{a+1} &: \qquad \alpha_a=\alpha_{a-1}\,, \qquad\,\,\, \,\beta_a=\beta_{a-1} \,, \qquad\,\,\,\,\,  \gamma_a = \sigma_{a}\,\gamma\,,
 \\
\sigma_a=-\sigma_{a+1} &:\qquad \alpha_a=-\beta_{a-1}\,, \qquad \beta_a=-\alpha_{a-1} \,, \qquad  \gamma_a = 0\,,
\end{aligned}
\end{equation}
with $\alpha_a$, $\beta_a$ and $\gamma$ obeying $\alpha_a+\beta_a+\sigma_{a+1} \gamma=0$. 

\begin{figure}[htbp]
\begin{minipage}{0.5\linewidth}
\centering
\begin{tikzpicture}[scale=1]
\node[state]  [regular polygon, regular polygon sides=4, draw=blue!50, very thick, fill=blue!10] (a1) at (0,2.236)  {$1$};
\node[state]  [regular polygon, regular polygon sides=4, draw=blue!50, very thick, fill=blue!10] (a2) at (2.126,0.691)  {$2$};
\node[state]  [regular polygon, regular polygon sides=4, draw=blue!50, very thick, fill=blue!10] (a3) at (1.134,-1.809)  {$3$};
\node[state]  [regular polygon, regular polygon sides=4, draw=blue!50, very thick, fill=blue!10] (a4) at (-1.314,-1.809)  {$4$};
\node[state]  [regular polygon, regular polygon sides=4, draw=blue!50, very thick, fill=blue!10] (a5) at (-2.127,0.691)  {$5$};
\path[->] 
(a1) edge [in=60, out=120, loop, thin, above] node {$h_3$} ()
(a2) edge [in=348, out=48, loop, thin, right] node {$h_3$} ()
(a4) edge [in=204, out=264, loop, thin, left] node {$-h_3$} ()
(a1) edge   [thin, bend left]  node [right] {$h_{1}$} (a2)
(a2) edge   [thin, bend left]  node [left]{$h_{2}$} (a1)
(a2) edge   [thin, bend left]  node [right] {$h_{1}$} (a3)
(a3) edge   [thin, bend left]  node [left]{$h_{2}$} (a2)
(a3) edge   [thin, bend right]  node [above] {$-h_{2}$} (a4)
(a4) edge   [thin, bend right]  node [below]{$-h_{1}$} (a3)
(a4) edge   [thin, bend right]  node [right] {$-h_{2}$} (a5)
(a5) edge   [thin, bend right]  node [left]{$-h_{1}$} (a4)
(a5) edge   [thin, bend left]  node [left] {$h_{1}$} (a1)
(a1) edge   [thin, bend left]  node [right]{$h_{2}$} (a5)
;
\end{tikzpicture}
\end{minipage}
\hfill
\begin{minipage}{0.5\linewidth}
\centering
\begin{tikzpicture}[scale=1]
\node[state]  [regular polygon, regular polygon sides=4, draw=blue!50, very thick, fill=blue!10] (a1) at (0,2.236)  {$1$};
\node[state]  [regular polygon, regular polygon sides=4, draw=blue!50, very thick, fill=blue!10] (a2) at (2.126,0.691)  {$2$};
\node[state]  [regular polygon, regular polygon sides=4, draw=blue!50, very thick, fill=blue!10] (a3) at (1.134,-1.809)  {$3$};
\node[state]  [regular polygon, regular polygon sides=4, draw=blue!50, very thick, fill=blue!10] (a4) at (-1.314,-1.809)  {$4$};
\node[state]  [regular polygon, regular polygon sides=4, draw=blue!50, very thick, fill=blue!10] (a5) at (-2.127,0.691)  {$5$};
\path[->] 
(a1) edge [in=60, out=120, loop, thin, above] node {$h_3$} ()
(a1) edge   [thin, bend left]  node [right] {$h_{1}$} (a2)
(a2) edge   [thin, bend left]  node [left]{$h_{2}$} (a1)
(a2) edge   [thin, bend right]  node [left] {-$h_{2}$} (a3)
(a3) edge   [thin, bend right]  node [right]{-$h_{1}$} (a2)
(a3) edge   [thin, bend left]  node [below] {$h_{1}$} (a4)
(a4) edge   [thin, bend left]  node [above]{$h_{2}$} (a3)
(a4) edge   [thin, bend right]  node [right] {-$h_{2}$} (a5)
(a5) edge   [thin, bend right]  node [left]{-$h_{1}$} (a4)
(a5) edge   [thin, bend left]  node [left] {$h_{1}$} (a1)
(a1) edge   [thin, bend left]  node [right]{$h_{2}$} (a5) 
;
\end{tikzpicture}
\end{minipage}
\caption{The quiver diagram for the choices  $\sigma=\{+1, +1,  +1,-1,  -1\}$ and $\sigma=\{+1, +1,  -1,+1,  -1\}$ }
\label{fig.gc_quiverv}
\end{figure}
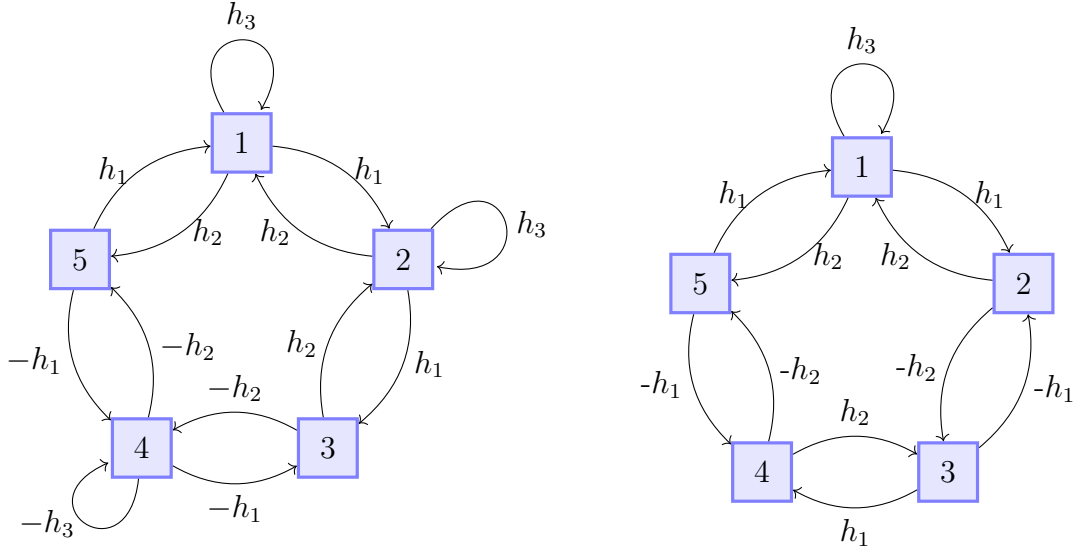

Now we can give the charge assignment in the presence of both loop and vertex constraints.
First,  define $\gamma\equiv h_3$.
Then without loss of generality, we can define, for an arbitrary vertex $a$,
\begin{equation}
\alpha_{a-1}\equiv \sigma_{a}\,h_1\,, \qquad \beta_{a-1}\equiv \sigma_{a}\, h_2\,, \qquad \gamma_a\equiv \left(\frac{\sigma_a+\sigma_{a+1}}{2}\right)\, h_3\,,
\end{equation}
where $(h_1,h_2,h_3)$ satisfy $h_1+h_2+h_3=0$.
Then applying the constraints (\ref{eq-constraint-glmn-v}) iteratively starting from vertex $a$, we have 
the general rule for the charge assignment with the vertex constraint  
(see Figure \ref{fig:gc_hv}):
\begin{itemize}
\item The arrow in the clockwise direction (vertex $a$ to  vertex $a+1$) has a weight $\alpha_a=\sigma_{a+1} h_1$ or $\sigma_{a+1} h_2$, where the choice of $h_1$ versus $h_2$ flips whenever we cross the odd quiver vertex $a$. 
\item Similarly, the arrow in the clockwise direction (vertex $a+1$ to  vertex $a$) has a weight $\beta_a=\sigma_{a+1} h_1$ or $\sigma_{a+1} h_2$, where again the choice of $h_1$ versus $h_2$ flips whenever we cross the odd quiver vertex $a$. 
\item When $\sigma_a=\sigma_{a+1}$ we have an arrow starting and ending at the vertex $a$, to which we assign a charge $\sigma_a h_3$.
\end{itemize}
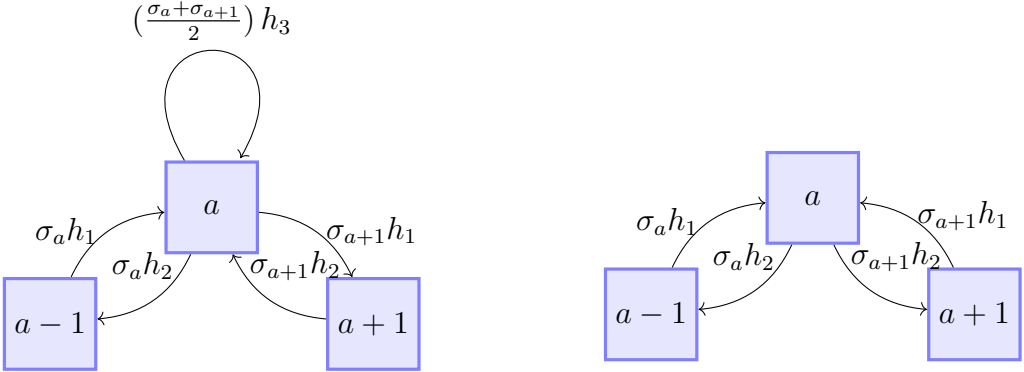
\begin{figure}[htbp]
\begin{center}
\begin{minipage}{0.48\linewidth}
\centering
\begin{tikzpicture}[scale=1]
\node[state][regular polygon, regular polygon sides=4, draw=blue!50, very thick, fill=blue!10, label=center:$a$,minimum width=17mm] (a1) at (0,2.236)  {};
\node[state][regular polygon, regular polygon sides=4, draw=blue!50, very thick, fill=blue!10, label=center:$a+1$,minimum width=17mm] (a2) at (2.126,0.691)  {};
\node[state][regular polygon, regular polygon sides=4, draw=blue!50, very thick, fill=blue!10, label=center:$a-1$,minimum width=17mm] (a5) at (-2.127,0.691)  {};
\path[->] 
(a1) edge [in=60, out=120, loop, thin, above] node {$(\frac{\sigma_a+\sigma_{a+1}}{2})\, h_3$} ()
(a1) edge   [thin, bend left]  node [right] {$\sigma_{a+1} h_1$} (a2)
(a2) edge   [thin, bend left]  node [right, pos=0.9]{$\sigma_{a+1} h_2$} (a1)
(a5) edge   [thin, bend left]  node [left] {$\sigma_{a} h_1$} (a1)
(a1) edge   [thin, bend left]  node [left, pos=0.1]{$\sigma_{a} h_2$} (a5) 
;
\end{tikzpicture}
\end{minipage}
\hfill
\begin{minipage}{0.48\linewidth}
\centering
\vspace{2.0cm}
\begin{tikzpicture}[scale=1]
\node[state][regular polygon, regular polygon sides=4, draw=blue!50, very thick, fill=blue!10, label=center:$a$,minimum width=17mm] (a1) at (0,2.236)  {};
\node[state][regular polygon, regular polygon sides=4, draw=blue!50, very thick, fill=blue!10, label=center:$a+1$,minimum width=17mm] (a2) at (2.126,0.691)  {};
\node[state][regular polygon, regular polygon sides=4, draw=blue!50, very thick, fill=blue!10, label=center:$a-1$,minimum width=17mm] (a5) at (-2.127,0.691)  {};
\path[->] 
(a1) edge   [thin, bend right]  node [right, pos=0.1] {$\sigma_{a+1} h_2$} (a2)
(a2) edge   [thin, bend right]  node [right, pos=0.6]{$\sigma_{a+1} h_1$} (a1)
(a5) edge   [thin, bend left]  node [left] {$\sigma_{a} h_1$} (a1)
(a1) edge   [thin, bend left]  node [left, pos=0.1]{$\sigma_{a} h_2$} (a5) 
;
\end{tikzpicture}
\end{minipage}
\end{center}
\caption{The charge assignment to the bifundamental/adjoint chiral multiplets around a vertex $a$. 
Note we have $\sigma_a=\sigma_{a+1}$ on the left and $\sigma_a=-\sigma_{a+1}$ on the right.}
\label{fig:gc_hv}
\end{figure}
For example, consider the case of $m=3, n=2$.
We show in Figure \ref{fig.gc_quiverv} quiver diagrams for two choices $\sigma_1=\{+1, +1,  +1,-1,  -1\}$ and for $\sigma_2=\{+1, +1,  -1,+1,  -1\}$.

Given the charge assignment, and the fact that the exchange of $h_1, h_2$ and the simultaneous flip of the orientation of all the arrows preserve the weights of the quiver, we can write the bond factor (\ref{eq-charge-function-glmn-1}) $\varphi^{a\Rightarrow b}(u)$ as
\begin{align}
\varphi^{a\Rightarrow b}(u)&= \frac{u+((Q^{+})_{a,b} h_1 +(Q^{-})_{a,b} h_2) }{u-((Q^{+})_{a,b} h_2 +(Q^{-})_{a,b} h_1) }  \;,
\end{align}
where $Q^{+}$ and $Q^{-}$ are matrices such that $(Q^{+})_{a, a}=(Q^{-})_{a, a}$ for all vertices $a$. 

The explicit expression for $Q^{+}$ and $Q^{-}$ are (recall the charge assignments in Figure \ref{fig:gc_hv} as well as the relation $h_1+h_2+h_3=0$): 
\begin{equation}\label{eq-Qpm}
\begin{aligned}
Q^{+}_{a,b}&=-\frac{\sigma_a+\sigma_{a+1}}{2}\,\delta_{a,b} +\frac{\sigma_{a+1}-1}{2}\, \,\delta_{a+1,b}+\frac{\sigma_a+1}{2} \,\delta_{a-1,b}   \;,\\
Q^{-}_{a,b}&= -\frac{\sigma_a+\sigma_{a+1}}{2}\,\delta_{a,b} +\frac{\sigma_{a+1}+1}{2}\, \,\delta_{a+1,b}+\frac{\sigma_a-1}{2} \,\delta_{a-1,b}  \;.
\end{aligned}
\end{equation}

It turns out that the algebra defined from the function $\varphi^{a\Rightarrow b}$ coincides with the relations of the affine Yangian of $\mathfrak{gl}_{m|n}$, up to Serre relations which we come to momentarily.

This is slightly easier if we define a symmetric matrix $A$ and an anti-symmetric matrix $M$ by 
\begin{align}
A:=-Q^{+}-Q^{-}  \;, \quad
M:=-Q^{+}+Q^{-} \;,
\end{align}
so that 
\begin{equation}\label{eq-AM}
\begin{aligned}
A_{a,b}&=(\sigma_a+\sigma_{a+1})\,\delta_{a,b} -\sigma_{a+1} \,\delta_{a+1,b}-\sigma_a \,\delta_{a-1,b} \;, \\
M_{a,b}&= \delta_{a+1,b}-\delta_{a-1,b} \;.
\end{aligned}
\end{equation}
We can then write
\begin{align}\label{varphi_glmn}
\varphi^{a\Rightarrow b}(u)
=\frac{u+(M_{a,b} \, (h_2-h_1)+ A_{a,b} \, h_3)/2 }{u+(M_{a,b} \, (h_2-h_1)- A_{a,b} \, h_3)/2} \;.
\end{align}
This coincides with the same function for the affine Yangian for the Lie superalgebra $\mathfrak{gl}_{m|n}$, which in turn arises from the rational reduction of the quantum toroidal $\mathfrak{gl}_{m|n}$ algebra constructed recently in 
\cite{Bezerra:2019dmp, bezerra2019braid}\footnote{
In the notations of \cite{Bezerra:2019dmp}, the quantum-toroidal counterpart of the function \eqref{varphi_glmn} is given by
\begin{align}
\varphi_{\textrm{trig.}}^{a\Rightarrow b}(u)
=\left(\frac{d^{\hat{M}_{a,b}}  z- q^{-1}w}{d^{\hat{M}_{a,b}} q^{-1} z-w}\right)^{\hat{A}_{a,b}} \;,
\end{align}
where $q_1=d q^{-1}, q_2=q^2, q_3=d^{-1} q^{-1}$, $\hat{A}_{a,b}=A_{a,b}$, and $\hat{M}_{a,b}=-\sigma_{a+1} \,\delta_{a+1,b}+ \sigma_a \,\delta_{a-1,b}$. 
After taking the trigonometric limit, using the fact that $\hat{A}_{a,b}\hat{M}_{a,b}=M_{a,b}$, and finally identifying their $(h_1, h_2,h_3)$ with our $(h_2, h_3,h_1)$, one obtains \eqref{varphi_glmn}.}
(which generalizes the case of quantum toroidal $\mathfrak{gl}_{n}$ constructed earlier in \cite{Feigin:2013fga}).
In particular, the symmetric matrix $A$ is nothing but the Dynkin diagram for the Lie superalgebra $\widehat{\mathfrak{gl}_{m|n}}$. 
In this language, different choices of the signs $\sigma$ are interpreted as different choices of the simple roots and of the Dynkin diagram for the Lie superalgebra $\mathfrak{gl}_{m|n}$.
(It is known that the choice of the Dynkin diagram is not unique, see e.g.\  \cite{Frappat:1996pb} for review of Lie superalgebras).
The ambiguity of the quiver gauge theory, which as we have seen corresponds to the ambiguity in the choice of the resolution of the toric diagram, now is identified precisely with the ambiguity of the Dynkin diagram of the Lie superalgebra. 
Moreover the boson/fermion statistics of the generators as we derived from the quiver diagram coincides with the even/odd nature of the Lie superalgebra generators.

The quantum toroidal algebra in principle depends on the choice of the sign $\sigma$, however it has recently been shown that algebras with different choices of the signs are related by toroidal braid groups \cite{bezerra2019braid}. 
This is the mathematical manifestation of the physical statement that different quiver gauge theories describe the same geometry.\footnote{
The special choice $s=\{ +1, +1, \dots, +1, -1, -1, \dots, -1 \}$ corresponds to the so-called distinguished Cartan matrix in the Lie superalgebra literature.}

Note that for the identification for the affine Lie superalgebra, it is crucial that both our quiver and the superpotential are invariant under a cyclic permutation of the signs $\sigma$. 
This existence of the affine Weyl group symmetry was noticed before, and also appears in the chamber structure of the K\"ahler moduli space when we consider BPS wall crossing phenomena \cite{MR2999994, Nagao:2009rq}.

The bond factor (\ref{varphi_glmn}) is only non-trivial when $b=a, a\pm 1$:
\begin{equation}
\begin{aligned}
\varphi^{a\Rightarrow a}(u)&=\frac{u+(\sigma_a+\sigma_{a+1})\,h_3/2}{u-(\sigma_a+\sigma_{a+1})\,h_3/2}  \;,\\
\varphi^{a\Rightarrow a+1}(u)&=\frac{u+(h_2-h_1-\sigma_{a+1}\, h_3)/2}{u+(h_2-h_1+\sigma_{a+1}\, h_3)/2} \;,\\
\varphi^{a\Rightarrow a-1}(u)&=\frac{u+(h_1-h_2-\sigma_{a}\, h_3)/2}{u+(h_1-h_2+\sigma_{a}\, h_3)/2} \;,
\end{aligned}
\end{equation}
 which gives the algebra
\begin{tcolorbox}[ams align]\label{eq-QY-glmn}
&\textrm{OPE:}\quad\begin{cases}\begin{aligned}
\psi^{(a)}(z)\, \psi^{(b)}(w)&\sim \psi^{(b)}(w)\, \psi^{(a)}(z) \;,\\
\psi^{(a)}(z)\, e^{(a)}(w)   &\sim \tfrac{\Delta+(\sigma_{a}+\sigma_{a+1})h_3/2 }{\Delta-(\sigma_{a}+\sigma_{a+1})h_3/2 } \, e^{(a)}(w)\, \psi^{(a)}(z)  \;,\\ 
e^{(a)}(z)\, e^{(a)}(w) & \sim  (-1)^{|a|}\tfrac{\Delta+(\sigma_{a}+\sigma_{a+1})h_3/2 }{\Delta-(\sigma_{a}+\sigma_{a+1})h_3/2 }  \, e^{(a)}(w)\, e^{(a)}(z) \;, \\
\psi^{(a)}(z)\, f^{(a)}(w) &  \sim\tfrac{\Delta-(\sigma_{a}+\sigma_{a+1})h_3/2 }{\Delta+(\sigma_{a}+\sigma_{a+1})h_3/2 }  \, f^{(a)}(w)\, \psi^{(a)}(z) \;, \\
 f^{(a)}(z)\, f^{(a)}(w) &  \sim  (-1)^{|a|} \tfrac{\Delta-(\sigma_{a}+\sigma_{a+1})h_3/2 }{\Delta+(\sigma_{a}+\sigma_{a+1})h_3/2 }  \, f^{(a)}(w)\, f^{(a)}(z)  \;,\\
 \psi^{(a+1)}(z)\, e^{(a)}(w)   &\sim \tfrac{\Delta+(h_2-h_1-\sigma_{a+1}\, h_3)/2}{\Delta+(h_2-h_1+\sigma_{a+1}\, h_3)/2} \, e^{(a)}(w)\, \psi^{(a+1)}(z)  \;,\\ 
   \psi^{(a-1)}(z)\, e^{(a)}(w)   &\sim \tfrac{\Delta+(h_1-h_2-\sigma_{a}\, h_3)/2}{\Delta+(h_1-h_2+\sigma_{a}\, h_3)/2}\, e^{(a)}(w)\, \psi^{(a-1)}(z) \;, \\ 
e^{(a+1)}(z)\, e^{(a)}(w) & \sim  (-1)^{|a||a+1|}\tfrac{\Delta+(h_2-h_1-\sigma_{a+1}\, h_3)/2}{\Delta+(h_2-h_1+\sigma_{a+1}\, h_3)/2}\,e^{(a)}(w) \,  e^{(a+1)}(z)  \;,\\
\psi^{(a+1)}(z)\, f^{(a)}(w) &  \sim \tfrac{\Delta+(h_2-h_1+\sigma_{a+1}\, h_3)/2}{\Delta+(h_2-h_1-\sigma_{a+1}\, h_3)/2}\, f^{(a)}(w)\, \psi^{(a+1)}(z)\;,\\
\psi^{(a-1)}(z)\, f^{(a)}(w) &  \sim  \tfrac{\Delta+(h_1-h_2+\sigma_{a}\, h_3)/2}{\Delta+(h_1-h_2-\sigma_{a}\, h_3)/2} \,f^{(a)}(w) \, \psi^{(a-1)}(z)  \;,\\
 f^{(a+1)}(z)\, f^{(a)}(w) &  \sim (-1)^{|a||a+1|}\tfrac{\Delta+(h_2-h_1+\sigma_{a+1}\, h_3)/2}{\Delta+(h_2-h_1-\sigma_{a+1}\, h_3)/2}\, f^{(a)}(w)\, f^{(a+1)}(z) \;,\\
\psi^{(b)}(z)\, e^{(a)}(w)   &\sim  \, e^{(a)}(w)\, \psi^{(b)}(z)\qquad\qquad (b\ne a, a\pm 1)   \;,\\ 
e^{(b)}(z)\, e^{(a)}(w) & \sim  \, (-1)^{|a||b|} e^{(a)}(w)\, e^{(b)}(z) \quad (b\ne a, a\pm 1)  \;,\\
\psi^{(b)}(z)\, f^{(a)}(w) &  \sim  \, f^{(a)}(w)\, \psi^{(b)}(z)  \qquad\qquad (b\ne a, a\pm 1) \;,\\
 f^{(b)}(z)\, f^{(a)}(w) &  \sim   \,(-1)^{|a||b|}   f^{(a)}(w)\, f^{(b)}(z)  \quad (b\ne a, a\pm 1) \;,\\
[e^{(a)}(z)\,, f^{(b)}(w) \}   &\sim  - \delta^{a,b}\, \frac{\psi^{(a)}(z) - \psi^{(b)}(w)}{z-w}  \;, 
\end{aligned}
\end{cases}\\
&\textcolor{black}{\textrm{Initial:}}\quad\begin{cases}
\begin{aligned}
&[\psi^{(a)}_0,e^{(a)}_m] = (\sigma_a+\sigma_{a+1})\,h_3\, e^{(a)}_m \;,\\
&[\psi^{(a)}_0,f^{(a)}_m] =-(\sigma_a+\sigma_{a+1})\, h_3\,f^{(a)}_m \;,\\
&[\psi^{(a+1)}_0,e^{(a)}_m] =-\sigma_{a+1}\,h_3\,e^{(a)}_m \;,\\
&[\psi^{(a+1)}_0,f^{(a)}_m] =\sigma_{a+1}\,h_3\,f^{(a)}_m  \;,\\
&[\psi^{(a-1)}_0,e^{(a)}_m] =-\sigma_{a}\,h_3\,e^{(a)}_m \;,\\
&[\psi^{(a-1)}_0,f^{(a)}_m] =\sigma_{a}\,h_3\,f^{(a)}_m  \;,\\
&[\psi^{(b)}_0,e^{(a)}_m] =[\psi^{(b)}_0,f^{(a)}_m] =0\;\quad ( b\neq a\,,\, a\pm 1)\,,
\end{aligned}
\end{cases}\\
&\textrm{Serre}:\quad
\begin{aligned}
\begin{cases}
&\begin{cases}
&\textrm{Sym}_{z_1,z_2}\, \left[ e^{(a)}(z_1)\,, \left[ e^{(a)}(z_2)\,, e^{(a\pm1)}(w)\right]\right] \sim 0  \;,\\
&\textrm{Sym}_{z_1,z_2}\, \left[ f^{(a)}(z_1)\,, \left[ f^{(a)}(z_2)\,, f^{(a\pm1)}(w)\right]\right] \sim 0  \;,
\end{cases}  \\ & \qquad \qquad \qquad \qquad  \qquad \qquad \qquad \qquad \qquad  \qquad \qquad \qquad (|a|=0) \\
&\begin{cases}
&\textrm{Sym}_{z_1,z_2}\, \left[e^{(a)}(z_1)\,, \left[  e^{(a+1)}(w_1)\,, \left[ e^{(a)}(z_2)\,,  e^{(a-1)}(w_2)\right\}\right\}\right\} \sim 0  \;,\\
&\textrm{Sym}_{z_1,z_2}\, \left[ f^{(a)}(z_1)\,, \left[  f^{(a+1)}(w_1)\,, \left[  f^{(a)}(z_2)\,,  f^{(a-1)}(w_2)\right\} \right\}\right\} \sim 0  \;,
\end{cases} \\ &\qquad \qquad \qquad \qquad  \qquad \qquad \qquad \qquad \qquad  \qquad \qquad \qquad (|a|=1) 
\end{cases}
\end{aligned}
\end{tcolorbox}

\bigskip

Let us remark that the appearance of the Lie superalgebra $\mathfrak{gl}_{m|n}$ for generalized conifold geometries was noticed in \cite{Rapcak:2019wzw}, which discussed W-algebras associated with $\mathfrak{gl}_{m|n}$. While we expect relations between our results and the results of \cite{Rapcak:2019wzw}, any such relation will likely require a non-trivial change of the generators of the algebra (see \cite{ueda2020affine} for a related discussion). It is also the case that the BPS state counting in \cite{Rapcak:2019wzw} is in a particular chamber of the K\"ahler moduli space. Our discussion by contrast applies to general chambers,
which are known to have crystal-melting description.

\subsubsection{\texorpdfstring{Conifold and Affine Yangian of $\mathfrak{gl}_{1|1}$}{Conifold and Affine Yangian of gl(1|1)}}
\label{sec:conifold}

\subsubsubsection{Conifold}

The toric diagram and its dual graph for the conifold $\mathcal{O}(-1)\times \mathcal{O}(-1)\rightarrow \mathbb{P}^1$
are 
\begin{equation}\label{fig-toric-conifold}
\begin{tikzpicture} 
\filldraw [red] (0,0) circle (2pt); 
\filldraw [red] (0,1) circle (2pt); 
\filldraw [red] (1,1) circle (2pt); 
\filldraw [red] (1,0) circle (2pt); 
\node at (-.5,-.5) {(0,0)}; 
\node at (-.5,1.5) {(0,1)}; 
\node at (1.5,1.5) {(1,1)}; 
\node at (1.5,-0.5) {(1,0)}; 
\draw (0,0) -- (0,1); 
\draw (0,1) -- (1,1); 
\draw (0,0) -- (1,0); 
\draw (1,0) -- (1,1); 
\end{tikzpicture}
\qquad \qquad \qquad
\begin{tikzpicture}[scale=0.6] 
\draw[->] (0,0) -- (-1,0); 
\draw[->] (0,0) -- (0,-1); 
\draw (0,0) -- (1,1); 
\draw[->] (1,1) -- (2,1); 
\draw[->] (1,1) -- (1,2); 
\node at (-1.5,0) {3}; 
\node at (2.5,1) {$\hat{3}$}; 
\node at (0,-1.5) {1}; 
\node at (1,2.5) {$\hat{1}$}; 
\end{tikzpicture}
\end{equation}
Its associated quiver diagram is similar to the one for the orbifold $(\mathbb{C}^2/\mathbb{Z}_2)$$\times$$\mathbb{C}$:
\begin{equation}\label{figure-quiver-conifold}
\begin{tikzpicture}[scale=1]
\node[state]  [regular polygon, regular polygon sides=4, draw=blue!50, very thick, fill=blue!10] (a1) at (-2,0)  {$1$};
\node[state]  [regular polygon, regular polygon sides=4, draw=blue!50, very thick, fill=blue!10] (a2) at (2,0)  {$2$};
\path[->] 
(a1) edge   [thin, bend left]  node [above] {$(A_1,\alpha_{1}), \,\, (B_2,\beta_{2})$} (a2)
(a2) edge   [thin, bend left]  node [below]{$(B_1,\beta_1), \,\, (A_2,\alpha_{2})$} (a1);
\end{tikzpicture}
\end{equation}
with super-potential 
\begin{equation}
W=\textrm{Tr}[-A_1B_1B_2A_2+A_1A_2B_2B_1]  \;.
\end{equation}
Since there is no self-loop for either vertex $1$ or $2$, both vertices are fermionic:
\begin{equation}
|a|=1\,, \qquad a=1,2\,,
\end{equation}
to be compared with the case of $(\mathbb{C}^2/\mathbb{Z}_n)$$\times$$\mathbb{C}$ shown in (\ref{figure-quiver-Z2}), where both vertices are bosonic, i.e.\ $|a|=0$.

The periodic quiver is shown in the left picture of the following
\begin{equation}\label{tilling-conifold}
\begin{aligned}
&
\begin{tikzpicture}[scale=0.8]
\filldraw[mygreen] (-3,0)--  (0, -3) -- (3,0) -- (0,3) -- cycle; 
\node[state]  [regular polygon, regular polygon sides=4, draw=blue!50, very thick, fill=blue!10] (a1) at (0,0)  {$1$};
\node[state]  [regular polygon, regular polygon sides=4, draw=blue!50, very thick, fill=blue!10] (a21) at (3,0)  {$2$};
\node[state]  [regular polygon, regular polygon sides=4, draw=blue!50, very thick, fill=blue!10] (a22) at (-3,0)  {$2$};
\node[state]  [regular polygon, regular polygon sides=4, draw=blue!50, very thick, fill=blue!10] (a41) at (0,3)  {$2$};
\node[state]  [regular polygon, regular polygon sides=4, draw=blue!50, very thick, fill=blue!10] (a42) at (0,-3)  {$2$};
\node[state]  [regular polygon, regular polygon sides=4, draw=blue!50, very thick, fill=blue!10] (a31) at (3,3)  {$1$};
\node[state]  [regular polygon, regular polygon sides=4, draw=blue!50, very thick, fill=blue!10] (a32) at (3,-3)  {$1$};
\node[state]  [regular polygon, regular polygon sides=4, draw=blue!50, very thick, fill=blue!10] (a34) at (-3,-3)  {$1$};
\node[state]  [regular polygon, regular polygon sides=4, draw=blue!50, very thick, fill=blue!10] (a33) at (-3,3)  {$1$};
\path[->] 
(a21) edge   [thick, red]   node [above] {$\alpha_2$} (a1)
(a22) edge   [thick, red]   node [above] {$\beta_1$} (a1)
(a31) edge   [thick, red]   node [right] {$\alpha_1$} (a21)
(a32) edge   [thick, red]   node [right] {$\beta_2$} (a21)
(a33) edge   [thick, red]   node [left] {$\alpha_1$} (a22)
(a34) edge   [thick, red]   node [left] {$\beta_2$} (a22)
(a41) edge   [thick, red]   node [above] {$\beta_1$} (a31)
(a41) edge   [thick, red]   node [above] {$\alpha_2$} (a33)
(a42) edge   [thick, red]   node [above] {$\alpha_2$} (a34)
(a42) edge   [thick, red]   node [above] {$\beta_1$} (a32)
(a1) edge   [thick, red]   node [right] {$\beta_2$} (a41)
(a1) edge   [thick, red]   node [right] {$\alpha_1$} (a42)
;
\end{tikzpicture}
\qquad
\begin{tikzpicture}[scale=0.8]
\filldraw[mygreen] (-3,0)--  (0, -3) -- (3,0) -- (0,3) -- cycle; 
\node[state]  [regular polygon, regular polygon sides=4, draw=blue!50, very thick, fill=blue!10] (a1) at (0,0)  {$1$};
\node[state]  [regular polygon, regular polygon sides=4, draw=blue!50, very thick, fill=blue!10] (a21) at (3,0)  {$2$};
\node[state]  [regular polygon, regular polygon sides=4, draw=blue!50, very thick, fill=blue!10] (a22) at (-3,0)  {$2$};
\node[state]  [regular polygon, regular polygon sides=4, draw=blue!50, very thick, fill=blue!10] (a41) at (0,3)  {$2$};
\node[state]  [regular polygon, regular polygon sides=4, draw=blue!50, very thick, fill=blue!10] (a42) at (0,-3)  {$2$};
\node[state]  [regular polygon, regular polygon sides=4, draw=blue!50, very thick, fill=blue!10] (a31) at (3,3)  {$1$};
\node[state]  [regular polygon, regular polygon sides=4, draw=blue!50, very thick, fill=blue!10] (a32) at (3,-3)  {$1$};
\node[state]  [regular polygon, regular polygon sides=4, draw=blue!50, very thick, fill=blue!10] (a34) at (-3,-3)  {$1$};
\node[state]  [regular polygon, regular polygon sides=4, draw=blue!50, very thick, fill=blue!10] (a33) at (-3,3)  {$1$};
\path[->] 
(a21) edge   [thick, red]   node [above] {$-h_2$} (a1)
(a22) edge   [thick, red]   node [above] {$h_2$} (a1)
(a31) edge   [thick, red]   node [right] {$h_1$} (a21)
(a32) edge   [thick, red]   node [right] {$-h_1$} (a21)
(a33) edge   [thick, red]   node [left] {$h_1$} (a22)
(a34) edge   [thick, red]   node [left] {$-h_1$} (a22)
(a41) edge   [thick, red]   node [above] {$h_2$} (a31)
(a41) edge   [thick, red]   node [above] {$-h_2$} (a33)
(a42) edge   [thick, red]   node [above] {$-h_2$} (a34)
(a42) edge   [thick, red]   node [above] {$h_2$} (a32)
(a1) edge   [thick, red]   node [right] {$-h_1$} (a41)
(a1) edge   [thick, red]   node [right] {$h_1$} (a42)
;
\end{tikzpicture}
\end{aligned}
\end{equation}
where the fundamental regions of the torus are shown as shaded regions.
Note its relation to the one for the orbifold $(\mathbb{C}^2/\mathbb{Z}_2)$$\times$$\mathbb{C}$, shown in (the right picture in) Figure~\ref{fig-periodic-quiver-gl2}.
Starting from the right picture in Figure~\ref{fig-periodic-quiver-gl2}, if one removes all the diagonal arrows, which correspond to the self-arrows in the quiver, and further flip the directions of arrows as one passes each vertex as one moves along either $x_1$ or $x_2$ direction, one then obtains the periodic quiver shown in (\ref{tilling-conifold}).
As we will see later, this is a general pattern relating the periodic quivers for the orbifold $(\mathbb{C}^2/\mathbb{Z}_n)$$\times$$\mathbb{C}$ and the generalized conifold with the same rank, resulting in the relation between affine Yangians of $\mathfrak{gl}_{m+n}$ and $\mathfrak{gl}_{m|n}$.

The loop constraint (\ref{eq-loop-constraint-toric}) translates to
\begin{equation}\label{eq-ConstraintL-conifold}
\alpha_1+\alpha_2+\beta_1+\beta_2=0\,.
\end{equation}
Again, the central condition (\ref{eq-central-condition}) is guaranteed by the loop constraint (\ref{eq-ConstraintL-conifold}).
One can then immediately read off the bond factors from the periodic quiver (\ref{tilling-conifold}) by the definition (\ref{eq-charge-atob})
\begin{equation}\label{eq-charge-function-conifold}
 \varphi^{a\Rightarrow a}(u)=1\,,\qquad \varphi^{a\Rightarrow a+1}(u)=\frac{(u+\alpha_{a+1})(u+\beta_a)}{(u-\alpha_a)(u-\beta_{a+1})} \;,
\end{equation}
where the indices are understood as mod $2$ and the four charges $(\alpha_{1,2}, \beta_{1,2})$ satisfy (\ref{eq-ConstraintL-conifold}). 
Accordingly, the resulting algebra is 
\begin{tcolorbox}[ams align]\label{eq-algebra-shifted-conifold}
&\textrm{OPE:}\quad
\begin{cases}
\begin{aligned}
\psi^{(a)}(z)\, \psi^{(b)}(w)&\sim \psi^{(b)}(w)\, \psi^{(a)}(z) \;,\\
\psi^{(a)}(z)\, e^{(a)}(w)   &\sim   e^{(a)}(w)\, \psi^{(a)}(z)  \;,\\ 
e^{(a)}(z)\, e^{(a)}(w) & \sim  - e^{(a)}(w)\, e^{(a)}(z)  \;,\\
\psi^{(a)}(z)\, f^{(a)}(w) &  \sim  f^{(a)}(w)\, \psi^{(a)}(z)  \;,\\
 f^{(a)}(z)\, f^{(a)}(w) &  \sim  - f^{(a)}(w)\, f^{(a)}(z) \;, \\
 \psi^{(a+1)}(z)\, e^{(a)}(w)   &\sim \tfrac{(\Delta+\alpha_{a+1})(\Delta+\beta_{a})}{(\Delta-\alpha_{a})(\Delta-\beta_{a+1})}  \, e^{(a)}(w)\, \psi^{(a+1)}(z)  \;,\\ 
e^{(a+1)}(z)\, e^{(a)}(w) &\sim - \tfrac{(\Delta+\alpha_{a+1})(\Delta+\beta_{a})}{(\Delta-\alpha_{a})(\Delta-\beta_{a+1})} \, e^{(a)}(w)\, e^{(a+1)}(z)  \;,\\
\psi^{(a+1)}(z)\, f^{(a)}(w) &  \sim \tfrac{(\Delta-\alpha_{a})(\Delta-\beta_{a+1})}{(\Delta+\alpha_{a+1})(\Delta+\beta_{a})}  \, f^{(a)}(w)\, \psi^{(a+1)}(z)  \;,\\
 f^{(a+1)}(z)\, f^{(a)}(w) &  \sim - \tfrac{(\Delta-\alpha_{a})(\Delta-\beta_{a+1})}{(\Delta+\alpha_{a+1})(\Delta+\beta_{a})}\,f^{(a)}(w)\, f^{(a+1)}(z) \;,\\
\{e^{(a)}(z)\,, f^{(b)}(w)\}   &=  - \delta^{a,b}\, \frac{\psi^{(a)}(z) - \psi^{(a)}(w)}{z-w}  \;.
\end{aligned}
\end{cases}
\end{tcolorbox}
\noindent Note that these relations form a subset of the relations for $(\mathbb{C}^2/\mathbb{Z}_n)$$\times$$\mathbb{C}$, given in (\ref{eq-algebra-shifted-gl2-1}), up to statistics factors in the $e-e$ and $f-f$ relations.
Correspondingly, the initial conditions are also a subset of those in (\ref{eq-algebra-shifted-gl2-2}). 
But the presence of the stronger constraint (\ref{eq-ConstraintL-conifold}) simplify the expressions:
\begin{tcolorbox}[ams align]\label{eq-algebra-shifted-conifold-re}
&\textcolor{black}{\textrm{Initial:}}\quad\begin{cases}
\begin{aligned}
&\begin{aligned}
&[\psi^{(a)}_0,e^{(a)}_m] =[\psi^{(a)}_1,e^{(a)}_m] =[\psi^{(a)}_0,f^{(a)}_m] =[\psi^{(a)}_1,f^{(a)}_m] =0\,,\\
& [\psi^{(a+1)}_0,e^{(a)}_m] =0 \;,\\
& [\psi^{(a+1)}_0,f^{(a)}_m] =0 \;, \\
&[\psi^{(a+1)}_1,e^{(a)}_m] =( \alpha_{a+1}\beta_{a}-\alpha_{a}\beta_{a+1})\,e^{(a)}_m 
\;,\\
&[\psi^{(a+1)}_1,f^{(a)}_m] =  -(\alpha_{a+1}\beta_{a}-\alpha_{a}\beta_{a+1})\,f^{(a)}_m
\;.
\\
\end{aligned}
\end{aligned}
\end{cases}
\end{tcolorbox}
\noindent From the initial conditions one can check that the combination $\psi_0\equiv \psi^{(1)}_0+\psi^{(2)}_0$ is indeed a central term.

\subsubsubsection{Truncation}

Consider the path from the origin to an atom $\sqbox{$1$}$ of color $1$, at which the growth of the crystal stops. 
The coordinate function of $\sqbox{$1$}$ is a special case of (\ref{eq-truncation-glmn-1}), with $\gamma_{a}=0$:
\begin{equation}\label{conifold_h}
h(\sqbox{$1$})=N_1(\alpha_1+\beta_1)+N_2(\alpha_2+\beta_2)+
N_{\alpha} (\alpha_1+\alpha_2)+N_{\beta}(\beta_1+\beta_2) \;,
\end{equation}
where $N_a$ is the number of segments $a\rightarrow a+1\rightarrow a$ with charge $\alpha_a$ and then $\beta_a$, etc.
Imposing the loop constraints (\ref{eq-ConstraintL-conifold}) gives the truncation condition:
\begin{equation}\label{conifold_truncate}
(N_1-N_2)(\alpha_1+\beta_1)+(N_{\alpha}-N_{\beta}) (\alpha_1+\alpha_2)+\sum^2_{a=1}\psi^{(a)}_0=0 \;,
\end{equation}
namely the truncation can be characterized by the two integer coefficients multiplying $(\alpha_1+\beta_1)$ and $(\alpha_1+\alpha_2)$. 

\begin{figure}[htbp]
\centering\includegraphics[scale=0.25]{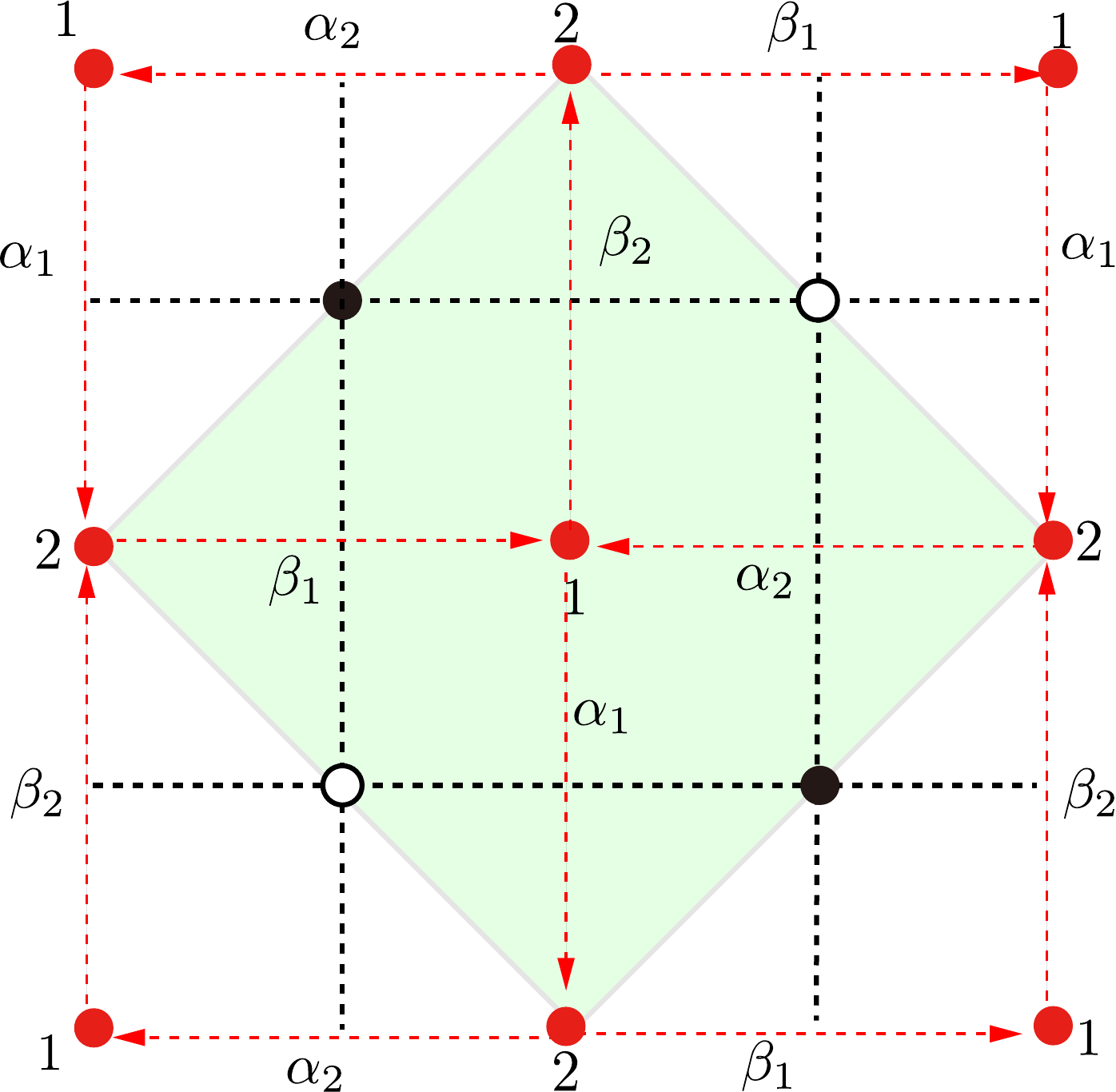}
\caption{The bipartite graph for the conifold geometry.}
\label{fig.conifoldbipartite}
\end{figure}

\begin{figure}[htbp]
\centering\includegraphics[scale=0.22]{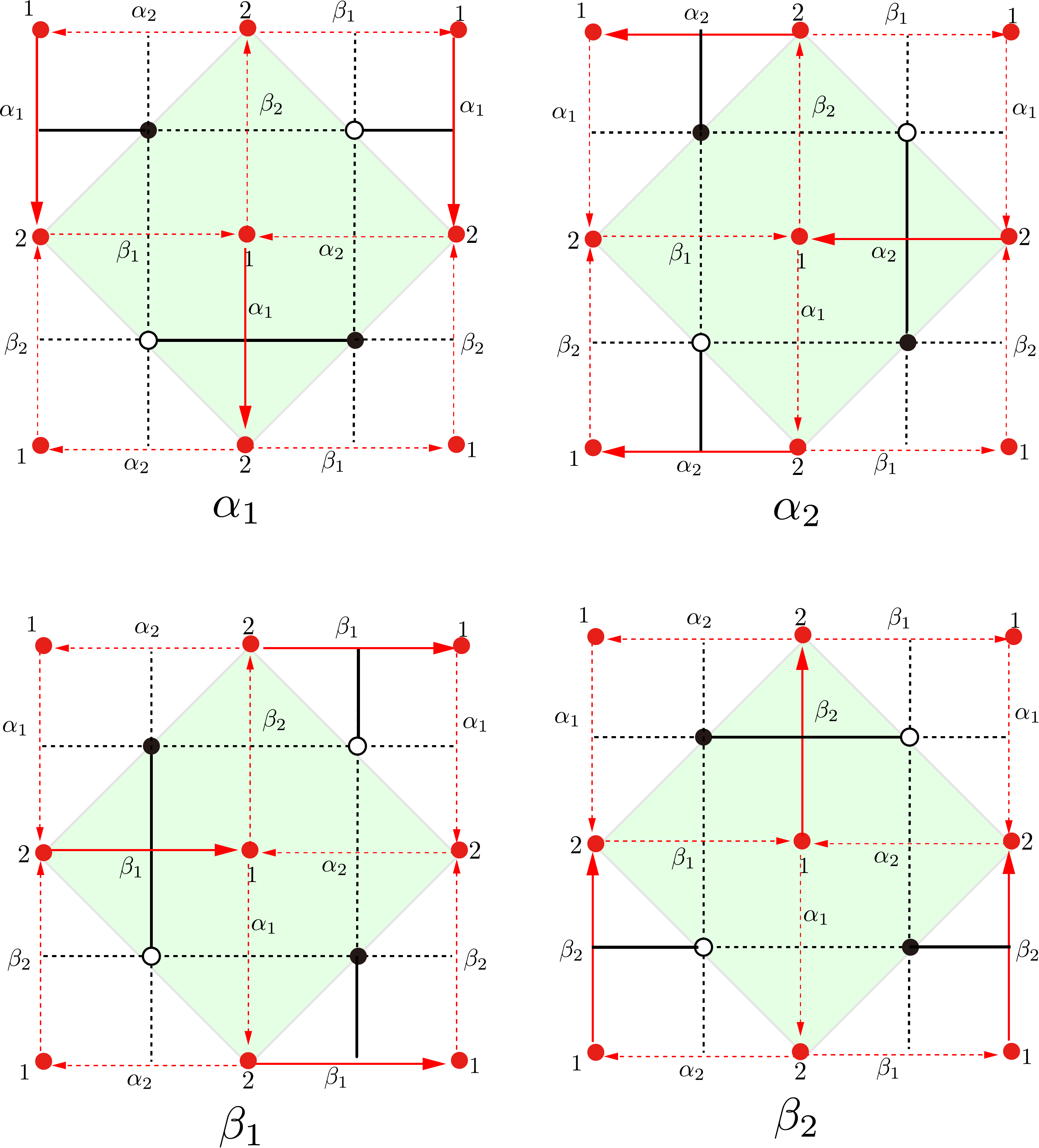}
\caption{The $4$ perfect matchings for the conifold geometry.
They correspond to the four parameters $\alpha_1, \alpha_2, \beta_1, \beta_2$.}
\label{fig.conifoldPM}
\end{figure}

One can compare this result with the expectation from D4-branes.
Starting with the bipartite graph in Figure \ref{fig.conifoldbipartite},
one obtains that the four perfect matchings (shown in Figure~\ref{fig.conifoldPM}) give
the linear combinations
\begin{align}\label{conifold_basis}
\alpha_1\;, \quad \alpha_2\;,\quad \beta_1 \;, \quad \beta_2=-(\alpha_1+\alpha_2+\beta_1) \;.
\end{align}
One might therefore conclude that we
obtain linear combination of the first three elements with non-negative integer coefficient.
This is more general than the previous result \eqref{conifold_truncate}, which suggests that 
there should be more general representations than those in this paper.

\subsubsubsection{Affine Yangian of $\mathfrak{gl}_{1|1}$}\label{sec.Ygl11}

For the periodic quiver (\ref{tilling-conifold}), the vertex constraint (\ref{eq-vertex-constraint-toric}) translates to
\begin{equation}\label{eq-ConstraintV-conifold}
\alpha_1+\beta_2=\alpha_2+\beta_2 \;.
\end{equation}
Together with the loop constraint (\ref{eq-ConstraintL-conifold}), it reduces the four parameters $(\alpha_{1,2},\beta_{1,2})$ to two independent parameters
\begin{equation}
\alpha_1=-\beta_2=h_1  \qquad \textrm{and}\qquad \beta_1=-\alpha_2=h_2\,.
\end{equation}
We have drawn the period quiver with both loop and vertex constraints imposed in the right figure of (\ref{tilling-conifold}).
With both the loop and vertex constraints imposed, the bond factor (\ref{eq-charge-function-conifold}) becomes
\begin{equation}
\begin{aligned}
&\varphi^{1\Rightarrow 1}(u)=\varphi^{2\Rightarrow 2}(u)=1 \;,\\
&\varphi^{1\Rightarrow 2}(u)=\frac{(u+h_2)(u-h_2)}{(u-h_1)(u+h_1)} \;,\\ 
&\varphi^{2\Rightarrow 1}(u)=\frac{(u+h_1)(u-h_1)}{(u-h_2)(u+h_2)} \;.
\end{aligned}
\end{equation}
The resulting algebra is
\begin{tcolorbox}[ams align]
&\textrm{OPE:}\quad
\begin{cases}
\begin{aligned}
\psi^{(a)}(z)\, \psi^{(b)}(w)&\sim \psi^{(b)}(w)\, \psi^{(a)}(z) \;,\\
\psi^{(a)}(z)\, e^{(a)}(w)   &\sim   e^{(a)}(w)\, \psi^{(a)}(z)  \;,\\ 
e^{(a)}(z)\, e^{(a)}(w) & \sim  - e^{(a)}(w)\, e^{(a)}(z)  \;,\\
\psi^{(a)}(z)\, f^{(a)}(w) &  \sim  f^{(a)}(w)\, \psi^{(a)}(z)  \;,\\
 f^{(a)}(z)\, f^{(a)}(w) &  \sim  - f^{(a)}(w)\, f^{(a)}(z) \;, \\
 \psi^{(2)}(z)\, e^{(1)}(w)   &\sim \tfrac{(\Delta+h_2)(\Delta-h_2)}{(\Delta+h_1)(\Delta-h_1)} \, e^{(1)}(w)\, \psi^{(2)}(z)  \;,\\ 
 \psi^{(1)}(z)\, e^{(2)}(w)   &\sim \tfrac{(\Delta+h_1)(\Delta-h_1)}{(\Delta+h_2)(\Delta-h_2)} \, e^{(2)}(w)\, \psi^{(1)}(z)  \;,\\ 
e^{(2)}(z)\, e^{(1)}(w) & \sim  - \tfrac{(\Delta+h_2)(\Delta-h_2)}{(\Delta+h_1)(\Delta-h_1)} \, e^{(1)}(w)\, e^{(2)}(z) \;, \\
\psi^{(2)}(z)\, f^{(1)}(w) &  \sim  \tfrac{(\Delta+h_1)(\Delta-h_1)}{(\Delta+h_2)(\Delta-h_2)} \, f^{(1)}(w)\, \psi^{(2)}(z)  \;,\\
\psi^{(1)}(z)\, f^{(2)}(w) &  \sim  \tfrac{(\Delta+h_2)(\Delta-h_2)}{(\Delta+h_1)(\Delta-h_1)} \, f^{(2)}(w)\, \psi^{(1)}(z)  \;,\\
 f^{(2)}(z)\, f^{(1)}(w) &  \sim -\tfrac{(\Delta+h_1)(\Delta-h_1)}{(\Delta+h_2)(\Delta-h_2)}\,f^{(1)}(w)\, f^{(2)}(z) \;,\\
\{e^{(a)}(z)\,, f^{(b)}(w)\}   &\sim  - \delta^{a,b}\, \frac{\psi^{(a)}(z) - \psi^{(b)}(w)}{z-w}  \;,
\end{aligned}
\end{cases}
\\
&\textcolor{black}{\textrm{Initial:}}\quad\begin{cases}
\begin{aligned}
&\begin{aligned}
&[\psi^{(a)}_0,e^{(a)}_m] =[\psi^{(a)}_1,e^{(a)}_m] =[\psi^{(a)}_0,f^{(a)}_m] =[\psi^{(a)}_1,f^{(a)}_m] =0\,,\\
& [\psi^{(a+1)}_0,e^{(a)}_m] =0 \;,\\
& [\psi^{(a+1)}_0,f^{(a)}_m] =0 \;, \\
&[\psi^{(a+1)}_1,e^{(a)}_m] = (-1)^a\, (h_2^2-h_1^2)\,e^{(a)}_m \;,\\
&[\psi^{(a+1)}_1,f^{(a)}_m] = -  (-1)^a\,(h_2^2-h_1^2)\,f^{(a)}_m \;,\\
\end{aligned}
\end{aligned}
\end{cases}\\
&\textrm{Serre}:\quad\begin{cases}\begin{aligned}
&\textrm{Sym}_{z_1,z_2}\, \left\{ e^{(a)}(z_1)\,, \left[  e^{(a+1)}(w_1)\,, \left\{  e^{(a)}(z_2)\,,  e^{(a+1)}(w_2)\right\}\right]\right\} \sim 0  \;,\\
&\textrm{Sym}_{z_1,z_2}\, \left\{ f^{(a)}(z_1)\,, \left[  f^{(a+1)}(w_1)\,, \left\{  f^{(a)}(z_2)\,,  f^{(a+1)}(w_2)\right\} \right]\right\} \sim 0  \;,\\
\end{aligned}
\end{cases}
\end{tcolorbox}
\noindent where $a,b=1,2$.

Here we adopted the Serre relation from \cite{Bezerra:2019dmp}.\footnote{
The paper \cite{Bezerra:2019dmp} strictly speaking does not deal with affine Yangians of $\mathfrak{gl}_{1|1}$, or 
more generally $\mathfrak{gl}_{n|n}$.}
Note that $e^{(a)}$ and $f^{(a)}$ are fermionic generators, and hence we have both commutators $[-,-]$ and 
anti-commutators $\{ -, - \}$ in the Serre relations.

While we do not have a top-down understanding of the Serre relations of the quiver Yangian in general,
we seem to be finding some pattern here.
Namely, the Serre relation for the fermionic generators $e^{(a)}$ involve
the $e^{(a)}, e^{(a+1)}, e^{(a)}, e^{(a+1)}$ from left to right in that order,
and this seems to correspond to the superpotential term
$\textrm{Tr}(\Phi_{a,a+1} \Phi_{a+1,a} \Phi_{a,a-1} \Phi_{a-1,a})$
in \eqref{eq.gc_W}. Similarly, 
we have a cubic Serre relation for the bosonic generators $e^{(a)}$
for the affine Yangian for $\mathfrak{gl}_{m|n}$ \cite{Bezerra:2019dmp}
(recall also the cubic Serre relation 
for the $\mathbb{C}^3$ geometry in \eqref{box-serre-gl1}),
and this corresponds naturally to the 
cubic superpotential term $\textrm{Tr}(\Phi_{a,a} \Phi_{a,a+1} \Phi_{a+1,a})$
in \eqref{eq.gc_W}. It is tempting to speculate that 
this is a general pattern and that the Serre relations can be
identified from the data of the superpotential.

\subsubsection{\texorpdfstring{Suspended Pinched Point and Affine Yangian of $\mathfrak{gl}_{2|1}$}{Suspended Pinched Point and Affine Yangian of gl(2|1)}}

Let us close this subsection with another special case of $m=2, n=1$. 
This is the Suspended Pinched Point geometry discussed in section \ref{sec:review_crystal}.
With both the loop constraints and the vertex constraint imposed, the corresponding algebra is the affine Yangian of $\mathfrak{gl}_{2|1}$. 

The quiver diagram for SPP is 
\begin{equation}\label{quiver-SPP}
\begin{tikzpicture}[scale=0.8]
\node[state]  [regular polygon, regular polygon sides=4, draw=blue!50, very thick, fill=blue!10] (a1) at (0,0)  {$1$};
\node[state]  [regular polygon, regular polygon sides=4, draw=blue!50, very thick, fill=blue!10] (a2) at (2,-3.4641)  {$2$};
\node[state]  [regular polygon, regular polygon sides=4, draw=blue!50, very thick, fill=blue!10] (a3) at  (-2,-3.4641)  {$3$};
\path[->] 
(a1) edge [in=60, out=120, loop, thin, above] node {$\gamma$} ()
(a1) edge   [thin, bend left]  node [right] {$\alpha_{1}$} (a2)
(a2) edge   [thin]  node [left]{$\beta_{1}$} (a1)
(a2) edge   [thin]  node [above] {$\alpha_2$} (a3)
(a3) edge   [thin, bend right]  node [below]{$\beta_2$} (a2)
(a3) edge   [thin, bend left]  node [left] {$\alpha_3$} (a1)
(a1) edge   [thin]  node [right]{$\beta_3$} (a3)
;
\end{tikzpicture}
\qquad \qquad\qquad
\begin{tikzpicture}[scale=0.8]
\node[state]  [regular polygon, regular polygon sides=4, draw=blue!50, very thick, fill=blue!10] (a1) at (0,0)  {$1$};
\node[state]  [regular polygon, regular polygon sides=4, draw=blue!50, very thick, fill=blue!10] (a2) at (2,-3.4641)  {$2$};
\node[state]  [regular polygon, regular polygon sides=4, draw=blue!50, very thick, fill=blue!10] (a3) at  (-2,-3.4641)  {$3$};
\path[->] 
(a1) edge [in=60, out=120, loop, thin, above] node {$h_3$} ()
(a1) edge   [thin, bend left]  node [right] {$h_{1}$} (a2)
(a2) edge   [thin]  node [left]{$h_{2}$} (a1)
(a2) edge   [thin]  node [above] {$-h_{2}$} (a3)
(a3) edge   [thin, bend right]  node [below]{$-h_{1}$} (a2)
(a3) edge   [thin, bend left]  node [left] {$h_{1}$} (a1)
(a1) edge   [thin]  node [right]{$h_{2}$} (a3)
;
\end{tikzpicture}
\end{equation}
where in the left one the charges are before any constraints are imposed; whereas in the right one, we have imposed both the loop constraints
\begin{equation}\label{eq-loop-constraint-SPP}
\begin{aligned}
-\gamma=\alpha_1+\beta_1=\alpha_3+\beta_3= -\alpha_2-\beta_2 \;,
\end{aligned}
\end{equation}
and the vertex constraint
\begin{equation}\label{eq-vertex-constraint-SPP}
\alpha_1-\beta_1=\alpha_2-\beta_2=\alpha_3-\beta_3\;.
\end{equation}
The solutions to the two sets of constraints are denoted by the three parameters $(h_1,h_2,h_3)$ satisfying
$h_1+h_2+h_3=0$.
The corresponding periodic quivers are
\begin{equation}\label{tilling-gl2-square}
\begin{tikzpicture}[scale=0.6]
\filldraw[mygreen] (-3,3)--  (-3, -3) -- (0,-3) -- (0,0) -- cycle; 
\filldraw[mygreen] (3,-3)--  (3, 3) -- (0,3) -- (0,0) -- cycle; 
\node[state]  [regular polygon, regular polygon sides=4, draw=blue!50, very thick, fill=blue!10] (a1) at (0,0)  {$1$};
\node[state]  [regular polygon, regular polygon sides=4, draw=blue!50, very thick, fill=blue!10] (a21) at (3,0)  {$3$};
\node[state]  [regular polygon, regular polygon sides=4, draw=blue!50, very thick, fill=blue!10] (a22) at (-3,0)  {$2$};
\node[state]  [regular polygon, regular polygon sides=4, draw=blue!50, very thick, fill=blue!10] (a41) at (0,3)  {$3$};
\node[state]  [regular polygon, regular polygon sides=4, draw=blue!50, very thick, fill=blue!10] (a42) at (0,-3)  {$2$};
\node[state]  [regular polygon, regular polygon sides=4, draw=blue!50, very thick, fill=blue!10] (a31) at (3,3)  {$2$};
\node[state]  [regular polygon, regular polygon sides=4, draw=blue!50, very thick, fill=blue!10] (a32) at (3,-3)  {$1$};
\node[state]  [regular polygon, regular polygon sides=4, draw=blue!50, very thick, fill=blue!10] (a34) at (-3,-3)  {$3$};
\node[state]  [regular polygon, regular polygon sides=4, draw=blue!50, very thick, fill=blue!10] (a33) at (-3,3)  {$1$};
\path[->] 
(a1) edge   [thick, red]   node [above] {$\beta_3$} (a21)
(a22) edge   [thick, red]   node [above] {$\beta_1$} (a1)
(a21) edge   [thick, red]   node [right] {$\beta_2$} (a31)
(a21) edge   [thick, red]   node [right] {$\alpha_3$} (a32)
(a33) edge   [thick, red]   node [left] {$\alpha_1$} (a22)
(a34) edge   [thick, red]   node [left] {$\beta_2$} (a22)
(a31) edge   [thick, red]   node [above] {$\alpha_2$} (a41)
(a33) edge   [thick, red]   node [above] {$\beta_3$} (a41)
(a42) edge   [thick, red]   node [above] {$\alpha_2$} (a34)
(a42) edge   [thick, red]   node [above] {$\beta_1$} (a32)
(a41) edge   [thick, red]   node [right] {$\alpha_3$} (a1)
(a1) edge   [thick, red]   node [right] {$\alpha_1$} (a42)
(a1) edge   [thick, red]   node [right] {$\gamma$} (a33)
(a32) edge   [thick, red]   node [right] {$\gamma$} (a1)
;
\end{tikzpicture}
\qquad \qquad 
\begin{tikzpicture}[scale=0.6]
\filldraw[mygreen] (-3,3)--  (-3, -3) -- (0,-3) -- (0,0) -- cycle; 
\filldraw[mygreen] (3,-3)--  (3, 3) -- (0,3) -- (0,0) -- cycle; 
\node[state]  [regular polygon, regular polygon sides=4, draw=blue!50, very thick, fill=blue!10] (a1) at (0,0)  {$1$};
\node[state]  [regular polygon, regular polygon sides=4, draw=blue!50, very thick, fill=blue!10] (a21) at (3,0)  {$3$};
\node[state]  [regular polygon, regular polygon sides=4, draw=blue!50, very thick, fill=blue!10] (a22) at (-3,0)  {$2$};
\node[state]  [regular polygon, regular polygon sides=4, draw=blue!50, very thick, fill=blue!10] (a41) at (0,3)  {$3$};
\node[state]  [regular polygon, regular polygon sides=4, draw=blue!50, very thick, fill=blue!10] (a42) at (0,-3)  {$2$};
\node[state]  [regular polygon, regular polygon sides=4, draw=blue!50, very thick, fill=blue!10] (a31) at (3,3)  {$2$};
\node[state]  [regular polygon, regular polygon sides=4, draw=blue!50, very thick, fill=blue!10] (a32) at (3,-3)  {$1$};
\node[state]  [regular polygon, regular polygon sides=4, draw=blue!50, very thick, fill=blue!10] (a34) at (-3,-3)  {$3$};
\node[state]  [regular polygon, regular polygon sides=4, draw=blue!50, very thick, fill=blue!10] (a33) at (-3,3)  {$1$};
\path[->] 
(a1) edge   [thick, red]   node [above] {$h_2$} (a21)
(a22) edge   [thick, red]   node [above] {$h_2$} (a1)
(a21) edge   [thick, red]   node [right] {$-h_1$} (a31)
(a21) edge   [thick, red]   node [right] {$h_1$} (a32)
(a33) edge   [thick, red]   node [left] {$h_1$} (a22)
(a34) edge   [thick, red]   node [left] {$-h_1$} (a22)
(a31) edge   [thick, red]   node [above] {$-h_2$} (a41)
(a33) edge   [thick, red]   node [above] {$h_2$} (a41)
(a42) edge   [thick, red]   node [above] {$-h_2$} (a34)
(a42) edge   [thick, red]   node [above] {$h_2$} (a32)
(a41) edge   [thick, red]   node [right] {$h_1$} (a1)
(a1) edge   [thick, red]   node [right] {$h_1$} (a42)
(a1) edge   [thick, red]   node [right] {$h_3$} (a33)
(a32) edge   [thick, red]   node [right] {$h_3$} (a1)
;
\end{tikzpicture}
\end{equation}
with the fundamental regions of the torus shown as shaded regions.
 (In these figures the fundamental regions are split into two to save space in this figure; in each case the understanding 
 is that the trapezoids are meant to be glued together along edges with labels $\alpha_2$ and $-h_2$ to obtain a parallelogram.)
The vertex $1$ is even while the vertices $2$ and $3$ are odd:
\begin{equation}
|1|=0\,, \qquad \qquad |2|=|3|=1\,.
\end{equation}

The algebra before the vertex constraints (\ref{eq-vertex-constraint-SPP}) are imposed is given by
\begin{equation}\label{eq-charge-function-SPP}
\begin{aligned}
&\varphi^{a\Rightarrow a+1}(u)=\frac{u+\beta_a}{u-\alpha_a}\,,   \qquad \varphi^{a\Rightarrow a-1}(u)=\frac{u+\alpha_{a-1}}{u-\beta_{a-1}}\,,\\
&\varphi^{1\Rightarrow 1}(u)=\frac{u+ \gamma}{u- \gamma}\,, \qquad \varphi^{2\Rightarrow 2}(u)=\varphi^{3\Rightarrow 3}(u)=1\,,
\end{aligned}
\end{equation}
where $a=1,2,3$, and the $\beta_{a}$ is fixed in terms of $\alpha_a$ and $\gamma$ by the loop constraints (\ref{eq-loop-constraint-SPP}).
The algebra relations and the initial conditions for the SPP geometry can then be obtained by plugging the choice $(\sigma_1,\sigma_2,\sigma_3)=(+,+,-)$ and the bond factors (\ref{eq-charge-function-SPP}) into the general formulae  (\ref{eq-OPE-generalizedO}) and (\ref{eq-initial-generalizedO}).

\subsubsubsection{Truncation}

The truncation condition can be obtained by taking the general  formula (\ref{eq-truncation-glmn-final}) and plugging in $(\sigma_1,\sigma_2,\sigma_3)=(+,+,-)$. This gives
\begin{equation}\label{eq-truncation-glmn-final-re}
\gamma\, \left(N_{\gamma}-N_1+N_2-N_3-N_{\beta}\right) +(\alpha_1+\alpha_2+\alpha_3)(N_{\alpha}-N_{\beta})+\sum^3_{a=1}\psi^{(a)}_0=0\,,
\end{equation}
namely the truncation can be characterized by the two integer coefficients multiplying $\gamma$ and $(\alpha_1+\alpha_2+\alpha_3)$.

\begin{figure}[htbp]
\centering\includegraphics[scale=0.25]{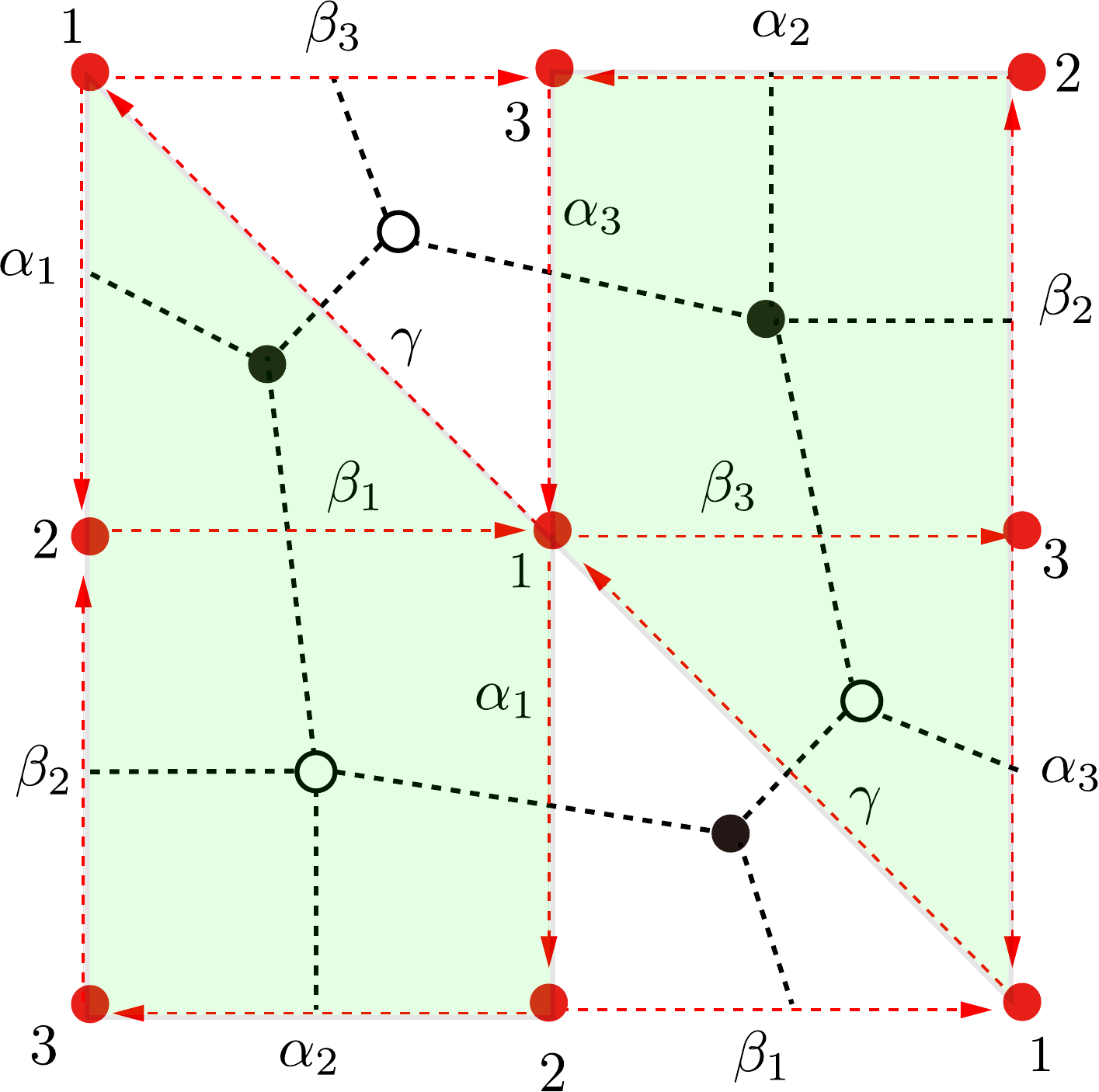}
\caption{The bipartite graph for the Suspended Pinched Point geometry.}
\label{fig.SPPbipartite_square}
\end{figure}

\begin{figure}[htbp]
\centering\includegraphics[scale=0.22]{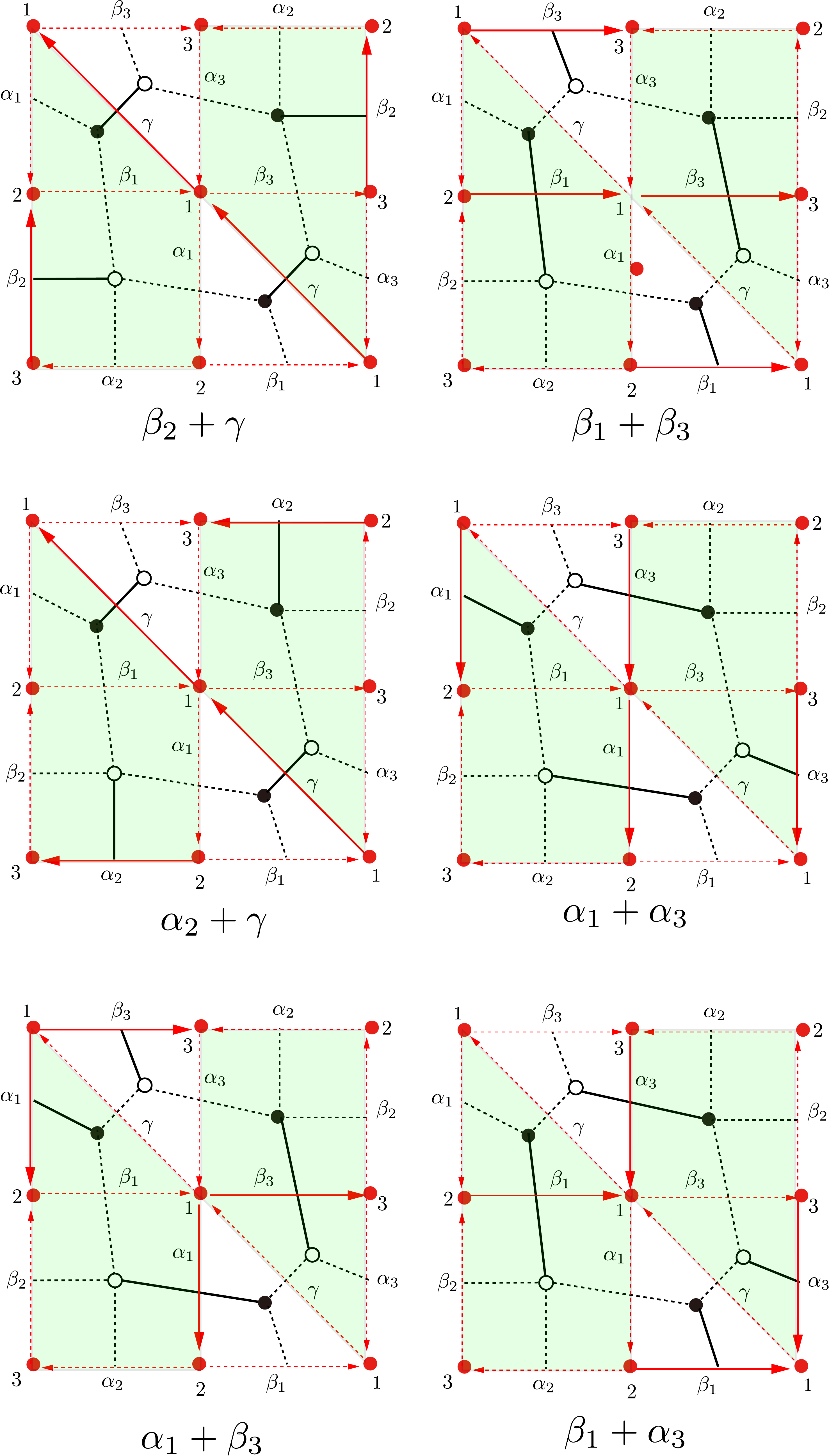}
\caption{The $6$ perfect matchings for the Suspended Pinched Point geometry.
They correspond to the combinations
$\beta_2+\gamma$, $\alpha_2+\gamma$, $\alpha_1+\alpha_3$, $\beta_1+\beta_3$, $\alpha_1+\beta_3$, $\alpha_3+\beta_1$.
}
\label{fig.SPPPM_square}
\end{figure}

Let us re-derive this result from perfect matchings.
We have already shown the bipartite graph and the perfect matchings in Figures \ref{fig.SPPbipartite} and \ref{fig.SPPPM}.
Now we reproduce them in slightly different-looking (albeit equivalent) forms in Figures \ref{fig.SPPbipartite_square} and \ref{fig.SPPPM_square}, to make the comparison with the quiver in \eqref{tilling-gl2-square} easier.

There are six perfect matchings as show in Figure \ref{fig.SPPPM_square},
giving rise to linear combinations
\begin{align}
\alpha_2+\gamma \;, \quad \beta_2+\gamma \;, \quad \alpha_1+\alpha_3 \;, \quad \beta_1+\beta_3 \;, \quad \alpha_1+\beta_3 \;, \quad \beta_1+\alpha_3\;.
\end{align}
When we impose the vertex constraints, these reduce to 
\begin{align}
-h_1-2h_2 \;, \quad -2h_1-h_2 \;, \quad 2 h_1 \;, \quad  2 h_2 \;, \quad  h_1+h_2 \;, \quad h_1+h_2 \;.
\end{align}
From this we find that only the first four perfect matchings are corner perfect matchings:
\begin{align}
\alpha_2+\gamma \;, \quad \beta_2+\gamma =2\gamma-\alpha_2\;, \quad \alpha_1+\alpha_3 \;, \quad \beta_1+\beta_3=-2\gamma-(\alpha_1+\alpha_3) \;, \end{align}
which by a change of basis can be replaced by
\begin{align}
\gamma \;, \quad \alpha_2\;, \quad \alpha_1+\alpha_3 \;.
\end{align}
This is (as in the conifold example) more general than the expectations from the representation theory.

\subsubsubsection{Affine Yangian of $\mathfrak{gl}_{2|1}$}

On top of the loop constraints (\ref{eq-loop-constraint-SPP}), if we further impose the vertex constraints (\ref{eq-vertex-constraint-SPP}), the charge assignment on the quiver is given in the right figure of (\ref{quiver-SPP}).
Accordingly, the matrices defined in (\ref{eq-Qpm}) are 
\begin{align}
Q^{+}
=\left(\begin{array}{ccc}
-1 & 0 & 1\\
1 & 0 & -1\\
0 & 0 & 0
\end{array}
\right)   \qquad\textrm{and} \qquad
Q^{-}
=\left(\begin{array}{ccc}
-1 & 1 & 0\\
0 & 0 & 0\\
1 & -1 & 0
\end{array}
\right), 
\end{align}
which gives
\begin{align}
A=\left(\begin{array}{ccc}
2 & -1 & -1\\
-1 & 0 & 1\\
-1 & 1 & 0
\end{array}
\right) \qquad\textrm{and}\qquad
M=\left(\begin{array}{ccc}
0 & 1 & -1\\
-1 & 0 & 1\\
1 & -1 & 0
\end{array}
\right)\,.
\end{align}
The bond factors are
\begin{equation}\label{eq-charge-function-gl21}
\begin{aligned}
&\varphi^{1\Rightarrow 1}(u)=\frac{u+h_3}{u-h_3} \,, \qquad\qquad \varphi^{2\Rightarrow 2}(u)=\varphi^{3\Rightarrow 3}(u)=1 \;,\\
 &\varphi^{1\Rightarrow 2}(u)=\varphi^{3\Rightarrow 1}(u)=\frac{u+h_2}{u-h_1}\,, \qquad  \varphi^{2\Rightarrow 3}(u)=\frac{u-h_1}{u+h_2} \;,\\
 &  \varphi^{2\Rightarrow 1}(u) =\varphi^{1\Rightarrow 3}(u)=\frac{u+h_1}{u-h_2} \,, \qquad  \varphi^{3\Rightarrow 2}(u)=\frac{u+h_1}{u-h_2}\,.
\end{aligned}
\end{equation}
One can check that the resulting algebra (from plugging the bond factors (\ref{eq-charge-function-gl21}) into the general formulae (\ref{eq-OPE-toric}), (\ref{eq-psi-e-f-initial-0}) and (\ref{eq-psi-e-f-initial-1})) agrees with the affine Yangian of $\mathfrak{gl}_{2|1}$ obtained by plugging  $(\sigma_1,\sigma_2,\sigma_3)=(+,+,-)$ and taking $a=1,2,3$ in the general formula (\ref{eq-QY-glmn}).

\section{\texorpdfstring{Examples: Calabi-Yau Threefolds with Compact $4$-Cycles}{Examples: Calabi-Yau Threefolds with Compact 4-Cycles}}
\label{sec:general4cycle}

In the previous section we have restricted ourselves to the toric Calabi-Yau threefolds without compact $4$-cycles.
Our discussion of the BPS quiver Yangian, however, works for
arbitrary toric Calabi-Yau threefolds, most of which have compact four-cycles and goes beyond the examples discussed in the previous section. 
This is in contrast with other existing approaches in the literature, where there seems to be technical problems associated with such generalizations.
We will discuss in detail examples of the canonical bundles overs $\mathbb{P}^1\times \mathbb{P}^1$ and $\mathbb{P}^2$ in the next subsections.

\subsection{Quiver Yangians for \texorpdfstring{$K_{\mathbb{P}^2}$}{Quiver Yangians for K(P2)}}

\subsubsection{Quiver and superpotential}

Let us consider the geometry $K_{\mathbb{P}^2}$, the canonical bundle over $\mathbb{P}^2$.
The geometry coincides with $\mathbb{C}^3/\mathbb{Z}_3$, 
where the action of $\mathbb{Z}_3$ is
$(z_1, z_2, z_3) \to (\omega z_1, \omega z_2, \omega z_3)$ with $\omega^3=1$.
The toric diagram and its dual graph are 
\begin{equation}\label{fig-toric-P2}
\begin{tikzpicture} 
\filldraw [red] (0,0) circle (2pt); 
\filldraw [red] (0,-1) circle (2pt); 
\filldraw [red] (1,1) circle (2pt); 
\filldraw [red] (-1,0) circle (2pt); 
\node at (0,0.8) {(0,0)}; 
\node at (-1.7,0) {(-1,0)}; 
\node at (1.5,1.5) {(1,1)}; 
\node at (0,-1.5) {(0,-1)}; 
\draw (0,0) -- (1,1); 
\draw (0,0) -- (-1,0); 
\draw (0,0) -- (0,-1); 
\draw (-1,0) -- (1,1); 
\draw (-1,0) -- (0,-1); 
\draw (0,-1) -- (1,1); 
\end{tikzpicture}
\qquad \qquad \qquad
\begin{tikzpicture}[scale=0.6] 
\draw (0,0) -- (1,0); 
\draw (0,0) -- (0,1); 
\draw (1,0) -- (0,1); 
\draw[->] (0,0) -- (-1,-1); 
\draw[->] (1,0) -- (3,-1); 
\draw[->] (0,1) -- (-1,3); 
\end{tikzpicture}
\end{equation}
Note that this is different from the $(\mathbb{C}^2/\mathbb{Z}_3)\times \mathbb{C}$
geometry discussed in section \ref{sec:C2Zn}.

The quiver diagram is the McKay quiver \cite{reid1997mckay} for the $\mathbb{Z}_3$-action
\begin{equation}\label{quiver-P2}
\begin{tikzpicture}[scale=0.8]
\node[state]  [regular polygon, regular polygon sides=4, draw=blue!50, very thick, fill=blue!10] (a1) at (0,0)  {$1$};
\node[state]  [regular polygon, regular polygon sides=4, draw=blue!50, very thick, fill=blue!10] (a2) at (2,-3.4641)  {$2$};
\node[state]  [regular polygon, regular polygon sides=4, draw=blue!50, very thick, fill=blue!10] (a3) at  (-2,-3.4641)  {$3$};
\path[->>>] 
(a1) edge   [thick, red]  node [right] {$(X^{(1)}_i\! , \alpha_i^{(1)})_{i=1,2,3}$} (a2)
(a2) edge   [thick, red]  node [below, pos=0.4] {$(X^{(2)}_i\! , \alpha_i^{(2)})_{i=1,2,3}$} (a3)
(a3) edge   [thick, red]  node [left] {$(X^{(3)}_i \! , \alpha_i^{(3)})_{i=1,2,3}$} (a1)
; 
\end{tikzpicture}
\end{equation}
with the superpotential
\begin{align}
W=-\sum_{i,j,k=1}^3 \varepsilon^{ijk}\textrm{Tr}(X^{(1)}_i X^{(2)}_j X^{(3)}_k) \;,
\end{align}
with the totally antisymmetric tensor $\varepsilon^{ijk}$.

The loop constraint (\ref{eq-loop-constraint-toric}) from the superpotential is 
\begin{align}
\alpha^{(1)}_i+\alpha^{(2)}_j+\alpha^{(3)}_k=0 \qquad \textrm{for} \quad  \{i,j,k\} \in \{ 1,2,3\} \;,
\end{align}
which reduces the $9$ parameters to $E+2I-1=4$:
\begin{equation}\label{eq-P2-a}
\alpha^{(a)}_{i}=h_i+g^{(a)}\,,
\end{equation}
with
\begin{equation}\label{eq-hg}
h_1+h_2+h_3=0 \qquad \textrm{and}\qquad  g^{(1)}+g^{(2)}+g^{(3)}=0 \,.
\end{equation}
The periodic quiver is given in the left figure of the following:
\begin{equation}\label{fig-periodic-quiver-P2}
\begin{aligned}
&
\begin{tikzpicture}[scale=0.8]
\filldraw[mygreen] (-3,3)--  (0, 6) -- (0,-3) -- (-3,-6) -- cycle; 
\node[state]  [regular polygon, regular polygon sides=4, draw=blue!50, very thick, fill=blue!10] (C) at (0,0)  {$3$};
\node[state]  [regular polygon, regular polygon sides=4, draw=blue!50, very thick, fill=blue!10] (E) at (3,0)  {$1$};
\node[state]  [regular polygon, regular polygon sides=4, draw=blue!50, very thick, fill=blue!10] (W) at (-3,0)  {$2$};
\node[state]  [regular polygon, regular polygon sides=4, draw=blue!50, very thick, fill=blue!10] (N) at (0,3)  {$2$};
\node[state]  [regular polygon, regular polygon sides=4, draw=blue!50, very thick, fill=blue!10] (S) at (0,-3)  {$1$};
\node[state]  [regular polygon, regular polygon sides=4, draw=blue!50, very thick, fill=blue!10] (a31) at (3,3)  {$3$};
\node[state]  [regular polygon, regular polygon sides=4, draw=blue!50, very thick, fill=blue!10] (a32) at (3,-3)  {$2$};
\node[state]  [regular polygon, regular polygon sides=4, draw=blue!50, very thick, fill=blue!10] (a34) at (-3,-3)  {$3$};
\node[state]  [regular polygon, regular polygon sides=4, draw=blue!50, very thick, fill=blue!10] (a33) at (-3,3)  {$1$};
\path[->] 
(C) edge   [thick, red]   node [above] {$\alpha^{(3)}_2$} (E)
(W) edge   [thick, red]   node [above] {$\alpha^{(2)}_2$} (C)
(a31) edge   [thick, red]   node [right] {$\alpha^{(3)}_1$} (E)
(E) edge   [thick, red]   node [right] {$\alpha^{(1)}_1$} (a32)
(a33) edge   [thick, red]   node [left] {$\alpha^{(1)}_1$} (W)
(W) edge   [thick, red]   node [left] {$\alpha^{(2)}_1$} (a34)
(N) edge   [thick, red]   node [above] {$\alpha^{(2)}_2$} (a31)
(a33) edge   [thick, red]   node [above] {$\alpha^{(1)}_2$} (N)
(a34) edge   [thick, red]   node [above] {$\alpha^{(3)}_2$} (S)
(S) edge   [thick, red]   node [above] {$\alpha^{(1)}_2$} (a32)
(N) edge   [thick, red]   node [right] {$\alpha^{(2)}_1$} (C)
(C) edge   [thick, red]   node [right] {$\alpha^{(3)}_1$} (S)
(C) edge   [thick, red]   node [right] {$\alpha^{(3)}_3$} (a33)
(a32) edge   [thick, red]   node [right] {$\alpha^{(2)}_3$} (C)
(E) edge   [thick, red]   node [right] {$\alpha^{(1)}_3$} (N)
(S) edge   [thick, red]   node [right] {$\alpha^{(1)}_3$} (W)
;
\end{tikzpicture}
\qquad 
\begin{tikzpicture}[scale=0.8]
\filldraw[mygreen] (-3,3)--  (0, 6) -- (0,-3) -- (-3,-6) -- cycle; 
\node[state]  [regular polygon, regular polygon sides=4, draw=blue!50, very thick, fill=blue!10] (C) at (0,0)  {$3$};
\node[state]  [regular polygon, regular polygon sides=4, draw=blue!50, very thick, fill=blue!10] (E) at (3,0)  {$1$};
\node[state]  [regular polygon, regular polygon sides=4, draw=blue!50, very thick, fill=blue!10] (W) at (-3,0)  {$2$};
\node[state]  [regular polygon, regular polygon sides=4, draw=blue!50, very thick, fill=blue!10] (N) at (0,3)  {$2$};
\node[state]  [regular polygon, regular polygon sides=4, draw=blue!50, very thick, fill=blue!10] (S) at (0,-3)  {$1$};
\node[state]  [regular polygon, regular polygon sides=4, draw=blue!50, very thick, fill=blue!10] (a31) at (3,3)  {$3$};
\node[state]  [regular polygon, regular polygon sides=4, draw=blue!50, very thick, fill=blue!10] (a32) at (3,-3)  {$2$};
\node[state]  [regular polygon, regular polygon sides=4, draw=blue!50, very thick, fill=blue!10] (a34) at (-3,-3)  {$3$};
\node[state]  [regular polygon, regular polygon sides=4, draw=blue!50, very thick, fill=blue!10] (a33) at (-3,3)  {$1$};
\path[->] 
(C) edge   [thick, red]   node [above] {$h_2$} (E)
(W) edge   [thick, red]   node [above] {$h_2$} (C)
(a31) edge   [thick, red]   node [right] {$h_1$} (E)
(E) edge   [thick, red]   node [right] {$h_1$} (a32)
(a33) edge   [thick, red]   node [left] {$h_1$} (W)
(W) edge   [thick, red]   node [left] {$h_1$} (a34)
(N) edge   [thick, red]   node [above] {$h_2$} (a31)
(a33) edge   [thick, red]   node [above] {$h_2$} (N)
(a34) edge   [thick, red]   node [above] {$h_2$} (S)
(S) edge   [thick, red]   node [above] {$h_2$} (a32)
(N) edge   [thick, red]   node [right] {$h_1$} (C)
(C) edge   [thick, red]   node [right] {$h_1$} (S)
(C) edge   [thick, red]   node [right] {$h_3$} (a33)
(a32) edge   [thick, red]   node [right] {$h_3$} (C)
(E) edge   [thick, red]   node [right] {$h_3$} (N)
(S) edge   [thick, red]   node [right] {$h_3$} (W)
;
\end{tikzpicture}
\end{aligned}
\end{equation}
where we have shown the fundamental regions of the torus as shaded regions.
Since there is no self-loop in the quiver diagram (\ref{quiver-P2}), all vertices are fermionic:
\begin{equation}
|a|=1\,, \qquad \qquad a=1,2,3\,.
\end{equation}

\subsubsection{Algebra}

From the quiver (\ref{quiver-P2}) with charge assignment (\ref{eq-P2-a}), one can read off the bond factors $\varphi^{a\Rightarrow b}(u)$:\begingroup
\setcounter{savefootnote}{\value{footnote}}
\setcounter{footnote}{0}
\renewcommand{\thefootnote}{\fnsymbol{footnote}}%
\footnote[7]{Note added for version 3: In this subsection we ignore the sign factors $\textrm{sign}(a,b)$ in the bond factors. 
One choice would be to have a ``$-1$" sign for $\varphi^{a\Rightarrow a+1}(u)$ (see footnote \ref{footnote11} on page 18); then 
$\Psi^{(2)}_{\Kappa}$ in \eqref{eq911}, $\Psi^{(2,3)}_{\Kappa}$ in \eqref{eq912}, and $\varphi^{a\Rightarrow a+1}$ together with $\varphi_{-}$ in 
\eqref{eq-charge-function-C3Z3-re} would acquire a  minus sign.}
\setcounter{footnote}{\value{savefootnote}}
\endgroup
\begin{equation}\label{eq-charge-function-C3Z3}
\begin{aligned}
 &\varphi^{a\Rightarrow a}(u)= 1\,, \\
 &\varphi^{a\Rightarrow a+1}(u)= \frac{1}{\prod_{i=1,2,3} \left(u-h_i-g^{(a)}\right)}\,, \\
 & \varphi^{a\Rightarrow a-1}(u)= \prod_{i=1,2,3} \left(u+h_i+g^{(a-1)}\right) \;.
\end{aligned}
\end{equation}

Using (\ref{eq-charge-function-C3Z3}), one can then write down the charge functions $\Psi^{(a)}_{\Kappa}(u)$ for any crystal $K$.
Here we give $\Psi^{(a)}_{\Kappa}(u)$ for the first few $|\Kappa\rangle$ as examples.
For the vacuum:
\begin{equation}
|\Kappa\rangle=|\emptyset\rangle \,: \quad 
\begin{cases}
\begin{aligned}
\Psi^{(1)}_{\Kappa}(u)&=1+\frac{C}{u}\,,\\
\Psi^{(2)}_{\Kappa}(u)&=\Psi^{(3)}_{\Kappa}(u)=1\,.
\end{aligned}
\end{cases}
\end{equation}
For the state with only the first atom $\sqbox{$1$}$:
\begin{equation}\label{eq911}
|\Kappa\rangle=|\sqbox{1}\rangle \,: \quad 
\begin{cases}
\begin{aligned}
\Psi^{(1)}_{\Kappa}(u)&=1+\frac{C}{u}\,,\\
\Psi^{(2)}_{\Kappa}(u)&=\frac{1}{\prod_{k=1,2,3} \left(u-h_k-g^{(1)}\right)}\,,\\
\Psi^{(3)}_{\Kappa}(u)&=\prod_{k=1,2,3} \left(u+h_k+g^{(3)}\right) \,.
\end{aligned}
\end{cases}
\end{equation}
For the state with the first atom $\sqbox{$1$}$ and one atom $\sqbox{$2$}$ (at the direction $i$, with $i=1,2,3$) at the level-$2$:
\begin{equation}\label{eq912}
|\Kappa\rangle=|\sqbox{1}\sqbox{2}_i\rangle \,: \quad 
\begin{cases}
\begin{aligned}
\Psi^{(1)}_{\Kappa}(u)&=\left(1+\frac{C}{u}\right)\prod_{k=1,2,3} \left(u+h_k-h_i\right) \,,\\
\Psi^{(2)}_{\Kappa}(u)&=\frac{1}{\prod_{k=1,2,3} \left(u-h_k-g^{(1)}\right)}\,,\\
\Psi^{(3)}_{\Kappa}(u)&=\prod_{k=1,2,3} \frac{\left(u+h_k+g^{(3)}\right)}{ \left(u-h_k+g^{(3)}-h_i\right)}\,.
\end{aligned}
\end{cases}
\end{equation}
One can thus proceed iteratively, and write down the charge function $\Psi^{(a)}_{\Kappa}(u)$ for all states $\Kappa$.
At each step, one can check the each pole $u^{*}$ of $\Psi^{(a)}_{\Kappa}(u)$ corresponds to the position of either an atom $\sqbox{$a$}$ (of color $a$) that can be added to $\Kappa$ or an atom $\sqbox{$a$}$ that can be removed from $\Kappa$.
Since the $\varphi^{a\Rightarrow b}(u)$ in (\ref{eq-charge-function-C3Z3}) is not homogeneous, generically the charge functions $\Psi^{(a)}_{\Kappa}(u)$ is also not homogeneous.
\bigskip

We can now  write down the quiver Yangian
\begin{tcolorbox}[ams align]\label{eq-OPE-C3Z3}
&\textrm{OPE:}\quad\begin{cases}\begin{aligned}
\psi^{(a)}(z)\, \psi^{(b)}(w)&= \psi^{(b)}(w)\, \psi^{(a)}(z)\;,\\
 \psi^{(a)}(z)\, e^{(a)}(w)   &\simeq   e^{(a)}(w)\, \psi^{(a)}(z) \;,\\ 
e^{(a)}(z)\, e^{(a)}(w) & \sim - e^{(a)}(w)\, e^{(a)}(z) \;,\\
\psi^{(a)}(z)\, f^{(a)}(w) &  \simeq f^{(a)}(w)\, \psi^{(a)}(z) \;,\\
 f^{(a)}(z)\, f^{(a)}(w) &  \sim  - f^{(a)}(w)\, f^{(a)}(z) \\
 \psi^{(a\pm 1)}(z)\, e^{(a)}(w)   &\simeq \varphi^{a\Rightarrow a\pm 1}(\Delta) \, e^{(a)}(w)\, \psi^{(a\pm 1)}(z) \;,\\ 
e^{(a+1)}(z)\, e^{(a)}(w) & \sim -  \varphi^{a\Rightarrow a+1}(\Delta)\, e^{(a)}(w)\, e^{(a+1)}(z)\;, \\
\psi^{(a\pm 1)}(z)\, f^{(a)}(w) &  \simeq  \varphi^{a\Rightarrow a\pm 1}(\Delta)^{-1}\, f^{(a)}(w)\, \psi^{(a\pm 1)}(z) \;,\\
 f^{(a+1)}(z)\, f^{(a)}(w) &  \sim - \varphi^{a\Rightarrow a+1}(\Delta)^{-1}\,f^{(a)}(w)\, f^{(a+1)}(z)\;,\\
\{e^{(a)}(z)\,, f^{(b)}(w)\}  &\sim  - \delta^{a,b}\, \frac{\psi^{(a)}(z) - \psi^{(b)}(w)}{z-w} \;,
\end{aligned}
\end{cases}
\end{tcolorbox}
\noindent where $a=1,2,3 \in \mathbb{Z}_3$. 

%
%
%

\subsubsection{Truncation}

Let us consider the truncation induced by the truncation of the crystal at an atom of color $1$.
The path starting and ending at the same vertex $1$ 
goes around the loop of the quiver diagram.
Each loop has the total weight of the form
$\alpha_i^{(1)}+\alpha_j^{(2)}+\alpha_k^{(3)}$ where $i,j,k$ runs separately from $1$ to $3$.
Using the loop constraints \eqref{eq-P2-a} and \eqref{eq-hg},
this is computed to be $h_i+h_j+h_k$.
This means that the coordinate function
at the location of the truncation takes the form
\begin{equation}\label{eq-truncation-C3Z3-1}
h(\sqbox{$1$})=\sum^3_{i,j,k=1}\, N_{i,j,k} (h_i+h_j+h_k)\;, \quad N_{i,j,k}\in \mathbb{Z}_{\ge 0} \;.
\end{equation}
Since $h_1+h_2+h_3=0$, we have integer linear combination of
$h_1$ and $h_2$, so that we have a truncation condition
\begin{equation}
N_1 \, h_1+N_2 \, h_2+C=0\;, \quad N_1, N_2 \in \mathbb{Z} \;.
\end{equation}

\begin{figure}[htbp]
\centering\includegraphics[scale=0.25]{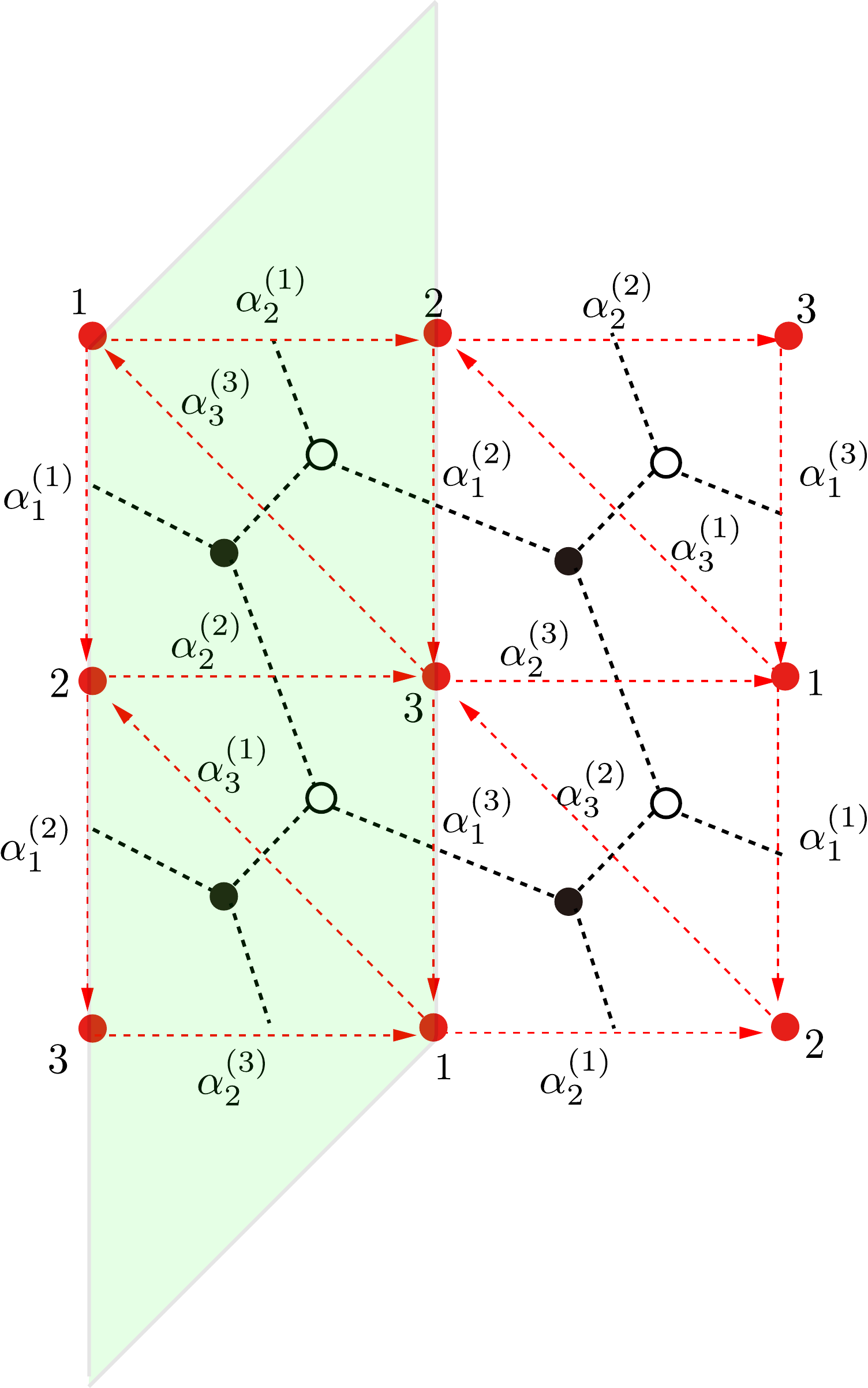}
\caption{The bipartite graph for the $K_{\mathbb{P}^2}$ geometry.}
\label{fig.P2bipartite}
\end{figure}

\begin{figure}[htbp]
\centering\includegraphics[scale=0.18]{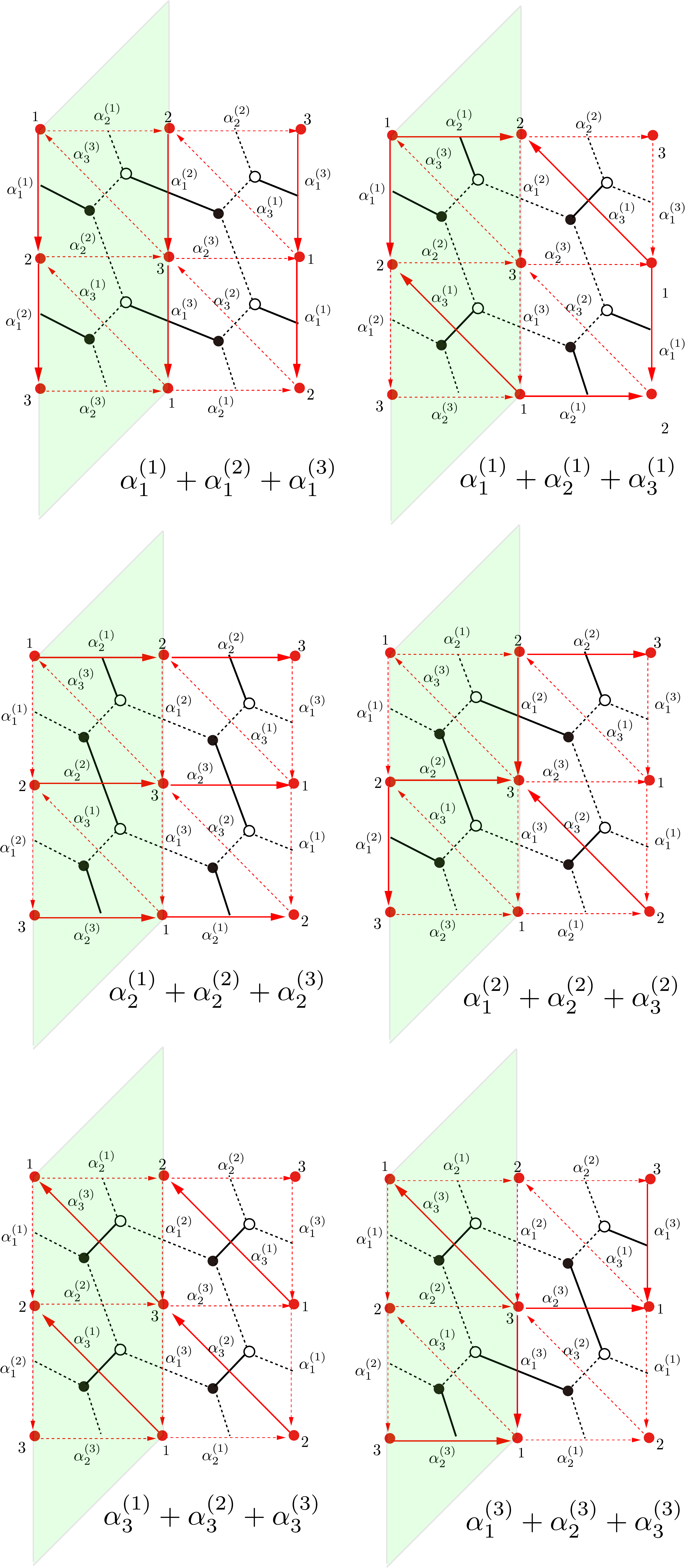}
\caption{The $6$ perfect matchings for the $K_{\mathbb{P}^2}$ geometry.
They correspond to the combinations
$\alpha_1^{(i)}+\alpha_2^{(i)}+\alpha_3^{(i)}$ ($i=1,2,3$)
and $\alpha_i^{(1)}+\alpha_i^{(2)}+\alpha_i^{(3)}$ ($i=1,2,3$).
}
\label{fig.P2PM}
\end{figure}

We can check this result from the perfect matching prescription introduced earlier,
as worked out in Figures \ref{fig.P2bipartite} and \ref{fig.P2PM}.
There are six perfect matchings as shown in Figure \ref{fig.P2PM}, corresponding to the linear combinations
\begin{equation}\label{P2_list}
\begin{aligned}
&\alpha_i^{(1)}+\alpha_i^{(2)}+\alpha_i^{(3)}=3 h_i \qquad (i=1,2,3)\;,\\
&\alpha_1^{(a)}+\alpha_2^{(a)}+\alpha_3^{(a)}=3 g^{(a)} \qquad (a=1,2,3) \;.
\end{aligned}
\end{equation}
When we further impose the vertex constraints, these reduce to 
\begin{equation}
3h_1\;,\quad 3h_2\;, \quad 3h_3=-3h_1-3h_2\;,\quad  0\;, \quad 0\;,\quad 0 \;,
\end{equation}
and when divided by a factor $3$ this matches with the lattice points 
\begin{equation}
(-1,0) \;, \quad (0,-1)\;, \quad (1,1)\;, \quad (0,0)\;,\quad  (0,0) \;,\quad (0,0) \;,
\end{equation}
of the toric diagram (\ref{fig-toric-P2}). 

When we take the corner lattice points from the list \eqref{P2_list} 
one obtains non-negative integer linear combinations of 
\begin{equation}
3h_1\;,\quad 3h_2 \;, \quad 3h_3=- 3h_1- 3h_2 \;,
\end{equation}
so that we have integer linear combination of $3h_1$ and $3h_2$.
This matches with the analysis of the truncation above up to a rescaling of $C$ by a factor of $3$.

\clearpage

\subsubsection{Vertex Constraint}

The vertex constraint (\ref{eq-vertex-constraint-toric}) for this case is
\begin{equation}
\sum^3_{i=1} \alpha^{(a)}_i = \sum^3_{i=1} \alpha^{(a+1)}_i \qquad \textrm{for} \quad a=1,2,3\,,
\end{equation}
which reduces the number of parameters to two, given by the triple $(h_1, h_2, h_3)$:
\begin{align}\label{eq.P2_final}
\alpha^{(1)}_i=\alpha^{(2)}_i=\alpha^{(3)}_i=h_i \quad (i=1,2,3) \;, 
\qquad h_1+h_2+h_3=0 \;.
\end{align}
We have also drawn the periodic quiver with the charge assignment (\ref{eq.P2_final}) in the right figure of (\ref{fig-periodic-quiver-P2}).

The bond factors (\ref{eq-charge-function-C3Z3}) reduces to 
\begin{equation}\label{eq-charge-function-C3Z3-re}
\begin{aligned}
 &\varphi^{a\Rightarrow a}(u)= 1\,, \\
 &\varphi^{a\Rightarrow a+1}(u)= \frac{1}{\prod_{i=1,2,3} \left(u-h_i\right)} \equiv \varphi_{-}(u)\,, \\
 & \varphi^{a\Rightarrow a-1}(u)= \prod_{i=1,2,3} \left(u+h_i\right) \equiv \varphi_{+}(u)\;.
\end{aligned}
\end{equation}
Accordingly the (reduced) quiver Yangian can be obtained by setting $g^{(a)}=0$ for $a=1,2,3$ in (\ref{eq-OPE-C3Z3}).

\subsection{\texorpdfstring{Quiver Yangian for $K_{\mathbb{P}^1\times \mathbb{P}^1}$}{Quiver Yangian for K(P1xP1)}}
\label{sec.P1P1}

\subsubsection{Quiver and superpotential}
The toric diagram and its dual graph for $K_{\mathbb{P}^1\times \mathbb{P}^1}$ are 
\begin{equation}\label{fig-toric-P1P1}
\begin{tikzpicture} 
\filldraw [red] (0,0) circle (2pt); 
\filldraw [red] (0,-1) circle (2pt); 
\filldraw [red] (1,0) circle (2pt); 
\filldraw [red] (0,1) circle (2pt); 
\filldraw [red] (-1,0) circle (2pt); 
\node at (0,0.3) {(0,0)}; 
\node at (0,-1.5) {(0,-1)}; 
\node at (1.7,0) {(1,0)}; 
\node at (0,1.5) {(0,1)}; 
\node at (-1.7,0) {(-1,0)}; 
\draw (0,0) -- (0,-1); 
\draw (0,0) -- (1,0); 
\draw (0,0) -- (0,1); 
\draw (0,0) -- (-1,0); 
\draw (0,-1) -- (1,0); 
\draw (1,0) -- (0,1); 
\draw (0,1) -- (-1,0); 
\draw (-1,0) -- (0,-1); 
\end{tikzpicture}
\qquad \qquad \qquad
\begin{tikzpicture}[scale=0.6] 
\draw (0,0) -- (1,0); 
\draw (0,0) -- (0,1); 
\draw (1,0) -- (1,1); 
\draw (0,1) -- (1,1); 
\draw[->] (0,0) -- (-1,-1); 
\draw[->] (1,0) -- (2,-1); 
\draw[->] (0,1) -- (-1,2); 
\draw[->] (1,1) -- (2,2); 
\end{tikzpicture}
\end{equation}
The quiver for $\mathbb{P}^1\times \mathbb{P}^1$ is shown in the left figure of (\ref{quiver-P1P1})
\begin{equation}\label{quiver-P1P1}
\begin{aligned}
&
\begin{tikzpicture}[scale=0.77]
\node[state]  [regular polygon, regular polygon sides=4, draw=blue!50, very thick, fill=blue!10] (a1) at (0,0)  {$1$};
\node[state]  [regular polygon, regular polygon sides=4, draw=blue!50, very thick, fill=blue!10] (a41) at (3,0)  {$4$};
\node[state]  [regular polygon, regular polygon sides=4, draw=blue!50, very thick, fill=blue!10] (a21) at (0,3)  {$2$};
\node[state]  [regular polygon, regular polygon sides=4, draw=blue!50, very thick, fill=blue!10] (a31) at (3,3)  {$3$};
\path[->] 
(a1) edge   [->>,thick, red]   node [left] {$\alpha^{(1)}_{1,2}$} (a21)
(a21) edge   [->>,thick, red]   node [above] {$\alpha^{(2)}_{1,2}$}  (a31)
(a31) edge   [->>,thick, red]   node [right] {$\alpha^{(3)}_{1,2}$}  (a41)
(a41) edge   [->>,thick, red]   node [below] {$\alpha^{(4)}_{1,2}$}  (a1)
;
\end{tikzpicture}
\qquad
\begin{tikzpicture}[scale=0.77]
\node[state]  [regular polygon, regular polygon sides=4, draw=blue!50, very thick, fill=blue!10] (a1) at (0,0)  {$1$};
\node[state]  [regular polygon, regular polygon sides=4, draw=blue!50, very thick, fill=blue!10] (a41) at (5,0)  {$4$};
\node[state]  [regular polygon, regular polygon sides=4, draw=blue!50, very thick, fill=blue!10] (a21) at (0,3)  {$2$};
\node[state]  [regular polygon, regular polygon sides=4, draw=blue!50, very thick, fill=blue!10] (a31) at (5,3)  {$3$};
\path[->] 
(a1) edge   [->>,thick, red]   node [left] {$h_1 + \delta_1, \,\,   h_3+ \delta_1$} (a21)
(a21) edge   [->>,thick, red]   node [above] {$h_2+\delta_2, \,\, h_4+\delta_2$} (a31)
(a31) edge   [->>,thick, red]   node [right] {$h_1- \delta_1, \,\,  h_3 -\delta_1$} (a41)
(a41) edge   [->>,thick, red]   node [below] {$h_2-\delta_2,\,\, h_4-\delta_2$} (a1)
;
\end{tikzpicture}
\end{aligned}
\end{equation}
The superpotential is 
\begin{equation}
\begin{aligned}
W=&-X^{(1)}_1 X^{(2)}_1 X^{(3)}_1 X^{(4)}_1 -X^{(1)}_2 X^{(2)}_2 X^{(3)}_2 X^{(4)}_2 \\
&+ X^{(1)}_1 X^{(2)}_2 X^{(3)}_1 X^{(4)}_2+ X^{(1)}_2 X^{(2)}_1 X^{(3)}_2 X^{(4)}_1\,.
\end{aligned}
\end{equation}
The corresponding periodic quiver is shown in the left figure of (\ref{periodic-quiver-P1P1}) 
\begin{equation}\label{periodic-quiver-P1P1}
\begin{aligned}
&
\begin{tikzpicture}[scale=0.77]
\filldraw[mygreen] (-3,3)--  (-3, -3) -- (3,-3) -- (3,3) -- cycle; 
\node[state]  [regular polygon, regular polygon sides=4, draw=blue!50, very thick, fill=blue!10] (a1) at (0,0)  {$1$};
\node[state]  [regular polygon, regular polygon sides=4, draw=blue!50, very thick, fill=blue!10] (a21) at (3,0)  {$4$};
\node[state]  [regular polygon, regular polygon sides=4, draw=blue!50, very thick, fill=blue!10] (a22) at (-3,0)  {$4$};
\node[state]  [regular polygon, regular polygon sides=4, draw=blue!50, very thick, fill=blue!10] (a41) at (0,3)  {$2$};
\node[state]  [regular polygon, regular polygon sides=4, draw=blue!50, very thick, fill=blue!10] (a42) at (0,-3)  {$2$};
\node[state]  [regular polygon, regular polygon sides=4, draw=blue!50, very thick, fill=blue!10] (a31) at (3,3)  {$3$};
\node[state]  [regular polygon, regular polygon sides=4, draw=blue!50, very thick, fill=blue!10] (a32) at (3,-3)  {$3$};
\node[state]  [regular polygon, regular polygon sides=4, draw=blue!50, very thick, fill=blue!10] (a34) at (-3,-3)  {$3$};
\node[state]  [regular polygon, regular polygon sides=4, draw=blue!50, very thick, fill=blue!10] (a33) at (-3,3)  {$3$};
\path[->] 
(a21) edge   [thick, red]   node [above]  {$\alpha^{(4)}_1$} (a1)
(a22) edge   [thick, red]   node [above]  {$\alpha^{(4)}_2$} (a1)
(a31) edge   [thick, red]   node [right]  {$\alpha^{(3)}_2$} (a21)
(a32) edge   [thick, red]   node [right]  {$\alpha^{(3)}_1$} (a21)
(a33) edge   [thick, red]   node [left]  {$\alpha^{(3)}_2$} (a22)
(a34) edge   [thick, red]   node [left]  {$\alpha^{(3)}_1$} (a22)
(a41) edge   [thick, red]   node [above]  {$\alpha^{(2)}_1$} (a31)
(a41) edge   [thick, red]   node [above]  {$\alpha^{(2)}_2$} (a33)
(a42) edge   [thick, red]   node [below]  {$\alpha^{(2)}_2$} (a34)
(a42) edge   [thick, red]   node [below]  {$\alpha^{(2)}_1$} (a32)
(a1) edge   [thick, red]   node [right]  {$\alpha^{(1)}_2$} (a41)
(a1) edge   [thick, red]   node [right] {$\alpha^{(1)}_1$} (a42)
;
\end{tikzpicture}
\qquad
\begin{tikzpicture}[scale=0.77]
\filldraw[mygreen] (-3,3)--  (-3, -3) -- (3,-3) -- (3,3) -- cycle; 
\node[state]  [regular polygon, regular polygon sides=4, draw=blue!50, very thick, fill=blue!10] (a1) at (0,0)  {$1$};
\node[state]  [regular polygon, regular polygon sides=4, draw=blue!50, very thick, fill=blue!10] (a21) at (3,0)  {$4$};
\node[state]  [regular polygon, regular polygon sides=4, draw=blue!50, very thick, fill=blue!10] (a22) at (-3,0)  {$4$};
\node[state]  [regular polygon, regular polygon sides=4, draw=blue!50, very thick, fill=blue!10] (a41) at (0,3)  {$2$};
\node[state]  [regular polygon, regular polygon sides=4, draw=blue!50, very thick, fill=blue!10] (a42) at (0,-3)  {$2$};
\node[state]  [regular polygon, regular polygon sides=4, draw=blue!50, very thick, fill=blue!10] (a31) at (3,3)  {$3$};
\node[state]  [regular polygon, regular polygon sides=4, draw=blue!50, very thick, fill=blue!10] (a32) at (3,-3)  {$3$};
\node[state]  [regular polygon, regular polygon sides=4, draw=blue!50, very thick, fill=blue!10] (a34) at (-3,-3)  {$3$};
\node[state]  [regular polygon, regular polygon sides=4, draw=blue!50, very thick, fill=blue!10] (a33) at (-3,3)  {$3$};
\path[->] 
(a21) edge   [thick, red]   node [above] {$h_4-\delta_2$} (a1)
(a22) edge   [thick, red]   node [above] {$h_2-\delta_2$} (a1)
(a31) edge   [thick, red]   node [right] {$h_1-\delta_1$} (a21)
(a32) edge   [thick, red]   node [right] {$h_3-\delta_1$} (a21)
(a33) edge   [thick, red]   node [left] {$h_1-\delta_1$} (a22)
(a34) edge   [thick, red]   node [left] {$h_3-\delta_1$} (a22)
(a41) edge   [thick, red]   node [above] {$h_2+\delta_2$} (a31)
(a41) edge   [thick, red]   node [above] {$h_4+\delta_2$} (a33)
(a42) edge   [thick, red]   node [below] {$h_4+\delta_2$} (a34)
(a42) edge   [thick, red]   node [below] {$h_2+\delta_2$} (a32)
(a1) edge   [thick, red]   node [right] {$h_3+\delta_1$} (a41)
(a1) edge   [thick, red]   node [right] {$h_1+\delta_1$} (a42)
;
\end{tikzpicture}
\end{aligned}
\end{equation}
where the fundamental regions are the shaded regions in the figure.
The loop constraint (\ref{eq-loop-constraint-toric}) translates to
\begin{equation}
\alpha^{(1)}_1+\alpha^{(3)}_1=\alpha^{(1)}_2+\alpha^{(3)}_2=-(\alpha^{(2)}_1+\alpha^{(4)}_1)=-(\alpha^{(2)}_2+\alpha^{(4)}_2)\,,
\end{equation}
whose solutions (shown in the right figure of (\ref{quiver-P1P1})) are
\begin{equation}\label{eq-loop-constraint-P1P1}
\begin{aligned}
\alpha^{(1)}_1=h_1+\delta_1\,, \qquad \alpha^{(1)}_2=h_3+\delta_1 \,;\qquad \alpha^{(2)}_1=h_2+\delta_2 \,,\qquad \alpha^{(2)}_2=h_4+\delta_2\,;\\
\alpha^{(3)}_1=h_1-\delta_1 \,,\qquad \alpha^{(3)}_2=h_3-\delta_1 \,;\qquad \alpha^{(4)}_1=h_2-\delta_2\,, \qquad \alpha^{(4)}_2=h_4-\delta_2\,;\\
\end{aligned}
\end{equation}
with
\begin{equation}\label{eq-ConstraintL-P1P1}
h_1+h_2+h_3+h_4=0 \;.
\end{equation}
Namely, the number of parameters is $E+2I-1=5$.

We can further impose the vertex constraint. One then obtains
\begin{align}
h_1=-h_3 \;, \quad h_2=-h_4\;, \quad
\delta_1=\delta_2=0  \;.
\end{align}

The periodic quiver is shown in the right figure of (\ref{periodic-quiver-P1P1}).
All the vertices are fermionic
\begin{equation}
|a|=1\,, \qquad \qquad a=1,2,3,4\,.
\end{equation}

The only non-trivial bond factors are
\begin{equation}\label{eq-charge-function-P1P1}
\begin{aligned}
&\varphi^{1\Rightarrow 2}(u)=\frac{1}{(u-h_1-\delta_1)(u-h_3-\delta_1)} \;,\quad
\varphi^{3\Rightarrow 4}(u)=\frac{1}{(u-h_1+\delta_1)(u-h_3+\delta_1)}  \;,
\\
&\varphi^{2\Rightarrow 3}(u)=\frac{1}{(u-h_2-\delta_2)(u-h_4-\delta_2)}\;,
\quad \varphi^{4\Rightarrow 1}(u)=\frac{1}{(u-h_2+\delta_2)(u-h_4+\delta_2)}\;,
\\
&\varphi^{2\Rightarrow 1}(u)=(u+h_1+\delta_1)(u+h_3+\delta_1)\;,
\quad \varphi^{4\Rightarrow 3}(u)=(u+h_1-\delta_1)(u+h_3-\delta_1)\;,
\\
&\varphi^{3\Rightarrow 2}(u)=(u+h_2+\delta_2)(u+h_4+\delta_2)\;,
\quad \varphi^{1\Rightarrow 4}(u)=(u+h_2-\delta_2)(u+h_4-\delta_2)\;.
\\
\end{aligned}
\end{equation}
It is now straightforward to write down the relations for the quiver Yangian.
Since the resulting commutation relations are rather lengthy and require several pages for the general case, we will only write down the algebra when the vertex constraints are imposed.

\subsubsection{Truncation}

Let us consider the truncation induced by the truncation of the crystal at an atom of color $1$.
When we have a closed path starting and ending at the quiver vertex $1$, we go around the quiver diagram. 
In each loop we obtain one of the $2^4=16$ possible weights:
\begin{align}
 \genfrac{\{ }{\} }{0pt}{}{h_1+\delta_1}{h_3+\delta_1}  +   \genfrac{\{ }{\} }{0pt}{}{h_2+\delta_2}{h_4+\delta_2}    + \genfrac{\{ }{\} }{0pt}{}{h_1-\delta_1}{h_3-\delta_1}  +   \genfrac{\{ }{\} }{0pt}{}{h_2-\delta_2}{h_4-\delta_2} \;.  
\end{align}
The factors of $\delta_1, \delta_2$ cancel out. 
Moreover, we need to impose the loop constraint \eqref{eq-ConstraintL-P1P1}, so that we obtain a linear combination of the following with non-negative integer coefficients:
\begin{align}
\begin{split}
&2(h_1+h_2) \;, \quad 2(h_1+h_4) \;, \quad 2(h_3+h_2) \;, \quad 2(h_3+h_4) \;, \\
&\pm (h_1-h_3) \;, \quad \pm (h_2-h_4) \;. 
\end{split}
\end{align}
The coordinate function can then be written as
\begin{align}\label{eq-truncation-P1P1-1}
\begin{split}
h(\sqbox{$1$})& =2N_1(h_1+h_2) +2N_2(h_1+h_4) +2N_3(h_3+h_2) +2N_4(h_3+h_4) \\
  & \qquad \qquad \qquad \qquad +N_5 (h_1-h_3) +N_6 (h_2-h_4)  \\
& =(2N_1-2N_2+2N_3-2 N_4) (h_1+h_2) +(2N_2-2N_3+N_5-N_6)(h_1-h_3) \;,
\end{split}
\end{align}
where in the last line we eliminated $h_4$ via \eqref{eq-ConstraintL-P1P1}.
This implies the truncation condition
\begin{equation}\label{P1P1_truncation}
2M_1 (h_1+h_2)+M_2 (h_1-h_3) +C=0 \;,
\end{equation}
for integers $M_1, M_2$.

\bigskip

\begin{figure}[htbp]
\centering\includegraphics[scale=0.25]{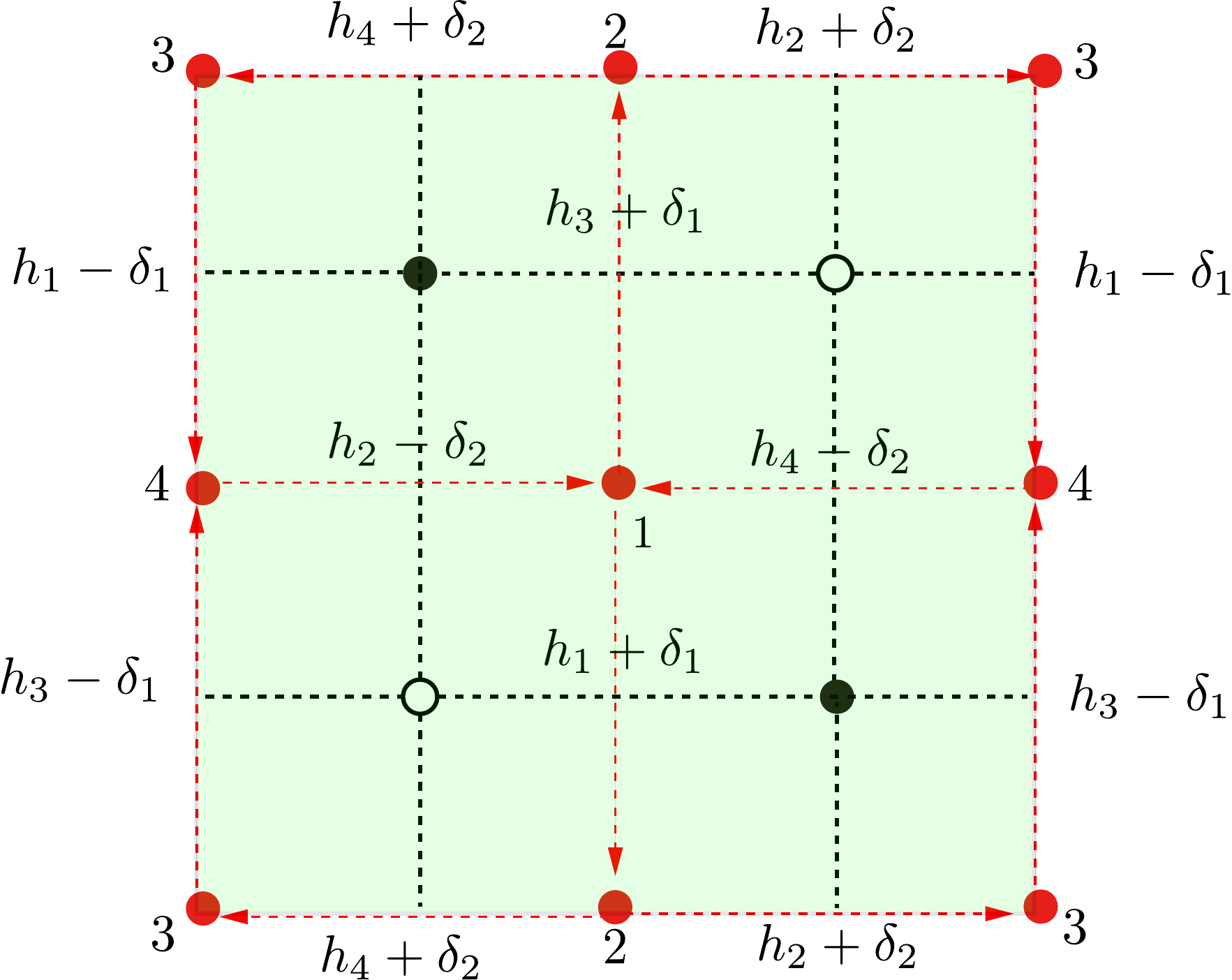}
\caption{The bipartite graph for the $K_{\mathbb{P}^1\times \mathbb{P}^1}$ geometry.}
\label{fig.P1P1bipartite}
\end{figure}

\begin{figure}[htbp]
\centering\includegraphics[scale=0.18]{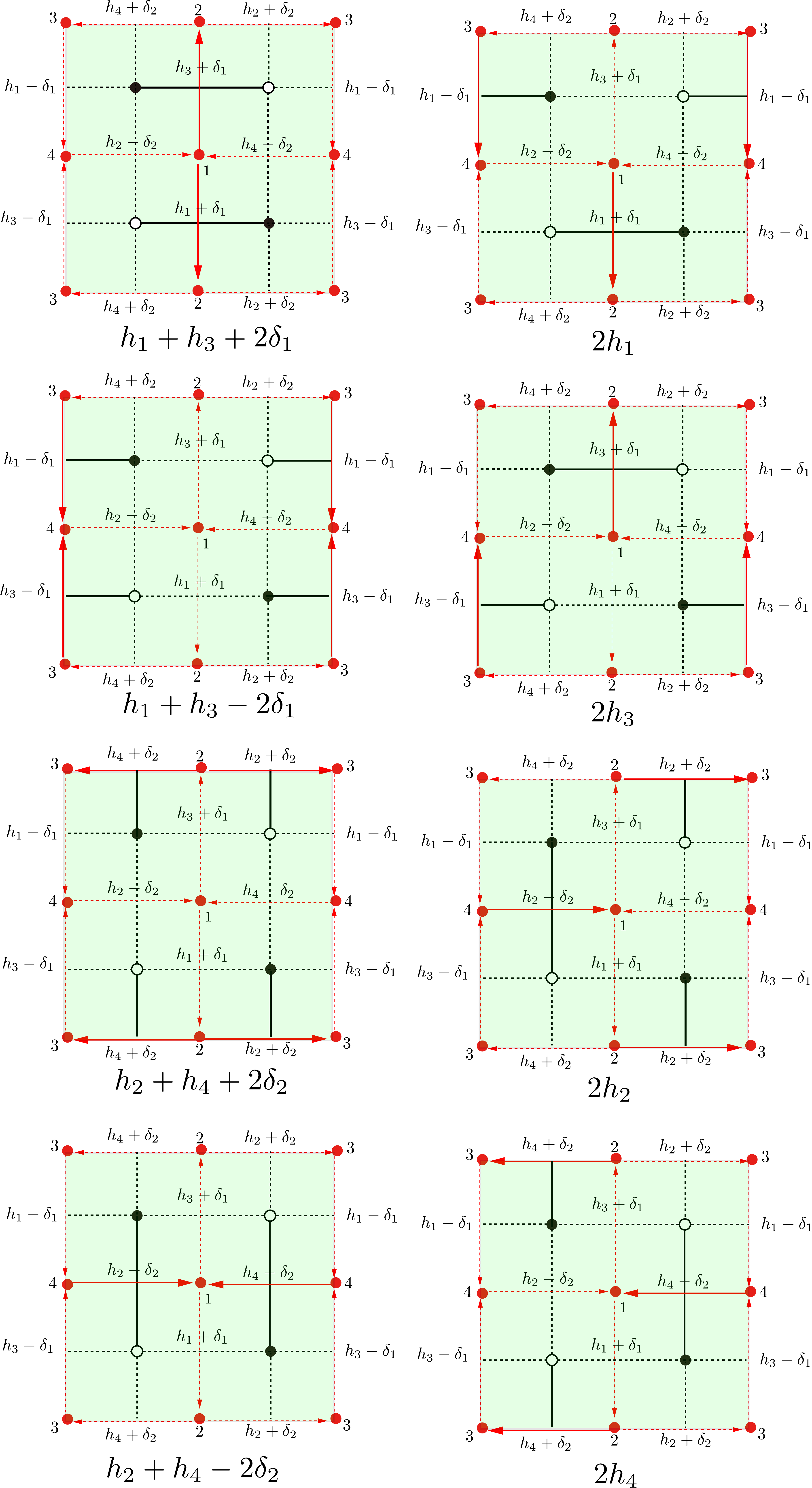}
\caption{The eight perfect matchings for the $K_{\mathbb{P}^1\times \mathbb{P}^1}$ geometry.
They correspond to the eight parameters $h_1+h_3+2\delta_1, h_1+h_3-2\delta_1, h_2+h_4+2\delta_2, h_2+h_4-2\delta_2$,
$2 h_1, 2h_2, 2h_3, 2h_4$.}
\label{fig.P1P1PM}
\end{figure}

Let us next consider truncations of the algebra corresponding to D4-branes.
The bipartite graph and the perfect matchings are shown in
Figures \ref{fig.P1P1bipartite} and \ref{fig.P1P1PM}.
There are eight perfect matchings, and they correspond to the
linear combinations 
\begin{align}\label{P1P1_list}
\begin{split}
&2h_1 \;, \quad 2h_2\;, \quad 2h_3 \;, \quad 2h_4 \;, \\
&h_1+h_3+2 \delta_1\;, \quad h_1+h_3-2 \delta_1\;, \quad  h_2+h_4+2 \delta_2\;, \quad h_2+h_4-2 \delta_2 \;.
\end{split}
\end{align}
When we further impose the vertex constraint, they reduce to 
\begin{align} \label{P1P1_h1h1}
\begin{split}
&2h_1 \;, \quad 2h_2 \;, \quad -2h_1 \;, \quad -2h_2 \;, \\
&0\;, \quad 0\;, \quad  0\;, \quad 0 \;.
\end{split}
\end{align}
and (after rescaling by a factor to $2$) are identified with the 
lattice points of the toric diagram (\ref{fig-toric-P1P1}):
\begin{align}\label{P1P1_lattice}
\begin{split}
&(1,0) \;, \quad (0,1) \;, \quad (-1,0) \;, \quad (0,-1) \;, \\
&(0,0)\;, \quad (0,0)\;, \quad  (0,0)\;, \quad (0,0)\;.
\end{split}
\end{align}
Note that the internal lattice point $(0,0)$ has multiplicity four.

For the comparison with the truncation analysis,
one needs to choose perfect matchings corresponding to the corner lattice points
$(\pm 1,0), (0,\pm1)$. In the list \eqref{P1P1_list}
these are 
\begin{align}
2h_1 \;, \quad 2h_2\;, \quad 2h_3 \;, \quad 2h_4=-2(h_1+h_2+h_3) \;.
\end{align}

\clearpage

\subsubsection{Vertex constraint}

The vertex constraint (\ref{eq-vertex-constraint-toric}) imposes
\begin{equation}\label{eq-ConstraintV-P1P1}
h_1+h_3=h_2+h_4 \;, \quad \delta_1=\delta_2=0 \;.
\end{equation}
These two constraints, when imposed together, leave two independent parameters
\begin{equation}\label{eq.h_P1P1_reduced}
h_1 \,, \qquad h_2  \,, \qquad h_3=-h_1 \,, \qquad h_4=-h_2 \;. 
\end{equation}
The charge assignment on the quiver becomes
\begin{equation}
\begin{aligned}
\begin{tikzpicture}[scale=0.8]
\filldraw[mygreen] (-3,3)--  (-3, -3) -- (3,-3) -- (3,3) -- cycle; 
\node[state]  [regular polygon, regular polygon sides=4, draw=blue!50, very thick, fill=blue!10] (a1) at (0,0)  {$1$};
\node[state]  [regular polygon, regular polygon sides=4, draw=blue!50, very thick, fill=blue!10] (a21) at (3,0)  {$4$};
\node[state]  [regular polygon, regular polygon sides=4, draw=blue!50, very thick, fill=blue!10] (a22) at (-3,0)  {$4$};
\node[state]  [regular polygon, regular polygon sides=4, draw=blue!50, very thick, fill=blue!10] (a41) at (0,3)  {$2$};
\node[state]  [regular polygon, regular polygon sides=4, draw=blue!50, very thick, fill=blue!10] (a42) at (0,-3)  {$2$};
\node[state]  [regular polygon, regular polygon sides=4, draw=blue!50, very thick, fill=blue!10] (a31) at (3,3)  {$3$};
\node[state]  [regular polygon, regular polygon sides=4, draw=blue!50, very thick, fill=blue!10] (a32) at (3,-3)  {$3$};
\node[state]  [regular polygon, regular polygon sides=4, draw=blue!50, very thick, fill=blue!10] (a34) at (-3,-3)  {$3$};
\node[state]  [regular polygon, regular polygon sides=4, draw=blue!50, very thick, fill=blue!10] (a33) at (-3,3)  {$3$};
\path[->] 
(a21) edge   [thick, red]   node [above] {$-h_2$} (a1)
(a22) edge   [thick, red]   node [above] {$h_2$} (a1)
(a31) edge   [thick, red]   node [right] {$h_1$} (a21)
(a32) edge   [thick, red]   node [right] {$-h_1$} (a21)
(a33) edge   [thick, red]   node [left] {$h_1$} (a22)
(a34) edge   [thick, red]   node [left] {$-h_1$} (a22)
(a41) edge   [thick, red]   node [above] {$h_2$} (a31)
(a41) edge   [thick, red]   node [above] {$-h_2$} (a33)
(a42) edge   [thick, red]   node [above] {$-h_2$} (a34)
(a42) edge   [thick, red]   node [above] {$h_2$} (a32)
(a1) edge   [thick, red]   node [right] {$-h_1$} (a41)
(a1) edge   [thick, red]   node [right] {$h_1$} (a42)
;
\end{tikzpicture}
\end{aligned}
\end{equation}
Accordingly the (non-trivial) bond factors (\ref{eq-charge-function-P1P1}) are reduced to
\begin{equation}
\begin{aligned}
&\varphi^{2\Rightarrow 1}(u)=\varphi^{1\Rightarrow 2}(u)^{-1}=\varphi^{4\Rightarrow 3}(u)=\varphi^{3\Rightarrow 4}(u)^{-1}=(u+h_1)(u-h_1) \equiv \varphi_1(u)\;,\\
&\varphi^{3\Rightarrow 2}(u)=\varphi^{2\Rightarrow 3}(u)^{-1}=\varphi^{1\Rightarrow 4}(u)=\varphi^{4\Rightarrow 1}(u)^{-1}=(u+h_2)(u-h_2) \equiv \varphi_2(u)\;,\\
\end{aligned}
\end{equation}
which give the algebra
\begin{tcolorbox}[ams align]\label{eq-OPE-P2-1}
&\textrm{OPE:}\quad
\begin{cases}
\begin{aligned}
\psi^{(a)}(z)\, \psi^{(b)}(w)&= \psi^{(b)}(w)\, \psi^{(a)}(z)\;,\\
\psi^{(a)}(z)\, e^{(a)}(w)   &\simeq   e^{(a)}(w)\, \psi^{(a)}(z)\;, \\ 
e^{(a)}(z)\, e^{(a)}(w) & \sim - e^{(a)}(w)\, e^{(a)}(z) \;,\\
\psi^{(a)}(z)\, f^{(a)}(w) &  \simeq f^{(a)}(w)\, \psi^{(a)}(z) \;,\\
 f^{(a)}(z)\, f^{(a)}(w) &  \sim  - f^{(a)}(w)\, f^{(a)}(z)\;, \\
 \psi^{(k)}(z)\, e^{(l)}(w)   &\simeq  \varphi_1(\Delta) \, e^{(l)}(w)\, \psi^{(k)}(z) \;,\\ 
 \psi^{(l)}(z)\, e^{(k)}(w)   &\simeq  \varphi_1(-\Delta)^{-1} \, e^{(k)}(w)\, \psi^{(l)}(z)\;, \\ 
 e^{(k)}(z)\, e^{(l)}(w) & \sim   -\varphi_1(\Delta) \, e^{(l)}(w)\, e^{(k)}(z) \;,\\
 \psi^{(k)}(z)\, f^{(l)}(w) &  \simeq  \varphi_1(\Delta)^{-1} \, f^{(l)}(w)\, \psi^{(k)}(z) \;,\\
 \psi^{(l)}(z)\, f^{(k)}(w)   &\simeq \varphi_1(-\Delta)\, f^{(k)}(w)\, \psi^{(l)}(z) \;,\\ 
f^{(k)}(z)\, f^{(l)}(w) &  \sim  -\varphi_1(\Delta)^{-1} \,f^{(l)}(w)\, f^{(k)}(z)\;,\\
 \psi^{(m)}(z)\, e^{(n)}(w)   &\simeq  \varphi_2(\Delta) \, e^{(n)}(w)\, \psi^{(m)}(z) \;,\\ 
 \psi^{(n)}(z)\, e^{(m)}(w)   &\simeq  \varphi_2(-\Delta)^{-1} \, e^{(m)}(w)\, \psi^{(n)}(z)\;, \\ 
 e^{(m)}(z)\, e^{(n)}(w) & \sim   -\varphi_2(\Delta) \, e^{(n)}(w)\, e^{(m)}(z) \;,\\
 \psi^{(m)}(z)\, f^{(n)}(w) &  \simeq  \varphi_2(\Delta)^{-1} \, f^{(n)}(w)\, \psi^{(m)}(z) \;,\\
 \psi^{(n)}(z)\, f^{(m)}(w)   &\simeq \varphi_2(-\Delta)\, f^{(m)}(w)\, \psi^{(n)}(z) \;,\\ 
f^{(m)}(z)\, f^{(n)}(w) &  \sim  -\varphi_2(\Delta)^{-1} \,f^{(n)}(w)\, f^{(m)}(z)\;,\\
  \psi^{(a+2)}(z)\, e^{(a)}(w)   &\simeq  \, e^{(a)}(w)\, \psi^{(a+2)}(z)   \;,\\ 
e^{(a+2)}(z)\, e^{(a)}(w) & \sim  \, - e^{(a)}(w)\, e^{(a+2)}(z)   \;,\\
\psi^{(a+2)}(z)\, f^{(a)}(w) &  \simeq  \, f^{(a)}(w)\, \psi^{(a+2)}(z)   \;,\\
 f^{(a+2)}(z)\, f^{(a)}(w) &  \sim   \, -   f^{(a)}(w)\, f^{(a+2)}(z)  \;,\\ 
\{e^{(a)}(z)\,, f^{(b)}(w)\}   &\sim  - \delta^{a,b}\, \frac{\psi^{(a)}(z) - \psi^{(b)}(w)}{z-w}\;, 
\end{aligned}
\end{cases}
\end{tcolorbox}
\noindent 
where $a,b=1,2,3, 4$, $\{k,l\}=\{1,2\}  \textrm{ or }  \{3,4\} $, $\{m,n\}=\{2,3\} \, \textrm{ or } \, \{4,1\}$.

\subsubsection{Dual Algebra}

\subsubsubsection{Dual Quiver Diagram}

The geometry $K_{\mathbb{P}^1\times \mathbb{P}^1}$ has another quiver description, related to the one above by Seiberg duality.
The quiver, obtained by Seiberg duality at vertex $4$, is given by
\begin{align}\label{quiver-P1P1-dual}
\begin{aligned}
&
\begin{tikzpicture}[scale=0.8]
\node[state]  [regular polygon, regular polygon sides=4, draw=blue!50, very thick, fill=blue!10] (a1) at (0,0)  {$1$};
\node[state]  [regular polygon, regular polygon sides=4, draw=blue!50, very thick, fill=blue!10] (a21) at (9,0)  {$2$};
\node[state]  [regular polygon, regular polygon sides=4, draw=blue!50, very thick, fill=blue!10] (a41) at (0,4)  {$4$};
\node[state]  [regular polygon, regular polygon sides=4, draw=blue!50, very thick, fill=blue!10] (a31) at (9,4)  {$3$};
\path[->] 
(a1) edge   [->>,thick, red]   node [above] {$h_1 +\delta_1, \,\, h_3+\delta_1$} (a21)
(a21) edge   [->>,thick, red]   node [right] {$h_2 + \delta_2, \,\, h_4 +\delta_2$} (a31)
(a41) edge   [->>,thick, red]   node [above] {$-h_1+\delta_1, \,\, -h_3+\delta_1$} (a31)
(a1) edge   [->>,thick, red]   node [left] {$-h_2+\delta_2,\,\,-h_4+\delta_2$} (a41)
(a31) edge   [->>>>,thick, red]   node [align=center] {$h_1+h_3-\delta_1-\delta_2 \;, h_1+h_4-\delta_1-\delta_2  $\\$h_2+h_3-\delta_1-\delta_2\;,
h_2+h_4-\delta_1-\delta_2$} (a1);
\end{tikzpicture}
\end{aligned}
\end{align}
which gives the periodic quiver 
\begin{equation}
\begin{aligned}
&
\begin{tikzpicture}[scale=0.8]
\node[state]  [regular polygon, regular polygon sides=4, draw=blue!50, very thick, fill=blue!10] (a1) at (0,0)  {$1$};
\node[state]  [regular polygon, regular polygon sides=4, draw=blue!50, very thick, fill=blue!10] (a21) at (6,0)  {$2$};
\node[state]  [regular polygon, regular polygon sides=4, draw=blue!50, very thick, fill=blue!10] (a22) at (-6,0)  {$2$};
\node[state]  [regular polygon, regular polygon sides=4, draw=blue!50, very thick, fill=blue!10] (a41) at (0,3)  {$4$};
\node[state]  [regular polygon, regular polygon sides=4, draw=blue!50, very thick, fill=blue!10] (a42) at (0,-3)  {$4$};
\node[state]  [regular polygon, regular polygon sides=4, draw=blue!50, very thick, fill=blue!10] (a31) at (6,3)  {$3$};
\node[state]  [regular polygon, regular polygon sides=4, draw=blue!50, very thick, fill=blue!10] (a32) at (6,-3)  {$3$};
\node[state]  [regular polygon, regular polygon sides=4, draw=blue!50, very thick, fill=blue!10] (a34) at (-6,-3)  {$3$};
\node[state]  [regular polygon, regular polygon sides=4, draw=blue!50, very thick, fill=blue!10] (a33) at (-6,3)  {$3$};
\path[->] 
(a1) edge   [thick, red]   node [above] {$h_1+\delta_1$} (a21)
(a1) edge   [thick, red]   node [above] {$h_3+\delta_1$} (a22)
(a21) edge   [thick, red]   node [right] {$h_2+\delta_2$} (a31)
(a21) edge   [thick, red]   node [right] {$h_4+\delta_2$} (a32)
(a22) edge   [thick, red]   node [left] {$h_2+\delta_2$} (a33)
(a22) edge   [thick, red]   node [left] {$h_4+\delta_2$} (a34)
(a41) edge   [thick, red]   node [above] {$-h_3+\delta_1$} (a31)
(a41) edge   [thick, red]   node [above] {$-h_1+\delta_1$} (a33)
(a42) edge   [thick, red]   node [above] {$-h_1+\delta_1$} (a34)
(a42) edge   [thick, red]   node [above] {$-h_3+\delta_1$} (a32)
(a1) edge   [thick, red]   node [align=center,yshift=-3pt] {$-h_4+\delta_2$} (a41)
(a1) edge   [thick, red]   node [align=center,yshift=3pt] {$-h_2+\delta_2$} (a42)
(a31) edge   [thick, red]   node [align=center,xshift=5pt,yshift=5pt] {$h_3+h_4-\delta_1-\delta_2$} (a1)
(a32) edge   [thick, red]   node [align=center,xshift=5pt,yshift=-5pt] {$h_2+h_3-\delta_1-\delta_2$} (a1)
(a33) edge   [thick, red]   node [align=center,xshift=-5pt,yshift=5pt] {$h_1+h_4-\delta_1-\delta_2$} (a1)
(a34) edge   [thick, red]   node [align=center,xshift=-5pt,yshift=-5pt] {$h_1+h_2-\delta_1-\delta_2$} (a1)
;
\end{tikzpicture}
\end{aligned}
\label{eq.P1P1-another-PQ}
\end{equation}

In the parametrization of the charge parameters as in \eqref{eq.P1P1-another-PQ}, the loop constraints  (\ref{eq-loop-constraint-toric}) and  the vertex constraint (\ref{eq-vertex-constraint-toric}) can be solved in exactly the same way as before, as in \eqref{eq-ConstraintL-P1P1} and \eqref{eq-ConstraintV-P1P1}.

\subsubsubsection{Dual Algebra}

The non-trivial bond factors are
\begin{equation}
\begin{aligned}
&\varphi^{1\Rightarrow 2}(u)=\frac{1}{(u-h_1-\delta_1)(u-h_3-\delta_1)} \;,\quad
\varphi^{3\Rightarrow 4}(u)  =(u-h_1+\delta_1)(u-h_3+\delta_1)\;,
\\
&\varphi^{2\Rightarrow 3}(u)=\frac{1}{(u-h_2-\delta_2)(u-h_4-\delta_2)}\;,
\quad \varphi^{4\Rightarrow 1}(u)=(u-h_2+\delta_2)(u-h_4+\delta_2)\;,
\\
&\varphi^{2\Rightarrow 1}(u)=(u+h_1+\delta_1)(u+h_3+\delta_1)\;,
\quad \varphi^{4\Rightarrow 3}(u)=\frac{1}{(u+h_1-\delta_1)(u+h_3-\delta_1)}\;,
\\
&\varphi^{3\Rightarrow 2}(u)=(u+h_2+\delta_2)(u+h_4+\delta_2)\;,
\quad \varphi^{1\Rightarrow 4}(u)=\frac{1}{(u+h_2-\delta_2)(u+h_4-\delta_2)}\;, \\
&\varphi^{1\Rightarrow 3}(u)=\prod_{i=1,2}\prod_{j=3,4} (u+h_i+h_j-\delta_1-\delta_2)\;, \\
&\varphi^{3\Rightarrow 1}(u)=\frac{1}{\prod_{i=1,2}\prod_{j=3,4}  (u-h_i-h_j+\delta_1+\delta_2)}\;.
\\
\end{aligned}
\end{equation}
When we impose the vertex constraint as in \eqref{eq-ConstraintV-P1P1}, these functions simplify as
\begin{equation}
\begin{aligned}
&\varphi^{2\Rightarrow 1}(u)=\varphi^{3\Rightarrow 4}(u)=\varphi^{1\Rightarrow 2}(u)^{-1}=\varphi^{4\Rightarrow 3}(u)^{-1}=(u+h_1)(u-h_1)\equiv \varphi_1(u)\;,\\
&\varphi^{3\Rightarrow 2}(u)= \varphi^{4\Rightarrow 1}(u)=\varphi^{2\Rightarrow 3}(u)^{-1}=\varphi^{1\Rightarrow 4}(u)^{-1}=(u+h_2)(u-h_2)\equiv \varphi_2(u)\;,\\
&\varphi^{1\Rightarrow 3}(u)=\varphi^{3\Rightarrow 1}(u)^{-1}=u^2(u+h_1-h_2)(u-h_1+h_2)\equiv \varphi_3(u)\;. \\
\end{aligned}
\end{equation}

It is again straightforward to write down relations for the quiver Yangian.
Since the resulting commutation relations are rather lengthy and require several lines for the general case, let us here write down the algebra only when the vertex constraints are imposed:
\begin{tcolorbox}[ams align]\label{eq-OPE-P2-2}
&\textrm{OPE:}\quad
\begin{cases}
\begin{aligned}
\psi^{(a)}(z)\, \psi^{(b)}(w)&= \psi^{(b)}(w)\, \psi^{(a)}(z)\;,\\
\psi^{(a)}(z)\, e^{(a)}(w)   &\simeq   e^{(a)}(w)\, \psi^{(a)}(z)\;, \\ 
e^{(a)}(z)\, e^{(a)}(w) & \sim  -e^{(a)}(w)\, e^{(a)}(z) \;,\\
\psi^{(a)}(z)\, f^{(a)}(w) &  \simeq f^{(a)}(w)\, \psi^{(a)}(z) \;,\\
 f^{(a)}(z)\, f^{(a)}(w) &  \sim   -f^{(a)}(w)\, f^{(a)}(z)\;, \\
 \psi^{(k)}(z)\, e^{(l)}(w)   &\simeq  \varphi_1(\Delta) \, e^{(l)}(w)\, \psi^{(k)}(z) \;,\\ 
 \psi^{(l)}(z)\, e^{(k)}(w)   &\simeq  \varphi_1(-\Delta)^{-1} \, e^{(k)}(w)\, \psi^{(l)}(z)\;, \\ 
 e^{(k)}(z)\, e^{(l)}(w) & \sim   -\varphi_1(\Delta) \, e^{(l)}(w)\, e^{(k)}(z) \;,\\
 \psi^{(k)}(z)\, f^{(l)}(w) &  \simeq  \varphi_1(\Delta)^{-1} \, f^{(l)}(w)\, \psi^{(k)}(z) \;,\\
 \psi^{(l)}(z)\, f^{(k)}(w)   &\simeq \varphi_1(-\Delta)\, f^{(k)}(w)\, \psi^{(l)}(z) \;,\\ 
f^{(k)}(z)\, f^{(l)}(w) &  \sim  -\varphi_1(\Delta)^{-1} \,f^{(l)}(w)\, f^{(k)}(z)\;,\\
 \psi^{(m)}(z)\, e^{(n)}(w)   &\simeq  \varphi_2(\Delta) \, e^{(n)}(w)\, \psi^{(m)}(z) \;,\\ 
 \psi^{(n)}(z)\, e^{(m)}(w)   &\simeq  \varphi_2(-\Delta)^{-1} \, e^{(m)}(w)\, \psi^{(n)}(z)\;, \\ 
 e^{(m)}(z)\, e^{(n)}(w) & \sim   -\varphi_2(\Delta) \, e^{(n)}(w)\, e^{(m)}(z) \;,\\
 \psi^{(m)}(z)\, f^{(n)}(w) &  \simeq  \varphi_2(\Delta)^{-1} \, f^{(n)}(w)\, \psi^{(m)}(z) \;,\\
 \psi^{(n)}(z)\, f^{(m)}(w)   &\simeq \varphi_2(-\Delta)\, f^{(m)}(w)\, \psi^{(n)}(z) \;,\\ 
f^{(m)}(z)\, f^{(n)}(w) &  \sim  -\varphi_2(\Delta)^{-1} \,f^{(n)}(w)\, f^{(m)}(z)\;,\\
%
%
\psi^{(3)}(z)\, e^{(1)}(w)   &\simeq  \varphi_3(\Delta)  \,e^{(1)}(w)\, \psi^{(3)}(z)\;, \\ 
\psi^{(1)}(z)\, e^{(3)}(w)   &\simeq  \varphi_3(-\Delta)^{-1} \,e^{(3)}(w)\, \psi^{(1)}(z)\;, \\ 
e^{(3)}(z)\, e^{(1)}(w) & \sim  -\varphi_3(\Delta)\,  e^{(1)}(w)\, e^{(3)}(z) \;,\\
\psi^{(3)}(z)\, f^{(1)}(w) &  \simeq \varphi_3(\Delta)^{-1} \,f^{(1)}(w)\, \psi^{(3)}(z) \;,\\
\psi^{(1)}(z)\, f^{(3)}(w) &  \simeq \varphi_3(-\Delta)\, f^{(3)}(w)\, \psi^{(1)}(z) \;,\\
 f^{(3)}(z)\, f^{(1)}(w) &  \sim   - \varphi_3(\Delta)^{-1}\, f^{(1)}(w)\, f^{(3)}(z)\;, \\
\psi^{(4)}(z)\, e^{(2)}(w)   &\simeq   e^{(2)}(w)\, \psi^{(4)}(z)\;, \\ 
\psi^{(2)}(z)\, e^{(4)}(w)   &\simeq   e^{(4)}(w)\, \psi^{(2)}(z)\;, \\ 
e^{(4)}(z)\, e^{(2)}(w) & \sim  -e^{(2)}(w)\, e^{(4)}(z) \;,\\
\psi^{(4)}(z)\, f^{(2)}(w) &  \simeq f^{(2)}(w)\, \psi^{(4)}(z) \;,\\
\psi^{(2)}(z)\, f^{(4)}(w) &  \simeq f^{(4)}(w)\, \psi^{(2)}(z) \;,\\
 f^{(4)}(z)\, f^{(2)}(w) &  \sim   -f^{(2)}(w)\, f^{(4)}(z)\;, \\
\{ e^{(a)}(z)\,, f^{(b)}(w) \}   &\sim  - \delta^{a,b}\, \frac{\psi^{(a)}(z) - \psi^{(b)}(w)}{z-w}\;, 
\end{aligned}
\end{cases}
\end{tcolorbox}
\noindent 
where $a,b=1,2,3, 4$, $\{k,l\}=\{1,2\}  \textrm{ or }  \{4,3\} $, $\{m,n\}=\{1,4\} \, \textrm{ or } \, \{2,3\}$. 

Our conjecture states that the algebra (\ref{eq-OPE-P2-2}) is equivalent to the algebra (\ref{eq-OPE-P2-1}), since its quiver (\ref{quiver-P1P1-dual}) is the Seiberg dual of the quiver (\ref{quiver-P1P1}) that the algebra (\ref{eq-OPE-P2-1}) is based on.

\subsection{Quiver Yangians for General Toric Calabi-Yau Threefolds}

We can repeat straightforwardly the analysis above for more general toric Calabi-Yau threefolds.
For example, when the toric diagram contains one internal lattice point, the geometry is a canonical bundle over a toric Fano surface: $\mathbb{P}^1\times \mathbb{P}^1$, $\mathbb{P}^2$, and their toric blow-ups (Hirzebruch surfaces $\mathbb{F}_{n=0,1,2}$ and del Pezzo surfaces $dP_{n=0,1,2,3}$).
All the data required for the quiver Yangian, including the periodic quiver and the charge assignments, are known and can be obtained by following the algorithms in the literature. It is more challenging to obtain the reduced quiver Yangian, 
which requires the Serre relations. It would be interesting to identify the Serre relations in general, and 
study the representation theory of the reduced quiver Yangian, see section \ref{sec:Serre} for a general discussion.

\section{Summary and Discussion}\label{sec:summary}

In this paper, we have proposed a general definition of an infinite-dimensional algebra, the BPS quiver Yangian $\mathsf{Y}_{(Q,W)}$, associated with a quiver $Q$ and a superpotential $W$. 
This algebra acts on configurations of BPS crystal melting model, which is also constructed by $Q$ and $W$.  
The pair $(Q,W)$ specifies a supersymmetric quantum mechanics dual to a toric Calabi-Yau threefold $X$, whose torus fixed points of the vacuum moduli space are classified by the configurations of BPS crystals.
Our algebra therefore acts on the (torus fixed points of) the BPS states in the type IIA compactifications on a toric Calabi-Yau threefold.

When the toric Calabi-Yau threefold has no compact $4$-cycles, the quiver Yangian $\mathsf{Y}_{(Q,W)}$, when supplemented with appropriate Serre relations, reproduces the affine Yangian for the Lie superalgebra $\mathfrak{gl}_{m|n}$.
More generally, our algebra seems to be new in the literature, but still acts on the configurations of the associated BPS crystals. 

The algebra depends on the set of charge parameters $h_I$.
We can consider a truncation of the algebra when the charger parameters are non-generic.
The resulting truncated algebra $\mathsf{Y}^{\vec{N}}_{(Q,W)}$ is labelled by $2$ integers.
We have discussed the relations of these integers to the numbers of D4-branes wrapping divisors.

The quiver/crystal-melting description of our algebra is rather powerful, and can naturally be adopted to discuss wall crossing phenomena of BPS states and open/closed BPS degeneracies.

We hope that the current paper uncovers only the tip of a huge iceberg, and we believe there are many interesting avenues for further research.
Let us conclude this paper by mentioning some of the problems for future investigation.

\begin{itemize}

\item In this paper we studied BPS state counting in a particular chamber of the moduli space.
Since BPS wall crossing for (closed/open) BPS state counting has been discussed in the literature
in terms of crystal melting \cite{Young:2008hn,MR2836398,MR2999994,Jafferis:2008uf,Chuang:2008aw,Nagao:2009ky,Nagao:2009rq,Sulkowski:2009rw,Aganagic:2010qr,Yamazaki:2010fz}, our discussion should generalize straightforwardly to other chambers.

\item A gluing construction for the affine Yangians has been worked out in \cite{Gaberdiel:2017hcn, Gaberdiel:2018nbs, Li:2019nna, Li:2019lgd}.
It would be interesting to compare the results of the current paper with those from the gluing approach in \cite{Gaberdiel:2017hcn, Gaberdiel:2018nbs, Li:2019nna, Li:2019lgd}. 
Similarly, the truncations of our algebra, as discussed in section \ref{sec:truncation}, should be related to another set of ``web of W-algebras'' obtained by gluing $\mathcal{W}_{1+\infty}$-algebras \cite{Gaiotto:2017euk, Prochazka:2017qum}.

\item Given a quiver and a superpotential one can define the cohomological Hall algebra \cite{MR2851153}.
We expect that the algebra $\mathsf{Y}_{(Q,W)}^{+}$ in the triangular decomposition \eqref{eq_triangular}, which we recall are generated by $e^{(a)}_n$'s, can be directly related to the shuffle algebra description of the cohomological Hall algebra. While cohomological Hall algebras for $\mathbb{C}^3$ are known \cite{Rapcak:2018nsl}, it seems to be difficult to generalize the discussion to a larger class of Calabi-Yau threefolds, and we hope that our work will shed some light on this problem (see  \cite{Kapranov:2019wri, zhao2019ktheoretic, MR4090584, MR4087863} for examples of recent studies of cohomological Hall algebras). 
Note also that the cohomological Hall algebra was recently discussed in the language of supersymmetric quiver quantum mechanics, which is closely related to the approach of this paper \cite{Galakhov:2018lta}.

\item As we discussed in section \ref{sec:Serre}, the question remains to identify the maximal set of Serre relation for the reduced quiver Yangian $\underline{\mathsf{Y}}_{(Q,W)}$, for a general toric quiver $(Q,W)$ (see the related discussion towards the end of section \ref{sec.Ygl11}). 
We have shown that for $(\mathbb{C}^2/\mathbb{Z}_n)\times \mathbb{C}$, the Serre relations are important in ensuring that the vacuum character of the reduced quiver Yangian $\underline{\mathsf{Y}}_{(Q,W)}$ reproduces the generating function of the colored crystals. One might try to use this criterion to help determine the Serre relations for the general toric quiver $(Q,W)$.

\item We expect that the definition of our quiver Yangian $\mathsf{Y}_{(Q,W)}$, as well as its representation in terms of crystal melting, can be lifted straightforwardly to the quiver quantum toroidal algebra $\mathsf{U}_{q, (Q,W)}$. The latter will contain the quantum toroidal algebras for $\mathfrak{gl}_n$ \cite{MR1324698} and $\mathfrak{gl}_{m|n}$ \cite{Bezerra:2019dmp} as special examples.
It seems that the crystal-melting representation for the $\mathfrak{gl}_{m|n}$ case was previously not known in the literature.

\item BPS crystal melting allows for a refinement (a one-parameter extension) \cite{Dimofte:2009bv, Nagao:2009ky, Nagao:2009rq}, which is natural in the context of wall crossing phenomena \cite{Kontsevich:2008fj}. Is there a corresponding refinement for our algebra?

\item It is known that the thermodynamic limit of the crystal melting model reproduces the geometry of the B-model
mirror Calabi-Yau geometry \cite{Ooguri:2009ri}. It is then natural to ask if our BPS algebra has anything to do
with the integrable hierarchies studied in the B-model geometry \cite{Aganagic:2003qj}. 

\item Recently a new approach to integrable models has been proposed based on a four-dimensional analogue of Chern-Simons theory \cite{Costello:2013zra, Costello:2018gyb, Costello:2017dso}, which in particular explains the Yangians of integrable models in terms of the algebra of loop operators.
It would be interesting to see if the quiver Yangians in this paper can be reproduced in a similar manner by a suitable Chern-Simons type gauge theory. 
This will in particular explain the geometrical origin of the spectral parameters, which are introduced as auxiliary parameters in the current discussion.

\end{itemize}

\section*{Acknowledgements}

We would like to thank Michio Jimbo, Matthias Gaberdiel, Dmitrii Galakhov, Hiraku Nakajima, Shigenori Nakatuka, Yoshihisa Saito, Yaping Yang, and Gufang Zhao for discussion.
We would like to thank Perimeter institute for hospitality, where this work was initiated.
WL is grateful for support from NSFC No.\ 11875064 and No.\ 11947302, the Max-Planck Partergruppen fund, and the Simons Foundation through the Simons Foundation Emmy Noether Fellows Program at Perimeter Institute.
The research of MY was supported in part by WPI Research Center Initiative, MEXT, Japan, and by the JSPS Grant-in-Aid for Scientific Research (No.\ 17KK0087, No.\ 19K03820 and No.\ 19H00689).

\appendix
\section{Serre Relations and State Counting}

\subsection{\texorpdfstring{$(\mathbb{C}^2/\mathbb{Z}_2)\times \mathbb{C}$, $2$-colored Plane Partitions, and Affine Yangian of $\mathfrak{gl}_{2}$}{(C2/Z2)x C, 2-colored Plane Partitions, and Affine Yangian of gl(2)}}
Recall that in the class of $(\mathbb{C}^2/\mathbb{Z}_n)\times \mathbb{C}$, the cases with $n=1$, $n=2$, and $n\geq 3$ need to be treated separately. 
The Serre relations for $n=1$ was discussed in section~\ref{sec:serre-gl1}, and the case of $n\geq 3$ was discussed in section~\ref{sec:serre-gl3}.
Let us now repeat this exercise for $n=2$.

The counting from the $2$-colored plane partitions gives \begin{equation}\label{eq-count-gl2}
\begin{aligned}
Z(q_1,q_2)&=\sum_{n_1,n_2}d(n_1,n_2)q^{n_1}_1\,q_2^{n_2}\\
&=1+q_1+(q_1^2+2\, q_1\, q_2)+(q_1^3+4\, q_1^2\,q_2+q_1\,q_2^2)+(q_1^4+4\, q_1^3\,q_2+8\,q_1^2\,q_2^2)+\dots\,,
\end{aligned}
\end{equation}
where we have grouped the terms with the same number of atoms $N=n_1+n_2$.
The goal is to reproduce this counting by the vacuum character of the reduced quiver Yangian algebra. 
We will do this level by level.
For simplicity, we impose the vertex constraint, which does not change the essence of argument but make the expressions shorter. 

\subsubsection{Vacuum}
There is one state at $(n_1,n_2)=(0,0)$, i.e.\ vacuum:
\begin{equation}
\textrm{vacuum:}\qquad |\emptyset\rangle \,.
\end{equation}

\subsubsection{One Atom}
Since we have assumed that the first atom in the crystal has color $a=1$, we have
\begin{equation}
e^{(1)}(z)|\emptyset\rangle=\frac{\#}{z}|\sqbox{1}\rangle\qquad \textrm{and}\qquad e^{(2)}(z)|\emptyset\rangle =0\,,
\end{equation}
which give 
\begin{equation}
e^{(1)}_{0}|\emptyset\rangle=\#|\sqbox{1}\rangle\quad \textrm{and}\quad e^{(1)}_{n\geq 1}|\emptyset\rangle =e^{(2)}_{n\geq 0}|\emptyset\rangle =0\,,
\end{equation}
when translated in terms of modes. 
Namely, there is only one state of the form $e^{(a)}_n|\emptyset\rangle$, which is
\begin{equation}\label{eq-gl2-N1}
(n_1,n_2)=(1,0): \qquad e^{(1)}_0  |\emptyset\rangle\,.
\end{equation}
This reproduce the counting in (\ref{eq-count-gl2}) at $N=1$.

\subsubsection{Two Atoms}

As shown by the counting (\ref{eq-count-gl2}) directly from the $2$-colored plane partitions, there are three states with $2$ atoms, one with $(n_1,n_2)=(2,0)$ and two with  $(n_1,n_2)=(1,1)$.
We need to reproduce this counting by enumerating independent states of the form $e^{(b)}_m\,e^{(a)}_n|\emptyset\rangle$.

The mode version of the OPE relations (\ref{eq-charge-function-Z2}) is
\begin{equation}\label{eq-mode-gl2}
\begin{aligned}
&[e^{(a)}_{n+1}, e^{(a)}_{m} ]-[e^{(a)}_n, e^{(a)}_{m+1} ]=h_3 \{e^{(a)}_{n}, e^{(a)}_{m} \}\,,\\
&( [e^{(a+1)}_{n+2}, e^{(a)}_{m} ]-2\,[e^{(a+1)}_{n+1}, e^{(a)}_{m+1} ]+[e^{(a+1)}_n, e^{(a)}_{m+2}])+h_1h_2\, [e^{(a+1)}_{n}, e^{(a)}_{m} ]\\
&\qquad \qquad \qquad \qquad =-h_3\, (\{ e^{(a+1)}_{n+1}, e^{(a)}_{m} \}- \{ e^{(a+1)}_{n}, e^{(a)}_{m+1} \})\,.
\end{aligned}
\end{equation}
Applying the $e^{(a)}_n$ with $n\in\mathbb{Z}_{\geq 0}$ on the unique $N=1$ state in (\ref{eq-gl2-N1}) and imposing the relations (\ref{eq-mode-gl2}), we obtain $3$ independent states, which can be chosen as
\begin{equation}\label{eq-states-gl2-N2}
\begin{aligned}
(n_1,n_2)=(2,0): 
&\quad e^{(1)}_0\,e^{(1)}_0  |\emptyset\rangle\,,
\\
(n_1,n_2)=(1,1): 
&\quad e^{(2)}_0\,e^{(1)}_0  |\emptyset\rangle\,,
\quad e^{(2)}_1\,e^{(1)}_0  |\emptyset\rangle\,.
\end{aligned}
\end{equation}
This match exactly with the counting (\ref{eq-mode-gl2}) at $N=2$.

\subsubsection{Three Atoms}

As shown by the counting (\ref{eq-count-gl2}) directly from the $2$-colored plane partitions, there are $6$ states with $3$ atoms, one with $(n_1,n_2)=(3,0)$, four with  $(n_1,n_2)=(2,1)$, and one with $(n_1,n_2)=(1,2)$.
We need to reproduce this counting by enumerating independent states of the form $e^{(c)}_\ell\,e^{(b)}_m\,e^{(a)}_n|\emptyset\rangle$.

First, applying $e^{(a)}_n$ on the three states (\ref{eq-states-gl2-N2}) from $N=2$ and using (\ref{eq-mode-gl2}) to eliminate dependent ones, we get
\begin{equation}
\begin{aligned}
(n_1,n_2)=(3,0): 
&\quad e^{(1)}_0\,e^{(1)}_0\,e^{(1)}_0  |\emptyset\rangle\,,
\\
(n_1,n_2)=(2,1): 
&\quad e^{(2)}_0\,e^{(1)}_0\,e^{(1)}_0\,  |\emptyset\rangle\,,
\quad e^{(2)}_1\,e^{(1)}_0\,e^{(1)}_0\,  |\emptyset\rangle\,,\\
&\quad e^{(1)}_0\,e^{(2)}_0\,e^{(1)}_0\,  |\emptyset\rangle\,,
\quad e^{(1)}_1\,e^{(2)}_0\,e^{(1)}_0\,  |\emptyset\rangle\,,
\quad e^{(1)}_2\,e^{(2)}_0\,e^{(1)}_0\,  |\emptyset\rangle\,,\\
&\quad e^{(1)}_0\,e^{(2)}_1\,e^{(1)}_0\,  |\emptyset\rangle\,,
\quad e^{(1)}_1\,e^{(2)}_1\,e^{(1)}_0\,  |\emptyset\rangle\,,
\quad e^{(1)}_2\,e^{(2)}_1\,e^{(1)}_0\,  |\emptyset\rangle\,,\\
(n_1,n_2)=(1,2): 
&\quad e^{(2)}_0\,e^{(2)}_0\,e^{(1)}_0\,  |\emptyset\rangle\,,
\quad e^{(2)}_0\,e^{(2)}_1\,e^{(1)}_0\,  |\emptyset\rangle\,,
\quad e^{(2)}_1\,e^{(2)}_1\,e^{(1)}_0\,  |\emptyset\rangle\,,
\end{aligned}
\end{equation}
which contain more states for $(n_1,n_2)=(2,1)$ and $(1,2)$ than the counting (\ref{eq-count-gl2}) at $N=3$ gives.
Therefore we need ``cubic" Serre relations, i.e.\ relations that  contain only terms of the form $e^{(b)}_\ell\,e^{(b)}_m\,e^{(a)}_n $ and $e^{(b)}_\ell\,e^{(a)}_n \,e^{(b)}_m$ with $b\neq a$.

The known Serre relation for the affine Yangian of $\mathfrak{gl}_2$ are quartic:
\begin{equation}
\textrm{Sym}_{z_1,z_2,z_3}\, \left[ e^{(a)}(z_1)\,, \left[ e^{(a)}(z_2)\,, \left[e^{(a)}(z_3)\,, e^{(a\pm1)}(w)\right]\right]\right] \sim 0\,,
\end{equation}
whose mode version is
 \begin{equation}\label{eq-serre-gl2-a}
\textrm{Sym}_{n_1,n_2,n_3}\, \left[ e^{(a)}_{n_1}\,, \left[ e^{(a)}_{n_2}\,, \left[e^{(a)}_{n_3}\,, e^{(a\pm1)}_{m}\right]\right]\right] = 0\,.
\end{equation}
Therefore, in order to use the Serre relations to reduce the number of states to the one dictated by the counting of the $2$-colored plane partitions, we need to derive a cubic version of the Serre relations using the quadratic relations (\ref{eq-mode-gl2}) and the quartic relations (\ref{eq-serre-gl2-a}).
We leave this to future work.

\bibliographystyle{nb}
\bibliography{yangian}

\end{document}